%% file: draft.tex
\documentclass[a4paper,11pt]{article}
\pdfoutput=1 

\usepackage{tikz-cd} 
\usepackage{amsmath} 
\usepackage{mathtools}
\allowdisplaybreaks
\usepackage{jheppub} 
\usepackage{empheq}
\usepackage{bm}
\usepackage{slashed}
\usepackage[T1]{fontenc} 
\usepackage[all]{xy}
\usepackage{rotating}
\usepackage{mathtools}
\usepackage{bbm}
\usepackage{dcolumn}
\usepackage{multirow}
\usepackage{mathrsfs}
\usepackage{arydshln}
\usepackage{verbatim}
\usepackage{graphics}
\usepackage{subcaption}
\usepackage{longtable}
\usepackage{tcolorbox}
\usepackage{cleveref}

\def\diag{\mathop{\rm diag}\nolimits}

\DeclareMathOperator{\Tr}{Tr}

\usepackage{slashed}

\def\a{\alpha}
\def\b{\beta}

\def\CA{{\cal A}}

\def\CF{{\cal F}}
\def\CG{{\cal G}}
\def\CH{{\cal H}}
\def\CI{{\cal I}}

\def\CK{{\cal K}}

\def\CM{{\cal M}}
\def\CN{{\cal N}}
\def\CO{{\cal O}}

\def\CT{{\cal T}}

\def\CW{{\cal W}}

\def\CZ{{\cal Z}}

\def\BI{{\mathbb I}}

\def\BT{{\mathbb T}}

\def\BZ{{\mathbb Z}}

\def\FF{{\mathfrak F}}

\def\FM{{\mathfrak M}}

\def\beq#1\eeq{\begin{align}#1\end{align}}

 \let\m=\mu \let\n=\nu  \let\p=\pi 
   \let\f=\phi \let\c=\chi 
    \let\D=\Delta  
    \let\F=\Phi

\newcommand{\be}{\begin{equation}}
	\newcommand{\ee}{\end{equation}}
\newcommand{\ba}{\begin{align}}
	\newcommand{\ea}{\end{align}}

\newcommand{\bi}{\begin{itemize}}
	\newcommand{\ei}{\end{itemize}}

\makeatletter
\newcommand*{\rom}[1]{\expandafter\romannumeral #1}
\makeatother

\newcommand{\paren}[1]{\left( {#1} \right)}

\def\FF{{\mathfrak F}}

\def\FM{{\mathfrak M}}

{\preprint{}}

\title{\boldmath Torus Knots and Minimal Models Revisited : Rational VOA characters from non-hyperbolic knots}

\abstract{In 2003, Hikami and Kirillov uncovered an intriguing connection between torus knots 
$\mathcal{K}_{(P,Q)}$ and Virasoro minimal models $\mathcal{M}(P,Q)$ by relating the 
Kashaev invariants of the knots to the characters of the corresponding minimal models. 
In this work, we recover and extend this connection by combining the 
3D--3D correspondence with a bulk--boundary correspondence. More concretely, we study 
the 3D $\mathcal{N}=2$ gauge theories associated with torus-knot complements via the 
Dimofte--Gaiotto--Gukov construction and show that, in the infrared, these theories 
either flow to a unitary TQFT (when $|P-Q| = 1$), whose boundary chiral algebra 
reproduces that of the associated unitary minimal model, or to a 3D $\mathcal{N}=4$ 
rank-0 SCFT (when $|P-Q| > 1$), which realizes the corresponding non-unitary chiral 
minimal model at the boundary after an appropriate topological twist. 
This framework yields new Nahm-sum--like expressions for the characters of Virasoro 
minimal models and other related rational conformal field theories, providing a 
systematic algorithm for constructing characters of rational VOAs directly from the 
combinatorial data of an ideal triangulation of a non-hyperbolic knot complement.
}
	
\author[a]{Dongmin Gang,}
\author[a]{Byoungyoon Park,}
\author[a]{Huijoon Sohn}

\affiliation[a]{
    Department of Physics and Astronomy $\&$ Center for Theoretical Physics,
    \\
    Seoul National University, 1 Gwanak-ro, Seoul 08826, Korea}

\emailAdd{arima275@snu.ac.kr}
\emailAdd{pby9707@snu.ac.kr}
\emailAdd{hjson99@snu.ac.kr}

\begin{document} 
    \maketitle
    \flushbottom

\section{Introduction and Summary}
A hallmark of research in string/M-theory and quantum field theory is the
discovery of unexpected—and often profound—connections between seemingly
distant areas of physics and mathematics. In some cases, such connections first arise from physical reasoning and
are only later recognized as precise mathematical statements.
In others, relations uncovered within mathematics find their natural
home in a physical framework, where an underlying physical picture not
only clarifies their origin but also points toward systematic
generalizations beyond the original setting.

The present work belongs to the latter category. In 2003, Hikami and Kirillov uncovered a remarkable relation between torus
knots $\mathcal{K}_{(P,Q)}$ and Virasoro minimal models
$\mathcal{M}(P,Q)$ by observing that the Kashaev invariants of torus knots
are intimately related to the characters of the corresponding minimal
models~\cite{Hikami:2003tb}, thereby building on earlier insights of
Zagier and Lawrence~\cite{zagier1999modular}.
 This striking observation strongly hints at an underlying physical structure that has yet to be fully elucidated. In this paper, we aim to understand and systematically extend the Hikami–Kirillov relation from the perspective of quantum field theory and M-theory. Our approach is based on the {\it 3D--3D correspondence}, which associates
three-dimensional $\mathcal{N}=2$ supersymmetric gauge theories to knot
complements, together with a {\it bulk--boundary correspondence} that relates bulk
three-dimensional topological quantum field theories to rational vertex
operator algebras (r-VOA) living at two-dimensional boundaries.  A schematic summary of our proposal is shown in \ref{fig : summary}. 
\begin{figure}[h]
\centering
\begin{equation*}
\begin{aligned}
\mathcal{K}_{(P,Q)}
\;&\xrightarrow{\;\;\text{3D--3D }\;\;}\; T[S^3\backslash \CK_{(P,Q)}] \simeq 
\begin{cases}
\text{unitary TQFT}, & |P-Q|=1, \\[0.3em]
\text{3D } \mathcal{N}=4 \text{ rank-}0 \text{ SCFT}, & |P-Q|>1,
\end{cases}
\\[0.5em]
&   \xrightarrow{\;\;\;\;\text{bulk--boundary }\;\;\;\;}\;
\begin{cases}
\text{unitary VOA of } \mathcal{M}(P,Q) \textrm{ or other unitary r-VOAs},\\[0.3em]
\text{non-unitary VOA of } \mathcal{M}(P,Q) \textrm{ or other non-unitary r-VOAs}.
\end{cases}
\end{aligned}
\end{equation*}
\caption{
Schematic summary of the main proposal of this work.
The 3D $\mathcal{N}=2$ supersymmetric gauge theory
$T[S^{3}\backslash \mathcal{K}_{(P,Q)}]$, associated with a torus-knot
complement via the 3D--3D correspondence, exhibits two distinct infrared
behaviors.
For $|P-Q|=1$, the theory flows to a unitary TQFT, whereas for $|P-Q|>1$
it flows to a three-dimensional $\mathcal{N}=4$ rank-$0$ superconformal
field theory.
The subsequent bulk--boundary correspondence then realizes, depending on
the choice of boundary conditions, the rational vertex operator algebra
of the corresponding Virasoro minimal model $\mathcal{M}(P,Q)$ or other
closely related rational conformal field theories
at the boundary.
}
\label{fig : summary}
\end{figure}

The 3D--3D correspondence \cite{Terashima:2011qi,Dimofte:2011ju} has its origin in the six-dimensional $A_{1}$ $(2,0)$
theory, which describes the low-energy effective world-volume dynamics of two
coincident M5-branes in M-theory.
A twisted compactification of this theory on a three-manifold $M$ yields a
three-dimensional $\mathcal{N}=2$ theory $T[M]$; see
\eqref{6d viewpoint of T_DGG} for a detailed discussion in the case where
$M = S^{3}\backslash \mathcal{K}$ is a knot complement.
While this six-dimensional picture provides the conceptual foundation,
a concrete and computable description of $T[M]$ is furnished by the
bottom-up construction of Dimofte--Gaiotto--Gukov (DGG)~\cite{Dimofte:2011ju},
in which $T[M]$ is described as the infrared fixed point of a
supersymmetric abelian Chern--Simons--matter theory.
The validity of this bottom-up approach is supported by the 3D--3D relation,
which equates supersymmetric partition functions of $T[M]$ with those of
$SL(2,\mathbb{C})$ Chern--Simons theory on $M$;  this relation
provided the primary motivation for proposing the bottom-up construction,
together with parallel developments in supersymmetric localization
for three-dimensional $\mathcal{N}=2$ gauge theories
\cite{Kim:2009wb,Kapustin:2009kz,Imamura:2011su,Jafferis:2010un,Hama:2010av,Hama:2011ea}
and state-integral models for complex Chern--Simons theory
\cite{HIKAMI_2001,Dimofte:2009yn,Dimofte:2011gm,EllegaardAndersen:2011vps,
Dimofte:2012qj,Dimofte:2014zga}.

The bulk--boundary correspondence between three-dimensional topological field theories
 and two-dimensional rational conformal field theories can be traced back to
Witten’s seminal work on Chern--Simons theory \cite{zbMATH04092352}. 
In this unitary setting, the correspondence between bulk topological
data and boundary rational vertex operator algebras is by now well
established.
More recently, however, it has been realized that this paradigm can be
substantially generalized beyond unitary TQFTs.
In particular, three-dimensional $\mathcal{N}=4$ rank-$0$
superconformal field theories \cite{Gang:2018huc,Gang:2021hrd} provide a natural framework in which
a non-unitary version of the bulk--boundary correspondence emerges,
realizing non-unitary rational vertex operator algebras at the boundary after topological twistings \cite{Gang:2023rei,Gang:2022kpe,Ferrari:2023fez}.\footnote{
Upon relaxing the rationality assumption, one may consider a more general
bulk--boundary correspondence \cite{Costello:2018fnz,Costello:2020ndc} associates boundary vertex operator
algebras to a wider class of supersymmetric three-dimensional gauge theories.
} Rank-$0$ theories, characterized by trivial Coulomb and Higgs branches,
are a recently discovered class of exotic yet simple
$\mathcal{N}=4$ superconformal field theories and have attracted growing
attention from various perspectives \cite{dedushenko:2018bpp,Gang:2023ggt,Dedushenko:2023cvd,Creutzig:2024ljv,ArabiArdehali:2024ysy,Gaiotto:2024ioj,ArabiArdehali:2024vli,Kim:2024dxu,Gang:2024loa,Go:2025ixu,Jeong:2025xid,Kim:2025klh,Kim:2025rog,Nishinaka:2025ytu}.

 From a direct field-theoretic analysis, we find that the three-dimensional
$\mathcal{N}=2$ supersymmetric gauge theories associated with torus-knot
complements $S^{3}\backslash \mathcal{K}_{(P,Q)}$, constructed via the explicit
algorithm of DGG \cite{Dimofte:2011ju}, exhibit two qualitatively distinct
infrared behaviors depending on the values of $(P,Q)$.
When $|P-Q|=1$, the theory develops a mass gap and flows in the infrared to
a unitary topological quantum field theory.
In contrast, when $|P-Q|>1$, the theory undergoes a non-trivial enhancement of
supersymmetry and flows to a three-dimensional $\mathcal{N}=4$ rank-$0$
superconformal field theory.
Although this infrared fixed point is unitary, it admits topologically twisted
sectors that play a central role in our analysis. These two distinct behaviors mirror precisely the properties of the associated Virasoro
minimal models.
For $|P-Q|=1$, the corresponding minimal model is unitary, and it is therefore
natural that the bulk theory is gapped and topological in the infrared.
For $|P-Q|>1$, on the other hand, the associated minimal model is non-unitary,
and the bulk theory is instead realized as a rank-$0$ superconformal field
theory, from which the relevant non-unitary topological data emerge only after
an appropriate topological twist. Accordingly, we expect that for $|P-Q|=1$ the bulk theory supports, upon
imposing suitable supersymmetric boundary conditions, the vertex operator
algebra of the corresponding unitary minimal model at the boundary.
For $|P-Q|>1$, the bulk theory is instead expected to support the vertex
operator algebra of the corresponding non-unitary minimal model only after an
appropriate topological twist of the bulk theory, together with a suitable
choice of boundary condition.

To probe the boundary rational vertex operator algebras, we make use of the
half-index \cite{Gadde:2013wq,Yoshida:2014ssa,Dimofte:2017tpi} , defined as the supersymmetric partition function on
$D_{2}\times S^{1}$.
For appropriate supersymmetric boundary conditions, the half-index is expected
to compute the vacuum character of the boundary rational VOA.
In principle, one could fix a given bulk UV gauge-theory description and
systematically scan over all admissible supersymmetric boundary conditions in
order to identify those that realize a rational boundary VOA.
Instead, we adopt the opposite strategy.
We fix the boundary condition to the simplest Dirichlet boundary condition and
instead vary the duality frame of the bulk theory.
In a suitable duality frame, this Dirichlet boundary condition becomes precisely
the boundary condition that supports a rational VOA, and the corresponding
half-index reproduces its vacuum character.
One of the key advantages of the Dimofte–Gaiotto–Gukov (DGG) construction is that
it provides an explicit and systematic prescription for how the gauge-theory
description transforms under changes of duality frame, see \eqref{polarization/duality}, rendering this approach
both practical and computationally efficient.
This strategy allows us not only to recover the characters of Virasoro minimal
models, but also to compute the characters of other closely related rational
conformal field theories.
\paragraph{Summary for Mathematicians}
Motivated by the physical considerations described above, we propose a
systematic procedure for constructing candidates for characters of rational
vertex operator algebras (r-VOAs), realized as half-indices, starting from the
combinatorial data of an ideal triangulation of a torus-knot complement.
See the conjecture in~\eqref{Main conjecture}.
\begin{figure}[h]
\centering
\begin{equation*}
\begin{aligned}
S^3\backslash \mathcal{K}_{(P,Q)}
\;&\xrightarrow{\;\;\text{ideal triangulation}\;\;}\;
\text{extended NZ data }(A,B,\boldsymbol{\nu}_x;\mathbf{Q})
\ \text{and}\ 
(A_{\rm r},B_{\rm r},\boldsymbol{\nu}_{x,{\rm r}};\mathbf{Q}_{\rm r})
\\[0.6em]
&\xrightarrow{\;\;\text{half-index}\;\;}
\text{ candidates for characters of r-VOAs}
\end{aligned}
\end{equation*}
\caption{Schematic summary of the construction of r-VOA characters from the
combinatorial data of an ideal triangulation of
$S^3\backslash \mathcal{K}_{(P,Q)}$.}
\label{fig:summary-math}
\end{figure}
\\
The construction proceeds as follows.
First, we define \emph{refined} Neumann--Zagier data $(A,B,\boldsymbol{\nu}_x)$
associated with a given ideal triangulation.
Here ``refined'' means that the internal edges are decomposed into ``easy'' and
``hard'' edges, as defined in~\eqref{def : Easy/hard edge}, and that the
corresponding Neumann--Zagier matrices $(A,B)$ are constructed using the
relations in~\eqref{NZ matrices}.
See also~\eqref{reduced NZ matrices} for the reduced Neumann--Zagier data
$(A_{\rm r},B_{\rm r},\boldsymbol{\nu}_{x,{\rm r}})$. Next, we introduce an additional integer-valued $r\times r$ matrix
$\mathbf{Q}$, subject to the conditions listed in~\eqref{NQB relation}.
Given these data, we define a family of half-indices in
\eqref{half-indices when |P-Q|=1},
\eqref{half-indices at A/B-twistings}, and
\eqref{half-indices at AA/AB/BA/BB-twistings}. We conjecture that, up to an overall prefactor of the form $q^{\mathbb{Q}}$,
these half-indices provide candidates for characters of rational VOAs
whenever the additional conditions
\eqref{half-indices when |P-Q|=1-2},
\eqref{half-indices at A/B-twistings-2}, and
\eqref{half-indices at AA/AB/BA/BB-twistings-2}
are satisfied.
This conjecture is supported by a large number of explicit examples in Appendix \ref{Appendix Examples}.

The rest of the paper is organized as follows.
In Section~\ref{sec : DGG review}, we review the DGG construction, with
several clarifications and refinements, focusing in particular on the
role of splitting internal edges into \emph{easy} and \emph{hard} ones
and on the associated flavor symmetries. We also introduce a convenient
modification of the construction that allows us to identify a UV
gauge-theory description suitable for efficient computations of the
half-index.
In particular, for torus-knot complements, we  propose a simpler
gauge-theory description,
$T^{\rm (r)}[S^{3}\backslash \mathcal{K}_{(P,Q)}]$.
In Section~\ref{sec : IR phases and BPS ptns}, we propose the infrared
phases of $T[S^{3}\backslash \mathcal{K}_{(P,Q)}]$ and explain how these
phases can be verified through computations of BPS partition functions.
Section~\ref{sec : RVOAs from Half-indices} constitutes the main part of
the paper, where we present explicit formulas for the half-indices of
$T[S^{3}\backslash \mathcal{K}_{(P,Q)}]$, which are completely determined
by the combinatorial data of ideal triangulations, together with the
main conjecture of the paper stated in~\eqref{Main conjecture}.
Finally, in the appendices, we provide numerous examples supporting the
conjecture, along with reviews of the state-integral models for
$SL(2,\mathbb{C})$ Chern--Simons theories and character formulas for
simple rational VOAs relevant to the examples.

\input{section_TfromTorusKnot}

\input{section_BPSpftnsIRphases}

\input{section_rVOAfromHI}

\section{Discussions and Future Directions} \label{sec: Discussion}

\paragraph{Choices of $\mathbf{Q}$} 
There may exist multiple admissible choices of $\mathbf{Q}$ satisfying the relation~\eqref{NQB relation}. 
If two such choices, say $\mathbf{Q}_1$ and $\mathbf{Q}_2$, have the same image in $\mathbb{Z}^r$, i.e.
\begin{align}
\mathbf{Q}_1 \cdot \mathbb{Z}^r = \mathbf{Q}_2 \cdot \mathbb{Z}^r \,,
\end{align}
then they yield identical half-indices and therefore the same characters. 
For a fixed pair $(A,B)$, there can be several inequivalent choices of $\mathbf{Q}$ with distinct image sets, 
potentially leading to different half-indices. 
It would be interesting to classify all admissible $\mathbf{Q}$ that produce genuinely different half-indices.

\paragraph{Relation to the work by Gang-Kang-Kim \cite{Gang:2024tlp}} In the paper, they argue that
\begin{align}
&(\textrm{Bulk field theory for $\mathcal{M}(P,Q)$}) = T[S^2 ((P,P-R),(Q,S),(3,1))]\;.
\end{align}
Here \( S^{2}\big((p_{1},q_{1}), (p_{2},q_{2}), (p_{3},q_{3})\big) \) denotes the Seifert-fibered space over \( S^{2} \) with three singular fibers.
Two integers $R$ and $S$ are determined by the relation $PS-QR=1$. We will claim that
\begin{align}
T[S^2 ((P,P-R),(Q,S),(3,1))] \simeq T[S^3\backslash \mathcal{K}_{P,Q}], \label{equivalence between two T[M]s}
\end{align}
where $\simeq$ denotes the IR equivalence. In general, the  theory $T[S^3\backslash \CK]$ theory has $SU(2)_m$ flavor symmetry associated with the knot $\CK$. For torus knots $\CK= \CK_{(P,Q)}$, the $SU(2)$ flavor symmetry decouples in the IR and it only appears as the background Chern-Simons terms with CS level $-P Q$, which can be seen from the 3D-index of the torus knot compelements in \eqref{3D index for torus knots}. Thus, using the Dehn filling operation in 3D-3D correspondence \cite{Gang:2018wek,Choi:2022dju}, we have
\begin{align}
\begin{split}
&T[(S^3\backslash \mathcal{T}_{(P,Q)})_{k}]  
\\
&= 
\begin{cases}
\left(\frac{T[S^3\backslash \mathcal{K}_{(P,Q)}]}{SU(2)_k}  \right)/\mathbb{Z}^{[1]}_2 =  \frac{\left(T[S^3\backslash \mathcal{K}_{(P,Q)}]\otimes SU(2)_{k-PQ}  \right)}{\mathbb{Z}^{[1]}_2 }, \quad k \in 2\mathbb{Z}
\\
\left( \frac{T[S^3\backslash \mathcal{K}_{(P,Q)}]}{SU(2)_k}\otimes U(1)_{\pm 2} \right)/\mathbb{Z}^{[1]}_2 =  \frac{\left( T[S^3\backslash \mathcal{K}_{(P,Q)}]\otimes SU(2)_{k-PQ}\otimes U(1)_{\pm 2} \right)}{\mathbb{Z}^{[1]}_2}, \quad k \in 2\mathbb{Z}+1
\end{cases}
\end{split}
\end{align}
Here $(S^{3}\backslash \mathcal{K})_{k}$ denotes the closed three-manifold obtained from the knot complement $S^{3}\backslash \mathcal{K}$ by performing Dehn filling with integer slope $k$. The tensoring with $U(1)_{\pm 2}$ and the gauging of the $\mathbb{Z}_2$ one-form symmetry are required in order to match the squashed three-sphere partition function with the $SL(2,\mathbb{C})$ Chern--Simons partition function on the resulting three-manifold, in accordance with the 3D--3D correspondence.
 
 When $k=PQ+3$, we have
\begin{align}
\begin{split}
&T[(S^3\backslash \mathcal{K}_{(P,Q)})_{k=PQ+3}]  
\\
&= 
\begin{cases}
  \frac{\left(T[S^3\backslash \mathcal{K}_{(P,Q)}]\otimes U(1)_2  \right)}{\mathbb{Z}^{[1]}_2 }, \quad PQ \in 2\mathbb{Z}+1
\\
\left( \frac{T[S^3\backslash \mathcal{K}_{(P,Q)}]}{SU(2)_k}\otimes U(1)_{\pm 2} \right)/\mathbb{Z}^{[1]}_2 =  \frac{\left( T[S^3\backslash \mathcal{K}_{(P,Q)}]\otimes U(1)_2 \otimes U(1)_{\pm 2} \right)}{\mathbb{Z}^{[1]}_2}, \quad PQ \in 2\mathbb{Z}
\end{cases}
\end{split}
\end{align}
Here we use the fact that (pure $\mathcal{N}=2 \;SU(2)_3$) $\simeq SU(2)_1 \simeq U(1)_2$. Topologically, on the other hand, it is known that 
\begin{align}
(S^3\backslash \CK_{(P,Q)})_{k=PQ+3}  = S^2 ((P, -R),(Q,S),(3,1))\;.
\end{align}
Thus, to confirm the equivalence in \eqref{equivalence between two T[M]s}, we need to check that 
\begin{align}
\begin{split}
&T[S^3 ((P,-R),(Q,S),(3,1))]  
\\
&= 
\begin{cases}
  \frac{\left(T[S^3 ((P,P-R),(Q,S),(3,1))]\otimes U(1)_2  \right)}{\mathbb{Z}^{[1]}_2 }, \quad PQ \in 2\mathbb{Z}+1
\\
\frac{\left( T[S^3 ((P,P-R),(Q,S),(3,1))]\otimes U(1)_2 \otimes U(1)_{\pm 2} \right)}{\mathbb{Z}^{[1]}_2}, \quad PQ \in 2\mathbb{Z}
\end{cases}
\end{split}
\end{align}
One can verify the above relation case by case at the level of twisted partition functions, using the 3D--3D relation for twisted partition functions in \cite{Cho:2020ljj,Gang:2024tlp}.

\paragraph{Other non-hyperbolic knots and links and higher $N$ (number of M5-branes)}
We expect that our analysis of torus knots can be extended to a broader class of
non-hyperbolic knots and links.
It would be interesting to determine which rational vertex operator algebras
arise from these more general examples.
Another natural direction is to consider three-dimensional $\mathcal{N}=2$
gauge theories obtained from the compactification of the six-dimensional
$A_{N-1}$ $(2,0)$ theory, with $N \geq 3$, on torus-knot complements.
A concrete field-theoretic construction of such theories is available via the
algorithm developed in~\cite{Dimofte:2013iv}.
Motivated by the results of~\cite{Kanade:2022wtm,Baek:2025uev}, we expect that these
three-dimensional theories admit boundary rational VOAs described by
$W_N$ minimal models, and that their characters can be constructed directly
from the combinatorial data encoded in the $PGL(N,\mathbb{C})$ gluing data \cite{Garoufalidis:2015}
of the torus-knot complement.

\paragraph{Haagerup-like non-unitary VOAs from once-punctured torus bundles}
Another class of three-manifolds that give rise to rank-$0$ SCFTs via the
3D--3D correspondence is provided by once-punctured torus bundles \cite{Terashima:2011qi,Gang:2013sqa}, which are
hyperbolic.  
Unlike the non-hyperbolic three-manifolds studied in this work, for these
examples the geometric R-symmetry $R_{\mathrm{geo}}$ in \eqref{R_geom} coincides with the
superconformal R-charge, rather than with R-symmetry at A-twisting or B-twisting point. In this case, the theory possesses a $U(1)$ symmetry associated with the
puncture on the fiber, which does not decouple in the infrared and instead
corresponds to the $U(1)_A$ symmetry of the rank-$0$ SCFT.
For certain classes of once-punctured torus bundles whose monodromy matrix is
of the form $ST^{k}$, the corresponding rational VOA characters have been
studied in \cite{Gang:2022kpe,Gang:2023ggt}.
An intriguing feature of these examples is that the modular data of the
associated rational VOAs provide a non-unitary generalization of the modular
data of the Haagerup TQFT \cite{Evans:2010yr}. It would be interesting to extend this analysis to more general classes of
once-punctured torus bundles and to identify new families of
Haagerup-like non-unitary rational VOAs.

\paragraph{Cobordisms between torus knots and domain walls in 3D TQFTs/2D RCFTs}
An interesting direction is to interpret cobordisms between torus-knot
complements as domain walls in the associated three-dimensional TQFTs.
Via the 3D--3D correspondence, such cobordisms define interfaces between the
corresponding three-dimensional theories and descend, in the presence of
boundaries, to conformal defects between the associated two-dimensional RCFTs.
From this perspective, mathematical results on cobordisms \cite{feller2015optimal,feller2021genus,baader2024minimal} may be leveraged to
understand domain walls in RCFTs, including their action on rational VOA
modules, fusion rules, and modular data.

\paragraph{Refined 3D-index}
As seen in the case of torus knots, symmetries that are not manifest from the
six-dimensional viewpoint can emerge in the infrared effective theory.
Exploiting such additional symmetries, one can introduce new refinements of the
superconformal index of $T[S^{3}\backslash \mathcal{K}]$, beyond the parameters
associated with the $U(1)_m$ symmetry.
While the 3D index itself has been extensively studied from various perspectives
\cite{Dimofte:2011py,Garoufalidis:2016ckn,Gang:2018gyt,Garoufalidis:2020xec,Celoria:2025lqm,Agarwal:2022cdm,Duan:2022ijp,Garoufalidis:2024hoe},
the refined extensions provide a natural starting point for
further exploration \cite{Refined3DIndex:ToAppear}.

\paragraph{Relation to the ``3d modularity''}
Our work is motivated by ideas similar in spirit to those of
\cite{Cheng:2018vpl,Cheng:2024vou}, in that both aim to relate vertex operator algebras
associated with three-manifolds via the 3D--3D correspondence and the
half-index.
However, the precise relationship between the two approaches is not yet clear. In the works, they do not rely on an explicit
three-dimensional field-theoretic analysis.
Instead, they directly relate quantum invariants of three-manifolds, known as
the $\widehat{Z}$-invariants \cite{Gukov:2016gkn,Gukov:2017kmk,Gukov:2019mnk}, to half-indices and study their nontrivial modular
properties.
The VOAs appearing in their construction are typically non-rational, even for
non-hyperbolic three-manifolds.
This is in contrast to our DGG-based construction, where non-hyperbolic
three-manifolds typically give rise to rational VOAs.
Accordingly, the connection between these VOAs and the rational VOAs obtained
in our framework remains unclear.

	\section*{Acknowledgements}
    We are grateful to Kibok Jeong, Soochang Lee, and Heesu Kang for valuable discussions related to this work. In particular, this project was initiated through insightful conversations with Kazuhiro Hikami during the workshop “Number Theory in Field and String Theory” held at the Institute of Mathematics Academia Sinica in September 2025, to whom we would like to express our special thanks. We also thank the organizers of the workshop for providing a stimulating environment that facilitated these discussions. This work was supported in part by the National Research Foundation of Korea (NRF) grant NRF-2022R1C1C1011979. We also acknowledge support from the National Research Foundation of Korea (NRF) grant RS-2024-00405629.

\newpage 

\appendix

\input{section_Examples}
	
\section{Partition functions of  $SL(2,\mathbb{C})$ CS theory } 
\paragraph{$k=1$ (state-integral model)}  When $k=1$ and $\sigma = \frac{1-b^2}{1+b^2}$, the partition function  on a knot complement $M=S^3\backslash \CK $ can be computed using the following finite-dimensional integral \cite{Dimofte:2012qj,Dimofte:2014zga}:
\begin{align}
\begin{split}
&Z^{SL(2,\mathbb{C})_{k=1}}_{M= S^3\backslash \CK} [X_m] =\frac{1}{\sqrt{\det B}}  \int \frac{d^r \boldsymbol{Z}}{(2\pi \hbar)^{r/2}} \exp \left( \frac{ \mathbf{Z}^T B^{-1}A  \mathbf{Z}}{2\hbar } + \frac{ \mathbf{Z}^T B^{-1}  (\mathbf{X} + (i \pi +\frac{\hbar}2) \boldsymbol{\nu}_x)}\hbar \right)\prod_{i=1}^r \psi_\hbar (Z_i )\;,
\\
& \mathbf{X}= (X_m, 0,\ldots,0)^T\;. \label{state-integral}
\end{split}
\end{align}
The special function $
\psi_\hbar (Z)$ is called quantum dilogarithm (Q.D.L) function, which is defined by \cite{Faddeev:1993rs} 
\begin{align}
\psi_\hbar (Z):= \begin{cases} 
\prod_{r=1}^{\infty} \frac{1-q^r e^{-Z}}{1-\widetilde{q}^{-r+1} e^{-\widetilde{Z}}}\;, \quad \textrm{if $|q|<1$}
\\
\prod_{r=1}^{\infty} \frac{1-\widetilde{q}^r e^{-\widetilde{Z}}}{1-q^{-r+1} e^{-Z}}\;, \quad \textrm{if $|q|>1$} \label{Q.D.L}
\end{cases}
\end{align}
with
\begin{align} 
q=e^{2\pi i b^2}, \quad \widetilde{q}:=e^{2\pi i b^{-2}}, \quad \widetilde{Z}= Z/b^2\;.
\end{align}
In the state-integral model,  we omit the overall factor of the form
\begin{align}
\exp \left( \big{(}\frac{\pi^2}{6\hbar}- \frac{\hbar}{24} \big{)} \mathbb{Z}  + \frac{i\pi }4 \mathbb{Z}+ \frac{(2\pi i  +\hbar)X_m }\hbar \mathbb{Q} +\frac{X_m^2}{\hbar } \mathbb{Q} \right)\;.
\end{align}
In the 3D-3D relation \eqref{3D-3D relation},   such overall factors can be interpreted as contributions from the gravitational Chern–Simons term and from mixed Chern–Simons terms for the background gauge fields coupled to the $U(1)_R$ and  $U(1)_m$ symmetries of the $T[M]$ theory. Because these background CS terms do not affect the IR dynamics of $T[M]$ theory  and can be ignored.

\paragraph{$k=0$ (3D index)} When $k=0$, the complex Chern–Simons partition function produces a quantum invariant known as the 3D index, $\CI^{SL(2,\mathbb{C})_{k=0}}_{S^3\backslash \CK}(m,e;q)$ \cite{Dimofte:2011py}. To define the 3D index from an ideal triangulation of $S^3\backslash \CK$, we need the expressions for the internal edges $\{C_I\}_{I=1}^r$, the meridian variable $M$, and the longitude variable $L$ in terms of the edge variables, as in \eqref{C, M in terms of edges} for the $C_I$ and $M$ variables. In addition to the symplectic structure among $\{C_I\}$ and $M$ given in \eqref{symplectic structure in gluing}, the longitude variable $L$ satisfies the following Poisson brackets: 
\begin{align} \begin{split} &\{ L, C_I \}_{\rm P.B.} = 0 ,\qquad \forall\, I = 1,\ldots,r ,\\[4pt] &\{ M, L \}_{\rm P.B.} = 2 \;. \end{split} 
\end{align}
Using the gluing data and a choice of polarization $(\mathbf{X},\mathbf{P})$, one can introduce an $Sp(2r,\mathbb{Z})$ matrix
\[
g=\begin{pmatrix} A & B \\ C & D \end{pmatrix},
\]
together with $r$-vectors $\boldsymbol{\nu}_{x}$ and $\boldsymbol{\nu}_{p}$, defined through the relation
\begin{align}
\left(
\begin{array}{c}
M \\[4pt]
C_1 - 2\pi i \\
\vdots \\
C_{r-1} - 2\pi i \\[2pt]
\hline
L/2 \\[2pt]
\Gamma_1 \\
\vdots \\
\Gamma_{r-1}
\end{array}
\right)
=
\begin{pmatrix}
A & B \\
C & D
\end{pmatrix}
\begin{pmatrix}
\mathbf{X} \\
\mathbf{P}
\end{pmatrix}
+ i\pi
\left(
\begin{array}{c}
\boldsymbol{\nu}_{x} \\
\boldsymbol{\nu}_{p}
\end{array}
\right)
\;.
\end{align}
Using the relations in the affine gluing space \eqref{gluing affine space}, the variable $L/2$ can always be rewritten as a linear combination of the edge variables with integer coefficients.  
The variables $\{\Gamma_I\}_{I=1}^{r-1}$ are the conjugate variables to $\{C_I\}_{I=1}^{r-1}$, satisfying
\begin{align}
\begin{split}
&\{C_I, \Gamma_J\}_{\rm P.B.} = \delta_{IJ}, \qquad
\{\Gamma_I, \Gamma_J\}_{\rm P.B.} = 0, \qquad
\{M, \Gamma_I\}_{\rm P.B.} = \{L, \Gamma_I\}_{\rm P.B.} = 0,
\\
&\forall I,J = 1,\ldots, r-1 \;.
\end{split}
\end{align}
Using the extended NZ data $(g, \boldsymbol{\nu}_x, \boldsymbol{\nu}_p)$, the 3D index in charge basis is defined as
\begin{align}
\begin{split}
&\CI^{SL(2,\mathbb{C})_{k=0}}_{S^3\backslash \CK}(m,e;q) 
\\
&:= \sum_{(e_2, \ldots, e_r) \in \mathbb{Z}^{r-1}}
\left(
(-q^{1/2})^{\mathbf{m} \cdot \boldsymbol{\nu}_p-\mathbf{e}\cdot \boldsymbol{\nu}_x}
\prod_{i=1}^r
\CI_\Delta \!\left( (g^{-1}\boldsymbol{\gamma})_{i},\, (g^{-1}\boldsymbol{\gamma})_{r+i} \right)
\right)
\bigg|_{\;m_1 \to m,\; e_1 \to e,\; m_{I>1}\to 0 }
\\[4pt]
&\text{where}\quad
\mathbf{m} := (m_1,\ldots,m_r)^T,\quad
\mathbf{e} := (e_1,\ldots,e_r)^T,\quad
\boldsymbol{\gamma} := (m_1,\ldots,m_r,\, e_1,\ldots,e_r)^T \;. \label{3D-index}
\end{split}
\end{align}
Here the tetrahedron index $\CI_\Delta (m,e)$ is defined as \cite{Dimofte:2011py}
\begin{align}
\begin{split}
&\CI_\Delta (m,u) := \prod_{r=0}^\infty \frac{1-q^{r-\frac{1}2 m+1}u^{-1}}{1-q^{r- \frac{1}2 m} u} =\sum_{e \in \mathbb{Z}}\CI_\Delta (m, e) u^e,  \;\; 
\\
&\textrm{where  } \CI_\Delta (m,e) = \sum_{n=\lfloor e \rfloor }^\infty \frac{(-1)^n q^{\frac{1}2 n (n+1)-(n+\frac{1}2 e)m}}{(q)_n (q)_{n+e}}\;. \label{tetrahedron index}
\end{split}
\end{align}
The 3D-index in fugacity basis is defined as
\begin{align}
&\CI^{SL(2,\mathbb{C})_{k=0}}_{S^3\backslash \CK}(m,u;q) := \sum_{e\in \mathbb{Z}}\CI^{SL(2,\mathbb{C})_{k=0}}_{S^3\backslash \CK}(m,e;q)u^e\;.  
\end{align}
For the case in which $M = S^{3}\backslash \CK_{(P,Q)}$ is the complement of the
$(P,Q)$ torus knot $\CK_{(P,Q)}$, the  3D index takes a remarkably simple form \cite{Celoria:2025lqm}:
\begin{align}
\begin{split}
&\CI^{SL(2,\mathbb{C})_{k=0}}_{S^{3}\backslash \CK_{(P,Q)}}(m,e;q)
= \delta_{\,e+\frac{PQ}{2} m,\,0}\;, \;\; \text{(in charge basis)}
\\
&\CI^{SL(2,\mathbb{C})_{k=0}}_{S^{3}\backslash \CK_{(P,Q)}}(m,u;q)
= u^{-\frac{PQ}{2}m}\;, \;\; \text{(in fugacity basis)}
\end{split}
\label{3D index for torus knots}
\end{align}

\section{Characters and Modular data of simple rational VOAs}

\paragraph{Virasoro Minimal model $\CM(P,Q)$}
The integers $P$ and $Q$ labeling $\CM(P,Q)$ should be coprime.
A primary operator of $\CM(P,Q)$ is labeled by two integers $1\leq r<P$ and $1\leq r < Q$, and its character is
\begin{equation}
\begin{gathered}
    \chi^{M(P,Q)}_{(r,s)} = \chi^{M(P,Q)}_{(P-r,Q-s)} = \frac{q^{\Delta^{(P,Q)}_{(r,s)}}}{(q)_\infty}
    \sum_{n\in \mathbb{Z}} \left( q^{n^2 PQ + n(Qr-Ps)} - q^{(nP+r)(nQ+s)}\right)\;, \\
    h^{(P,Q)}_{(r,s)} = \frac{(Qr-Ps)^2-(P-Q)^2}{4PQ}\;, \quad c(P,Q) = 1-\frac{6(P-Q)^2}{PQ}\,, \\
    \Delta^{(P,Q)}_{(r,s)} = h^{(P,Q)}_{(r,s)}-\frac{1}{24}c(P,Q)\,.
\end{gathered}
\end{equation}
The $S$-matrix of $\CM(P,Q)$ is
\begin{equation}
    S^{M(P,Q)}_{(r,s),{(r',s')}} = \sqrt{\frac{8}{PQ}} (-1)^{1+rs'+sr'} 
    \sin\paren{\pi \frac{Q}{P} r r'} \sin\paren{\pi \frac{P}{Q} s s'}\,.
\end{equation}

\paragraph{Supersymmetric Virasoro Minimal model $SM(P,Q)$}
The integers $P$ and $Q$ labeling $SM(P,Q)$ should satisfy two conditions:
\begin{align}
    P\equiv Q \textrm{ (mod 2)}\,, \quad {\rm gcd}(P,\frac{P-Q}{2})=1\,.
\end{align}
Again a primary operator of $SM(P,Q)$ is labeled by two integers $1\leq r<P$ and $1\leq r < Q$. $(r,s)$ where $r-s\in 2\BZ$ corresponds to NS sector, and otherwise R sector. The character is
\begin{equation}
\begin{gathered}
\chi^{SM(P,Q)}_{(r,s)} = q^{\Delta^{(P,Q)}_{(r,s)}} \frac{(-q^t;q)_\infty}{(q)_\infty}
\sum_{n\in \mathbb{Z}} \left( q^{(n^2 PQ + n(Qr-Ps))/2} - q^{(nP+r)(nQ+s)/2} \right)\,, \\
h^{(P,Q)}_{(r,s)} = \frac{(Qr-Ps)^2-(P-Q)^2}{8PQ}+\frac{2t-1}{16}\,, \quad c(P,Q) = \frac{3}{2}\left( 1-\frac{2(P-Q)^2}{PQ}\right)\,, \\
\Delta^{(P,Q)}_{(r,s)} = h^{(P,Q)}_{(r,s)}-\frac{1}{24}c(P,Q)\,, \quad t=\begin{cases} 1/2, & r-s\in2\BZ\\
                1, & r-s\in2\BZ+1\,.     \end{cases}    
\end{gathered}
\end{equation}
The $S$-matrix for the NS characters of $SM(P,Q)$ is
\begin{align}
\begin{split}
    S^{SM(P,Q)}_{(r,s),(r',s')} &= \frac{2}{\sqrt{PQ}}
    \biggr( \cos\left( \frac{\pi}{2PQ}(Ps-Qr)(Ps'-Qr') \right) \\
    & \quad - \cos\left( \frac{\pi}{2PQ}(Ps+Qr)(Ps'-Qr') \right) \biggr)\,.
\end{split}
\end{align}

\paragraph{Nahm--sum like expressions}
We review the known Nahm--sum like expressions for the characters of minimal models. It is known that a large class of $2d$ RCFT characters can be written as fermionic sum, or \textit{Nahm sum formula} \cite{Nahm:2004ch, Zagier:2007knq, Berkovich_1996, melzer1994susyanalogGAid, Bytsko_1999, Kedem:1993ze, Berkovich:1994es, welsh2005fermionic}:
\begin{equation} \label{nahmsum}
    \chi(q) = \sum_{m\in \BZ^r} \frac{q^{ \frac{1}{2} m^T A m + B^T m + C }}{ (q)_{m_1}(q)_{m_2} \cdots (q)_{m_k} }\,,
\end{equation}
where $A$, $B$, and $C$ are some $k\times k$ matrix, length-$k$ vector, and scalar each.
\\
The characters of $\CM(2,2k+3)$ can be represented exactly as the form of \eqref{nahmsum} \cite{melzer1994susyanalogGAid}:
\begin{equation}
\begin{gathered}
    \chi^{\CM(2,2k+3)}_{(1,s)}(q) = q^{\Delta^{(2,2k+3)}_{(1,s)}} \sum_{m \in \BZ^k_{\geq0}}
    \frac{ q^{ \frac{1}{2} m^TAm + B^T m } }{ (q)_{m_1}(q)_{m_2} \cdots (q)_{m_k} }\, \quad (1\leq s \leq k+1), \\
    \textrm{where } (A)_{ij} = 2\,\mathrm{min}(i,j), \quad B = (\mathbf{0}_{s-1},\, 1,\, 2,\, \cdots,\, k-s+1 )^T \,.
\end{gathered}
\label{Nahmsum2,k}
\end{equation}
\\
For $\CM(3,l)$ \cite{Bytsko_1999},
\begin{align} \label{Nahmsum3,l}
    \chi^{\CM(3,l)}_{(1,s)}(q) = q^{\Delta^{(3,l)}_{(1,s)}} \sum_{m \in \BZ^k_{\geq0}}
    \frac{ q^{ \frac{1}{2} m^TA^{(l)}m + m^T B^{(l)}_s } }{ (q)_{m_1}(q)_{m_2} \cdots (q)_{m_{k-1}} (q)_{2m_k} }\,,
\end{align}
where $A^{(l)}$ and $B^{(l)}$ are defined as
\begin{align}
\begin{split}
    (A^{(3k+1)})_{ij} &= (A^{(3k+1)})_{ji} = 2\min(i,j)\,, \qquad 1 \le i,j \le k-1\,, \\
    (A^{(3k+1)})_{jk} &= (A^{(3k+1)})_{kj} = 2j + (1-k)\,\delta_{jk}\,, \qquad 1\le j\le k-1\,, \\
    (A^{(3k+1)})_{kk} &= 2k+2\,, \\
    (A^{(3k+2)})_{ij} &= (A^{(3k+1)})_{ij} - 2\,\delta_{ik}\delta_{jk}\,, \\
\end{split}
\end{align}
and
\begin{align}
\begin{split}
    (B^{(3k+1)}_s)_j &= \max(j-s+1,0) + (j-k-1)\,\delta_{jk}\,, \\
    (B^{(3k+2)}_s)_j &= (B^{(3k+1)}_s)_j + \,\delta_{jk}\,.
\end{split}
\end{align}

\paragraph{Abelian Wess-Zumino-Witten model} For a Wess-Zumino-Witten (WZW) model with gauge group $U(1)^r$ and level matrix $K\in \BZ^r$, there are $|\det K|$ primary operators $\CO_{\boldsymbol{\a}\in \BZ^r}$ with the following equivalence relation
\begin{align}
    \CO_{\boldsymbol{\a}}\sim \CO_{\boldsymbol{\a}+K\cdot\BZ^r}\;.
\end{align}
The central charge of $U(1)^r_K$ WZW model and the conformal weight of primary operator $\CO_{\boldsymbol{\a}}$ are \cite{Delmastro:2019vnj}
\begin{align}
    c=r,\quad h_{\boldsymbol{\a}}=\frac{1}{2}\boldsymbol{\a}^T\cdot K^{-1}\cdot\boldsymbol{\a}\;.
\end{align}
The character is
\begin{align}
    \c_{\boldsymbol{\a}}^{U(1)^r_K}(q)=\frac{q^{h_{\boldsymbol{\a}}-c/24}}{(q)_\infty^r}\sum_{\mathbf{m}\in \BZ^r} q^{\frac{1}{2}\mathbf{m}^T\cdot K \cdot \mathbf{m}-\boldsymbol{\a}^T\cdot\mathbf{m}}\;.
\end{align}
The modular $S$-matrix is
\begin{align}
    S_{\boldsymbol{\a},\boldsymbol{\b}}=\frac{1}{\sqrt{|\det K|}} \exp\paren{2\p i \boldsymbol{\a}^T\cdot K^{-1}\cdot\boldsymbol{\b}}\;.
\end{align}
For $r=1$, there is Nahm--sum like expression of the character\cite{Gang:2024loa}
\begin{align}
    \c_\a^{U(1)_k}(q)=q^{h_\a-1/24}\sum_{\mathbf{m}\in (\BZ_{\geq 0})^2} \frac{q^{\frac{1}{2}\mathbf{m}^T\cdot J_k\cdot\mathbf{m}} (-q^{1/2})^{-2 \a (1,-1)\cdot\mathbf{m}}}{(q)_{m_1} (q)_{m_2}}\;,
    \label{U(1) Nahm sum}
\end{align}
where
\begin{align}
    J_k=\begin{pmatrix}
        k & 1-k\\
        1-k & k
    \end{pmatrix}\;.
\end{align}
\paragraph{Free fermionic RCFT} The central charge of the free fermionic RCFT is $\frac{1}{2}$, and its character is \cite{DiFrancesco:1997nk}
\begin{align}
    \c_F(q)=q^{-1/48} (-q^{1/2};q)_{\infty}\;.
\end{align}
$\c_F(q)$ is a modular form itself.

\bibliographystyle{ytphys}
\bibliography{ref}

\end{document}

%% file: section_TfromTorusKnot.tex
\section{$T[M]$ for torus knot complement $M=S^3\backslash \mathcal{K}_{(P,Q)}$} \label{sec : DGG review}

In this section, we review the field theoretic construction of 3D $\mathcal{N}=2$ gauge theory $T[M]$ labeled by 3-manifold $M$ proposed by Dimofte-Gaiotto-Gukov (DGG) \cite{Dimofte:2011ju}. In the paper, we consider the case when the 3-manifold is torus knot complement $M= S^3\backslash \mathcal{K}_{(P,Q)}$ and will review the construction for general knot complement on $S^3$, $M= S^3\backslash \mathcal{K}$.  

\paragraph{$T[M]$ from a twisted compactification of 6D theory} The 3D theory can be regarded as an effective field theory obtained from a twisted compactification of the 6D $A_1$ $(2,0)$ theory:
\begin{align}
\left(
\begin{aligned}
&\text{6D $A_1$ $(2,0)$ theory on $\mathbb{R}^{1,2}\times S^3$}
\\
&\text{with a BPS codimension-two defect along $\mathbb{R}^{1,2}\times \mathcal{K}$}
\end{aligned}
\right)
\xrightarrow{\quad \text{size}(S^3)\to 0 \quad}
T[S^3\backslash \mathcal{K}] \; .
\label{6d viewpoint of T_DGG}
\end{align}
A partial topological twist is performed along $S^3$ using an $SO(3)$ subgroup of the $SO(5)$ R-symmetry of the 6D theory. 
This twist preserves four supercharges, and the resulting 3D theory is expected to possess $\mathcal{N}=2$ supersymmetry. The theory has a distinguished $U(1)_R$ symmetry, often referred to as the \emph{geometrical} R-symmetry, whose charge $R_{\rm geo}$ can be identified as
\begin{align}
\begin{split}
R_{\rm geo}
=\;& \text{the charge of } SO(2)\subset SO(2)\times SO(3)
\\
&\hspace{5.6em}\subset SO(5)\text{ R-symmetry of the 6D theory}\; .
\end{split}
\label{R_geom}
\end{align}
A characteristic property of this
geometric $U(1)_R$ symmetry is that its charge is quantized,
\begin{align}
R_{\rm geo} \in \mathbb{Z},
\label{quantization of R_{geo}}
\end{align}
since it arises from a $U(1)$ subgroup of $SO(5)$.
In addition, the theory enjoys an $SU(2)$ flavor symmetry, denoted by $SU(2)_m$, which originates from the codimension-two defect:
\begin{align}
\exists \;SU(2)_m \text{ symmetry in } T[S^3\backslash \mathcal{K}] \; .
\label{su(2)m}
\end{align}
\paragraph{Brief Sketch of the DGG construction} The construction of $T[S^3\backslash \mathcal{K}]$  can be summarized as follows
\begin{align}
\begin{split}
&(\textrm{an ideal triangulation $\mathcal{T}$ of $S^3\backslash \mathcal{K}$})+(\textrm{a choice of `polarizations'})
\\
& \rightarrow \textrm{Neunmann-Zagier matrices }(A,B) \textrm{ and }\ g = \begin{pmatrix} A & B \\ C & D \end{pmatrix} \in Sp(2r, \mathbb{Z}) 
\\
& \rightarrow T[S^3\backslash \mathcal{K}] = \left( g \cdot [(T_\Delta)^{\otimes r}] \textrm{ with superpotential }\CW = \sum_{E_I  : \textrm{easy internal edge}} \CO_{E_I} \right) \;.
\end{split} \label{sketch of DGG}
\end{align}
In Section \ref{sec: NZ matrices from ideal triangulation}, we review how to obtain the Neumann–Zagier  matrices (NZ matrices) from an ideal triangulation of a knot complement together with a choice of polarizations on the triangulation.  Here $T_\Delta$ denotes a free chiral theory, and 
\begin{align}
g\cdot [(T_\Delta)^{\otimes r}]
\end{align}
denotes the theory obtained by acting with an element $g\in Sp(2r, \mathbb{Z})$ action on the free theory of $r$ chirals   $[(T_\Delta)^{\otimes r}]$,  using its $U(1)^r$  flavor symmetry. For $r=1$, the $Sp(2,\mathbb{Z}) = SL(2,\mathbb{Z})$ action is the supersymmetric version of the Witten's $SL(2,\mathbb{Z})$ action \cite{Witten:2003ya}, and it extends naturally  to general $r \geq 1$. The theory $g\cdot [(T_\Delta)^{\otimes r}]$ flows to an $\CN=2$ SCFT which contains 1/2 BPS chiral primaries $\CO_{E_I}$ associated with `easy internal edges' $\{E_I\}$, which will be defined in \eqref{def : Easy/hard edge} and \eqref{O_{E}}. By turning on superpotential deformations using these
chiral primaries, one obtains  the theory $T[S^3\backslash \mathcal{K}]$. A beautiful aspect of this construction is that the IR physics of 
$T[S^3 \backslash \mathcal{K}]$ 
does not depend on the auxiliary choices---such as the ideal triangulation, the choice of polarizations, or the completion of the NZ matrices into an element of $Sp(2r,\mathbb{Z})$.  
Different choices lead to different UV descriptions, but all of them are related by IR dualities.  
In particular, different choices of polarization yield UV field theories that are related to one another by a sequence of the following basic $\CN=2$ duality \cite{Aharony:1997bx}:
\begin{align}
\begin{split}
\text{Basic $\CN=2$ duality :}\;\;
&\text{(a free theory of a single chiral multiplet)}
\\
&\simeq\;
\left(
U(1)_{\frac12}
\;\text{coupled to a single chiral multiplet of charge $+1$}
\right)\;.
\label{mirror duality}
\end{split}
\end{align}
The most nontrivial step in the construction is to obtain a field-theoretic description of $ g\cdot [(T_\Delta)^{\otimes r}]$. In principle,  one can decompose the element  $g \in Sp(2r, \mathbb{Z})$ into a product of the basic generators of $Sp(2r,\mathbb{Z})$-the called "S-type", "T-type" and "GL-type" transformations defined in \cite{Dimofte:2011ju}. However, such decompositions quickly become complicated as $r$ increases, and the resulting gauge theories are often far from simple. When $|\det B|=1$, the theory admits a much simpler description:
\begin{align}
\begin{split}
&g\cdot [(T_\Delta)^{\otimes r}]  
\\
&= \bigg{(}U(1)^r_{K_{\rm eff}}  \;\CN=2\;\textrm{ CS theory coupled to $r$ chirals of charge matrix $\mathbf{Q} = \mathbb{I}$} 
\\
& \qquad \; \textrm{with UV effective mixed CS level }K_{\rm eff} = B^{-1} A - \mathbb{I}/2\bigg{)}\;.
\end{split}
\end{align}
When $|\det B|>1$, the gauge-theory description above is no longer well-defined, since the matrix $K:=B^{-1}A$  is not integer-valued and therefore the mixed Chern–Simons level  fails to be properly quantized.\footnote{Here \(K\) denotes the bare mixed Chern-Simons level in the "$(U(1)_{-\frac{1}{2}}$ quantization", whose entries are all integers.
The matrix
$
K_{\rm eff} := K - \frac{1}{2}\,\mathbf{Q}^{T}\mathbf{Q}
$ denotes the UV effective Chern-Simons levels.
} In Section \ref{subsec : construction of DGG}, we will address how to make the simple gauge-theory description sensible in this case. 

\paragraph{A 3D--3D relation}
One useful guideline for identifying the correct low-energy field-theory description is provided by the following 3D--3D relation \cite{Terashima:2011qi,Dimofte:2011ju,Dimofte:2014zga,Cordova:2013cea}:
\begin{align}
\begin{split}
&Z^{T[M]}_{S^3_b}[X_m] := \left(\textrm{BPS partition function of $T[M]$ on a squashed 3-sphere $S^3_b$ \cite{Hama:2011ea}}\right)
\\
&=
\\
&Z^{SL(2,\mathbb{C})_{k=1}}_{M= S^3\backslash \CK} [X_m] := \left(SL(2,\mathbb{C})_{k=1, \sigma = \frac{1-b^2}{1+b^2}} \textrm{ CS theory partition function on } M\right)\;. \label{3D-3D relation}
\end{split}
\end{align}
The $SL(2,\mathbb{C})$ Chern--Simons partition function on a knot complement
$M=S^3\backslash \mathcal{K}$ depends on a complex parameter $X_m$, which may be interpreted as the boundary meridian holonomy variable.
In the 3D--3D correspondence, this variable is mapped to the holomorphic mass parameter for a $U(1)$ flavor symmetry of the theory $T[S^3\backslash \CK]$. As we will see in Section \eqref{subsec : construction of DGG}, the theory $T[S^3\backslash \CK]$ always possesses a $U(1)$ symmetry, which we denote by $U(1)_m$.
This symmetry can be identified with the Cartan subgroup of the $SU(2)_m$ flavor symmetry introduced in \eqref{su(2)m}.
More explicitly, the holomorphic parameter $X_m$ is given by ($\hbar:=2\pi i b^2$)
\begin{align}
X_m = \zeta_m + \Bigl(i\pi + \frac{\hbar}{2}\Bigr)\mu_m \; ,
\label{3D-3D relation 2}
\end{align}
where $\zeta_m$ is the rescaled real mass parameter, $\zeta_m = b \times (\text{real mass})$, for the $U(1)_m$ symmetry, and $\mu_m$ parametrizes the mixing of the $U(1)_m$ symmetry with the R-symmetry. The R-charge at mixing parameter $\mu_m$ is then given by
\begin{align}
R_{\mu_m} = R_{\rm geo} + \mu_m T_m \; ,
\label{3D-3D relation 3}
\end{align}
where
\begin{align}
T_m \; \text{is the Cartan generator of the $U(1)_m \subset SU(2)_m$ symmetry.}
\label{T_m}
\end{align}
Here $R_{\rm geo}$ denotes the distinguished geometric $U(1)$ $R$-charge
defined in \eqref{R_geom}, which originates from the $SO(5)$ $R$-symmetry
of the parent six-dimensional theory.

\paragraph{Accidental and IR-decoupled symmetries in $T[S^3\backslash \CK]$}
The $U(1)_R$ symmetry and the $SU(2)_m$ flavor symmetry of
$T[S^3\backslash \mathcal{K}]$ are the symmetries expected from the
six-dimensional compactification picture.
In general, however, $T[S^3\backslash \mathcal{K}]$ may possess additional
symmetries beyond these, as emphasized in \cite{Gang:2018wek}.
From the six-dimensional viewpoint, such symmetries can be interpreted as
emergent accidental symmetries arising in the infrared.

In the bottom-up construction of the theory $T[S^3\backslash \mathcal{K}]$ in \eqref{sketch of DGG},
these accidental symmetries arise from the fact that the ideal
triangulation may not contain a sufficient number of \emph{easy edges},
defined in \eqref{def : Easy/hard edge}.
Each easy edge gives rise to a superpotential term that breaks one Cartan
generator of the flavor symmetry.
Consequently, when there are too few easy edges, some flavor symmetries
remain unbroken, leading to additional flavor symmetries beyond $SU(2)_m$.
The rank of the ultraviolet flavor symmetry of $T[S^3\backslash \CK]$
is given in \eqref{rank(F_UV) of T_{DGG}}.

This, however, is not the end of the story for flavor symmetry counting.
Some of the ultraviolet flavor symmetries of $T[S^3\backslash \CK]$
may decouple in the infrared, acting trivially on all IR observables.
To identify the faithful infrared symmetry, one must therefore mod out
such decoupled flavor symmetries, as in \eqref{F[T]_IR}.
Both the accidental symmetries and the IR-decoupled symmetries will play
crucial roles in identifying the correct infrared phases of
$T[S^3\backslash \CK]$ for the torus knot case,
i.e.\ $\CK = \CK_{(P,Q)}$, as we will see in
Section~\ref{subsect : T-DGG for torus knot}.

\subsection{Symplectic structure in ideal triangulation} \label{sec: NZ matrices from ideal triangulation}
\paragraph{Gluing equations} We consider an ideal triangulation $\mathcal{T}$ of a knot complement \cite{thurston2022geometry}:
\begin{align}
\mathcal{T} \;:\; S^3 \backslash \mathcal{K}
   = \left( \bigcup_{i=1}^r \Delta_i \right) / \sim \;.
\end{align}
For each ideal tetrahedron $\Delta_i$, we assign the boundary phase space $P(\partial \Delta_i)$ \cite{Dimofte:2011gm}
\begin{align}
\begin{split}
&P(\partial \Delta_i)
= \{ Z_i, Z_i', Z_i'' \;:\; Z_i + Z_i' + Z_i'' = i\pi \}, 
\\
&\text{with symplectic structure } \;\;
\{ Z_i, Z_i'' \}_{\rm P.B.}
 = \{ Z_i', Z_i \}_{\rm P.B.}
 = \{ Z_i'', Z_i' \}_{\rm P.B.}
 = 1 \;.
\end{split}
\end{align}
The choice of {\it polarization} specifies the position variable $X_i$ and the momentum variable $P_i$
for each tetrahedron. For the polarization choice, one may take
\begin{align}
\text{polarization choices :}\;\;
\begin{pmatrix} X_i \\[2pt] P_i \end{pmatrix}
=
\begin{pmatrix} Z_i \\ Z_i'' \end{pmatrix},
\;\; \text{or} \;\; 
\begin{pmatrix} Z'_i \\ Z_i \end{pmatrix},
\;\; \text{or} \;\;
\begin{pmatrix} Z''_i \\ Z'_i \end{pmatrix}. \label{polarization choices}
\end{align}
The gluing equations consist of the internal-edge equations and the boundary equation:
\begin{align}
\begin{split}
\textrm{internal edges : }C_I &= \sum_{i=1}^r \left( G_{I i} Z_i + G'_{I i} Z'_i + G''_{I i} Z''_i \right),
\qquad I = 1, \ldots, r,
\\
\textrm{meridian : }M &= \sum_{i=1}^r \left( \alpha_i Z_i + \beta_i Z'_i + \gamma_i Z''_i \right).
\label{C, M in terms of edges}
\end{split}
\end{align}
%
 Among the $r$ internal edges, only $(r-1)$ are linearly independent due to the relation
\begin{align}
\sum_{I=1}^r C_I = 2 \sum_{i=1}^r (Z_i + Z_i' + Z_i'')  = 2\pi i r\;.
\label{linear relation of Cs}
\end{align}
These gluing equations are known to satisfy the following symplectic relations \cite{neumann1985volumes}:
\begin{align}
\{ C_I, C_J \}_{\rm P.B.} = 0,
\qquad
\{ C_I, M \}_{\rm P.B.} = 0,
\qquad
\forall\, I, J = 1, \ldots, r \;. \label{symplectic structure in gluing}
\end{align}
\paragraph{Hard/easy internal edges and "refined" NZ matrices} We will generalize the notion of internal edges and  we call
\begin{align}
C = \sum_{i=1}^r (n_i Z_i + n'_i Z_i' + n''_i Z''_i) \textrm{ with $n_i, n'_i, n''_i \in \mathbb{Z}_{\ge 0}$}
\end{align}
 an \emph{internal edge} if
\begin{align}
\begin{split}
&C\big|_{\mathcal{G}[\mathcal{T}]} = 2\pi i,\quad
\{C, M\}_{\rm P.B.} = \{C, C_I\}_{\rm P.B.} = 0 \quad (I=1,\ldots,r)\;.
\end{split}
\end{align}
Here the gluing affine space $\mathcal{G}[\mathcal{T}]$ is defined as
\begin{align}
\mathcal{G}[\mathcal{T}] 
= \Big\{ (Z_i, Z_i', Z_i'')_{i=1}^{r} \;:\; 
(Z_i + Z'_i + Z''_i = i\pi)\big|_{i=1}^{r},\; 
(C_I = 2\pi i)\big|_{I=1}^{r} \Big\} \;. \label{gluing affine space}
\end{align}
Note that every internal edge $\{ C_I \}_{I=1}^{r}$ satisfies the conditions. More generally, the internal edge can be written as
\begin{align}
C
&= \sum_{I=1}^{r-1} N_{I} C_I 
 + \sum_{i=1}^r L_i (Z_i + Z_i' + Z_i'') \;.
\end{align}
For later use, we also define
\begin{align}
\textrm{internal edge $C$ is }
\begin{cases}
\textrm{easy},  \quad  \textrm{at most one of $(n_i, n'_i, n''_i)$ is nonzero, for each $i$.}
\\
\textrm{hard},  \quad \textrm{otherwise}
\end{cases} \label{def : Easy/hard edge}
\end{align}
Let
\begin{align}
\begin{split}
&\{E_I\}_{I=1}^{\sharp_E}\;
  :\; \text{the maximal set of linearly independent easy internal edges}, 
\\[0.3em]
&\{H_I\}_{I=1}^{\sharp_H := r-1-\sharp_E}\;
  :\; \text{a linearly independent set of hard edges} 
\\
&\hspace{8.3em}
\text{that are linearly independent of the set }\{E_I\}\;.
\end{split}
\label{sets of easy/hard edges}
\end{align}
Fixing a choice of polarizations, the {\it refined} Neumann–Zagier matrices $(A,B ;\boldsymbol{\nu}_x)$ are defined by the linear relation
\begin{align}
\begin{pmatrix}
M \\[4pt]
H_1 - 2\pi i \\
\vdots
\\
H_{\sharp_H}-2\pi i 
\\
E_1 - 2\pi i \\
\vdots \\
E_{\sharp_E} - 2\pi i
\end{pmatrix}
=
\begin{pmatrix}
A & B
\end{pmatrix}
\begin{pmatrix}
\mathbf{X} \\
\mathbf{P}
\end{pmatrix}
+ i\pi\, \boldsymbol{\nu}_{x} \;. \label{NZ matrices}
\end{align}
We assume that the matrix $B$ is invertible, which can always be achieved by choosing an appropriate set of polarizations \cite{Dimofte:2012qj}.

\paragraph{Example : $\CK = \CK_{(P,Q)=(2,7)}$} The torus knot complement can be ideally triangulated with $r=4$ tetrahedra with following gluing data, see \eqref{(2,7) gluing}
\begin{align}
\begin{split}
    & C_1=2 Z_1''+Z_2'+Z_3'+Z_4'', \quad C_2=Z_2''+Z_4',\\
    & C_3=2 Z_1+Z_1'+2 Z_2+2 Z_3+2 Z_4,\\
    & C_4=Z_1'+Z_2'+Z_2''+Z_3'+2 Z_3''+Z_4'+Z_4'',\\
    & M=Z_2'+Z_3''-Z_4\;. \label{Example : (2,7)}
\end{split}
\end{align}
Then,
\begin{align}
\{E_I\} = \{C_2,C_1\}, \quad \{H_I\} = \{ C_3\}\;.
\end{align}
When we choose the polarizations as follows
\begin{align}
    \mathbf{X}=(Z_1',Z_2,Z_3'',Z_4)^T \textrm{ and } \mathbf{P}=(Z_1,Z_2'',Z_3',Z_4'')^T
\end{align}
the NZ matrices are
\begin{align}
    A=\begin{pmatrix}
        0 & -1 & 1 & -1\\
        1 & 2 & -2 & 2\\
        0 & 0 & 0 & -1\\
        -2 & -1 & 0 & 0
    \end{pmatrix},\quad B=\begin{pmatrix}
        0 & -1 & 0 & 0\\
        2 & 0 & -2 & 0 \\
        0 & 1 & 0 & -1 \\
        -2 & -1 & 1 & 1
    \end{pmatrix}, \quad {\boldsymbol{\n}}_x=(1,0,-1,1)^T \label{NZ matrices for (2,7)}
\end{align}

\subsection{Field theoretic construction of $T[S^3\backslash \mathcal{K}]$} \label{subsec : construction of DGG}
Using the Neumann--Zagier matrices $(A,B)$ together with an additional
integer-valued matrix $\mathbf{Q}\in M_r(\mathbb{Z})$, we define the theory
\begin{align}
\begin{split}
\mathbb{T}[A,B;\mathbf{Q}]
:=\;&
\left(
U(1)^r_{K_{\rm eff}}
\ \text{coupled to $r$ chiral multiplets } \Phi s
\ \text{with charge matrix } \mathbf{Q}
\right),
\\[4pt]
\text{with}\quad
K_{\rm eff}
:=\;&
K - \tfrac{1}{2}\,\mathbf{Q}^{T}\mathbf{Q}
\;=\;
\mathbf{Q}^{T}\!\left(B^{-1}A - \tfrac{1}{2}\,\mathbb{I}\right)\!\mathbf{Q}\;.
\end{split}
\label{tT[A,B]}
\end{align}
We define the charge matrix by
\begin{align}
(\mathbf{Q})_{ij}
= \text{charge of the $i$-th chiral under the $j$-th } U(1)\;.
\end{align}
The matrix $\mathbf{Q}$ is chosen such that
\begin{align}
\begin{split}
\text{i)}\;& 
K := \mathbf{Q}^{T} B^{-1} A\, \mathbf{Q} \;\in\; M_{r}(\mathbb{Z}), 
\\[2pt]
\text{ii)}\;&
N := \left| \frac{\det \mathbf{Q}}{\sqrt{\det B}} \right|^{2}
\quad \text{is a square-free natural number}.
\\[2pt]
\end{split}
\label{NQB relation}
\end{align}
We assume that such a choice of $\mathbf{Q}$ exists.  
In all examples considered in this work, we are able to find a matrix $\mathbf{Q}$ satisfying these conditions.

Let  $\Gamma^{[1]}$ denotes the 1-form symmetry of $\mathbb{T}[A,B;\mathbf{Q}]$, which is given by
\begin{align}
\Gamma^{[1]}
= \frac{\mathbb{Z}^r}{K \cdot \mathbb{Z}^r + \mathbf{Q} \cdot \mathbb{Z}^r}\;.
\label{1-form symmetry of tT}
\end{align}
Here the numerator $\mathbb{Z}^r$ represents the Wilson-line charge lattice of the $U(1)^r$ gauge fields,  
while $K \cdot \mathbb{Z}^r$ and $\mathbf{Q}\cdot \mathbb{Z}^r$ correspond to the charges screened by the mixed Chern--Simons couplings and the charged matter fields, respectively. The $\mathbb{Z}_N$ is non-anomalous when
\begin{align}
\begin{split}
&q([v]):=\frac{N}{2} v^T K^{-1} v  \;(\textrm{mod 1})\; \;\textrm{is well-defined for $[v] \in \mathbb{Z}_N\in \Gamma^{[1]} (K, \mathbf{Q})$} \;,
\\
&\textrm{and $q([v])=0$ for a generator $[v]$ of $\mathbb{Z}_N$}\;.
\end{split}
\end{align}
We assume that the theory $\mathbb{T}[A,B;\mathbf{Q}]$ develops an anomalous
$\mathbb{Z}_N$ 1-form symmetry in the infrared, whose generating line operator
has topological spin $\tfrac{p}{2N}$.
This $\mathbb{Z}_N$ symmetry may descend from a UV 1-form symmetry
(i.e.\ $\mathbb{Z}_N \subset \Gamma^{[1]}$), or it may instead be an emergent
symmetry appearing only in the infrared. The motivation for this assumption is as follows.
As we will see below, the theory $\mathbb{T}[A,B;\mathbf{Q}]$ shares the same
supersymmetric partition function on the squashed three-sphere $S^3_b$
\cite{Hama:2011ea} as the theory $g \cdot (T_\Delta)^{\otimes r}$ appearing in
\eqref{sketch of DGG}, up to an overall factor of $1/\sqrt{N}$:
\begin{align}
\CZ_{S^3_b}\!\left[\mathbb{T}[A,B;\mathbf{Q}]\right]
=
\frac{1}{\sqrt{N}}\,
\CZ_{S^3_b}\!\left[g \cdot (T_\Delta)^{\otimes r}\right].
\end{align}
This relation is highly nontrivial, since the partition function depends
nontrivially on the squashing parameter as well as on multiple real mass
parameters. The factor $1/\sqrt{N}$ can be interpreted as the contribution of a decoupled
topological field theory $\mathcal{A}^{N,p}$, which is the minimal TQFT
carrying the anomalous $\mathbb{Z}_N$ 1-form symmetry~\cite{Hsin:2018vcg}.\footnote{
When $N$ is not square-free, say $N = N_1^{\,2} N_2$, the theory
$\mathbb{T}[A,B;\mathbf{Q}]$ is expected to possess a non-anomalous
$\mathbb{Z}_{N_1}$ 1-form symmetry.
Upon gauging this symmetry, the resulting theory is infrared equivalent to
$g \cdot (T_\Delta)^{\otimes r}$ up to a decoupled
$\mathcal{A}^{N_2,p_2}$ sector.
In this case, the local operator spectrum of
$\mathbb{T}[A,B;\mathbf{Q}]$ generally differs from that of
$g \cdot (T_\Delta)^{\otimes r}$, since gauging a 1-form symmetry—unlike
tensoring with a TQFT—modifies the spectrum of local operators.
In particular, certain monopole operators present in
$g \cdot (T_\Delta)^{\otimes r}$ are absent in
$\mathbb{T}[A,B;\mathbf{Q}]$.
Consequently, one should not expect the existence of a chiral primary operator
$\mathcal{O}_E$ in $\mathbb{T}[A,B;\mathbf{Q}]$ associated with the easy internal
edge in \eqref{sketch of DGG}.
For this reason, we restrict our attention to square-free values of $N$.
}
The infrared 1-form symmetry can be analyzed systematically using
supersymmetric localization.
Upon removing the decoupled topological sector, we define the theory $T[A,B]$ by
\begin{align}
T[A,B] \otimes \mathcal{A}^{N,p}
\;:=\;
\mathbb{T}[A,B;\mathbf{Q}] \, .
\label{T[A,B]}
\end{align}
The theory $T[A,B]$ is expected to be infrared equivalent to
$g \cdot (T_\Delta)^{\otimes r}$, possibly up to decoupled invertible TQFTs,
since the two theories share the same partition function on the squashed
three-sphere.


\paragraph{Easy edges and superpotential deformations} Now we turn to a superpotential deformation of the $T[A,B]$ theory.  We first need to identify the spectrum of chiral primary operators in $T[A,B]$. Interestingly, to each \emph{easy} internal edge $E_I$ of the ideal triangulation, one can associate a corresponding chiral primary operator $\mathcal{O}_{E_I}$.
Let $\{E_I\}$ denote the set of easy internal edges.  
Each $E_I$ can be written as
\begin{align}
E_I 
&= \sum_{i=1}^r \big( g_{Ii} X_i + g'_{Ii} R_i + g''_{Ii} P_i \big) \;\; \textrm{ with } g_{Ii}, g'_{Ii}, g''_{Ii} \in \mathbb{Z}_{\geq 0}\;,\label{E in terms of g}
\end{align}
where for each $i$, at most one of $(g_{Ii}, g'_{Ii}, g''_{Ii})$ is nonzero.  
Here $R_i$ is whichever one among $\{Z_i, Z_i', Z_i''\}$ is not equal to $X_i$ or $P_i$. In the first line, the index $J$ runs only up to $J = r - 1$, because only $r-1$ of the $C_I$ are independent due to the relation in \eqref{linear relation of Cs}.
For each easy internal edge $E_I$, there exists a corresponding gauge-invariant chiral primary operator $\mathcal{O}_{E_I}$ in $T[A,B]$:
\begin{align}
\begin{split}
\mathcal{O}_{E_I} 
&= \left( \prod_{i=1}^r \phi_i^{\, g_{Ii}} \right) V_{\mathbf{m}(E_I)}, 
\\
\mathbf{m}(E_I) 
&:= \mathbf{Q}^{-1} \cdot (g'_{I1} - g''_{I1}, \ldots, g'_{Ir} - g''_{Ir})^{T} 
= \mathbf{Q}^{-1} \cdot (\mathbf{g}'_I - \mathbf{g}''_I) \;. \label{O_{E}}
\end{split}
\end{align}
Here $V_{\mathbf{m}}$ denotes the $\tfrac12$-BPS bare monopole operator of monopole flux 
$\mathbf{m} = (m_1, \ldots, m_r)^T$.
Its gauge charge $q_i$ under the $i$-th $U(1)$ gauge symmetry is
\begin{align} \label{echargeofVm}
\begin{split}
&q_i (V_{\mathbf{m}}) = \left(\mathbf{Q}^T \cdot (B^{-1}A -\mathbb{I}/2) \cdot \mathbf{Q} \cdot \mathbf{m} \right)_i - \frac{1}2 \sum_j (\mathbf{Q}^T)_{ij} |(\mathbf{Q}\cdot \mathbf{m})_j|\;,
\\
&\textrm{or equivalently,}
\\
& \mathbf{q} (V_{\mathbf{m}}) = \left(\mathbf{Q}^T \cdot (B^{-1}A - \mathbb{I}/2) \cdot \mathbf{Q} \cdot \mathbf{m} \right) - \frac{1}2 \mathbf{Q}^T \cdot  |\mathbf{Q}\cdot \mathbf{m}|
\end{split}
\end{align}
$\phi_i$ denotes the scalar in the $i$-th chiral multiplet. Then,
\begin{align}
\mathbf{q} \left(\big{(}\prod_{i=1}^r\phi^{g_{Ii}}_i \big{)} \right) = \mathbf{Q}^T \cdot \mathbf{g}_I\;.
\end{align}
Using these expressions, one can check that $\CO_{E_I}$ is indeed gauge-invariant:
\begin{align}
\begin{split}
 \mathbf{q}(\CO_{E_I})
 &= \mathbf{Q}^T \mathbf{g}_I
 + \mathbf{Q}^T B^{-1}\!\left(
 A(\mathbf{g}'_I-\mathbf{g}''_I)
 - \frac12\, B\big(\mathbf{g}'_I-\mathbf{g}''_I+\lvert \mathbf{g}'_I-\mathbf{g}''_I\rvert\big)
 \right)
\\
 &= \mathbf{Q}^T B^{-1}\!\left(
 A(\mathbf{g}'_I-\mathbf{g}''_I) + B(\mathbf{g}_I-\mathbf{g}'_I)
 \right)
\\
 &= \mathbf{Q}^T B^{-1}\!
 \begin{pmatrix}
 \{E_I,M\}_{\rm P.B.}\\
 \{E_I,H_1\}_{\rm P.B.}\\
 \vdots\\
 \{E_I,E_{\sharp_E}\}_{\rm P.B.}
 \end{pmatrix}
 = \mathbf{0}\,.
\end{split}
\end{align}
Furthermore, the dressed monopole operator $\CO_{E_I}$ has spin zero, since
\begin{align}
g_{Ii}\,(\mathbf{Q}\cdot \mathbf{m}(E_I))_i = 0
\qquad \forall\, i=1,\ldots,r \, .
\end{align}
Using the 1/2 BPS monopole operators, we finally define
\begin{align}
T[S^3\backslash \mathcal{K}]
:= \Big( T[A,B] \ \text{deformed by the superpotential}\ 
\CW = \sum_{E_I: \text{easy internal edges}} \mathcal{O}_{E_I} \Big)\;.
\label{T-DGG}
\end{align}
Equivalently, the theory can be defined by the relation
\begin{align}
\begin{split}
&T[S^3\backslash \mathcal{K}] \otimes \mathcal{A}^{N,p}
= \mathbb{T}[A,B;\mathbf{Q},\{E_I\}] \,,
\\
&\mathbb{T}[A,B;\mathbf{Q},\{E_I\}]
:= \Big( \mathbb{T}[A,B;\mathbf{Q}]\
\text{deformed by the superpotential } 
\sum \mathcal{O}_{E_I} \Big)\;.
\label{T-DGG-2}
\end{split}
\end{align}
The relations among the four theories are summarized in the following diagram:
\[
\begin{tikzcd}[column sep=9em, row sep=4em]
\fbox{$\mathbb{T}[A,B;\mathbf{Q}]$}
  \arrow[r, "{\quad \text{removing the } \mathcal{A}^{N,p} \quad}"]
  \arrow[d, "{\CW = \sum_I \mathcal{O}_{E_I}}"']
&
\fbox{$T[A,B]$}
  \arrow[d, "{\CW = \sum_I \mathcal{O}_{E_I}}"]
\\
\fbox{$\mathbb{T}[A,B;\mathbf{Q},\{E_I\}]$}
  \arrow[r, "{\quad \text{removing the } \mathcal{A}^{N,p} \quad}"']
&
\fbox{$T[S^3\backslash \mathcal{K}]$}
\end{tikzcd}
\]
We expect that the infrared behavior of the theory $T[S^{3}\backslash \mathcal{K}]$
does not depend on auxiliary choices, such as the choice of polarization
\eqref{polarization choices} or the choice of $\mathbf{Q}$ in
\eqref{NQB relation}. In particular, different choices of polarization lead to different UV
descriptions, which are related to one another by a sequence of elementary
$\mathcal{N}=2$ dualities in
\eqref{mirror duality}:
\begin{align}
(\text{different choices of polarization})
\;\longleftrightarrow\;
(\text{different duality frames related by \eqref{mirror duality}})\, .
\label{polarization/duality}
\end{align}
Different choices of polarization affect the Neumann--Zagier matrices in
\eqref{NZ matrices}, and consequently modify the admissible charge matrix
$\mathbf{Q}$ as well as the Chern--Simons levels, $K = \mathbf{Q}^{T} B^{-1} A \mathbf{Q}$, of the UV gauge theory

\paragraph{Flavor symmetry of $T[S^3\backslash \CK]$}
The rank of the flavor symmetry $F_{\rm UV}$ of the UV gauge theory
$T[S^3\backslash \mathcal{K}]$ is given by
\begin{align}
\begin{split}
\mathrm{rank}\,(F_{\mathrm{UV}})
&= r \;(\text{number of tetrahedra})
 - \sharp_E \;(\text{number of easy internal edges})
\\
&= \sharp_H + 1 \;.
\label{rank(F_UV) of T_{DGG}}
\end{split}
\end{align}
Here $r$ is the rank of the gauge group $U(1)^r$, which is equal to the number of
topological $U(1)_{\rm top}$ symmetries in the UV description.
Each easy internal edge gives rise to a superpotential term involving a chiral
primary operator, which explicitly breaks one Cartan generator of the flavor
symmetry. We define the vector space
\begin{align}
\mathfrak{F}[\mathcal{T}]
:=
\Big\{
\boldsymbol{v}\in\mathbb{R}^{r}
\;\big|\;
\mathbf{m}(E_I)\!\cdot\!\boldsymbol{v}=0,
\quad
\forall\ \text{easy internal edges } E_I
\Big\}.
\label{F[T]}
\end{align}
This space can be identified with the Cartan subalgebra of the UV flavor symmetry
$F_{\rm UV}$,
\begin{align}
\big\{
\boldsymbol{v}\!\cdot\!\mathbf{T}
\;\big|\;
\boldsymbol{v}\in\mathfrak{F}[\mathcal{T}]
\big\}
=
\big\{
\text{Cartan subalgebra of the UV flavor symmetry of }
T[S^{3}\backslash\mathcal{K}]
\big\}.
\nonumber
\end{align}
Here
\begin{align} 
\textrm{$\mathbf{T} = (T_1, \ldots, T_r)^T$ denotes the charges of the symmetry $U(1)^r_{\rm top}$ in the $T[A,B]$ theory}\;. \label{topological charges} 
\end{align} 
A convenient basis of $\mathfrak{F}[\mathcal{T}]$ may be chosen as
\begin{align}
\begin{split}
\mathfrak{F}[\mathcal{T}]
&=
\mathrm{Span}
\big\{
\boldsymbol{v}_{m},\boldsymbol{v}_{1},\ldots,\boldsymbol{v}_{\sharp_{H}}
\big\},
\\[4pt]
\boldsymbol{v}_{m}
&:=
\mathbf{Q}^{T} B^{-1}
\begin{pmatrix}
1 \\ \mathbf{0}_{r-1}
\end{pmatrix},
\qquad
\boldsymbol{v}_{1}
:=
\mathbf{Q}^{T} B^{-1}
\begin{pmatrix}
0 \\ 1 \\ \mathbf{0}_{r-2}
\end{pmatrix},
\\
&\hspace{2cm}\ldots,
\\
\boldsymbol{v}_{\sharp_{H}}
&:=
\mathbf{Q}^{T} B^{-1}
\begin{pmatrix}
\mathbf{0}_{\sharp_{H}} \\ 1 \\ \mathbf{0}_{r-1-\sharp_{H}}
\end{pmatrix}.
\end{split}
\label{basis of F[T]}
\end{align}
One can check that
 \begin{align}
 \begin{split}
 &\forall\; I = 1,\ldots, \sharp_E\;,
 \\
 &(\boldsymbol{v}_m \cdot \mathbf{T} \textrm{  of } \CO_{E_I}) =\boldsymbol{v}_m \cdot \mathbf{m}(E_I) \\
 & =  (1,0,\ldots, 0)\cdot (B^{-1})^T \cdot (\mathbf{g}'_I-\mathbf{g}''_I)= \sum_{j=1}^r (B^{-1})_{j 1} (-B)_{I+\sharp_H+1,j}=0\;,
 \\
 & (\boldsymbol{v}_i \cdot \mathbf{T} \textrm{  of } \CO_{E_I})|_{i=1}^{\sharp_H}=\boldsymbol{v}_i \cdot \mathbf{m}(E_I)\\
 & =(\mathbf{0}_i,1,\mathbf{0}_{r-i-1})\cdot (B^{-1})^T\cdot(\mathbf{g}'_I-\mathbf{g}''_I)= \sum_{j=1}^r (B^{-1})_{j,i+1} (-B)_{I+\sharp_H+1,j}=0\;.
 \end{split}
 \end{align}
In the $T[S^3\backslash  \mathcal{K}]$ theory,  there is always a $U(1)_m$ symmetry, which can be identified with the $U(1)_m$ in \eqref{T_m}, whose Cartan $T_m$ is given by 
\begin{align}
T_m = \boldsymbol{v}_m \cdot \mathbf{T} \; \textrm{with } \boldsymbol{v}_m:= \mathbf{Q}^T \cdot B^{-1}\cdot (1,0,\ldots, 0)^T\;. \label{v_m}
\end{align}
The $U(1)_m$ symmetry is in fact the Cartan subgroup of an $SU(2)_m$ symmetry 
associated with the regular codimension-two defect inserted along 
$\mathcal{K} \subset S^{3}$ in \eqref{6d viewpoint of T_DGG}. 
In addition to this $U(1)_m$, there are $\sharp_H$ additional flavor Cartans. 
These extra flavor symmetries are not visible in the 6D compactification picture 
\eqref{6d viewpoint of T_DGG}.
 
Some of the UV flavor symmetries may decouple in the IR, acting trivially on all 
IR observables. For torus knot cases, this decoupling always occurs. 
To analyze the structure of the decoupled flavor symmetry, we define
\begin{align}
\begin{split}
&(\mathfrak{F}_{\rm decouple}[\CT] \subset  \mathfrak{F} [\CT] )
\\
&:= \bigl( \text{the Cartan subalgebra of $F_{\rm UV}$ that decouples in the IR 
for } T[S^3\backslash \CK] \bigr)\;. \label{F_decouple}
\end{split}
\end{align}
As we will demonstrate in the next subsection, this decoupled 
symmetry can be analyzed via BPS partition functions.
Then the Cartan subalgebra of the faithful IR flavor symmetry\footnote{
More precisely, this refers to the part of the IR flavor symmetry that descends 
faithfully from $F_{\rm UV}$. There may also exist emergent IR flavor Cartans 
that are difficult to detect from the UV gauge theory description.
} can be identified as
\begin{align}
\mathfrak{F}_{\rm IR}[\CT] := \mathfrak{F}[\CT] \big/ \mathfrak{F}_{\rm decouple}[\CT]\;. \label{F[T]_IR}
\end{align}
\paragraph{Example : $(P,Q)=(2,7)$} Using the NZ-matrices  in \eqref{NZ matrices for (2,7)} and choosing $\mathbf{Q} = \textrm{diag} \{2,1,1,1 \}$, we obtain
\begin{align}
&K = \mathbf{Q}^T B^{-1} A \mathbf{Q} =  \left(
\begin{array}{cccc}
 6 & 0 & 2 & 0 \\
 0 & 1 & -1 & 1 \\
 2 & -1 & 2 & -1 \\
 0 & 1 & -1 & 2 \\
\end{array}
\right)\;, \quad N = \bigg{|}\frac{(\det \mathbf{Q})^2 }{\det B}\bigg{|} =2\;. \label{K and Q for (2,7)}
\end{align}
Using \eqref{T[A,B]} and \eqref{O_{E}}, we have
\begin{align}
\begin{split}
&\mathbb{T}[A,B;\mathbf{Q}, \{E_1, E_2\}] 
\\
&:= \bigg{(}U(1)^4_{K_{\rm eff}} \textrm{ $\CN=2$ gauge theory coupled to 4 chiral multiplets $\Phi$s in  $\mathbf{Q}$} 
\\
& \qquad \;\;  \textrm{with superpotential  $\CW = \CO_{E_1} +\CO_{E_2}= V_{\mathbf{m}= (0,-1,0,1)} + V_{\mathbf{m}= (1,1,-1,-1)}$} \bigg{)}
\\
&\textrm{where }K_{\rm eff} = \mathbf{Q}^T (B^{-1}A -\mathbb{I}/2) \mathbf{Q} = \left(
\begin{array}{cccc}
 4 & 0 & 2 & 0 \\
 0 & \frac{1}{2} & -1 & 1 \\
 2 & -1 & \frac{3}{2} & -1 \\
 0 & 1 & -1 & \frac{3}{2} \\
\end{array}
\right)\;.
\end{split}
\end{align}
The UV gauge theory has a flavor symmetry $F_{\rm UV}$, whose Cartan subalebra is given as
\begin{align}
\mathfrak{F}[\CT]=\bigg{\{} \boldsymbol{v}=(v_1,v_2,v_3,v_4)^T: \begin{pmatrix}
                0 & -1 & 0 & 1 \\
                1 & 1 & -1 & -1 
            \end{pmatrix}\cdot {\boldsymbol{v}}=\begin{pmatrix}
                0  \\
                0 
            \end{pmatrix}\bigg{\}}\;.
\end{align}
The vector space is spanned by $\boldsymbol{v}_m$ and $\boldsymbol{v}_1$ in \eqref{basis of F[T]}:
\begin{align}
\begin{split}
\boldsymbol{v}_m = \mathbf{Q}^T B^{-1}\cdot (1,0,0,0)^T = (0,-1,0,-1)^T\;,
\\
\boldsymbol{v}_1 = \mathbf{Q}^T B^{-1}\cdot (0,1,0,0)^T = (-1,0,-1,0)^T\;. \label{vm and v1 for (2,7)}
\end{split}
\end{align}
The theory has anomalous $\mathbb{Z}_{N=2}$ 1-form symmetry
\begin{align}
\Gamma^{[1]} = \frac{\mathbb{Z}^4}{K \cdot \mathbb{Z}^4+ \mathbf{Q} \cdot \mathbb{Z}^4} \simeq  \mathbb{Z}_2\;.
\end{align}
By removing the decoupled $\mathcal{A}^{N=2,p}$ from $\mathbb{T}[A,B;\mathbf{Q}]$, we obtain the $T[S^3\backslash \CK_{(2,7)}]$
\begin{align}
T[S^3\backslash \CK_{(2,7)}] \otimes \CA^{N=2,p} \simeq \mathbb{T}[A,B;\mathbf{Q},\{E_1, E_2\}]\;.
\end{align}

\subsection{$T[S^3\backslash \CK]$ for torus knots $\CK=\CK_{(P,Q)}$} \label{subsect : T-DGG for torus knot}
For torus knots, we find that the ideal triangulation of 
$S^3 \backslash \mathcal{K}_{(P,Q)}$, provided by the computer software SnapPy \cite{SnapPy},
always exhibits the following curious properties.\footnote{We have checked this for $(P,Q)=$ $\left.(2,2r-1)\right|_{2\leq r\leq 7}$, $\left.(3,3 r-2)\right|_{2\leq r\leq 5}$, $\left.(3,3 r-1)\right|_{2\leq r\leq 4}$, $\left.(4,4 r-3)\right|_{2\leq r\leq 4}$, $(4,7)$, $(4,11)$, $\left.(5,r)\right|_{6\leq r\leq 9}$, and $(7,9)$.}
\begin{align}
\begin{split}
&\bullet \textrm{There exists a special easy internal edge, say $E_{I=1}$, which is the sum of two edge parameters  }
\\
& \quad \;\textrm{from two ideal tetrahedra, } \textrm{i.e. $E_{I=1}= Z_i^{*}+Z_{j}^{**}$ \textrm{ with }$Z^*,Z^{**}\in \{Z, Z',Z''\}$}
\\
&\bullet \textrm{$r$ (number of tetrahedra)-$\sharp_E$(number of easy internal edges)}  
\\
& \quad  =
\begin{cases}
1, & |P - Q| = 1,\\[0.3em]
2, & \left(Q \equiv \pm 1\ (\mathrm{mod}\ P)\right) 
    \ \oplus\ 
    \left(P \equiv \pm 1\ (\mathrm{mod}\ Q)\right)\\[0.3em]
3, & \left(Q \not\equiv \pm 1 \ (\mathrm{mod}\ P)\right) 
    \ \land\ 
    \left(P \not\equiv \pm 1\ (\mathrm{mod}\ Q)\right),
\end{cases}
\end{split} \label{triangulation of torus knots}
\end{align}
For the $(P,Q)=(2,7)$ case in \eqref{Example : (2,7)}, $C_2=Z_2''+Z_4'$ is the special easy edge, $E_{I=1}= C_2$, and $r -\sharp_E =4-2=2$. Based on the general counting given in \eqref{rank(F_UV) of T_{DGG}}, we predict the rank of the UV flavor symmetry $F_{\rm UV}$ for the theory $T[S^3\backslash \mathcal{K}_{(P,Q)}]$ to be:
\begin{align}
\begin{split}
&\textrm{rank of $F_{\rm UV}$ of  $T[S^3\backslash \CK_{(P,Q)}]$}
\\
& = \begin{cases}
1, & |P - Q| = 1,\\[0.3em]
2, & \left(Q \equiv \pm 1\ (\mathrm{mod}\ P)\right) 
    \ \oplus\ 
    \left(P \equiv \pm 1\ (\mathrm{mod}\ Q)\right)\\[0.3em]
3, & \left(Q \not\equiv \pm 1 \ (\mathrm{mod}\ P)\right) 
    \; \land \; 
    \left(P \not\equiv \pm 1\ (\mathrm{mod}\ Q)\right),
\end{cases}
\end{split} \label{sharp_E and sharp_H for T[torus]}
\end{align}
More explicitly,
\begin{align}
\begin{split}
&(\textrm{Cartan subalgebra of $F_{\rm UV}$}) \simeq  \mathfrak{F}[\CT]
\\
 &= \textrm{Span} \begin{cases}
\{ \boldsymbol{v}_m\}, & |P - Q| = 1\\[0.3em]
\{ \boldsymbol{v}_m,\boldsymbol{v}_1\}, & \left(Q \equiv \pm 1\ (\mathrm{mod}\ P)\right) 
    \ \oplus\ 
    \left(P \equiv \pm 1\ (\mathrm{mod}\ Q)\right)\\[0.3em]
\{ \boldsymbol{v}_m,\boldsymbol{v}_1, \boldsymbol{v}_2\}, & \left(Q \not\equiv \pm 1 \ (\mathrm{mod}\ P)\right) 
    \; \land \; 
    \left(P \not\equiv \pm 1\ (\mathrm{mod}\ Q)\right)
\end{cases}
\end{split} \label{sharp_E and sharp_H for T[torus]-2}
\end{align}
The basis vectors are defined in \eqref{basis of F[T]} with \eqref{NZ matrices}.
However, it is not guaranteed that this counting works in the IR. Actually, one of the Cartan generators of the UV flavor symmetry decouples in the IR due to the special internal edge. 
The easy internal edge involving only two edge parameters plays a special role in the field theory $T[S^3\backslash \CK ]$.
If we choose a polarization such that $X_i = Z_i^{*}$ and $X_j = Z_j^{**}$, the superpotential operator $\CO_{E_1}$ becomes
\begin{align}
\CW = \Phi_i\Phi_j\;.
\end{align}
It is nothing but a mass term, and the two corresponding chiral multiplets can be integrated out.
This superpotential lifts both of the $U(1)$ flavor symmetries that rotate $\Phi_i$ and $\Phi_j$, 
whereas a generic single superpotential term lifts only a single $U(1)$.  
One may say that the $U(1)$ symmetry preserved by the superpotential 
(namely the $U(1)$ that rotates the two chiral multiplets with opposite charges) 
effectively decouples in the IR.  
Remarkably, this decoupled symmetry always turns out to be the $U(1)_m$ symmetry defined in \eqref{v_m}.
\begin{align}
\begin{split}
&\text{$U(1)_m$ in $T[S^3\backslash \CK]$ always decouples in the IR for }
\CK = \CK_{(P,Q)}, 
\\
&\Rightarrow\; \boldsymbol{v}_m \in \mathfrak{F}_{\rm decouple}[\CT]\;. \label{decoupling of U(1)_m}
\end{split}
\end{align}
From the BPS partition function computation, we will see that 
\begin{align}
\mathfrak{F}_{\rm decouple}[\CT] = \operatorname{Span}\{\boldsymbol{v}_m\}\;, \label{F-decouple for torus}
\end{align}
i.e. there is no additional decoupled symmetry other than the $U(1)_m$.
It implies that 
\begin{align}
(\mathfrak{F}_{\rm IR} [\CT] \textrm{ in \eqref{F[T]_IR}}) = \textrm{Span} \begin{cases}
\{ \emptyset\}, & |P - Q| = 1\\[0.3em]
\{ [\boldsymbol{v}_1]\}, & \left(Q \equiv \pm 1\ (\mathrm{mod}\ P)\right) 
    \ \oplus\ 
    \left(P \equiv \pm 1\ (\mathrm{mod}\ Q)\right)\\[0.3em]
\{ [\boldsymbol{v}_1], [\boldsymbol{v}_2]\}, & \left(Q \not\equiv \pm 1 \ (\mathrm{mod}\ P)\right) 
    \; \land \; 
    \left(P \not\equiv \pm 1\ (\mathrm{mod}\ Q)\right) \label{F[T]_{IR} for torus knots}
\end{cases}
\end{align}
\paragraph{Reduced UV description $T^{(\rm r)}[S^3\backslash \mathcal{K}_{(P,Q)}]$}
By integrating out the chiral multiplets $\Phi_i$ and $\Phi_j$ in the theory $T[S^3\backslash \mathcal{K}_{(P,Q)}]$,
one obtains a reduced UV description
$T^{(\rm r)}[S^3\backslash \mathcal{K}_{(P,Q)}]$, in which the decoupled
$U(1)_m$ sector has already been taken into account.
This procedure simplifies the UV gauge theory by removing redundant degrees
of freedom associated with the decoupled sector.

At the level of the gluing equations, integrating out the chiral multiplets
$\Phi_i$ and $\Phi_j$ corresponds to the following specialization:
\begin{align}
\begin{split}
& Z_i^{*},\, Z_j^{**} \;\longrightarrow\; \pi i \,,
\\
& \text{(all other edge parameters of the $i$- and $j$-th tetrahedra)}
\;\longrightarrow\; 0 \;. \label{reduction procedure}
\end{split}
\end{align}
In the 3D--3D correspondence, the edge parameter $Z_i^{*}$ is identified with
the complexified mass parameter~\cite{Dimofte:2011ju},
\[
Z_i^{*} = i\pi\, R(\Phi_i) + m_{\Phi_i},
\]
where $R(\Phi_i)$ denotes the $R$-charge and $m_{\Phi_i}$ the real mass of the
chiral multiplet $\Phi_i$~\cite{Dimofte:2011ju}.
Therefore, this specialization corresponds to setting
$R(\Phi_i)=R(\Phi_j)=1$ and turning off the real mass parameters.
This has the same effect as introducing superpotential mass terms for the two
chiral multiplets and integrating them out in the infrared. Geometrically, the imaginary part of $Z_i^{*}$ corresponds to the dihedral angle
between the two faces meeting at the internal edge.
Assigning it the value $\pi$ implies that the $i$-th ideal tetrahedron becomes
completely flat and therefore effectively disappears from the triangulation.

After this reduction, there remain $(r-2)$ linearly independent internal edges $\{C^{(\rm r)}_{I}\}_{I=1}^{r-2}$, which are linear combinations of the edge parameters of the remaining $(r-2)$ tetrahedra.
Interestingly, these reduced internal edges still satisfy
\begin{align}
\{C^{(\rm r)}_I,C^{(\rm r)}_J\}_{\rm P.B.}=0\;, \quad \forall I, J=1,\ldots, r-2\;,
\end{align}
i.e. the symplectic structure among internal edges is preserved after the integrating-out procedure.
Upon choosing a polarization $(\mathbf{X}_{\rm r}, \mathbf{P}_{\rm r})$ for the remaining $(r-2)$ tetrahedra, the reduced NZ matrices 
$(A_{\rm r}, B_{\rm r}, \boldsymbol{\nu}_{x;\mathrm{red}})$ 
are defined by
\begin{align}
\begin{pmatrix}
H^{(\rm r)}_1 - 2\pi i \\
\vdots \\
H^{(\rm r)}_{\sharp^{(\rm r)}_H} - 2\pi i \\
E^{(\rm r)}_1 -2\pi i \\
\vdots \\
E^{(\rm r)}_{\sharp^{(\rm r)}_E} - 2\pi i
\end{pmatrix}
=
\begin{pmatrix}
A_{\rm r} & B_{\rm r}
\end{pmatrix}
\begin{pmatrix}
\mathbf{X}_{\rm r} \\
\mathbf{P}_{\rm r}
\end{pmatrix}
+ i\pi\, \boldsymbol{\nu}_{x;{\rm r}} \;. \label{reduced NZ matrices}
\end{align}
Here $\{H_I^{(\rm r)}\}$ and $\{E_I^{(\rm r)}\}$ denote, respectively, the sets of hard and easy internal edges after the reduction. 
Then the theory $T^{(\rm r)}[S^3\backslash \mathcal{K}_{(P,Q)}]$ can be constructed from $(A_{\rm r}, B_{\rm r})$ in exactly the same way that one constructs 
$T[S^3\backslash \mathcal{K}]$ from the original NZ matrices $(A,B)$:
\begin{align}
\begin{split}
& \BT^{(\rm r)}[A_{\rm r}, B_{\rm r};\mathbf{Q}_{\rm r}, \{ \CO_{E_I^{(\rm r)}}\}] 
\\
&:=\;
\bigg{(}
U(1)^{r-2}_{K^{(\rm r)}_{\rm eff}}
\ \text{coupled to $r$ chirals }\Phi_i
\ \text{with charge matrix}\ \mathbf{Q}_{\rm r}
\\
& \quad \quad \;\; \textrm{with the UV effective mixed CS level $K^{(\rm r)}_{\rm eff}:= K_{\rm r} - 1/2 \mathbf{Q}_{\rm r}^T \mathbf{Q}_{\rm r}:=\mathbf{Q}_{\rm r}^T (B^{-1}_{\rm r}A_{\rm r} -  \mathbb{I}/2) \mathbf{Q}_{\rm r}$} \bigg{)}
\\
&  \quad \quad \;\; \textrm{+ superpotential }\CW = \sum  \CO_{E_I^{\rm (r)}}\;.
\\[0.2em]
&T^{(\rm r)}[S^3\backslash \mathcal{K}_{(P,Q)}] \otimes \CA^{N_{\rm r}, p_{\rm r}} :=   \mathbb{T}^{(\rm r)}[A_{\rm r}, B_{\rm r};\mathbf{Q}_{\rm r}, \{ \CO_{E_I^{(\rm r)}}\}]  \;. \label{T-red-DGG}
\end{split}
\end{align}
The chiral primary operator $\mathcal{O}_{E_I^{\rm (r)}}$ can be constructed in
the same way as $\mathcal{O}_{E_I}$ in \eqref{O_{E}}. The two theories, $T[S^{3}\backslash \mathcal{K}_{(P,Q)}]$ and $T^{(\mathrm{r})}[S^{3}\backslash \mathcal{K}_{(P,Q)}]$, are expected to be IR-equivalent, possibly up to a decoupled Abelian TQFT:\footnote{As we will see in concrete examples, this decoupled TQFT is trivial in many cases. For example, for $(P,Q)=(2,*)$, the theory $T[S^3\backslash \CK_{(P,Q)}]$ does not contain a decoupled Abelian TQFT factor, and hence the Abelian sector is trivial.}

\begin{align}
T[S^3\backslash \CK_{(P,Q)}]  \simeq  T^{\rm (r)}[S^3\backslash \CK_{(P,Q)}] \otimes (\textrm{Abelian TQFT})\;. \label{Relation between T and T^r}
\end{align}
The decoupled Abelian TQFT may originate from dynamical (possibly discrete)
Abelian gauge fields in \(T[S^{3}\backslash \mathcal{K}_{(P,Q)}]\) that carry
non-trivial Chern--Simons terms and under which only massive chiral multiplets
are charged.

Since $(r, \sharp_E)$ in \eqref{triangulation of torus knots} reduces to $(r-2, \sharp_E-1)$,  we expect that 
\begin{align}
\begin{split}
&(\sharp^{(\rm r)}_{H}, \sharp^{(\rm r)}_{E}) =  \begin{cases}
( 0, r-2), & |P - Q| = 1\\[0.3em]
(1,r-3), & \left(Q \equiv \pm 1\ (\mathrm{mod}\ P)\right) 
    \ \oplus\ 
    \left(P \equiv \pm 1\ (\mathrm{mod}\ Q)\right)\\[0.3em]
(2,r-4), & \left(Q \not\equiv \pm 1 \ (\mathrm{mod}\ P)\right) 
    \; \land \; 
    \left(P \not\equiv \pm 1\ (\mathrm{mod}\ Q)\right)
\end{cases}
\\
&\textrm{and }(\mathrm{rank \;of }\, F_{\rm UV}) = \sharp^{(\rm r)}_{H}\;.
\end{split}
\end{align}
More explicitly, the Cartan generators are given by 
\begin{align}
\mathfrak{F}^{(\rm r)} [\CT]  = \textrm{Span} \begin{cases}
\{ \emptyset\}, & |P - Q| = 1\\[0.3em]
\{ \boldsymbol{v}^{(\rm r)}_1\}, & \left(Q \equiv \pm 1\ (\mathrm{mod}\ P)\right) 
    \ \oplus\ 
    \left(P \equiv \pm 1\ (\mathrm{mod}\ Q)\right)\\[0.3em]
\{ \boldsymbol{v}^{(\rm r)}_1, \boldsymbol{v}^{(\rm r)}_2\}, & \left(Q \not\equiv \pm 1 \ (\mathrm{mod}\ P)\right) 
    \; \land \; 
    \left(P \not\equiv \pm 1\ (\mathrm{mod}\ Q)\right) \label{F[T]^{(red)} for torus knots}
\end{cases}
\end{align}
Here we define
\begin{align}
\boldsymbol{v}^{(\rm r)}_1 
:=  \mathbf{Q}_{\rm r}^{\,T} \cdot B_{\rm r}^{-1} \cdot (1,\mathbf{0}_{r-3})^{T}, \;\; \boldsymbol{v}^{(\rm r)}_2
:=  \mathbf{Q}_{\rm r}^{\,T} \cdot B_{\rm r}^{-1} \cdot (0,1,\mathbf{0}_{r-4})^{T}. \label{v_i-red}
\end{align}
\paragraph{Example : $(P,Q)=(2,5)$} The torus knot complement can be ideally triangulated with $r=3$ tetrahedra with following gluing data, see \eqref{(2,5) ge}
\begin{align}
\begin{split}    &C_1 = Z_1+Z_3, \quad C_2 = 2Z'_1+Z''_1+Z_2+2Z_2''+Z'_3+2 Z_3'',
\\
&C_3 = Z_1+Z_1''+Z_2+2Z_2'+Z_3+Z_3', \quad M = -Z_1''-Z_2'+Z_3''\;.
\end{split}
\end{align}
Note that the $C_1$ is the special easy internal edge in \eqref{triangulation of torus knots}. By taking the reduction procedure \eqref{reduction procedure} on the gluing equations, we have
\begin{align}
H_1^\mathrm{(r)}:=C_2^\mathrm{(r)}=Z_2+2 Z_2''\;.
\end{align}
There is no easy internal edge after the reduction. 
When we choose the following polarizations
\begin{align}
    \mathbf{X}_{\rm r}=(Z_2')\textrm{ and }\mathbf{P}_{\rm r}=(Z_2)
\end{align}
the reduced NZ matrices in \eqref{reduced NZ matrices} are
\begin{align}     
    A_\mathrm{r}=(-2),\quad B_\mathrm{r}=(-1),\quad \boldsymbol{\n}_{x;\mathrm{r}}=(0)
\end{align}
In the case, we choose $\mathbf{Q}_\mathrm{r}=\BI$. Then,
\begin{align}
\begin{split}
    K_\mathrm{r}= B_{\rm r}^{-1} A_{\rm r}=(2), \quad N_\mathrm{r}=\left|\frac{(\det \mathbf{Q})^2}{\det B} \right|=1, \quad \{E_I^{\rm {(r)}}\} = \emptyset\;.
    \label{reduced K and Q for (2,5)}
\end{split}
\end{align}
Using the above data, one can construct 
$T^{(\rm r)}[S^3\backslash \CK_{(2,5)}]= \mathbb{T}^{\rm (r)}[A_{\rm r}, B_{\rm r}; \mathbf{Q}_{\rm r}, \{E^{\rm (r)}_I\}]$
as defined in \eqref{T-red-DGG}.
This theory is precisely the UV gauge-theory description of the minimal
$\mathcal{N}=4$ rank-$0$ SCFT studied in \cite{Gang:2018huc}.

\paragraph{Example : $(P,Q)=(2,7)$} By taking the reduction procedure \eqref{reduction procedure} on the gluing equations in \eqref{Example : (2,7)}, we have
\begin{align}
H^{(\rm r)}_1 := C^{(\rm r)}_3 = 2Z_1+Z_1'+2Z_3, \quad E^{(\rm r)}_1=C^{(\rm r)}_1= 2Z_1''+Z_3'\;. \label{Example : (2,7)-2}
\end{align}
i) When we choose $\mathbf{X}_{\rm r} = (Z_1',Z_3')^T$ and $\mathbf{P}_{\rm r} = (Z_1,Z_3)^T$, the reduced NZ matrices are
\begin{align}
A_{\rm r} = \begin{pmatrix} 1 & 0 \\ -2 & 1 \end{pmatrix}, \quad  B_{\rm r} = \begin{pmatrix} 2 & 2 \\ -2 & 0 \end{pmatrix}, \quad \boldsymbol{\nu}_{x;{\rm r}} = (-2,0)^T\;.
\end{align}
In the case, we choose $\mathbf{Q}_{\rm r} = \textrm{diag} \{1,2 \}$. Then,
\begin{align}
K_{\rm r} = (\mathbf{Q}^T B^{-1} A \mathbf{Q})_{\rm r} = \left(
\begin{array}{cc}
 1 & -1 \\
 -1 & 2 \\
\end{array}
\right), \quad N_r= \left|\frac{(\det \mathbf{Q})^2}{\det B} \right|= 1, \quad  \CO_{E_1^{\rm(r)}} = \phi_2 V_{\mathbf{m}=(2,0)}\;.
\end{align}
Using the above information, one can construct $\mathbb{T}^{\rm (r)}[A_r, B_r;\mathbf{Q}_r, \{E_1^{\rm (r)}\}]$ in \eqref{T-red-DGG}, which is identical to $T^{\rm (r)}[S^3\backslash \CK_{(2,7)}]$ in the case since $N_r=1$.  The theory has a flavor symmetry whose Cartan subalgebra is 
\begin{align}
\mathfrak{F}^{\rm (r)}[\mathcal{T}] = \{\boldsymbol{v}=(v_1, v_2)^T \;:\; v_1=0 \}\;.
\end{align}
The one-dimensional space is spanned by a vector $\boldsymbol{v}^{\rm (r)}_1  = \mathbf{Q}^T B_r^{-1}\cdot (1,0)^T = (0,1)^T$. 
\\
\\
ii) When we choose $\mathbf{X}_{\rm r} = (Z_1,Z_3)^T$ and $\mathbf{P}_{\rm r} = (Z_1'',Z_3'')^T$, the reduced NZ matrices are
\begin{align}
A_{\rm r} = \begin{pmatrix} 1 & 2 \\ 0 & -1 \end{pmatrix}, \quad  B_{\rm r} = \begin{pmatrix} -1 & 0 \\ 2 & -1 \end{pmatrix}, \quad \boldsymbol{\nu}_{x;{\rm r}} = (-1,-1)^T \;. \label{reduced NZ for (2,7)}
\end{align}
In the case, we choose $\mathbf{Q}_{\rm r} = \mathbb{I}$. Then,
\begin{align}
K_{\rm r} = B^{-1}_{\rm r} A_{\rm r}= \left(
\begin{array}{cc}
 -1 & -2 \\
 -2 & -3 \\
\end{array}
\right), \quad N_{\rm r}= \left|\frac{(\det \mathbf{Q}_{\rm r})^2}{\det B_{\rm r}} \right|= 1\;, \quad \CO_{E_1^{\rm(r)}} = V_{\mathbf{m}=(-2,1)}\;. \label{reduced K and Q for (2,7)}
\end{align}
Using the above information, one can construct $\mathbb{T}^{\rm (r)}[A_r, B_r;\mathbf{Q}_r, \{E_1^{\rm (r)}\}]$ in \eqref{T-red-DGG}, which is equivalent to $T^{\rm (r)}[S^3\backslash \CK_{(2,7)}]$ since $N_r=1$.  
This theory coincides with the UV gauge-theory description of the $\CT_{r=2}$ theory introduced in~\cite{Gang:2023rei}, which flows to a rank-$0$ SCFT that serves as the bulk theory for $M(2,7)$, up to a parity transformation.
 The theory has a flavor symmetry whose Cartan subalgebra is 
\begin{align}
\mathfrak{F}^{\rm (r)}[\mathcal{T}] = \{\boldsymbol{v}=(v_1, v_2)^T \;:\; -2v_1+v_2=0 \}\;.
\end{align}
The one-dimensional space is spanned by a vector $\boldsymbol{v}^{\rm (r)}_1  = \mathbf{Q}^T B_r^{-1}\cdot (1,0)^T = (-1,-2)^T$.

%% file: section_BPSpftnsIRphases.tex
\section{BPS partition functions and IR phases of $T[S^3\backslash \mathcal{K}_{(P,Q)}]$} \label{sec : IR phases and BPS ptns}
\subsection{Proposed infrared behaviors of $T[S^3\backslash \mathcal{K}_{(P,Q)}]$}
Motivated by the intriguing observation of H.~Hikami and A.~N.~Kirillov \cite{Hikami:2003tb}, together with recent developments in non-unitary bulk--boundary correspondences \cite{Gang:2023rei,Gang:2023ggt,Baek:2024tuo,Gang:2024tlp}, we propose the following:
\begin{align}
\begin{split}
&T[S^{3} \backslash \mathcal{K}_{(P,Q)}]
\;\simeq\;
T^{(\mathrm{r})}[S^{3} \backslash \mathcal{K}_{(P,Q)}]
\otimes (\text{Abelian TQFT})
\\[0.3em]
&\xrightarrow{\;\mathrm{IR}\;}
\begin{cases}
\text{unitary TQFT}, 
& |P - Q| = 1,\\[0.45em]
\mathcal{N}=4 \text{ rank-0 SCFT}, 
& \left(Q \equiv \pm 1 \pmod{P}\right)
  \oplus
  \left(P \equiv \pm 1 \pmod{Q}\right),\\[0.45em]
\dfrac{\text{a product of two rank-0 SCFTs}}{\text{(1-form symmetry)}}, 
& \left(Q \not\equiv \pm 1 \pmod{P}\right)
  \land
  \left(P \not\equiv \pm 1 \pmod{Q}\right).
\end{cases}
\\[0.7em]
&\text{With an appropriate boundary condition, the IR theory supports the rational VOA of} 
\\
&
\CM(P,Q)\text{ on the boundary. For }  |P-Q|>1,\text{ the realization of the VOA further } 
\\
&\text{requires a suitable topological twisting.}
\end{split}
\label{IR phase of S3-K_{(P,Q)}}
\end{align}
There exist earlier proposals for bulk field theories corresponding to the Virasoro minimal models—namely, the $\CM(2,*)$ series in \cite{Gang:2023rei} and the general $\CM(P,Q)$ case in \cite{Gang:2024tlp}. We expect that the theory $T[S^{3}\backslash \CK_{(P,Q)}]$ is IR dual to these previous constructions. In particular, the IR behaviors proposed above follow from the analysis in \cite{Gang:2024tlp}. 

An advantage of our proposal is that we have access to many duality frames, and by choosing appropriate ones and computing half-indices, we can reproduce various rational VOA characters—including the Virasoro minimal model characters. These different duality frames have a geometric origin in the ideal–triangulated 3-manifolds, arising from choices of polarization. Moreover, these polarization choices are fully understood in our framework, enabling us to identify the appropriate duality frame for half-index computations and to compute the half-index directly from the geometric data of the ideal triangulation. We will examine the half-index computation in detail in Section~\ref{sec : RVOAs from Half-indices}. 

In contrast, within the field-theoretic descriptions of \cite{Gang:2024tlp}, it was not feasible to extract rational VOA characters from half-index computations due to the complexity of their UV gauge theories. A more direct relation between our $T[S^{3}\backslash \CK_{(P,Q)}]$ and the field theories proposed in \cite{Gang:2024tlp} will be discussed in Section~\ref{sec: Discussion}, using Dehn filling operations in the 3D--3D correspondence.

\paragraph{When $|P-Q|=1$}  The theory has a mass gap and the IR theory is described by a unitary TQFT, where the $U(1)_m$ symmetry and supersymmetry decouple.

\paragraph{When $\left(Q \equiv \pm 1 \pmod{P}\right) 
  \oplus 
  \left(P \equiv \pm 1 \pmod{Q}\right)$} Under the RG, the $U(1)_R \times U(1)_A$, where $\mathfrak{u}(1)_A$ can be identified with $\mathfrak{F}_{\rm IR}[\CT]$ in \eqref{F[T]_{IR} for torus knots} or $\mathfrak{F}^{(\rm r)}[\CT]$ in \eqref{F[T]^{(red)} for torus knots}, is expected  to be enhanced to  $SO(4)_R$ symmetry of IR $\CN=4$ rank-0 SCFT. The generator of $U(1)_R\times U(1)_A$ is embedded into $SO(4)_R$ as follows:
\begin{align}
\begin{split}
&R \simeq  \bigl(J_3^C + J_3^H \text{ of the IR SCFT}\bigr)\;,
\\
&T_A \simeq \bigl(J_3^C - J_3^H \text{ of the IR SCFT}\bigr)\,. \label{Embedding of T_a in SO(4)}
\end{split}
\end{align}
Here $J_3^C$ and $J_3^H$ are the Cartan generators of  
\(
SU(2)^C \times SU(2)^H \simeq SO(4)_R,
\)
normalized such that \( J_3\in \mathbb{Z}/2 \).   This is the standard supersymmetry-enhancement pattern that produces IR $\CN=4$ rank-0 SCFTs from UV $\mathcal{N}=2$ gauge theories \cite{Gang:2018huc,Gang:2021hrd}.

\paragraph{When $\left(Q \not\equiv \pm 1 \pmod{P}\right) 
  \land 
  \left(P \not\equiv \pm 1 \pmod{Q}\right)$} Under the RG, the $U(1)_R \times U(1)_{A_1}\times U(1)_{A_2}$, where $\mathfrak{u}(1)_{A_1} \oplus \mathfrak{u}(1)_{A_2}$ can be identified with  $\mathfrak{F}_{\rm IR}[\CT]$ in \eqref{F[T]_{IR} for torus knots} or $\mathfrak{F}^{(\rm r)}[\CT]$ in \eqref{F[T]^{(red)} for torus knots}, is expected to be enhanced to $SO(4)^{(1)}_{R}\times SO(4)^{(2)}_{R}$ symmetry of the product of two $\CN=4$ rank-0 SCFTs. The Cartan generators of $U(1)_{A_1} \times U(1)_{A_2}$ are embedded into the $SO(4)^{(1)}_R \times SO(4)^{(2)}_R$ as follows:
\begin{align}
\begin{split}
&R \simeq  \bigl(J^{(1)C}_3 + J^{(1)H}_3 + J^{(2)C}_3 + J^{(2)H}_3\text{ of the IR SCFT}\bigr)\;,
\\
&T_{A}^{(1)}\simeq \left(J^{(1)C}_3 - J^{(1)H}_3 \textrm{ of the IR SCFT}\right)\;, 
\\
&T_{A}^{(2)}\simeq \left(J^{(2)C}_3 - J^{(2)H}_3 \textrm{ of the IR SCFT}\right)\;.
\label{Embedding of T_a1 and T_a2 in SO(4)}
\end{split}
\end{align}
\\
\paragraph{Example: $(P,Q)=(5,7)$}
The knot complement can be ideally triangulated using five tetrahedra, and the
corresponding gluing equations are, see \eqref{(5,7) ge}
\begin{align}
\begin{split}
C_1 &= 2Z_1+Z_1''+Z_3+Z_3'+Z_4'+Z_4'', 
\\
C_2 &= Z_3''+Z_4,
\qquad
C_3 = 2Z_2+Z_3'+Z_4'+Z_5+2Z_5'',
\\
C_4 &= Z_2'+Z_3+Z_4''+2Z_5',
\qquad
C_5 = 2Z_1'+Z_1''+Z_2'+2Z_2''+Z_3''+Z_4+Z_5,
\\
M &=
-Z_1+Z_1'-Z_2+Z_2''+2Z_3'' \; .
\end{split}
\end{align}
Note that $C_2$ corresponds to a special easy internal edge in \eqref{triangulation of torus knots}.
After performing the reduction procedure \eqref{reduction procedure}, we obtain
\begin{align}
\begin{split}
H_1^{(r)} &= C_1^{(r)} = 2Z_1+Z_1'',
\\
H_2^{(r)} &= C_3^{(r)} = 2Z_2+Z_5+2Z_5'',
\\
E_1^{(r)} &= C_4^{(r)} = Z_2'+2Z_5' \; .
\end{split}
\end{align}
Here $C_1^{\rm (r)}$ denotes the internal edge $C_1$ after the reduction. Choosing $\mathbf{X}_{\rm r}=(Z_1,Z_2,Z_5'')$ and
$\mathbf{P}_{\rm r}=(Z_1'',Z_2'',Z_5')$, the reduced Neumann--Zagier data are
\begin{align}
A_{\rm r} =
\begin{pmatrix}
 2 & 0 & 0 \\
 0 & 2 & 1 \\
 0 & -1 & 0
\end{pmatrix},
\qquad
B_{\rm r} =
\begin{pmatrix}
 1 & 0 & 0 \\
 0 & 0 & -1 \\
 0 & -1 & 2
\end{pmatrix},
\qquad
\boldsymbol{\nu}_{x,{\rm r}} = (-2,-1,-1)^{T}. \label{reduced NZ for (5,7)}
\end{align}
Since $|\det B_{\rm r}|=1$, one may choose $\mathbf{Q}_{\rm r}=\mathbb{I}$, and
thus $N_{\rm r}=1$.
The reduced Chern--Simons matrix $K_{\rm r}$ and the corresponding
$\tfrac{1}{2}$-BPS operator $\mathcal{O}_{E_1^{(r)}}$ are
\begin{align}
K_{\rm r} := B_{\rm r}^{-1} A_{\rm r}
=
\begin{pmatrix}
 2 & 0 & 0 \\
 0 & -3 & -2 \\
 0 & -2 & -1
\end{pmatrix},
\qquad
\mathcal{O}_{E_1^{(r)}} = V_{\mathbf{m}=(0,1,-2)} \; . \label{reduced K and Q for (5,7)}
\end{align}
From the factorized structure of the Chern--Simons matrix, we find
\begin{align}
T^{(r)}[S^{3}\backslash \mathcal{K}_{(5,7)}]
\simeq
T^{(r)}[S^{3}\backslash \mathcal{K}_{(2,5)}]
\;\otimes\;
T^{(r)}[S^{3}\backslash \mathcal{K}_{(2,7)}] \; .
\end{align}
Here the theory $T^{(r)}[S^{3}\backslash \mathcal{K}_{(2,5)}]$ can be constructed
from the data in \eqref{reduced K and Q for (2,5)}, while
$T^{(r)}[S^{3}\backslash \mathcal{K}_{(2,7)}]$ follows from
\eqref{reduced K and Q for (2,7)}.
This result is compatible with the general claim in
\eqref{IR phase of S3-K_{(P,Q)}}.
The factorization can also be made manifest in a UV description of
$T[S^{3}\backslash \mathcal{K}_{(5,7)}]$ by an appropriate choice of
polarizations.
We leave the detailed analysis to the interested reader.

\subsection{Superconformal Index}
The superconformal index (SCI) \cite{Kim:2009wb,Imamura:2011su,Dimofte:2011py} is a basic tool for analyzing the infrared phases
of three-dimensional $\mathcal{N}=2$ gauge theories.
When the index is identically equal to $1$, this strongly suggests that the theory
develops a mass gap and flows to a unitary topological quantum field theory (TQFT)
in the infrared.
If instead the theory flows to a three-dimensional $\mathcal{N}=4$ rank-$0$
superconformal field theory (SCFT) with enhanced supersymmetry, the index becomes
trivial in certain limits, known as the A- and B-twisting limits.
This is because, in these limits, the SCI computes the Hilbert series of the
Coulomb or Higgs branch, which is trivial for rank-$0$ SCFTs.
Moreover, supersymmetry enhancement may leave characteristic imprints in the SCI,
contributions of extra supersymmetry
current multiplets.

The index for $\mathbb{T}[A,B;\mathbf{Q}]$ in \eqref{tT[A,B]}, which coincides with
the index of $T[A,B]$ in \eqref{T[A,B]} since the decoupled TQFT does not affect
the local operator spectrum, is 
 \begin{align}
 \begin{split}
 &\CI_{\rm sci}^{T[A,B]} [\boldsymbol{x},\boldsymbol{\mu}] := \textrm{Tr}_{\CH_{\rm rad}(S^2)} (-1)^{R_{\boldsymbol{\mu}}} q^{\frac{R_{\boldsymbol{\mu}}}2 +j_3} \prod_{i=1}^r x_i^{T_i}
 \\
 &= \sum_{\mathbf{m} \in \mathbb{Z}^r}  \prod_{i=1}^r  \oint_{|u_i|=1} \frac{du_i}{2\pi i u_i}  \prod_{i,j=1}^r u_i^{K_{ij}m_j}  \prod_{i=1}^r (x_i (-q^{1/2})^{\mu_i})^{m_i}\prod_{i=1}^r \CI_\Delta \big{(}\sum_{j=1}^r Q_{ij}m_j, \prod_{j=1}^r u_j^{Q_{ij}}\big{)}\;, \label{SCI for T[A,B]}
 \end{split}
 \end{align}
 where $K = \mathbf{Q}^T B^{-1} A\mathbf{Q}$ as defined in \eqref{NQB relation}. $\mathcal{H}_{\rm rad}(S^{2})$ denotes the radially quantized Hilbert space,
whose elements can be identified with local operators of the IR SCFT. Here  $\boldsymbol x = (x_1, \ldots, x_r)^T$ are the fugacity variables for the topological symmetry $U(1)^r$ ($U(1)^r_{\rm top}$) of the theory.  $ \boldsymbol \mu  = (\mu_1, \ldots, \mu_r)^T$ parameterize the mixing of the R-symmetry with the topological symmetry $U(1)^r_{\rm top}$
\begin{align}
R_{ \boldsymbol \mu } =  R_{ * } +  \boldsymbol \mu  \cdot \mathbf{T} \;,
\end{align}
where $ \mathbf{T}  = (T_1, \ldots, T_r)^T$ denotes the charges of the symmetry $U(1)^r_{\rm top}$ as in \eqref{topological charges}. In the superconformal index formula, the reference R-charge $R_*$ is chosen as%
\begin{align}
R_{*} (V_{\mathbf{m}}) =\frac{1}2 \sum_{i=1}^r \left((\mathbf{Q}\cdot \mathbf{m})_i+|(\mathbf{Q}\cdot \mathbf{m})_i|\right) \;. \label{reference R-charge}
\end{align}
The tetrahedron index  $\CI_\Delta(m,u)$, defined in \eqref{tetrahedron index}, computes the generalized superconformal index
of a free chiral multiplet with R-charge assignment $R_*(\Phi)=0$, in the presence of
background Chern--Simons couplings $k_{FF}=-\tfrac12$ and $k_{RF}=\tfrac12$.
Here $U(1)_F$ is the flavor symmetry under which $\Phi$ has charge $+1$, and
$(m,u)$ denote the background monopole flux and fugacity for $U(1)_F$.
Accordingly,
\[
\prod_{i=1}^r
\CI_\Delta\!\left(\sum_j Q_{ij} m_j,\; \prod_j u_j^{Q_{ij}}\right)
\]
computes the generalized superconformal index of $r$ free chiral multiplets
with $U(1)^r$ flavor symmetry and charge matrix $\mathbf{Q}$.
The induced background mixed Chern--Simons levels are
\[
K_{F_i F_j}=-\tfrac12 (\mathbf{Q}^T\mathbf{Q})_{ij},
\;\;
K_{R F_i}=\tfrac12 \sum_j Q_{ji}. \label{Background CS terms}
\]
The R-charge of a monopole operator $V_{\mathbf{m}}$ is then
\[
R_*(V_{\mathbf{m}})
=
\sum_i K_{R F_i} m_i
+
\sum_i \frac{1-R_* (\Phi_i)}{2}\,
\big|(\mathbf{Q}\cdot \mathbf{m})_i\big| \, ,
\]
which give the expression in \eqref{reference R-charge}. 
Note that the reference $R$-charge is quantized as
\begin{align}
R_* \in \mathbb{Z}\;.
\label{quantization of R*}
\end{align}
The superconformal index of $T[S^3\backslash \mathcal{K}_{(P,Q)}]$ is obtained from 
$\CI^{T[A,B]}_{\rm sci}[\boldsymbol{x},\boldsymbol{\mu}]$ by restricting the parameters 
$\{\boldsymbol{x},\boldsymbol{\mu}\}$ to satisfy
\begin{align}
\begin{split}
&R_{\boldsymbol{\mu}}(\mathcal{O}_{E_I})
= R_*(V_{\mathbf{m}(E_I)}) + \mathbf{m}(E_I)\!\cdot\!\boldsymbol{\mu} = 2 \;,
\\
&\text{and}
\\
&\prod_i x_i^{\,(\mathbf{m}(E_I))_i} = 1 \;, 
\qquad \forall\ \text{easy internal edges } E_I \;. \label{constraints on x and mu}
\end{split}
\end{align}
\paragraph{Affine space $\mathfrak{M}[\CT]$ for R-symmetry mixings} The constraints on $\boldsymbol{x}$ can be solved by 
\begin{align}
\boldsymbol{x}
= \boldsymbol{\eta}_{\boldsymbol{v}}
:= (\eta^{v_1},\eta^{v_2},\ldots,\eta^{v_r}) \; \textrm{ with }\boldsymbol{v}=(v_1,\ldots,v_r)^T \in \mathfrak{F}[\mathcal{T}]
\label{eta v}
\end{align}
The $\mathfrak{F}[\CT]$ is defined in \eqref{F[T]}.
To solve the constraints of $\boldsymbol{\mu}$, we introduce the affine linear space $\mathfrak{M}[\mathcal{T}]$:
\begin{align}
\mathfrak{M}[\mathcal{T}]
:= \Big\{\, \boldsymbol{\mu}\in\mathbb{R}^r \;:\;
R_{\boldsymbol{\mu}}(\mathcal{O}_{E_I}) = 2,\ 
\forall\ \text{easy internal edges } E_I \Big\} \;.
\label{M[T]}
\end{align}
There is a special point ${\boldsymbol \mu}_{\rm geo} $ in $\mathfrak{M}[\mathcal{T}] $, which is 
    \begin{align}
    {\boldsymbol \mu}_{\rm geo}  := \mathbf{Q}^T \cdot B^{-1}\cdot {\boldsymbol \nu}_{x}\in \mathfrak{M}[\CT]\;. \label{mu_geom}
    \end{align}
    One can check that\footnote{Because $C_I-2\pi i|_\CG=0$,
    $\mathbf{P}|_\CG=-(B^{-1} A)\cdot \mathbf{X}-\pi i B^{-1}\cdot{\boldsymbol{\nu}_x}+B^{-1}\cdot (\CM,0,\cdots,0)^T|_\CG$
    with an arbitrary number $\CM$. Then (\ref{E in terms of g}) can be written as
    \begin{align}
        \begin{split}
            & \paren{(\mathbf{g}_I-\mathbf{g}_I')^T+(\mathbf{g}_I'-\mathbf{g}_I'')^T\cdot (B^{-1} A)}\cdot\mathbf{X}+(\mathbf{g}_I''-\mathbf{g}_I')^T\cdot B^{-1}\cdot(\CM,0,\cdots,0)^T\\
            & \left.+\pi i \paren{\sum_{i=1}^{r}(\mathbf{g}_I')_i+(\mathbf{g}_I'-\mathbf{g}_I'')^T\cdot B^{-1}\cdot {\boldsymbol{\nu}_x}}\right|_\CG=2\pi i\;.
        \end{split}
    \end{align}
    As the first two terms vanish, it proves the identity. }
    \begin{align}
    \begin{split}
    \forall E_I, \quad R_{\boldsymbol{\mu}_{\rm geo} } (\CO_{E_I}) &= \frac{1}2 \sum_{i=1}^r (\mathbf{g}'_I-\mathbf{g}''_{I} +|\mathbf{g}'_I-\mathbf{g}''_{I}|)_i + \boldsymbol{\nu}_x^T (B^{-1})^T \cdot (\mathbf{g}'_I-\mathbf{g}''_{I})
    \\
    &= \sum_{i=1}^r (\mathbf{g}'_I)_i + \boldsymbol{\nu}_x^T (B^{-1})^T \cdot (\mathbf{g}'_I-\mathbf{g}''_{I}) =2\;.
    \end{split}
    \end{align}
    We will argue that
\begin{align}
R_{\boldsymbol{\mu}_{\rm geo}}
\;=\;\left(
R_{\rm geo}
\text{ defined in \eqref{R_geom}}\right),
\label{R_geom-2}
\end{align}
using the 3D--3D relation \eqref{3D-3D relation},
as will be demonstrated in \eqref{Check for 3D-3D for S3b}. Using the $\boldsymbol{\mu}_{\rm geo} $ as a reference point, the affine space can be given as
    \begin{align}
    &\mathfrak{M}[\CT] = \boldsymbol{\mu}_{\rm geo} + \mathfrak{F}[\CT]\;.
    \end{align}
Similarly, one defines
\begin{align}
\mathfrak{M}_{\rm IR}[\mathcal{T}]
:= \mathfrak{M}[\mathcal{T}] \,/\, \mathfrak{F}_{\rm decouple}[\mathcal{T}]
= \boldsymbol{\mu}_{\rm geo} + \mathfrak{F}_{\rm IR}[\mathcal{T}] \;,
\label{M[T]_IR}
\end{align}
which parametrizes all possible R-symmetry mixings with the faithful IR  flavor symmetry.  In practice, the decoupled IR symmetries can be identified using the superconformal index as follows:
\begin{align}
\begin{split}
\mathbf{d} \in \mathfrak{F}_{\rm decouple}[\mathcal{T}]
\;\; \Rightarrow  \;\;
&\mathcal{I}_{\rm sci}^{T[A,B]}
\big[\boldsymbol{\eta}_{ \mathbf{d}},\, \boldsymbol{\mu}_0 \big]\textrm{ is $\eta$-independent at a }\boldsymbol{\mu}_0 \in \mathfrak{M}[\mathcal{T}] \;.
\end{split}
\end{align}
Using the basis in \eqref{basis of F[T]}, we consider the following superconformal index for $T[S^3\backslash \CK]$:
\begin{align}
\begin{split}
&\CI^{T[S^3\backslash \CK]}_{\rm sci}[u, \boldsymbol{\eta} ;\mu_m, \boldsymbol{\nu} ]:= \CI_{\rm sci}^{T[A,B]}[\boldsymbol{x} ,\boldsymbol{\mu} ]\big{|}_* \quad \bigl(\boldsymbol{\eta}:=(\eta_1, \ldots, \eta_{\sharp_H})^T, \boldsymbol{\nu} = (\nu_1, \ldots, \nu_{\sharp_H})^T\bigr)
\\
&\textrm{with the replacement }\big{|}_* : \boldsymbol{\mu} \rightarrow  \boldsymbol{\mu}_{\rm geo} + \mu_m \boldsymbol{v}_m + \sum_{I=1}^{\sharp_H}\nu_I \boldsymbol{v}_I , \quad x_i \rightarrow u^{(\boldsymbol{v}_m )_i} \prod_{I=1}^{\sharp_H} \eta_I^{(\boldsymbol{v}_I )_i}\;. \label{SCI for T_{DGG}}
\end{split}
\end{align} 
Here $(u, \mu_m)$ is the (fugacity, R-symmetry mixing parameter) for $U(1)_m$ symmetry while  $(\boldsymbol{\eta}, \boldsymbol{\nu})$ are that for other flavor symmetries.  3D-3D relation for the superconformal index is \cite{Dimofte:2011py,Lee:2013ida,Yagi:2013fda}
\begin{align}
\CI^{T[S^3\backslash \CK]}_{\rm sci} [u, \boldsymbol{\eta}= \mathbf{1}; \mu_m=0, \boldsymbol{\nu}= \mathbf{0} ] = \CI_{S^3\backslash \CK}^{SL(2,\mathbb{C})_{k=0}} (m=0, u;q)\;. \label{3D-3D for SCI}
\end{align}
Here  $\CI^{SL(2,\mathbb{C})_{k=0}}_{S^3\backslash \CK} (m, u;q)$ is the 3D-index in fugacity basis, defined in \eqref{3D-index}, for the knot complement. 
\paragraph{Torus knot case}
For $\CK=\CK_{(P,Q)}$, the 3D index becomes trivial when $m=0$, see \eqref{3D index for torus knots}:
\begin{align}
\CI^{SL(2,\mathbb{C})_{k=0}}_{S^3\backslash \CK_{(P,Q)}}(m=0,u;q)=1\;. 
\end{align}
Thus the index is independent of $u$, as expected, because the $U(1)_m$ symmetry of $T[S^3\backslash \CK_{(P,Q)}]$ always decouples in the IR, as stated in \eqref{decoupling of U(1)_m}.  Indeed, for torus knots, one can check that the full superconformal index is completely independent of both $u$ and $\mu_m$:
\begin{align}
\CI^{T[S^3\backslash \CK_{(P,Q)}]}_{\rm sci}[u,\boldsymbol{\eta};\mu_m,\boldsymbol{\nu}]=
\CI^{T[S^3\backslash \CK_{(P,Q)}]}_{\rm sci}[\boldsymbol{\eta};\boldsymbol{\nu}] \;.
\end{align}
Furthermore, from explicit computations we have confirmed that, the index depends nontrivially on all fugacities $\boldsymbol{\eta}$ as well as on the mixing parameters $\boldsymbol{\mu}$.
This verifies the claim in \eqref{F-decouple for torus}. 

Using the basis $\{ \boldsymbol{v}_I\}$ of $\mathfrak{F}[\CT]$ in \eqref{F[T]_{IR} for torus knots}, the index is defined as
\begin{align}
\begin{split}
&\CI^{T[S^3\backslash \CK_{(P,Q)}]}_{\rm sci}[ \boldsymbol{\eta} ; \boldsymbol{\nu} ]
\\
&= \begin{cases}
\CI^{T[S^3\backslash \CK_{(P,Q)}]}_{\rm sci}, & |P - Q| = 1,\\[0.3em]
\CI^{T [S^3\backslash \CK_{(P,Q)}]}_{\rm sci}[\eta ;\nu ], & \left(Q \equiv \pm 1\ (\mathrm{mod}\ P)\right) 
    \ \oplus\ 
    \left(P \equiv \pm 1\ (\mathrm{mod}\ Q)\right)\\[0.3em]
\CI^{T[S^3\backslash \CK_{(P,Q)}]}_{\rm sci}[\eta_1, \eta_2 ;\nu_1, \nu_2], & \left(Q \not\equiv \pm 1 \ (\mathrm{mod}\ P)\right) 
    \; \land \; 
    \left(P \not\equiv \pm 1\ (\mathrm{mod}\ Q)\right).
\end{cases}
\\
&:= \begin{cases}
\textrm{Tr} (-1)^R q^{\frac{R}2+j_3}\;, \quad R:=R_{\rm geo} := R_{\boldsymbol{\mu}_{\rm (geo)}}
\\
\textrm{Tr} (-1)^{R(\nu)} q^{\frac{R(\nu)}2+j_3} \eta^{T},\; R(\nu):= R_{\rm geo} + \nu T,\; T:= \boldsymbol{v}_1\cdot \mathbf{T}
\\
\textrm{Tr} (-1)^{R(\nu_1, \nu_2)} q^{\frac{R(\nu_1, \nu_2)}2+j_3} \eta_1^{T_1} \eta_2^{T_2}, \; R(\nu_1, \nu_2):= R_{\rm geo} + \sum_{I=1}^2\nu_I T_I,\;  T_I:= \boldsymbol{v}_I\cdot \mathbf{T} \label{SCI for torus : case II}
\end{cases}
\end{split}
\end{align}
The superconformal index for the reduced UV field theory  $T^{(\rm r)}[S^3\backslash \CK_{(P,Q)}]$ in \eqref{T-red-DGG} is (note that $\sharp_H^{(\rm r)} = \sharp_{H}$)
\begin{align}
\begin{split}
&\CI^{T^{(\rm r)}[S^3\backslash \CK]}_{\rm sci}[ \boldsymbol{\eta} ; \boldsymbol{\nu} ]:= \CI_{\rm sci}^{T[A_{\rm r},B_{\rm r}]}[\boldsymbol{x} ,\boldsymbol{\mu} ]\big{|}_* \quad \bigl(\boldsymbol{\eta}:=(\eta_1, \ldots, \eta_{\sharp_H})^T, \boldsymbol{\nu} = (\nu_1, \ldots, \nu_{\sharp_H})^T\bigr)
\\
&\textrm{with the replacement }\big{|}_* : \boldsymbol{\mu} \rightarrow  \boldsymbol{\mu}^{\rm (r)}_{\rm geo}  + \sum_{I=1}^{\sharp_H}\nu_I \boldsymbol{v}^{(\rm r)}_I , \quad x_i \rightarrow \prod_{I=1}^{\sharp_H} \eta_I^{(\boldsymbol{v}^{(\rm r)}_I )_i}\;. \label{SCI for T-red-DGG}
\end{split}
\end{align} 
Here $\boldsymbol{v}^{(\rm r)}_{I}$ is defined in \eqref{v_i-red} and $\boldsymbol{\mu}^{\rm (r)}_{\rm geo}$ is defined as (cf. \eqref{mu_geom})
\begin{align}
\boldsymbol{\mu}^{\rm (r)}_{\rm geo} := \mathbf{Q}_{r}^T \cdot  B^{-1}  \cdot \boldsymbol{\nu}_{x;{\rm r}}\;. \label{mu_geom-r}
\end{align}
The two theories are related by integrating out two massive chiral multiplets and
are infrared equivalent up to an Abelian TQFT,
as summarized in \eqref{Relation between T and T^r}.
Accordingly, we expect their superconformal indices to coincide:
\begin{align}
\CI^{T^{(\rm r)}[S^3\backslash \CK_{(P,Q)}]}_{\rm sci}
\big[ \boldsymbol{\eta} \, ; \, \boldsymbol{\nu} \big]
=
\CI^{T[S^3\backslash \CK_{(P,Q)}]}_{\rm sci}
\big[ \boldsymbol{\eta} \, ; \, \boldsymbol{\nu} \big] \; .
\end{align}
When $|P-Q|=1$, the 3D index becomes just 1:
\begin{align}
\CI^{T[S^3\backslash \CK_{(P,Q)}]}_{\rm sci} (q)=1\;, \quad \textrm{when $|P-Q|=1$}\;.
\end{align}
It is expected from \eqref{3D-3D for SCI} and \eqref{3D index for torus knots}. A trivial superconformal index provides strong evidence for a mass gap, fully consistent with our main claim in \eqref{IR phase of S3-K_{(P,Q)}}.
\\ 
When 
\[
\left(Q \equiv \pm 1\;(\mathrm{mod}\;P)\right)
\ \oplus\ 
\left(P \equiv \pm 1\;(\mathrm{mod}\;Q)\right),
\]
we find that there always exist two special values of $\nu$, denoted $\nu_a$ and $\nu_b$, at which
\begin{align}
\CI_{\rm sci}^{T[S^3\backslash \CK_{(P,Q)}]}[\eta=1;\,\nu=\nu_a]
=
\CI_{\rm sci}^{T[S^3\backslash \CK_{(P,Q)}]}[\eta=1;\,\nu=\nu_b]
=1\;. \label{two special points}
\end{align}
One of these special values is $\nu_a = 0$, as anticipated from 
\eqref{3D-3D for SCI} and \eqref{3D index for torus knots}.  
The second special value is always given by $\nu_b = 2$ or $\nu_b = -2$.  
This is fully consistent with our main claim in \eqref{IR phase of S3-K_{(P,Q)}}, since the SCI of a 3D $\CN=4$ rank-0 SCFT is known to possess precisely two such distinguished limits: the A-twisting and B-twisting points. In these limits, the SCI takes the form
\begin{align}
\begin{split}
&\text{SCI at the A-twisting point :}\;\; 
\CI^A_{\rm sci}(q) := \Tr_{\CH_{\rm rad}(S^2)} (-1)^{R_A}\, q^{\frac{R_A}{2}+j_3}, 
\\
&\text{SCI at the B-twisting point :}\;\;
\CI^B_{\rm sci}(q) := \Tr_{\CH_{\rm rad}(S^2)} (-1)^{R_B}\, q^{\frac{R_B}{2}+j_3},
\\
&\text{where}\quad R_A = 2J_3^H,\qquad R_B = 2J_3^C\;. \label{SCI for rank-0}
\end{split}
\end{align}
At these twisting points, contributions to the index arise only from operators on the Coulomb or Higgs branch, along with their descendants.  
Because rank-0 SCFTs have trivial Coulomb and Higgs branches by definition, the superconformal index in both limits reduces to $1$.

Comparing \eqref{SCI for torus : case II} with \eqref{SCI for rank-0}, we obtain the identification
\begin{align}
R_{\rm geo} = R_A\;, \qquad
\frac{\nu_b}2 \times \left( T := \boldsymbol{v}_1 \cdot \mathbf{T} \right)=  (T_A := J_3^C - J_3^H) \label{symmetry embedding for Case II}
\end{align}
When 
\[
\left(Q \not\equiv \pm 1\,(\mathrm{mod}\,P)\right)
\;\land\;
\left(P \not\equiv \pm 1\,(\mathrm{mod}\,Q)\right),
\]
we instead find four special points,
\begin{align}
\begin{split}
&(\nu_1,\nu_2)=\boldsymbol{\nu}_{aa},\ \boldsymbol{\nu}_{ab},\ \boldsymbol{\nu}_{ba},\ \boldsymbol{\nu}_{bb},  
\\
&\textrm{satisfying }  \boldsymbol{\nu}_{ba} - \boldsymbol{\nu}_{aa}=   \boldsymbol{\nu}_{bb} - \boldsymbol{\nu}_{ab}, \quad  \boldsymbol{\nu}_{ab} - \boldsymbol{\nu}_{aa}=   \boldsymbol{\nu}_{bb} - \boldsymbol{\nu}_{ba}
\end{split}
\end{align}
at which
\begin{align}
\CI_{\rm sci}^{T[S^3\backslash \CK_{(P,Q)}]}
\big[\eta_1=\eta_2=1;\,(\nu_1,\nu_2)\big]
\Big|_{(\nu_1,\nu_2)=\boldsymbol{\nu}_{aa},\,\boldsymbol{\nu}_{ab},\,\boldsymbol{\nu}_{ba},\,\boldsymbol{\nu}_{bb}}
=1\;. \label{four special points}
\end{align}
One of these special values is $\boldsymbol{\nu}_{aa} = (0,0)$, as anticipated from 
\eqref{3D-3D for SCI} and \eqref{3D index for torus knots}. This is again consistent with our claim in \eqref{IR phase of S3-K_{(P,Q)}}, which identifies these four points with the 
$(A,A)$, $(A,B)$, $(B,A)$, and $(B,B)$ twisting limits of the product of two rank-0 SCFTs. Comparing \eqref{SCI for torus : case II} with \eqref{SCI for rank-0}, we obtain the identification
\begin{align}
\begin{split}
&R_{\rm geo} = R^{(1)}_A + R^{(2)}_A\;, \;\;
\frac{\boldsymbol{\nu}_{ba} \cdot (T_1, T_2)}2 = T_A^{(1)}, \;\;\frac{\boldsymbol{\nu}_{ab} \cdot (T_1, T_2)}2 = T_A^{(2)}
\\
&\textrm{where }T_I := \boldsymbol{v}_I\cdot \mathbf{T}\;. \label{symmetry embedding for Case III}
\end{split}
\end{align}
\paragraph{Example : $T[S^3\backslash \CK_{(P,Q)} ]$ with $(P,Q)=(2,7)$} Using the matrices $(K, \mathbf{Q})$ in \eqref{K and Q for (2,7)}, the index for $T[A,B]$ theory is given as
\begin{align}
\begin{split}
&\CI_{\rm sci}^{T[A,B]}[x_1, x_2, x_3, x_4; \mu_1, \mu_2, \mu_3, \mu_4] (q)
\\
&= \sum_{m_1, m_2, m_3, m_4 \in \mathbb{Z}} \prod_{i=1}^4 \oint_{|u_i|=1} \frac{du_i}{2\pi i u_i } u_1^{6 m_1+2m_3} u_2^{m_2-m_3+m_4} u_3^{2 m_1-m_2+2m_3-m_4} u_4^{m_2-m_3+2m_4} 
\\
&  \qquad \qquad  \qquad \times \left( \prod_{i=1}^4 (x_i (-q^{1/2})^{\mu_i})^{m_i} \right) \CI_{\Delta}(2m_1, u_1^2) \CI_{\Delta}(m_2, u_2)\CI_{\Delta}(m_3, u_3)\CI_{\Delta}(m_4, u_4)
\end{split}
\end{align}
Then, the index for $T[S^3\backslash \CK_{(2,7)}]$ is 
\begin{align}
\CI_{\rm sci}^{T[S^3\backslash \CK_{(2,7)}]} [u, \eta; \mu_m, \nu] = \CI_{\rm sci}^{T[A,B]}[\boldsymbol{x}; \boldsymbol{\mu}] (q) \bigg{|}_{*}
\end{align}
where the replacement is (using the NZ matrices in \eqref{NZ matrices for (2,7)})
\begin{align}
\begin{split}
&\boldsymbol{\mu} \rightarrow \boldsymbol{\mu}_{\rm geo} + \mu_m \boldsymbol{v}_m+ \nu  \boldsymbol{v}_1, \; x_i \rightarrow u^{(\boldsymbol{v}_m)_i}  \eta^{(\boldsymbol{v}_1)_i}  \textrm { with } 
\\
& \boldsymbol{\mu}_{\rm geo}  =\mathbf{Q}^T B^{-1}\cdot \boldsymbol{\nu}_x = (0,-1,0,0)^T\;,
\\
&\boldsymbol{v}_m = \mathbf{Q}^T B^{-1}\cdot (1,0,0,0)^T = (0,-1,0,-1)^T\;,
\\
&\boldsymbol{v}_1 = \mathbf{Q}^T B^{-1}\cdot (0,1,0,0)^T = (-1,0,-1,0)^T\;. 
\end{split}
\end{align}
Computing the index in $q$-expansion, we obtain
\begin{align}
\begin{split}
&\CI_{\rm sci}^{T[S^3\backslash \CK_{(2,7)}]} [u, \eta; \mu_m, \nu] = \bigg{(} 1+ (\eta-1)q+ \left(\eta+\frac{1}\eta -2\right)q^2+(\eta-1)q^3
\\
&\qquad \quad+\left(\eta^2-\eta- \frac{1}\eta+\frac{1}{\eta^2}\right)q^4+\left(4+\eta^2-3\eta -\frac{3}\eta + \frac{1}{\eta^2}\right)q^5+\ldots \bigg{)}\bigg{|}_{\eta \rightarrow \eta (-q^{1/2})^\nu} \label{SCI for (2,7)}
\end{split}
\end{align}
Note that the index is independent on $(u, \mu_m)$, which is compatible with that the $U(1)_m$ symmetry decouples in the IR. The index becomes trivial when
\begin{align}
\CI_{\rm sci}^{T[S^3\backslash \CK_{(2,7)}]} [u, \eta=1; \mu_m, \nu=0]=\CI_{\rm sci}^{T[S^3\backslash \CK_{(2,7)}]} [u, \eta=1; \mu_m, \nu=2]=1\;.
\end{align}
Two special points, $\nu=\nu_{a}=0$ and $\nu=\nu_{b}=2$, can be regarded as the
A-twisting and B-twisting limits, respectively.
At $\nu=\nu_{\rm con}=1$, the index computes the superconformal index using the
superconformal $R$-charge and coincides with the index of the
$\mathcal{T}_{r=2}$ theory studied in~\cite{Gang:2023rei}.
This theory is a rank-$0$ SCFT that supports the rational VOA
$\mathcal{M}(2,7)$ at the boundary.
At the superconformal point $\nu=1$, the index contains the following term:
\begin{align}
\CI_{\rm sci}^{T[S^3\backslash \CK_{(2,7)}]}[\eta,\nu=1]
\;\supset\;
-(\eta+\eta^{-1})\,q^{3/2}.
\end{align}
Such a contribution is consistent with the presence of conserved current
multiplets associated with enhanced supersymmetry, although by itself it should
be regarded as suggestive rather than definitive evidence \cite{Cordova:2016emh,Evtikhiev:2017heo}.

\paragraph{Example : $T^{\rm (r)}[S^3\backslash \CK_{(P,Q)} ]$ with $(P,Q)=(2,7)$}   Using the matrices $(K_{\rm r
}, \mathbf{Q}_{\rm r})$ in \eqref{reduced K and Q for (2,7)}, the index for $T[A_{\rm r},B_{\rm r}]$ theory is given as
\begin{align}
\begin{split}
&\CI_{\rm sci}^{T[A_{\rm r},B_{\rm r}]}[x_1, x_2; \mu_1, \mu_2] (q)
\\
&= \sum_{m_1, m_2 \in \mathbb{Z}} \prod_{i=1}^2 \oint_{|u_i|=1} \frac{du_i}{2\pi i u_i } u_1^{-m_1-2m_2} u_2^{-2m_1-3m_2}   \left( \prod_{i=1}^2 (x_i (-q^{1/2})^{\mu_i})^{m_i} \right) \CI_{\Delta}(m_1, u_1) \CI_{\Delta}(m_2, u_2)\;.
\end{split}
\end{align}
Then, the index for $T^{\rm (r)}[S^3\backslash \CK_{(2,7)}]$ is 
\begin{align}
\CI_{\rm sci}^{T^{\rm (r)}[S^3\backslash \CK_{(2,7)}]} [ \eta ; \nu] = \CI_{\rm sci}^{T[A_{\rm r},B_{\rm r}]}[\boldsymbol{x}; \boldsymbol{\mu}] (q) \bigg{|}_{*}
\end{align}
where the replacement is (using the NZ matrices in \eqref{reduced NZ for (2,7)})
\begin{align}
\begin{split}
&\boldsymbol{\mu} \rightarrow \boldsymbol{\mu}^{\rm (r)}_{\rm geo} + \nu  \boldsymbol{v}^{\rm (r)}_1, \; x_i  \rightarrow  \eta^{(\boldsymbol{v}^{\rm (r)}_1)_i}  \textrm { with } 
\\
& \boldsymbol{\mu}^{\rm (r)}_{\rm geo}  =B^{-1}_{\rm (r)}\cdot \boldsymbol{\nu}_{x;{\rm r}} = (1,3)^T\;,
\\
&\boldsymbol{v}^{\rm (r) }_1 =  B^{-1}_{\rm r}\cdot (1,0)^T = (-1,-2)^T\;. 
\end{split}
\end{align}
Computing the index in $q$-expansion, we obtain the same index in  \eqref{SCI for (2,7)}, as expected from \eqref{Relation between T and T^r} and the index becomes trivial when
\begin{align}
\CI_{\rm sci}^{T^{\rm (r)}[S^3\backslash \CK_{(2,7)}]} (u, \eta=1; \mu_m, \nu=0)=\CI_{\rm sci}^{T^{\rm (r)}[S^3\backslash \CK_{(2,7)}]} (u, \eta=1; \mu_m, \nu=2)=1\;.
\end{align}
Two two special points, $\nu=\nu_a =0$ and $\nu=\nu_b=2$, can be regarded as A-twisting and B-twisting limits. 
\paragraph{Example : $T^{\rm (r)}[S^3\backslash \CK_{(P,Q)} ]$ with $(P,Q)=(5,7)$} Using the matrices $(K_{\rm r
}, \mathbf{Q}_{\rm r})$ in \eqref{reduced K and Q for (5,7)}, the index for $T[A_{\rm r},B_{\rm r}]$ theory is given as
\begin{align}
\begin{split}
&\CI_{\rm sci}^{T[A_{\rm r},B_{\rm r}]}[x_1, x_2, x_3; \mu_1, \mu_2, \mu_3] (q)
\\
&= \sum_{m_1, m_2, m_3 \in \mathbb{Z}} \left(  \prod_{i=1}^3 \oint_{|u_i|=1} \frac{du_i}{2\pi i u_i } \right) u_1^{2m_1} u_2^{-3m_2-2m_3} u_3^{-2m_2 - m_3}   \left( \prod_{i=1}^3 (x_i (-q^{1/2})^{\mu_i})^{m_i} \right) 
\\
&\qquad \qquad \qquad \qquad \qquad \qquad \quad  \times   \CI_{\Delta}(m_1, u_1) \CI_{\Delta}(m_2, u_2) \CI_{\Delta}(m_3, u_3)\;.
\end{split}
\end{align}
Then, the index for $T^{\rm (r)}[S^3\backslash \CK_{(5,7)}]$ is 
\begin{align}
\CI_{\rm sci}^{T^{\rm (r)}[S^3\backslash \CK_{(5,7)}]} [ \eta_1, \eta_2 ; \nu_1, \nu_2] = \CI_{\rm sci}^{T[A_{\rm r},B_{\rm r}]}[\boldsymbol{x}; \boldsymbol{\mu}] (q) \bigg{|}_{*}
\end{align}
where the replacement is (using the reduced NZ matrices in \eqref{reduced NZ for (5,7)})
\begin{align}
\begin{split}
&\boldsymbol{\mu} \rightarrow \boldsymbol{\mu}^{\rm (r)}_{\rm geo} + \nu_1  \boldsymbol{v}^{\rm (r)}_1 +  \nu_2 \boldsymbol{v}^{\rm (r)}_2, \; x_i  \rightarrow  \eta_1^{(\boldsymbol{v}^{\rm (r)}_1)_i}  \eta_2^{(\boldsymbol{v}^{\rm (r)}_2)_i}  \textrm { with } 
\\
& \boldsymbol{\mu}^{\rm (r)}_{\rm geo}  =B^{-1}_{\rm (r)}\cdot \boldsymbol{\nu}_{x;{\rm r}} = (-2,3,1)^T\;,
\\
&\boldsymbol{v}^{\rm (r) }_1 = 
 B^{-1}_{\rm r}\cdot (1,0,0)^T = (1,0,0)^T, \quad  \boldsymbol{v}^{\rm (r) }_2
 = B^{-1}_{\rm r}\cdot (0,1,0)^T = (0,-2,-1)^T\;. 
\end{split}
\end{align}
Computing the index in $q$-expansion, we obtain
\begin{align}
\begin{split}
&\CI_{\rm sci}^{T[S^3\backslash \CK_{(5,7)}]} [\eta_1, \eta_2; \nu_1, \nu_2]  ( q) = 
\\
&  \bigg{(} 1+ \big{(}\eta_1-1\big{)}q+ \big{(}\eta_1+\frac{1}{\eta_1} -2\big{)}q^2+\big{(}\eta_1+\frac{1}{\eta_1} -2\big{)} q^3 + \big{(}\eta_1^2+\frac{1}{\eta_1} -2\big{)} q^4+ (\eta_1^2-\eta_1) q^5 +\ldots \bigg{)}
\\
&\times \bigg{(} 1+ (\eta_2-1)q+ \big{(}\eta_2+\frac{1}{\eta_2} -2\big{)}q^2+(\eta_2-1)q^3+\big{(}\eta_2^2-\eta_2- \frac{1}{\eta_2}+\frac{1}{\eta_2^2}\big{)}q^4
\\
&\qquad  +\big{(}4+\eta_2^2-3\eta_2 -\frac{3}{\eta_2} + \frac{1}{\eta_2^2}\big{)} q^5+\ldots \bigg{)}\bigg{|}_{\eta_1 \rightarrow \eta_1 (-q^{1/2})^{\nu_1},\; \eta_2 \rightarrow \eta_2 (-q^{1/2})^{\nu_2}} \label{SCI for (5,7)}
\end{split}
\end{align}
The index exhibits a factorization property, as expected from
\eqref{IR phase of S3-K_{(P,Q)}}. The index becomes trivial ($=1$) when $\boldsymbol{\eta}=\boldsymbol{1}$ and
$\boldsymbol{\nu}$ takes one of the following four values:
\begin{align}
\boldsymbol{\nu}_{aa}=(0,0), \qquad
\boldsymbol{\nu}_{ba}=(2,0), \qquad
\boldsymbol{\nu}_{ab}=(0,2), \qquad
\boldsymbol{\nu}_{bb}=(2,2)\;. \label{nu_{aa/ab/ba/bb} for reduced (5,7)}
\end{align}

\subsection{Squashed 3-sphere partition function} 
The squashed three-sphere partition function  \cite{Kapustin:2009kz,Hama:2011ea,Dimofte:2011ju,Gang:2019jut} is an important RG-invariant BPS
partition function of three-dimensional $\mathcal{N}=2$ gauge theories. It is, in many respects, a more refined invariant than the superconformal index,
since it is sensitive to decoupled topological sectors that may be invisible in
the index. At $b=1$, the partition function plays a central role in determining the
infrared superconformal $R$-charge via $F$-maximization \cite{Jafferis:2010un}.
Combined with the superconformal index computation, this information is crucial
for analyzing the structure of superconformal multiplets at the IR fixed point. Moreover, the squashed-sphere partition function provides a non-trivial
dictionary of the 3D--3D correspondence, which will be crucial for identifying the correct bottom-up
field-theoretic construction of the three-dimensional theory $T[M]$.

The partition function for $\mathbb{T}[A,B; \mathbf{Q}]$ and $T[A,B]$ is 
\begin{align}
\begin{split}
&\CZ^{\mathbb{T}[A,B;\mathbf{Q}]}_{S^3_b} [\boldsymbol{\xi}, \boldsymbol \mu] = \int \frac{d^r \mathbf{Z}}{(2\pi \hbar )^{r/2}}  \CI_\hbar (\mathbf{Z};\boldsymbol{\xi}, \boldsymbol{\mu}) \;,
\\
&\textrm{ with }  \CI_\hbar (\mathbf{Z};\boldsymbol{\zeta}, \boldsymbol{\mu}) := \exp \left(\frac{\mathbf{Z}^T\mathbf{Q}^T B^{-1}A \mathbf{Q}\mathbf{Z} +2 \mathbf{Z}^T \mathbf{W}}{2\hbar }\right)\prod_i \psi_\hbar (\sum_j Q_{ij}Z_j) \bigg{|}_{\mathbf{W} =\boldsymbol{\xi}+(i \pi +\frac{\hbar}2)  \boldsymbol \mu }
\\
&\CZ^{T[A,B]}_{S^3_b} [\boldsymbol{\xi},  \boldsymbol \mu ]  =\sqrt{N}  \CZ^{\mathbb{T}[A,B;\mathbf{Q}]}_{S^3_b} [\boldsymbol{\xi},\boldsymbol{\mu}]\;. \label{S3-b for tT[A,B;Q]}
\end{split}
\end{align}
Here we define $\hbar:=2\pi i b^2$ and  $\boldsymbol{\mu} =(\mu_1, \ldots, \mu_r)^T$. $\psi_\hbar$ denotes the quantum dilogarithm function defined in \eqref{Q.D.L}. The vector $\boldsymbol{\xi} = (\xi_1, \ldots, \xi_r)^T$ denotes the (rescaled) FI parameters for the $U(1)^r$ gauge symmetry, which are equivalent to real masses for the  $U(1)^r_{\rm top}$ symmetry. The partition function  $\CZ^{T[A,B]}_{S^3_b} [\boldsymbol{\xi},  \boldsymbol \mu ]$ is  equivalent to the partition function for $g\cdot (T_\Delta)^{\otimes r}$ in \eqref{sketch of DGG} \cite{Dimofte:2011ju,Dimofte:2012qj}. In the above, we use the fact that $\CZ_{S^3_b}[\CA^{N,p}] = 1/\sqrt{N}$. 

The partition function for $T[S^3\backslash \mathcal{K}]$ (resp. $\mathbb{T}[A,B;\mathbf{Q}, \{E_I\}]$) can be obtained from $\CZ_{S^3_b}^{T[A,B]}$ (resp. $\CZ_{S^3_b}^{\BT[A,B;\mathbf{Q}]}$) by specializing $\boldsymbol{\xi} \in \mathfrak{F}[\mathcal{T}]$ in \eqref{F[T]} and $\boldsymbol{\mu} \in \mathfrak{M}[\CT]$ in \eqref{M[T]}. Using the basis in \eqref{basis of F[T]}, the partition functions for  $T[S^3\backslash \CK]$ and $\mathbb{T}[A,B;\mathbf{Q}, \{E_I\}]$ can be given as: 
\begin{align}
\begin{split}
&\CZ^{T[S^3\backslash \CK]}_{S^3_b}[\zeta_m, \boldsymbol{\zeta} ;\mu_m, \boldsymbol{\nu} ]:= \CZ^{T[A,B]}_{S^3_b}[\boldsymbol{\xi} ,\boldsymbol{\mu} ]\big{|}_* \quad \bigl(\boldsymbol{\zeta}:=(\zeta_1, \ldots, \zeta_{\sharp_H})^T, \boldsymbol{\nu} = (\nu_1, \ldots, \nu_{\sharp_H})^T\bigr)\;,
\\
&\CZ^{\mathbb{T}[A,B;\mathbf{Q}, \{E_I\}]}_{S^3_b}[\zeta_m, \boldsymbol{\zeta} ;\mu_m, \boldsymbol{\nu} ]:= \CZ^{\mathbf{T}[A,B; \mathbf{Q}]}_{S^3_b}[\boldsymbol{\xi} ,\boldsymbol{\mu} ]\big{|}_*\;,
\\
&\textrm{with the replacement }\big{|}_* : \boldsymbol{\mu} \rightarrow  \boldsymbol{\mu}_{\rm geo} + \mu_m \boldsymbol{v}_m + \sum_{I=1}^{\sharp_H}\nu_I \boldsymbol{v}_I , \quad \boldsymbol{\xi} \rightarrow   \zeta_m \boldsymbol{v}_m  +\sum_{I=1}^{\sharp_H} \zeta_I \boldsymbol{v}_I \;. \label{S3b ptn for T_{DGG}}
\end{split}
\end{align}  
$\boldsymbol{\mu}_{\rm geo} \in \mathfrak{M}[\CT]$ is given in \eqref{mu_geom}. $(\zeta_m, \mu_m)$ can be regarded as the (rescaled real mass, R-symmetry mixing parameter) for the $U(1)_m$ symmetry while the $(\boldsymbol{\zeta}, \boldsymbol{\nu})$ are that for other flavor symmetries.  
Especially by turning off the all the parameters except for the $U(1)_m$ symmetry, we have  (after rescaling the integral variables $\mathbf{Z} \rightarrow \mathbf{Q}^{-1}\mathbf{Z}$)
    \begin{align}
    \begin{split}
    &\CZ^{T[S^3\backslash \mathcal{K}]}_{S^3_b}[\zeta_m, \boldsymbol{\zeta} = \mathbf{0}; \mu_m, \boldsymbol{\nu} = \mathbf{0}] = \CZ^{T[A,B]}_{S^3_b} \big{[}\boldsymbol{\xi} = \zeta_m \boldsymbol{v}_{\mathbf{m}}  ,  \boldsymbol \mu  = \boldsymbol{\mu }_{\rm geo} + \mu_m \boldsymbol{v}_{\mathbf{m}}\big{]} 
    \\
    &= \frac{\sqrt{N}}{\det \mathbf{Q}} \int \frac{d^r \mathbf{Z}}{(2\pi \hbar)^{r/2}} \exp \bigg{(} \frac{\mathbf{Z} B^{-1}A \mathbf{Z}+(2\pi i +\hbar)\mathbf{Z}^T B^{-1} \boldsymbol{\nu}_x}{2\hbar }
    \\
    &  \qquad  \qquad  \qquad \qquad \qquad \quad   +\frac{2 (\zeta_m+ (i \pi +\frac{\hbar}2)\mu_m) \mathbf{Z}^T B^{-1}(1,\ldots, 0)^T}{2\hbar }\bigg{)} \prod_{i=1}^r \psi_\hbar (Z_i)\;.
    \end{split}
    \end{align}
    Comparing with the state-integral expression in \eqref{state-integral}, and using the relation in \eqref{NQB relation}, we obtain 
    \begin{align}
    \CZ^{T[S^3\backslash \mathcal{K}]}_{S^3_b}[\zeta_m, \boldsymbol{\zeta} = \mathbf{0}; \mu_m, \boldsymbol{\nu} = \mathbf{0}]  =  Z^{SL(2,\mathbb{C})_{k=1}}_{M= S^3\backslash \mathcal{K}}[X_m = \zeta_m +(i \pi +\frac{\hbar}2) \mu_m ]\;. \label{Check for 3D-3D for S3b}
    \end{align}
   This agrees exactly with the 3D–3D relations in \eqref{3D-3D relation}, \eqref{3D-3D relation 2}, and \eqref{3D-3D relation 3}. Such a highly non-trivial consistency check justifies our definition of $T[S^3\backslash \CK]$ in \eqref{tT[A,B]}, \eqref{T[A,B]}, and \eqref{T-DGG}, as well as the identifications in \eqref{v_m} and \eqref{R_geom-2}.

   \paragraph{Torus knot case} For the case when $M= S^3\backslash \CK_{(P,Q)}$, the $U(1)_m$ symmetry decouples in the IR and the partition function is expected to be independent on the $(\zeta_m, \mu_m)$ modulo a phase factor:
   \begin{align}
   \CZ^{T[S^3\backslash \CK_{(P,Q)}]}_{S^3_b}[\zeta_m, \boldsymbol{\zeta} ;\mu_m, \boldsymbol{\nu} ] \simeq   \CZ^{T[S^3\backslash \CK_{(P,Q)}]}_{S^3_b}[ \boldsymbol{\zeta} ; \boldsymbol{\nu} ] \;.
   \end{align}
   The partition function for the simplified UV field theory  $T^{(\rm r)}_{\rm DGG
}[S^3\backslash \CK_{(P,Q)}]$ in \eqref{T-red-DGG} is (note that $\sharp_H^{(\rm r)} = \sharp_{H}$)
\begin{align}
\begin{split}
&\CZ^{T^{(\rm r)}[S^3\backslash \CK]}_{S^3_b}[ \boldsymbol{\zeta} ; \boldsymbol{\nu} ]:= \CZ_{S^3_b}^{T[A_{\rm r},B_{\rm r}]}[\boldsymbol{\xi} ,\boldsymbol{\mu} ]\big{|}_* 
\\
&\textrm{with the replacement }\big{|}_* : \boldsymbol{\mu} \rightarrow  \boldsymbol{\mu}^{\rm (r)}_{\rm geo}  + \sum_{I=1}^{\sharp_H}\nu_I \boldsymbol{v}^{(\rm r)}_I , \quad \boldsymbol{\xi} \rightarrow \sum_{I=1}^{\sharp_H}\zeta_I \boldsymbol{v}^{(\rm r)}_I \;. \label{Z3-b for T-red-DGG}
\end{split}
\end{align} 
Here $\boldsymbol{v}^{(\rm r)}_{I}$ and $\boldsymbol{\mu}^{\rm (r)}_{\rm geo}$ are defined in \eqref{v_i-red} and \eqref{mu_geom-r} respectively. Two theories are IR equivalent up to an decoupled TQFT \eqref{Relation between T and T^r}, we expect their partition functions should agree up to a contribution from the decoupled TQFT:
\begin{align}
\CZ^{T[S^3\backslash \CK_{(P,Q)}]}_{S^3_b}[ \boldsymbol{\zeta} ; \boldsymbol{\nu} ] \simeq \CZ^{T^{(\rm r)}[S^3\backslash \CK_{(P,Q)}]}_{S^3_b}[ \boldsymbol{\zeta} ; \boldsymbol{\nu} ] \times \CZ_{S^3}[\textrm{Abelian TQFT}] \;.
\end{align}
Using the basis $\{ \boldsymbol{v}_I\}$ of $\mathfrak{F}[\CT]$ in \eqref{F[T]_{IR} for torus knots}, the partition function is given as
\begin{align}
\begin{split}
&\CZ^{T[S^3\backslash \CK_{(P,Q)}]}_{S^3_b}[ \boldsymbol{\zeta} ; \boldsymbol{\nu} ] 
\\
&= \begin{cases}
\CZ^{T[S^3\backslash \CK_{(P,Q)}]}_{S^3_b}, & |P - Q| = 1,\\[0.3em]
\CZ^{T [S^3\backslash \CK_{(P,Q)}]}_{S^3_b}[\zeta; \nu], & \left(Q \equiv \pm 1\ (\mathrm{mod}\ P)\right) 
    \ \oplus\ 
    \left(P \equiv \pm 1\ (\mathrm{mod}\ Q)\right)\\[0.3em]
\CZ^{T[S^3\backslash \CK_{(P,Q)}]}_{S^3_b}[\zeta_1, \zeta_2 ;\nu_1, \nu_2], & \left(Q \not\equiv \pm 1 \ (\mathrm{mod}\ P)\right) 
    \; \land \; 
    \left(P \not\equiv \pm 1\ (\mathrm{mod}\ Q)\right).
\end{cases}
\end{split}
\end{align}
When $|P-Q|=1$, we found that the partition function is independent of the squashing parameter $b$, 
up to an overall phase, and its magnitude is given by  
\begin{align}
\CZ^{T[S^3\backslash \mathcal{K}_{(P,Q)}]}_{S^3_b}
   \;\simeq\; 
   S_{00}\big(\mathcal{M}(P,Q)\big)
   = \sqrt{\frac{8}{P Q}}\,
     \sin\!\left(\frac{\pi Q}{P}\right)
     \sin\!\left(\frac{\pi P}{Q}\right).
\end{align}
This result is compatible with the claim in
\eqref{IR phase of S3-K_{(P,Q)}}, together with the dictionary of the
bulk--boundary correspondence \cite{zbMATH04092352}.  Checking the $b$-independence is highly nontrivial, since performing the full localization integral is generically very difficult. 
Instead, one can verify the independence perturbatively by expanding the integral in powers of $(b-1)$ using the method in \cite{Gang:2019jut}, and checking the independence order by order. 

When 
\[
\left(Q \equiv \pm 1\;(\mathrm{mod}\;P)\right)
\oplus
\left(P \equiv \pm 1\;(\mathrm{mod}\;Q)\right),
\]
we find that there exist two special values of $\nu$, denoted $\nu=\nu_a$ and 
$\nu=\nu_b$, for which the partition function becomes $b$-independent when $\zeta=0$:
\begin{align}
\CZ_{S^3_b}^{T[S^3\backslash \mathcal{K}_{(P,Q)}]}
   [\zeta=0;\nu]\Big|_{\nu=\nu_a\ \mathrm{or}\ \nu_b}
   \quad\text{is $b$-independent modulo a phase.}
\end{align}
These two points coincide with the special values identified in the
superconformal index analysis of \eqref{two special points}:
one is always $\nu_a = 0$, while the other is $\nu_b = \pm 2$.
They correspond to the A- and B-twisting choices, respectively.
Accordingly, from the claim in \eqref{IR phase of S3-K_{(P,Q)}}, we expect
\begin{align}
\begin{split}
&\textrm{at }\nu = \nu_a =0 \textrm{ or }\nu= \nu_b , 
\\
&\CZ^{T[S^3\backslash \mathcal{K}_{(P,Q)}]}_{S^3_b}
   [\zeta=0;\nu]
   \;\simeq\;
   S_{00}\big(\mathcal{M}(P,Q)\big)
   = \sqrt{\frac{8}{P Q}}\,
     \sin\left(\frac{\pi Q}{P}\right)
     \sin\left(\frac{\pi P}{Q}\right).
\end{split}
\end{align}
If the point is at $\nu=\nu_a$ (respectively, $\nu_b$), we expect that the
$r$-VOA of $\CM(P,Q)$ is realized in the topologically A-twisted
(respectively, B-twisted) theory.
We further expect that, at $b=1$, the absolute value of the partition function is 
minimized at 
\[
\nu = \nu_{\rm con} := \frac{\nu_b}{2}.
\]
This follows from the fact that the squashed $S^3_b$ partition function is computed 
using the R-charge 
\[
R_{\nu} = R_A + \nu T= 2J_3^H + \frac{2\nu}{ \nu_b} \,(J_3^C - J_3^H),
\]
see \eqref{symmetry embedding for Case II}, and that 
$|\CZ_{S^3_{b=1}}|$ is minimized at the superconformal 
R-charge $R_{\rm con} = J_3^H + J_3^C$, according to the F-maximization principle 
\cite{Jafferis:2010un}.

When 
\[
\left(Q \not\equiv \pm 1\;(\mathrm{mod}\;P)\right)
\land
\left(P \not\equiv \pm 1\;(\mathrm{mod}\;Q)\right),
\]
we find that there exist four special values of 
$\boldsymbol{\nu} = (\nu_1,\nu_2)$, denoted 
$\boldsymbol{\nu}_{aa}, \boldsymbol{\nu}_{ab}, \boldsymbol{\nu}_{ba}, \boldsymbol{\nu}_{bb}$,
for which the partition function becomes $b$-independent when 
$\boldsymbol{\zeta}=\mathbf{0}$:
\begin{align}
\CZ_{S^3_b}^{T[S^3\backslash \mathcal{K}_{(P,Q)}]}
   [\boldsymbol{\zeta}=\mathbf{0};\,\boldsymbol{\nu}]
   \Big|_{\boldsymbol{\nu}=\boldsymbol{\nu}_{aa},\ \boldsymbol{\nu}_{ba},\ 
   \boldsymbol{\nu}_{ab},\ \boldsymbol{\nu}_{bb}}
   \quad\text{is $b$-independent modulo a phase.}
\end{align}
These four points coincide with the special values obtained from the superconformal 
index analysis in \eqref{four special points}; one of them is always 
$\boldsymbol{\nu}_{aa} = \mathbf{0}$.  
They correspond to the $(\mathrm{A},\mathrm{A})$-, $(\mathrm{B},\mathrm{A})$-, 
$(\mathrm{A},\mathrm{B})$-, and $(\mathrm{B},\mathrm{B})$-twisting choices of the two 
rank-0 SCFT factors. From \eqref{IR phase of S3-K_{(P,Q)}}, we expect that there exists a choice
\[
\boldsymbol{\nu} \in 
\{\boldsymbol{\nu}_{aa},\,\boldsymbol{\nu}_{ba},\,\boldsymbol{\nu}_{ab},\,\boldsymbol{\nu}_{bb}\}
\]
for which
\begin{align}
\CZ^{T[S^3\backslash \mathcal{K}_{(P,Q)}]}_{S^3_b}
   [\boldsymbol{\zeta}=\mathbf{0};\,\boldsymbol{\nu}]
   \;\simeq\;
   S_{00}\big(\mathcal{M}(P,Q)\big)
   =
   \sqrt{\frac{8}{P Q}}\,
   \sin\!\left(\frac{\pi Q}{P}\right)
   \sin\!\left(\frac{\pi P}{Q}\right).
\end{align}
Furthermore, we expect that at $b=1$ the absolute value of the partition function is 
minimized at
\[
\boldsymbol{\nu}_{\rm con} 
:= \frac{\boldsymbol{\nu}_{ba}+\boldsymbol{\nu}_{ab}}{2}.
\]
This follows from the fact that the squashed $S^3_b$ partition function is computed 
using the R-charge
\begin{align}
R_{\boldsymbol{\nu}}
= R^{(1)}_A + R^{(2)}_A + \nu_1 T_1 + \nu_2 T_2 = 2J_3^{(1)H} + 2J_3^{(2)H} + \nu_1 T_1 + \nu_2 T_2
\end{align}
and that the superconformal R-charge
\[
R_{\rm con} 
= J_3^{(1)H} + J_3^{(1)C} + J_3^{(2)H} + J_3^{(2)C}
\]
is equal to $R_{\boldsymbol{\nu}=\boldsymbol{\nu}_{\rm con}}$, 
see \eqref{symmetry embedding for Case III}.

\paragraph{Example : $T[S^3\backslash \CK_{(P,Q)} ]$ with $(P,Q)=(2,7)$} Using the matrices $(K, \mathbf{Q})$ in \eqref{K and Q for (2,7)}, the integrand  for $\mathbb{T}[A,B;\mathbf{Q}, \{E_I\}]$ theory is given as
\begin{align}
\begin{split}
& \CI^{\mathbb{T}[A, B; \mathbf{Q}, \{E_I\}]}_\hbar (\mathbf{Z};\zeta_m, \zeta; \mu_m, \nu) 
\\
&= \exp \left(\frac{6 Z_1^2 + Z_2^2 + 4 Z_1 Z_3 + 2 Z_4^2 + 2 (Z_2 - Z_3) (Z_4 - Z_3)+ 2\sum_{i=1}^4 Z_i \big{(} \xi_i +(\pi i +\hbar/2   )\mu_i \big{)} }{2\hbar} \right)
\\
& \quad \times \psi_\hbar (2Z_1) \psi_\hbar (Z_2) \psi_\hbar (Z_3) \psi_\hbar (Z_4) \bigg{|}_*\;, \label{Ih for (2,7)}
\end{split}
\end{align}
where the replacement is (using the NZ matrices in \eqref{NZ matrices for (2,7)})
\begin{align}
\begin{split}
&\boldsymbol{\mu} \rightarrow \boldsymbol{\mu}_{\rm geo} + \mu_m \boldsymbol{v}_m+ \nu  \boldsymbol{v}_1, \; \boldsymbol{\xi} \rightarrow  \zeta_m \boldsymbol{v}_m+ \zeta  \boldsymbol{v}_1  \textrm { with } 
\\
& \boldsymbol{\mu}_{\rm geo}  =\mathbf{Q}^T B^{-1}\cdot \boldsymbol{\nu}_x = (0,-1,0,0)^T\;,
\\
&\boldsymbol{v}_m = \mathbf{Q}^T B^{-1}\cdot (1,0,0,0)^T = (0,-1,0,-1)^T\;,
\\
&\boldsymbol{v}_1 = \mathbf{Q}^T B^{-1}\cdot (0,1,0,0)^T = (-1,0,-1,0)^T\;. 
\end{split}
\end{align}
A direct evaluation of the localization integral is difficult.
However, for $b=1$, corresponding to the round three-sphere,
the $S^3_b$ partition function can be computed using an alternative approach,
namely the Bethe-sum formula, which will be discussed in
Section~\ref{sec : twisted ptns}.
The resulting partition function is independent of
$(\zeta_m,\mu_m)$, as expected from the decoupling of $U(1)_m$.
The function evaluated at $\zeta=0$ is shown in
Figure~\ref{fig : ZS3 for (2,7)}.
\begin{figure}[h] 
    \centering
    \includegraphics[width=0.5\linewidth]{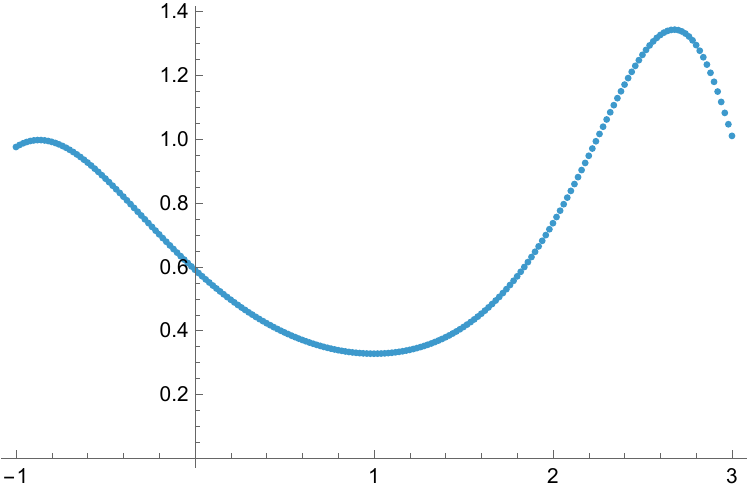}
    \caption{
        Absolute value of the round $S^3$ partition function for 
        $T[S^3\backslash \CK_{(P,Q)=(2,7)}]$, computed using the 
        Bethe-sum formula in~\eqref{Bethe-sum for S3}. 
        The partition function attains its minimum at 
        $\nu=\nu_{\rm con}=1$, where the value is 
        $\frac{2}{\sqrt{7}}\sin\!\big(\frac{6\pi}{7}\big) \approx 0.327985$. 
        At $\nu=0$, the value 
        $\frac{2}{\sqrt{7}}\sin\!\big(\frac{2\pi}{7}\big) \approx 0.591009$ 
        matches $|S_{00}|$ of the minimal model $\CM(2,7)$. 
        At $\nu=2$, the value is 
        $\frac{2}{\sqrt{7}}\sin\!\big(\frac{4\pi}{7}\big) \approx 0.736976$.
    }
    \label{fig : ZS3 for (2,7)}
\end{figure}

\paragraph{Example : $T^{\rm (r)}[S^3\backslash \CK_{(P,Q)} ]$ with $(P,Q)=(2,7)$}   Using the matrices $(K_{\rm r
}, \mathbf{Q}_{\rm r})$ in \eqref{reduced K and Q for (2,7)}, the integrand  for $\mathbb{T}^{\rm (r)}[A_{\rm r},B_{\rm r};\mathbf{Q}, \{E_I\}]$ theory is given as
\begin{align}
\begin{split}
& \CI^{\mathbb{T}^{\rm (r)}[A_{\rm r},B_{\rm r};\mathbf{Q}, \{E_I\}]}_\hbar (\mathbf{Z}; \zeta;  \nu) 
\\
&= \exp \left(\frac{-Z_1^2-3 Z_2^2 - 4 Z_1  Z_2+ 2\sum_{i=1}^2 Z_i \big{(} \xi_i +(\pi i +\hbar/2   )\mu_i \big{)} }{2\hbar} \right) \psi_\hbar (Z_1) \psi_\hbar (Z_2)  \bigg{|}_*\;,
\end{split}
\end{align}
where the replacement is (using the NZ matrices in \eqref{reduced NZ for (2,7)})
\begin{align}
\begin{split}
&\boldsymbol{\mu} \rightarrow \boldsymbol{\mu}^{\rm (r)}_{\rm geo} + \nu  \boldsymbol{v}^{\rm (r)}_1, \; \boldsymbol{\xi}  \rightarrow  \zeta \boldsymbol{v}^{\rm (r)}_1 \textrm { with } 
\\
& \boldsymbol{\mu}^{\rm (r)}_{\rm geo}  =B^{-1}_{\rm (r)}\cdot \boldsymbol{\nu}_{x;{\rm r}} = (1,3)^T\;,
\\
&\boldsymbol{v}^{\rm (r) }_1 =  B^{-1}_{\rm r}\cdot (1,0)^T = (-1,-2)^T\;. 
\end{split}
\end{align}

\subsection{Twisted Partition functions} \label{sec : twisted ptns}
Twisted partition functions \cite{Benini:2015noa,Benini:2016hjo,Closset:2016arn,Closset:2017zgf,Closset:2018ghr} provide very useful BPS invariants, especially in
situations where the theory flows in the infrared to a unitary TQFT or to a
three-dimensional $\mathcal{N}=4$ rank-$0$ superconformal field theory.
They allow one to probe basic data of the TQFTs appearing in the IR, including
both unitary TQFTs and non-unitary TQFTs arising after topological twisting. In particular, twisted partition functions can be used to extract part of the
modular data of the resulting topological field theories.
This information characterizes the modular transformation properties of the
characters of boundary rational vertex operator algebras (r-VOAs).

Here we compute the twisted partition functions for the theory $\mathbb{T}[A,B;\mathbf{Q},\{E_I\}]$ defined in
\eqref{T-DGG-2}.
This theory differs from $T[S^{3}\backslash \CK]$ only by a decoupled
topological quantum field theory,
\begin{align}
\mathbb{T}[A,B;\mathbf{Q},\{E_I\}]
\simeq
T[S^{3}\backslash \CK]\otimes\mathcal{A}^{N,p}\,,
\label{Decouled sector A^{N,p}}
\end{align}
where $\mathcal{A}^{N,p}$ is an Abelian TQFT.
In the special case $N=1$, equivalently
$\left|\det \mathbf{Q}/\sqrt{\det B}\right|^{2}=1$,
the decoupled sector is trivial, and the two theories coincide.

A convenient way to compute the twisted partition function is to consider
the asymptotic expansion of the integrand $\CI_\hbar$ of the squashed
three-sphere partition function in the limit
$\hbar := 2\pi i b^2 \to 0$~\cite{Gang:2019jut}. The integrand is given by
\begin{align}
\CI_\hbar (\mathbf{Z}; \zeta_m, \boldsymbol{\zeta}; \mu_m, \boldsymbol{\nu})
:=
\bigl(
\CI_\hbar (\mathbf{Z}; \boldsymbol{\xi}, \boldsymbol{\mu})
\ \text{in~\eqref{S3-b for tT[A,B;Q]}}
\bigr)\Big|_{*},
\end{align}
where the replacement $\big|_{*}$ is defined in
\eqref{S3b ptn for T_{DGG}}. Then, $\CW_0$ and $\CW_1$ are defined through the asymptotic expansion
\begin{align}
\log \CI_\hbar(\mathbf{Z}; \zeta_m, \boldsymbol{\zeta}; \mu_m, \boldsymbol{\nu})
\xrightarrow{\;\hbar \to 0\;}
\frac{\CW_0 (\mathbf{Z}; \zeta_m, \boldsymbol{\zeta}; \mu_m, \boldsymbol{\nu})}{\hbar}
+ \CW_1 (\mathbf{Z}; \zeta_m, \boldsymbol{\zeta}; \mu_m, \boldsymbol{\nu})
+ O(\hbar),
\end{align}
where
\begin{align}
\begin{split}
\CW_0 (\mathbf{Z}; \zeta_m, \boldsymbol{\zeta}; \mu_m, \boldsymbol{\nu})
&\simeq
\frac{1}{2}\,
\mathbf{Z}^T \mathbf{Q}^T B^{-1} A \mathbf{Q} \mathbf{Z}
+ \mathbf{Z}^T (\boldsymbol{\xi} + i \pi \boldsymbol{\mu})\Big|_{*}
+ \sum_i {\rm Li}_2 \!\left(e^{-\sum_j Q_{ij} Z_j}\right),
\\[4pt]
\CW_1 (\mathbf{Z}; \zeta_m, \boldsymbol{\zeta}; \mu_m, \boldsymbol{\nu})
&\simeq
\frac{1}{2}\,\mathbf{Z}^T \boldsymbol{\mu}\Big|_{*}
- \frac{1}{2} \sum_i \log \!\left(1 - e^{-\sum_j Q_{ij} Z_j}\right).
\label{W0 and W1}
\end{split}
\end{align}
We use the following asymptotic behavior of the quantum dilogarithm function:
\begin{align}
\log \psi_\hbar (Z)
\simeq
\frac{{\rm Li}_2 (e^{-Z})}{\hbar}
- \frac{1}{2} \log \!\left(1 - e^{-Z}\right)
+ O(\hbar),
\qquad \text{as } \hbar \to 0 .
\end{align}
Here the equivalence relation $\simeq$ is defined as
\begin{align}
\CW_0 &\simeq \CW_0 + 2\pi i\, \mathbf{Z}\cdot \mathbf{n} + 4\pi^2 l,
\qquad (\mathbf{n}\in \mathbb{Z}^r,\; l\in \mathbb{Z}),
\\
\CW_1 &\simeq \CW_1 + \pi i\, \mathbb{Z}.
\end{align}
We then define the Bethe vacua, fibering operator, and handle-gluing operator as
\begin{align}
\begin{split}
\text{Bethe vacua :}\quad
{\rm BV}
&=
\Bigl\{
\mathbf{z}=(z_1,\ldots,z_r)
\; \big|\;
\exp\!\left(\partial_{Z_i}\CW_0\right)\Big|_{\mathbf{Z}\to\log\mathbf{z}} = 1,
\; i=1,\ldots,r
\Bigr\},
\\[4pt]
\text{Fibering operator :}\quad
\CF
&=
\exp\!\left(
-\frac{
\CW_0
- \mathbf{Z}\cdot \partial_{\mathbf{Z}}\CW_0
- \zeta_m \partial_{\zeta_m}\CW_0
- \boldsymbol{\zeta}\cdot \partial_{\boldsymbol{\zeta}}\CW_0
}{2\pi i}
\right)\Bigg|_{\mathbf{Z}\to\log\mathbf{z}},
\\[6pt]
\text{Handle-gluing operator :}\quad
\CH
&=
\det_{i,j}\!\left(\partial_{Z_i}\partial_{Z_j}\CW_0\right)
\, e^{-2\CW_1}
\Big|_{\mathbf{Z}\to\log\mathbf{z}} .
\label{BV, F and H}
\end{split}
\end{align}
More explicitly, the Bethe-vacua equations take the form
\begin{align}
\begin{split}
\text{BV equations :}\quad
\prod_{j} z_j^{K_{ij}} \bigl(1 - z_j^{-Q_{ij}}\bigr)
&=
\exp\!\left(-\xi_i - i\pi \mu_i\right)\Big|_{*},
\qquad i=1,\ldots,r,
\\
\text{with}\;\;
K &= \mathbf{Q}^T B^{-1} A \mathbf{Q}.
\end{split}
\end{align}
Let $\mathbf{HF}$ denote the set
\begin{align}
\mathbf{HF}(\zeta_m, \boldsymbol{\zeta}; \mu_m, \boldsymbol{\nu})
:=
\bigcup_{\mathbf{z}\in {\rm BV}} (\CH,\CF).
\end{align}
Using this set, the twisted partition function on $\CM_{(g,p)}$—the degree-$p$
circle bundle over a genus-$g$ Riemann surface $\Sigma_g$—is given by
\cite{Closset:2017zgf}
\begin{align}
\CZ_{\CM_{(g,p)}}
=
\sum_{(\CH,\CF)\in \mathbf{HF}
}
\CH^{\,g-1}\,\CF^{\,p}.
\end{align}
In this construction, we turn on suitable magnetic fluxes along
$\Sigma_g$ for the $U(1)_R$ symmetry. Due to the Dirac quantization
condition, the set $\mathbf{HF}$ is well defined only when the
$R$-symmetry mixing parameters are integers, namely
\begin{align}
\boldsymbol{\mu} = \boldsymbol{\mu}_{\rm geo} + \mu_m \boldsymbol{v}_m + \sum_{I=1}^{\sharp_H}\nu_I \boldsymbol{v}_I  \in \mathbb{Z}^r \;.
\end{align}
The twisted partition function depends on the choice of spin structure
on $\CM_{(g,p)}$. The formula above corresponds to the spin structure
with anti-periodic boundary conditions along the $S^1$ fiber.
Such a spin structure exists only when $p$ is even, and therefore the
above expression is valid only for even $p$. For this reason, it is more natural to introduce the modified set
\begin{align}
\mathbf{HF}^{2}(\zeta_m, \boldsymbol{\zeta};\mu_m, \boldsymbol{\nu})
:= \bigcup_{\mathbf{z}\in {\rm BV}} (\CH, \CF^{2})\;,
\end{align}
which can be used to compute $\CZ_{\CM_{(g,p)}}$ for all even
$p\in 2\mathbb{Z}$. Alternatively, one may consider the set $\mathbf{HF}$ associated with
periodic boundary conditions along the $S^1$ fiber in order to compute
the twisted partition function for odd $p$.
It turns out that the sets $\mathbf{HF}$ defined using anti-periodic
and periodic boundary conditions coincide provided that the following
condition is satisfied:
\begin{align}
K_{ii} + \boldsymbol{\mu}_i \in 2\mathbb{Z},
\qquad \forall\, i=1,\ldots,r\;.
\end{align}
When this condition holds, the twisted partition function
$\CZ_{\CM_{(g,p)}}$ can be computed using $\mathbf{HF}$ for arbitrary
$p\in \mathbb{Z}$.
Otherwise, one must restrict to the set $\mathbf{HF^2}$, and the
partition function is well defined only for even values
$p\in 2\mathbb{Z}$.
\begin{align}
\begin{pmatrix}
\mathbf{HF}
\\
\mathbf{HF^2}
\end{pmatrix} \textrm{ is well-defined when }  
\begin{pmatrix}
K_{ii}+\mu_i \in 2\mathbb{Z}
\\
\mu_i \in \mathbb{Z} 
\end{pmatrix}\;\; \forall i=1,\ldots,r\;.\label{HF and HF2}
\end{align}
From \eqref{Decouled sector A^{N,p}}, the set $\mathbf{HF}$ (or $\mathbf{HF^2}$) is expected to be  factorized as
\begin{align}
\begin{split}
& (\mathbf{HF} \textrm{ of }  \mathbb{T}[A,B;\mathbf{Q}, \{E_I\}]) = (\mathbf{HF} \textrm{ of } T[S^3\backslash \CK]) \times (\mathbf{HF} \textrm{ of }  \mathcal{A}^{N,p})\;.
\end{split}
\end{align}
For topological field theories or rational conformal field theories,
the sets $\mathbf{HF}$ and $\mathbf{HF}^{2}$ are defined as
\begin{align}
\begin{split}
\mathbf{HF}(\text{TQFT or RCFT})
&\;\simeq\;
\bigcup_{\alpha \in \mathcal{I}}
\left(
S_{0\alpha}^{-2},\; T_{\alpha\alpha}
\right),
\\[4pt]
\mathbf{HF}^{2}(\text{TQFT or RCFT})
&\;\simeq\;
\bigcup_{\alpha \in \mathcal{I}}
\left(
S_{0\alpha}^{-2},\; T_{\alpha\alpha}^{2}
\right).
\end{split}
\end{align}
Here $\mathcal{I}$ labels the set of simple objects in the TQFT,
or equivalently the set of primary fields in the corresponding RCFT.
The matrices $(S,T)$ denote the modular $S$- and $T$-matrices,
and the label $\alpha=0$ corresponds to the identity operator.
The symbol $\simeq$ denotes equality up to an overall phase
in the identification of the $\mathcal{F}$s.
More precisely,
\begin{align}
\begin{split}
\bigl(
\mathbf{HF}
=
\{(\CH_\alpha,\CF_\alpha)\}
\bigr)
\;\simeq\;
\bigl(
\widetilde{\mathbf{HF}}
=
\{(\tilde{\CH}_\alpha,\tilde{\CF}_\alpha)\}
\bigr)
\end{split}
\end{align}
if and only if
\begin{align}
\begin{split}
\CH_\alpha = \tilde{\CH}_\alpha,
\qquad
\CF_\alpha = e^{i\delta}\,\tilde{\CF}_\alpha,
\qquad
\forall\,\alpha \in \mathcal{I},
\end{split}
\label{Equivalence in HF}
\end{align}
for some real phase $\delta$ independent of $\alpha$.

\paragraph{Torus knot case}
The set $\mathbf{HF}$ is independent of $(\zeta_m,\mu_m)$, as expected, because the 
$U(1)_m$ symmetry of $T[S^3\backslash \mathcal{K}_{(P,Q)}]$ always 
decouples in the IR, as stated in \eqref{decoupling of U(1)_m}:
\begin{align}
\mathbf{HF}(\zeta_m,\boldsymbol{\zeta};\mu_m,\boldsymbol{\nu})
\simeq 
\mathbf{HF}(\boldsymbol{\zeta};\boldsymbol{\nu})\;.
\end{align}
For the simplified UV gauge theory 
$T^{(\rm r)}[S^3\backslash \mathcal{K}_{(P,Q)}]$, the set 
$\mathbf{HF}$ can be computed in the same way using the integrand of the 
$S^3_b$ partition function in \eqref{Z3-b for T-red-DGG}.  
Since the the two theories are expected to be IR equivalent up to an decoupled TQFT \eqref{Relation between T and T^r}, we expect that 
\begin{align}
\begin{split}
&\bigl(\mathbf{HF}(\boldsymbol{\zeta};\boldsymbol{\nu})
\text{ of } T[S^3\backslash \mathcal{K}_{(P,Q)}]\bigr) 
\\
& \simeq\ 
\bigl(\mathbf{HF}(\boldsymbol{\zeta};\boldsymbol{\nu})
\text{ of } T^{(\rm r)}[S^3\backslash \mathcal{K}_{(P,Q)}]\bigr) \times (\mathbf{HF} \textrm{ of deocupled Abelian TQFT})\;.
\end{split}
\end{align}
Using the basis $\{\boldsymbol{v}_I\}$ of $\mathfrak{F}[\mathcal{T}]$ in 
\eqref{F[T]_{IR} for torus knots}, the set $\mathbf{HF}$ takes the form
\begin{align}
\begin{split}
&\bigl(\mathbf{HF}(\boldsymbol{\zeta};\boldsymbol{\nu})
\text{ of } T[S^3\backslash \mathcal{K}_{(P,Q)}]\bigr)
\\
&=
\begin{cases}
\mathbf{HF}, 
& |P-Q|=1,\\[0.3em]
\mathbf{HF}(\zeta;\nu),
& \left(Q \equiv \pm1\ (\mathrm{mod}\ P)\right)
  \ \oplus\
  \left(P \equiv \pm1\ (\mathrm{mod}\ Q)\right),\\[0.3em]
\mathbf{HF}(\zeta_1,\zeta_2;\nu_1,\nu_2),
& \left(Q \not\equiv \pm1\ (\mathrm{mod}\ P)\right)
  \ \land\
  \left(P \not\equiv \pm1\ (\mathrm{mod}\ Q)\right).
\end{cases}
\end{split}
\end{align}
When $|P-Q|=1$, using \eqref{IR phase of S3-K_{(P,Q)}} and
\eqref{HF and HF2}, together with the bulk--boundary dictionaries
\cite{Gang:2021hrd,Gang:2023rei}, we expect
\begin{align}
\begin{cases}
\bigl(\mathbf{HF}
\text{ of } T[S^3\backslash \mathcal{K}_{(P,Q)}]\bigr)
\simeq
\bigl(\mathbf{HF} \text{ or } \overline{\mathbf{HF}}
\text{ of } \mathcal{M}(P,Q)\bigr),
&
\text{if } K_{ii}+(\boldsymbol{\mu}_{\rm geom})_i \in 2\mathbb{Z}
\quad \forall\, 1\le i\le r,
\\[6pt]
\bigl(\mathbf{HF}^2
\text{ of } T[S^3\backslash \mathcal{K}_{(P,Q)}]\bigr)
\simeq
\bigl(\mathbf{HF}^2 \text{ or } \overline{\mathbf{HF}^2}
\text{ of } \mathcal{M}(P,Q)\bigr),
&
\text{otherwise}.
\end{cases}
\end{align}
Here we define
\begin{align}
\overline{\mathbf{HF}}
:=
\left(\text{the complex conjugate of }\mathbf{HF}\right).
\end{align}
This computes the $\mathbf{HF}$ associated with the parity-conjugate theory.
In the construction of $T[S^3\backslash \mathcal{K}]$, we are not careful about the
choice of orientation of the knot complement.
As a result, $\mathbf{HF}$ may be identified with its complex conjugate
$\overline{\mathbf{HF}}$, which corresponds to reversing the orientation of the
internal three-manifold.

When
\[
\bigl(Q \equiv \pm 1 \!\!\!\pmod{P}\bigr)
\ \oplus \
\bigl(P \equiv \pm 1 \!\!\!\pmod{Q}\bigr),
\]
we define
\begin{align}
\begin{split}
\mathbf{HF}_{\rm a}
&:= \mathbf{HF}(\zeta;\nu)\Big|_{\zeta=0,\ \nu=\nu_a},
\\
\mathbf{HF}_{\rm b}
&:= \mathbf{HF}(\zeta;\nu)\Big|_{\zeta=0,\ \nu=\nu_b},
\end{split}
\end{align}
where $\nu_a = 0$ and $\nu_b = \pm 2$, as determined by the criterion
in~\eqref{two special points}.
We then expect
\begin{align}
\begin{split}
&\text{if } K_{ii}+(\boldsymbol{\mu}_{\rm geom})_i \in 2\mathbb{Z}
\qquad \forall\, 1\le i\le r,
\\[4pt]
&\bigl(\mathbf{HF}_{\rm a} \text{ or } \mathbf{HF}_{\rm b}
\text{ of } T[S^3\backslash \mathcal{K}_{(P,Q)}]\bigr)
\simeq
\bigl(\mathbf{HF} \text{ or } \overline{\mathbf{HF}}
\text{ of } \mathcal{M}(P,Q)\bigr),
\\[6pt]
&\text{otherwise},
\\[4pt]
&\bigl(\mathbf{HF}^2_{\rm a} \text{ or } \mathbf{HF}^2_{\rm b}
\text{ of } T[S^3\backslash \mathcal{K}_{(P,Q)}]\bigr)
\simeq
\bigl(\mathbf{HF}^2 \text{ or } \overline{\mathbf{HF}^2}
\text{ of } \mathcal{M}(P,Q)\bigr).
\end{split}
\end{align}

When
\[
\bigl(Q \not\equiv \pm 1 \!\!\!\pmod{P}\bigr)
\ \land \
\bigl(P \not\equiv \pm 1 \!\!\!\pmod{Q}\bigr),
\]
we define
\begin{align}
\begin{split}
\mathbf{HF}_{\rm aa/ba/ab/bb}
:=
\mathbf{HF}(\boldsymbol{\zeta};\boldsymbol{\nu})
\Big|_{\boldsymbol{\zeta}=\mathbf{0},\ 
      \boldsymbol{\nu}=
      \boldsymbol{\nu}_{\rm aa}/
      \boldsymbol{\nu}_{\rm ba}/
      \boldsymbol{\nu}_{\rm ab}/
      \boldsymbol{\nu}_{\rm bb}} \; .
\end{split}
\end{align}
Here $\boldsymbol{\nu}_{\rm aa}=\mathbf{0}$, and the remaining
$\boldsymbol{\nu}_{*}$ are defined in~\eqref{four special points}.
We then expect
\begin{align}
\begin{split}
&\text{if } K_{ii}+(\boldsymbol{\mu}_{\rm geom})_i \in 2\mathbb{Z}
\qquad \forall\, 1\le i\le r,
\\[4pt]
&\bigl(\mathbf{HF}_{\rm aa}
\text{ or } \mathbf{HF}_{\rm ba}
\text{ or } \mathbf{HF}_{\rm ab}
\text{ or } \mathbf{HF}_{\rm bb}
\text{ of } T[S^3\backslash \mathcal{K}_{(P,Q)}]\bigr)
\simeq
\bigl(\mathbf{HF} \text{ or } \overline{\mathbf{HF}}
\text{ of } \mathcal{M}(P,Q)\bigr),
\\[6pt]
&\text{otherwise},
\\[4pt]
&\bigl(\mathbf{HF}^2_{\rm aa}
\text{ or } \mathbf{HF}^2_{\rm ba}
\text{ or } \mathbf{HF}^2_{\rm ab}
\text{ or } \mathbf{HF}^2_{\rm bb}
\text{ of } T[S^3\backslash \mathcal{K}_{(P,Q)}]\bigr)
\simeq
\bigl(\mathbf{HF}^2 \text{ or } \overline{\mathbf{HF}^2}
\text{ of } \mathcal{M}(P,Q)\bigr).
\end{split}
\end{align}
\paragraph{Computation of the round three-sphere partition function from the Bethe sum}
If
$K_{ii}+(\boldsymbol{\mu}_{\rm geom})_i \in 2\mathbb{Z}$
for all $i=1,\ldots,r$, the round three-sphere partition function
$\CZ_{S^3_b}$ at $b=1$ can be computed using the set $\mathbf{HF}$ as follows:
\begin{align}
\begin{split}
\CZ_{S^3_{b=1}}[\zeta_m,\boldsymbol{\zeta};\mu_m=0,\boldsymbol{\nu}=\mathbf{0}]
&\simeq
\CZ_{\CM_{(g=0,p=1)}}[\zeta_m,\boldsymbol{\zeta};\mu_m=0,\boldsymbol{\nu}=\mathbf{0}]
\\[0.3em]
&\simeq
\sum_{(\CH,\CF)\,\in\,\mathbf{HF}(\zeta_m,\boldsymbol{\zeta};\mu_m=0,\boldsymbol{\nu}=0)}
\CH^{-1}\,\CF ,
\\[0.4em]
\text{and}\;\;
\CZ_{S^3_{b=1}}[\zeta_m,\boldsymbol{\zeta};\mu_m,\boldsymbol{\nu}]
&=
\CZ_{S^3_{b=1}}
\bigl[\zeta_m+i\pi\mu_m,\;\boldsymbol{\zeta}+i\pi\boldsymbol{\nu};\;\mu_m=0,\;\boldsymbol{\nu}=\mathbf{0}\bigr].
\end{split}
\label{Bethe-sum for S3}
\end{align}
If instead
$K_{ii}+(\boldsymbol{\mu}_{\rm geo})_i \notin 2\mathbb{Z}$
for some $1\leq i\leq r$, the round three-sphere partition function can
still be computed by choosing a reference vector
$\boldsymbol{\mu}_0$ satisfying
$K_{ii}+(\boldsymbol{\mu}_0)_i \in 2\mathbb{Z}$
for all $i=1,\ldots,r$. One then proceeds as follows:
\begin{align}
\begin{split}
&\CZ_{S^3_{b=1}}[\zeta_m,\zeta;\mu_m,\nu]
=
\CZ_{S^3_{b=1}}[\boldsymbol{\xi},\boldsymbol{\mu}]
\Big|_{*}
\;
\text{with the replacement specified in
\eqref{S3b ptn for T_{DGG}}}\;,
\\[0.4em]
&\CZ_{S^3_{b=1}}[\boldsymbol{\xi},\boldsymbol{\mu}]
=
\CZ_{S^3_{b=1}}
\big[\boldsymbol{\xi}+ i\pi(\boldsymbol{\mu}-\boldsymbol{\mu}_0),
\boldsymbol{\mu}_0\big]\;,
\\[0.6em]
&\CZ_{S^3_{b=1}}[\boldsymbol{\xi},\boldsymbol{\mu}_0]
\simeq
\sum_{(\CH,\CF)\,\in\,\mathbf{HF}(\boldsymbol{\xi},\boldsymbol{\mu}_0)}
\CH^{-1}\,\CF \;.
\end{split}
\end{align}
Here $\mathbf{HF}(\boldsymbol{\xi},\boldsymbol{\mu}_0)$ denotes the set of
pairs $(\CH,\CF)$ computed from the asymptotic expansion coefficients
$\CW_0$ and $\CW_1$ of the function
$\CI_\hbar(\mathbf{Z};\boldsymbol{\xi},\boldsymbol{\mu})$
appearing in \eqref{S3-b for tT[A,B;Q]}, following the prescription
outlined in \eqref{BV, F and H}.

\paragraph{Example: $T[S^3\backslash \CK_{(P,Q)}]$ with $(P,Q)=(2,7)$}
Using the integrand of $\mathbb{T}^{[A,B;\mathbf{Q},\{E_I\}]}$ in \eqref{Ih for (2,7)} 
together with the general formula \eqref{BV, F and H}, one can compute the set 
$\mathbf{HF}(\zeta_m, \zeta; \mu_m, \nu)$. 
From the explicit computation, we find:

\begin{align}
\begin{split}
&{\rm i)}\quad 
\mathbf{HF} \text{ is independent of } (\zeta_m, \mu_m)\;, \\[0.4em]
&{\rm ii)}\quad 
\mathbf{HF}_a := \mathbf{HF}(\zeta_m, \zeta = 0; \mu_m = 0, \nu = 0) \\[0.2em]
&\qquad\simeq 
\bigg\{
\Big(\frac{7}{4\sin^2(\tfrac{2\pi}{7})},\; \mathbf{e}(0)\Big),\;
\Big(\frac{7}{4\sin^2(\tfrac{4\pi}{7})},\; \mathbf{e}(\tfrac{2}{7})\Big),\;
\Big(\frac{7}{4\sin^2(\tfrac{6\pi}{7})},\; \mathbf{e}(\tfrac{3}{7})\Big)
\bigg\} 
\\
&\qquad\times 
\bigg\{
(2,\mathbf{e}(0)),\;
\big(2,\mathbf{e}(\tfrac{3}{4})\big)
\bigg\}\;, \\[0.6em]
&{\rm iii)}\quad
\mathbf{HF}_b := \mathbf{HF}(\zeta_m, \zeta = 0; \mu_m = 0, \nu = 2) \\[0.2em]
&\qquad\simeq 
\bigg\{
\Big(\frac{7}{4\sin^2(\tfrac{4\pi}{7})},\; \mathbf{e}(0)\Big),\;
\Big(\frac{7}{4\sin^2(\tfrac{6\pi}{7})},\; \mathbf{e}(\tfrac{4}{7})\Big),\;
\Big(\frac{7}{4\sin^2(\tfrac{2\pi}{7})},\; \mathbf{e}(\tfrac{6}{7})\Big)
\bigg\} 
\\
&\qquad\times 
\bigg\{
(2,\mathbf{e}(0)),\;
\big(2,\mathbf{e}(\tfrac{3}{4})\big)
\bigg\}\;.
\end{split}
\end{align}
We define
\begin{align}
\mathbf{e}(h) := e^{2\pi i h}\;.
\end{align}
The property in (i) is consistent with the fact that the $U(1)_m$ symmetry decouples in 
the infrared.  
In both $\mathbf{HF}_a$ and $\mathbf{HF}_b$, the first factor corresponds to the 
$\mathbf{HF}$ set of $T[S^3\backslash \CK_{(2,7)}]$, while the second factor corresponds 
to the decoupled Abelian TQFT $\CA^{N=2,p}$.  
Therefore, we confirm that
\begin{align}
\big(\mathbf{HF}_a \text{ of } T[S^3\backslash \CK_{(2,7)}]\big)
\;\simeq\;
\big(\overline{\mathbf{HF}} \text{ of } \CM(2,7)\big)\;.
\end{align}
Using the Bethe-sum formula in \eqref{Bethe-sum for S3}, one can compute the 
round three-sphere partition function. 
Its absolute value, as a function of $\nu$, is shown in 
Figure~\ref{fig : ZS3 for (2,7)}.

%% file: section_rVOAfromHI.tex
\section{Rational VOA characters from Half-indices} \label{sec : RVOAs from Half-indices}

Now we compute the half-index \cite{Gadde:2013wq,Yoshida:2014ssa,Dimofte:2017tpi} of $T[S^3\backslash \mathcal{K}_{(P,Q)}]$, or more precisely, of $\mathbb{T}^{[A,B;\mathbf{Q},\{E_I\}]}$, which is identical to $T[S^3\backslash \mathcal{K}_{(P,Q)}]$ up to a decoupled TQFT $\mathcal{A}^{N,p}$. This will make the direct connection with the boundary rational VOAs more transparent.

Unlike the case of bulk BPS partition functions, the computation of the half-index is more subtle, since one must choose appropriate boundary conditions in order to realize a rational VOA at the boundary. Moreover, a single bulk theory may admit several distinct rational VOAs, depending on the choice of boundary condition. From the UV gauge-theory perspective, it is generally unclear which UV boundary condition flows to a given desired IR boundary condition. 

Furthermore, even when the bulk theory factorizes in the infrared as a product,
the half-index does not necessarily exhibit a corresponding factorization.
This is because the choice of boundary conditions can explicitly break the
factorization between the two sectors.
Since such factorization is not manifest in the UV description---and, in
particular, because our choice of Dirichlet boundary conditions on the UV fields
obscures the factorized structure of the IR theory---the decoupling of the Abelian
TQFT $\mathcal{A}^{N,p}$ in \eqref{Decouled sector A^{N,p}} is not automatic at the
level of the half-index computation. For this reason, we focus on the theory
$\mathbb{T}[A,B;\mathbf{Q},\{E_I\}]$ rather than directly on
$T[S^3 \backslash \mathcal{K}]$.
This perspective naturally explains how the bulk theory associated with a
minimal model can be realized as a product of rank-0 SCFTs, as proposed in
\eqref{IR phase of S3-K_{(P,Q)}}.

A naive approach to identifying appropriate boundary conditions is to systematically scan the supersymmetric boundary conditions available in a given UV description, compute the corresponding half-index for each choice, and then identify those boundary conditions for which the resulting half-index reproduces the characters of rational VOAs. In this section, however, we adopt a different strategy. We fix a simple boundary condition—essentially Dirichlet boundary conditions for both vector and chiral multiplets—in the UV gauge theory, and instead vary the duality frame to determine which frame yields half-indices that match rational VOA characters. 
In the construction of $T[S^{3}\backslash \CK]$, different choices of polarization correspond to different duality frames, related by the basic mirror dualities \eqref{mirror duality}. Distinct polarizations lead to different NZ matrices $(A,B)$. The Dirichlet half-index is well defined only when the matrix \(B^{-1}A\) is positive definite.
We therefore scan over the possible choices of polarization and identify those for which
\(B^{-1}A\) becomes positive definite.
We find that, in any duality frame for which $B^{-1}A$ is positive
definite, the Dirichlet half-index consistently yields sensible rational
VOA characters (or trivial results): some reproduce the expected
Virasoro minimal-model characters, while others correspond to different
rational VOAs. A similar strategy was employed in \cite{Gang:2023ggt}.
We now present the main claim of this paper:
\begin{align}
\begin{split}
\textbf{Conjecture:}\quad
&\text{The half-indices in }
\eqref{half-indices when |P-Q|=1-2},\;
\eqref{half-indices at A/B-twistings-2},\;
\eqref{half-indices at AA/AB/BA/BB-twistings-2} \text{ are either } 0,\;1,\;
\\
&
\text{or a character of an RCFT up to an overall factor of } \pm q^{\mathbb{Q}} \; .
\end{split} \label{Main conjecture}
\end{align}
We assume that the ideal triangulation of the torus-knot complement
satisfies the conditions stated in~\eqref{triangulation of torus knots}.
The half-indices compute the Dirichlet half-index of 
\(\mathbb{T}[A,B;\mathbf{Q},\{E_I\}]\) or 
\(\mathbb{T}^{\rm (r)}[A_{\rm r},B_{\rm r};\mathbf{Q}_{\rm r},\{E^{\rm (r)}_I\}]\)
evaluated at the topological twisting points.
The half-index can be \(0\) when the Dirichlet boundary conditions break supersymmetry,
and it can be \(1\) when the UV boundary condition becomes trivial topological boundary condition in the infrared.
We claim that, except for these exceptional cases, the theory supports a rational VOA
at the boundary, and the half-index computes its character.
We will explicitly check this claim through concrete examples in Appendix \ref{Appendix Examples}.

\paragraph{Half-index of $\mathbb{T}[A,B;\mathbf{Q},\{E_I\}]$ } For theory $\mathbb{T}[A,B; \mathbf{Q}]$ in \eqref{tT[A,B]}, the half-index in $(\mathcal{D}, D_c)$ boundary condition\footnote{
In terms of a 2D $\mathcal{N}=(0,2)$ subalgebra of the 3D $\mathcal{N}=2$ superconformal algebra, 
a 3D $\mathcal{N}=2$ chiral multiplet $\Phi$ decomposes into a 2D $(0,2)$ chiral multiplet 
$\Phi_{\rm 2d}$ and a 2D $(0,2)$ Fermi multiplet $\Psi_{\rm 2d}$. 
The deformed Dirichlet boundary condition $D_c$ for $\Phi$ corresponds to imposing
$\Phi_{\rm 2d}=c$ at the boundary, with a nonzero constant $c$. Similarly, a 3D $\mathcal{N}=2$ vector multiplet decomposes into a 2D $(0,2)$ vector multiplet 
and a 2D $(0,2)$ chiral multiplet under the same subalgebra.
The Dirichlet boundary condition $\mathcal{D}$ for the 3D vector multiplet is defined by setting
the 2D $(0,2)$ vector multiplet to zero at the boundary.
} is given as \cite{Dimofte:2017tpi,Jockers:2021omw,Gang:2023rei}

\begin{align}
\CI_{\rm half}^{\mathbb{T}[A,B;\mathbf{Q}]}
[\boldsymbol{x},\boldsymbol{\mu}]
=
\sum_{\substack{
\mathbf{m} = (m_1,\ldots,m_r)^T \in \mathbb{Z}^r \\
\mathbf{Q}\cdot\mathbf{m} \in (\mathbb{Z}_{\ge 0})^r
}}
\frac{
q^{\frac12\, \mathbf{m}^T K\, \mathbf{m}}
\,(-q^{1/2})^{-\mathbf{m}\cdot\boldsymbol{\mu}}
\prod_{i=1}^r x_i^{-m_i}
}{
\displaystyle \prod_{i=1}^r (q)_{[\mathbf{Q}\cdot\mathbf{m}]_i}
}\;.
\end{align}
Here $(q)_n$ denotes the $q$-Pochhammer symbol,
\begin{align}
(q)_n := \prod_{i=1}^{n} (1 - q^{i}) \; .
\end{align}
Here $K = \mathbf{Q}^{T} B^{-1} A \mathbf{Q}$, and the index is well-defined
when $B^{-1} A > 0$.
 The half-index for the $\mathbb{T} [A,B; \mathbf{Q}, \{E_I\}]$ theory is obtained from that of $\mathbb{T} [A,B;\mathbf{Q}] $ 
      by restricting the parameters $\{\boldsymbol{x},\boldsymbol{\mu}\}$ to satisfy conditions in \eqref{constraints on x and mu}. Using the basis in \eqref{basis of F[T]}, we consider the following half index
\begin{align}
\begin{split}
&\CI^{\mathbb{T}[A,B; \mathbf{Q}, \{E_I\}]}_{\rm half}[u, \boldsymbol{\eta} ;\mu_m, \boldsymbol{\nu} ]:= \CI_{\rm half}^{\mathbb{T}[A,B;\mathbf{Q}]}[\boldsymbol{x} ,\boldsymbol{\mu} ]\big{|}_*\;, \label{half-index for T_{DGG}}
\end{split}
\end{align} 
with the replacement $|_*$  given in \eqref{SCI for T_{DGG}},
\begin{align}
&\big{|}_* : \boldsymbol{\mu} \rightarrow  \boldsymbol{\mu}_{\rm geo} + \mu_m \boldsymbol{v}_m + \sum_{I=1}^{\sharp_H}\nu_I \boldsymbol{v}_I , \quad x_i \rightarrow u^{(\boldsymbol{v}_m )_i} \prod_{I=1}^{\sharp_H} \eta_I^{(\boldsymbol{v}_I )_i}\;.  \nonumber
\end{align}
\paragraph{Half-index of $\overline{\mathbb{T}[A,B;\mathbf{Q}, \{E_I\}]}$ } Let \(\overline{T}\) denote the theory obtained by applying parity conjugation to \(T\). In 3D-3D correspondence, $\overline{T[M]}= T[\overline{M}]$, where $\overline{M}$ denotes the orientation reversal of $M$. The half-index of $\overline{\mathbb{T}[A,B;\mathbf{Q}]}$ is given as
\begin{align}
\begin{split}
&\CI_{\rm half}^{\overline{\mathbb{T}[A,B;\mathbf{Q}]}} [\boldsymbol{x}, \boldsymbol{\mu}] =\sum_{\substack{
\mathbf{m} = (m_1,\ldots,m_r)^T \in \mathbb{Z}^r \\
\mathbf{Q}\cdot\mathbf{m} \in (\mathbb{Z}_{\ge 0})^r
}} \frac{q^{\frac{1}2 \mathbf{m}^T \cdot \overline{K} \cdot \mathbf{m} } (-q^{1/2})^{-\mathbf{m}\cdot  \overline{\boldsymbol{\mu}}} \prod_{i=1}^r x_i^{m_i}}{\prod_{i=1}^r (q)_{[\mathbf{Q}\cdot \mathbf{m}]_i}}\;,
\\
&\textrm{with } \overline{K} := \mathbf{Q}^T \mathbf{Q}-K = \mathbf{Q}^T(\mathbb{I}-B^{-1}A) \mathbf{Q}, \; \overline{\boldsymbol{\mu}} := -\mathbf{Q}^T \cdot (1,\ldots, 1)^T - \boldsymbol{\mu}\;.
\end{split}
\end{align}
Under the parity transformation, the effective Chern--Simons data flips sign:
\begin{align}
\overline{K}_{\rm eff} &= - K_{\rm eff}\,, \quad \textrm{where}\;\;
\overline{K}_{\rm eff} = \overline{K} - \frac{1}{2}\,\mathbf{Q}^{T}\mathbf{Q}\;\textrm{ and }
K_{\rm eff} = K - \frac{1}{2}\,\mathbf{Q}^{T}\mathbf{Q}\,.
\end{align}
Similarly,
\begin{align}
\overline{\boldsymbol{\mu}}_{\rm eff} = - \boldsymbol{\mu}_{\rm eff}\,, \;\; \textrm{where}\;\;
\overline{\boldsymbol{\mu}}_{\rm eff} 
= \overline{\boldsymbol{\mu}} + \frac{1}{2}\,\mathbf{Q}^{T}\cdot (1,\ldots,1)^{T} \textrm{ and }
\boldsymbol{\mu}_{\rm eff} 
= \boldsymbol{\mu} + \frac{1}{2}\,\mathbf{Q}^{T}\cdot (1,\ldots,1)^{T}\,.
\end{align}
See \eqref{Background CS terms} for the background mixed Chern--Simons levels, which account for the offset between \(K_{\rm eff}\) and \(K\).
Note that \(\boldsymbol{\mu}\) can be regarded as the mixed Chern--Simons coupling
between the \(U(1)_R\) symmetry and the dynamical \(U(1)^r\) gauge symmetry.

The half-index is well-defined  when $(\mathbb{I}-B^{-1}A)>0$. Then, the half-index for $\overline{\mathbb{T}[A,B;\mathbf{Q},\{E_I\}]}$ is given as
\begin{align}
\begin{split}
&\CI^{\overline{\mathbb{T}[A,B;\mathbf{Q},\{E_I\}]}}_{\rm half}[u, \boldsymbol{\eta} ;\mu_m, \boldsymbol{\nu} ]:= \CI_{\rm half}^{\overline{\mathbb{T}[A,B;\mathbf{Q}]}}[\boldsymbol{x} ,\boldsymbol{\mu} ]\big{|}_*\;, \label{half-index for bT_{DGG}}
\end{split}
\end{align} 
 with the replacement $|_*$  given in \eqref{SCI for T_{DGG}}.

\paragraph{Torus knot case} Using the basis
$\{\boldsymbol{v}_m,\boldsymbol{v}_I\}$ of $\mathfrak{F}[\mathcal{T}]$
 in~\eqref{sharp_E and sharp_H for T[torus]-2}, the half-indices can be given as
\begin{align}
\begin{split}
&  \CI^{\mathbb{T}[A,B;\mathbf{Q},\{E_I\}]}_{\rm half}
[ \boldsymbol{\eta} ; \mu_m, \boldsymbol{\nu} ]
=
\CI^{\mathbb{T}[A,B;\mathbf{Q},\{E_I\}]}_{\rm half}
[u=1,  \boldsymbol{\eta} ;\mu_m, \boldsymbol{\nu} ]\;,
\\
& \CI^{\overline{\mathbb{T}[A,B;\mathbf{Q},\{E_I\}]}}_{\rm half}
[ \boldsymbol{\eta} ;\mu_m,  \boldsymbol{\nu} ]
=
\CI^{\overline{\mathbb{T}[A,B;\mathbf{Q},\{E_I\}]}}_{\rm half}
[u=1,  \boldsymbol{\eta} ;\mu_m, \boldsymbol{\nu} ]\;.
\end{split}
\end{align}
 The $U(1)_m$ symmetry decouples in the infrared, and we turn off the corresponding fugacity, $u=1$. We nevertheless keep the parameter $\mu_m$, which parametrizes the mixing
between the $U(1)_R$ symmetry and the $U(1)_m$ symmetry.
All bulk partition functions are independent of $\mu_m$
(up to an overall phase factor), and one may therefore set $\mu_m=0$
at the level of bulk observables.
However, the half-index computation is more subtle.
Some imprint of the UV $U(1)_m$ symmetry can survive in the infrared,
especially when the IR theory contains a decoupled TQFT sector,
on which a discrete subgroup of $U(1)_m$ may act nontrivially. The parameter $\mu_m$ can be regarded as specifying a choice of the geometric
$R$-symmetry mixed with the $U(1)_m$ symmetry,
\begin{align}
R_{\rm geo}\;\longrightarrow\; R_{\rm geo} + \mu_m\, T_m\;,
\end{align}
where $T_m$ is the generator of $U(1)_m$.
The parameter $\mu_m$ cannot take arbitrary real values,
since $R_{\rm geo}$ must be properly quantized,
as discussed in~\eqref{quantization of R_{geo}}. The generator $T_m$ is embedded into the $SU(2)_m$ algebra as $J_3$, $T_m = J_3$\;,
with the normalization $J_3 \in \mathbb{Z}/2$ \cite{Gang:2018wek}.
Consequently, the admissible values of $\mu_m$ are restricted to
\begin{align}
\mu_m \in 2\mathbb{Z}\;.
\end{align}
In the following, we will compute the half-index with $\mu_m\in 2\mathbb{Z}$ turned on.

For the reduced UV gauge theory
$\mathbb{T}[A_{\rm r},B_{\rm r};\mathbf{Q}_{\rm r},\{E_I^{(\rm r)}\}]$ in~\eqref{T-red-DGG}, the half-index is given by
(note that $\sharp_H^{(\rm r)}=\sharp_H$)
\begin{align}
\begin{split}
\CI^{\mathbb{T}[A_{\rm r},B_{\rm r};\mathbf{Q}_{\rm r}, \{E_I^{(\rm r)}\}]}_{\rm half}
[ \boldsymbol{\eta} ; \boldsymbol{\nu} ]
&=
\CI^{\mathbb{T}[A_{\rm r},B_{\rm r};\mathbf{Q}_{\rm r}]}_{\rm half}
[ \boldsymbol{x} ; \boldsymbol{\mu} ]\Big|_* ,
\\
\text{with the replacement }\Big|_*:\qquad
\boldsymbol{\mu}
&\;\longrightarrow\;
\boldsymbol{\mu}_{\rm geo}
+ \sum_{I=1}^{\sharp_H}\nu_I\,\boldsymbol{v}^{(\rm r)}_I,
\qquad
x_i \;\longrightarrow\;
\prod_{I=1}^{\sharp_H}\eta_I^{(\boldsymbol{v}^{(\rm r)}_I)_i}\; .
\end{split}
\end{align}
Here $\boldsymbol{v}^{(\rm r)}_{I}$ is defined in~\eqref{v_i-red}.
Similarly, for  its parity conjugation
\begin{align}
\begin{split}
\CI^{\overline{\mathbb{T}[A_{\rm r},B_{\rm r};\mathbf{Q}_{\rm r}, \{E_I^{(\rm r)}\}]}}_{\rm half}
[ \boldsymbol{\eta} ; \boldsymbol{\nu} ]
&=
\CI^{\overline{\mathbb{T}[A_{\rm r},B_{\rm r};\mathbf{Q}_{\rm r}]}}_{\rm half}
[ \boldsymbol{x} ; \boldsymbol{\mu} ]\Big|_* ,
\\
\text{with the replacement }\Big|_*:\qquad
\boldsymbol{\mu}
&\;\longrightarrow\;
\boldsymbol{\mu}^{(\rm r)}_{\rm geo}
+ \sum_{I=1}^{\sharp_H}\nu_I\,\boldsymbol{v}^{(\rm r)}_I,
\qquad
x_i \;\longrightarrow\;
\prod_{I=1}^{\sharp_H}\eta_I^{(\boldsymbol{v}^{(\rm r)}_I)_i}\; .
\end{split}
\end{align}
\paragraph{The case when $|P-Q|=1$}
We introduce a shorthand notation for the four half-indices:
\begin{align}
\begin{split}
\CI_{\rm half}[\mu_m]
&:=
\sum_{\substack{
\mathbf{m}=(m_1,\ldots,m_r)^T \in \mathbb{Z}^r \\
\mathbf{Q}\cdot \mathbf{m}\in (\mathbb{Z}_{\ge 0})^r
}}
\frac{
q^{\frac12\, \mathbf{m}^T K \mathbf{m}}
\,(-q^{1/2})^{-\mathbf{m}^T(\boldsymbol{\mu}_{\rm geo}+\mu_m \boldsymbol{v}_m)}
}{
\prod_{i=1}^r (q)_{[\mathbf{Q}\cdot\mathbf{m}]_i}
}\;,
\\
&\hspace{1.5em}\text{with }\;
K:=\mathbf{Q}^T B^{-1}A\,\mathbf{Q}\;,
\;\;
\boldsymbol{\mu}_{\rm geo}+\mu_m \boldsymbol{v}_m
:=\mathbf{Q}^T B^{-1}\!\left(\boldsymbol{\nu}_x+(\mu_m,\mathbf{0}_{r-1})^T\right)\;,
\\[0.4em]
\overline{\CI_{\rm half}}[\mu_m]
&:=
\sum_{\substack{
\mathbf{m}\in \mathbb{Z}^r \\
\mathbf{Q}\cdot \mathbf{m}\in (\mathbb{Z}_{\ge 0})^r
}}
\frac{
q^{\frac12\, \mathbf{m}^T \overline{K}\,\mathbf{m}}
\,(-q^{1/2})^{-\mathbf{m}^T(\overline{\boldsymbol{\mu}}_{\rm geo}-\mu_m \boldsymbol{v}_m)}
}{
\prod_{i=1}^r (q)_{[\mathbf{Q}\cdot\mathbf{m}]_i}
}\;,
\\
&\hspace{1.5em}\text{with }\;
\overline{K}:=\mathbf{Q}^T\mathbf{Q}-K\;,
\;\;
\overline{\boldsymbol{\mu}}_{\rm geo}
:=-\mathbf{Q}^T(1,\ldots,1)^T-\boldsymbol{\mu}_{\rm geo}\;,
\\[0.6em]
\CI^{(\rm r)}_{\rm half}
&:=
\sum_{\substack{
\mathbf{m}\in \mathbb{Z}^{r-2} \\
\mathbf{Q}_{\rm r}\cdot \mathbf{m}\in (\mathbb{Z}_{\ge 0})^{r-2}
}}
\frac{
q^{\frac12\, \mathbf{m}^T K_{\rm r}\mathbf{m}}
\,(-q^{1/2})^{-\mathbf{m}^T \boldsymbol{\mu}^{(\rm r)}_{\rm geo}}
}{
\prod_{i=1}^{r-2} (q)_{[\mathbf{Q}_{\rm r}\cdot\mathbf{m}]_i}
}\;,
\\
&\hspace{1.5em}\text{with }\;
K_{\rm r}:=\mathbf{Q}_{\rm r}^T B_{\rm r}^{-1}A_{\rm r}\mathbf{Q}_{\rm r}\;,
\;\;
\boldsymbol{\mu}^{(\rm r)}_{\rm geo}
:=\mathbf{Q}_{\rm r}^T B_{\rm r}^{-1}\boldsymbol{\nu}_{x,{\rm r}}\;,
\\[0.4em]
\overline{\CI^{(\rm r)}_{\rm half}}
&:=
\sum_{\substack{
\mathbf{m}\in \mathbb{Z}^{r-2} \\
\mathbf{Q}_{\rm r}\cdot \mathbf{m}\in (\mathbb{Z}_{\ge 0})^{r-2}
}}
\frac{
q^{\frac12\, \mathbf{m}^T \overline{K}_{\rm r}\mathbf{m}}
\,(-q^{1/2})^{-\mathbf{m}^T \overline{\boldsymbol{\mu}}^{(\rm r)}_{\rm geo}}
}{
\prod_{i=1}^{r-2} (q)_{[\mathbf{Q}_{\rm r}\cdot\mathbf{m}]_i}
}\;,
\\
&\hspace{1.5em}\text{with }\;
\overline{K}_{\rm r}:=\mathbf{Q}_{\rm r}^T\mathbf{Q}_{\rm r}-K_{\rm r}\;,
\;\;
\overline{\boldsymbol{\mu}}^{(\rm r)}_{\rm geo}
:=-\mathbf{Q}_{\rm r}^T(1,\ldots,1)^T-\boldsymbol{\mu}^{(\rm r)}_{\rm geo}\;.
\label{half-indices when |P-Q|=1}
\end{split}
\end{align}
These are the half-indices for
$\mathbb{T}[A,B;\mathbf{Q},\{E_I\}]$,
$\overline{\mathbb{T}[A,B;\mathbf{Q},\{E_I\}]}$,
$\mathbb{T}[A_{\rm r},B_{\rm r};\mathbf{Q}_{\rm r},\{E^{(\rm r)}_I\}]$, and
$\overline{\mathbb{T}[A_{\rm r},B_{\rm r};\mathbf{Q}_{\rm r},\{E^{(\rm r)}_I\}]}$,
respectively.

We expect that the following half-indices, up to an overall prefactor of the form
$\pm q^{\mathbb{Q}}$, provide candidates for characters of rational VOAs:
\begin{align}
\begin{split}
&\CI_{\rm half}\,[\mu_m \in 2\mathbb{Z}]
     \quad \text{when } B^{-1}A>0 \;, 
\qquad
\overline{\CI_{\rm half}}\,[\mu_m \in 2\mathbb{Z}]
     \quad \text{when } (\mathbb{I}-B^{-1}A)>0 \;, 
\\
&\CI^{(\rm r)}_{\rm half}
     \quad \text{when } B_{\rm r}^{-1}A_{\rm r}>0 \;, 
\qquad
\overline{\CI^{(\rm r)}_{\rm half}}
     \quad \text{when } (\mathbb{I}-B_{\rm r}^{-1}A_{\rm r})>0 \;.
\label{half-indices when |P-Q|=1-2}
\end{split}
\end{align}
For a fixed ideal triangulation $\mathcal{T}$ of a torus-knot complement,
varying the choice of polarization \eqref{polarization choices} for each tetrahedron
may lead to Neumann--Zagier matrices $(A,B)$ that satisfy one of the above positivity
conditions. In the examples we have examined, there are typically several admissible
choices. Although $\mu_m$ can take arbitrary values in $2\mathbb{Z}$, the resulting
inequivalent half-indices form only a finite set.

\paragraph{The case when
$\big(Q \equiv \pm1\ (\mathrm{mod}\ P)\big)\oplus
\big(P \equiv \pm1\ (\mathrm{mod}\ Q)\big)$}
We define
\begin{align}
\begin{split}
\CI^{a/b}_{\rm half}[\mu_m]
&:=
\sum_{\substack{
\mathbf{m}=(m_1,\ldots,m_r)^T \in \mathbb{Z}^r \\
\mathbf{Q}\cdot\mathbf{m}\in (\mathbb{Z}_{\ge 0})^r
}}
\frac{
q^{\frac12\, \mathbf{m}^T K \mathbf{m}}
\,(-q^{1/2})^{-\mathbf{m}^T\!\left(\boldsymbol{\mu}_{\rm geo}+\mu_m \boldsymbol{v}_m+\nu \boldsymbol{v}_1\right)}
}{
\prod_{i=1}^r (q)_{[\mathbf{Q}\cdot\mathbf{m}]_i}
}\Bigg|_{\nu=\nu_{a/b}}\;,
\\
&\hspace{1.5em}\text{with }\;
K:=\mathbf{Q}^T B^{-1}A\,\mathbf{Q}\;,
\;\;
\boldsymbol{\mu}_{\rm geo}+\mu_m \boldsymbol{v}_m+\nu \boldsymbol{v}_1
:=\mathbf{Q}^T B^{-1}\!\left(\boldsymbol{\nu}_x+(\mu_m,\nu,\mathbf{0}_{r-2})^T\right)\;,
\\[0.4em]
\overline{\CI^{a/b}_{\rm half}}[\mu_m]
&:=
\sum_{\substack{
\mathbf{m}\in \mathbb{Z}^r \\
\mathbf{Q}\cdot\mathbf{m}\in (\mathbb{Z}_{\ge 0})^r
}}
\frac{
q^{\frac12\, \mathbf{m}^T \overline{K}\,\mathbf{m}}
\,(-q^{1/2})^{-\mathbf{m}^T\!\left(\overline{\boldsymbol{\mu}}_{\rm geo}-\mu_m \boldsymbol{v}_m-\nu \boldsymbol{v}_1\right)}
}{
\prod_{i=1}^r (q)_{[\mathbf{Q}\cdot\mathbf{m}]_i}
}\Bigg|_{\nu=\nu_{a/b}}\;,
\\
&\hspace{1.5em}\text{with }\;
\overline{K}:=\mathbf{Q}^T\mathbf{Q}-K\;,
\;\;
\overline{\boldsymbol{\mu}}_{\rm geo}
:=-\mathbf{Q}^T(1,\ldots,1)^T-\boldsymbol{\mu}_{\rm geo}\;,
\\[0.6em]
\CI^{a/b;(\rm r)}_{\rm half}
&:=
\sum_{\substack{
\mathbf{m}\in \mathbb{Z}^{r-2} \\
\mathbf{Q}_{\rm r}\cdot\mathbf{m}\in (\mathbb{Z}_{\ge 0})^{r-2}
}}
\frac{
q^{\frac12\, \mathbf{m}^T K_{\rm r}\mathbf{m}}
\,(-q^{1/2})^{-\mathbf{m}^T\!\left(\boldsymbol{\mu}^{(\rm r)}_{\rm geo}+\nu \boldsymbol{v}^{(\rm r)}_1\right)}
}{
\prod_{i=1}^{r-2} (q)_{[\mathbf{Q}_{\rm r}\cdot\mathbf{m}]_i}
}\Bigg|_{\nu=\nu_{a/b}}\;,
\\
&\hspace{1.5em}\text{with }\;
K_{\rm r}:=\mathbf{Q}_{\rm r}^T B_{\rm r}^{-1}A_{\rm r}\mathbf{Q}_{\rm r}\;,
\;\;
\boldsymbol{\mu}^{(\rm r)}_{\rm geo}
:=\mathbf{Q}_{\rm r}^T B_{\rm r}^{-1}\boldsymbol{\nu}_{x,{\rm r}}\;,
\qquad
\boldsymbol{v}^{(\rm r)}_1:=\mathbf{Q}_{\rm r}^T B_{\rm r}^{-1}(1,\mathbf{0}_{r-3})^T\;,
\\[0.4em]
\overline{\CI^{a/b;(\rm r)}_{\rm half}}
&:=
\sum_{\substack{
\mathbf{m}\in \mathbb{Z}^{r-2} \\
\mathbf{Q}_{\rm r}\cdot\mathbf{m}\in (\mathbb{Z}_{\ge 0})^{r-2}
}}
\frac{
q^{\frac12\, \mathbf{m}^T \overline{K}_{\rm r}\mathbf{m}}
\,(-q^{1/2})^{-\mathbf{m}^T\!\left(\overline{\boldsymbol{\mu}}^{(\rm r)}_{\rm geo}-\nu \boldsymbol{v}^{(\rm r)}_1\right)}
}{
\prod_{i=1}^{r-2} (q)_{[\mathbf{Q}_{\rm r}\cdot\mathbf{m}]_i}
}\Bigg|_{\nu=\nu_{a/b}}\;,
\\
&\hspace{1.5em}\text{with }\;
\overline{K}_{\rm r}:=\mathbf{Q}_{\rm r}^T\mathbf{Q}_{\rm r}-K_{\rm r}\;,
\;\;
\overline{\boldsymbol{\mu}}^{(\rm r)}_{\rm geo}
:=-\mathbf{Q}_{\rm r}^T(1,\ldots,1)^T-\boldsymbol{\mu}^{(\rm r)}_{\rm geo}\;.
\label{half-indices at A/B-twistings}
\end{split}
\end{align}
Here $\nu_a=0$ and $\nu_b=\pm2$ are the special values defined in~\eqref{two special points}.
These are the half-indices for the four theories evaluated at the $A$- and $B$-twisting limits.
We expect that the following half-indices, up to an overall prefactor of the form $\pm q^{\mathbb{Q}}$,
are candidates for characters of rational VOAs:

\begin{align}
\begin{split}
&\CI^{a/b}_{\rm half} [\mu_m \in 2\mathbb{Z}] \textrm{ when }B^{-1}A >0\;, \;\;\;\;\;\;\overline{\CI^{a/b}_{\rm half}}[\mu_m \in 2\mathbb{Z}] \textrm{ when }(\mathbb{I}-B^{-1}A) >0\;,
\\
&\CI^{a/b;{(\rm r)}}_{\rm half} \textrm{ with }B_{\rm r}^{-1}A_{\rm r} >0\;,\;\;\;\;\;\;\overline{\CI^{a/b;{(\rm r)}}_{\rm half}}\textrm{ with }(\mathbb{I}-B_{\rm r}^{-1}A_{\rm r}) >0\;. \label{half-indices at A/B-twistings-2}
\end{split}
\end{align}
 \paragraph{Example : $(P,Q)=(2,7)$} We use the reduced gluing equations in \eqref{Example : (2,7)-2} with  $\mathbf{X}_{\rm r} = (Z_1,Z_3)$ and $\mathbf{P}_{\rm r} = (Z_1'',Z_3'')$.  Then, $(B^{-1} A)_{\rm r} $ is not positive definite but  $(\mathbb{I}- B^{-1} A)_{\rm r} $ is positive definite. With $\mathbf{Q}_{\rm r}=\mathbb{I}$,\; 
 \begin{align}
 \begin{split}
 &\overline{K}_{\rm r} = \mathbb{I} -(B^{-1}A)_{\rm r} = \left(
\begin{array}{cc}
 2 & 2 \\
 2 & 4 \\
\end{array}
\right)\;,
\\
& \overline{\boldsymbol{\mu}}^{\rm (r)}_{\rm geo} = (-1,-1)^{T} - (B^{-1}\cdot \boldsymbol{\nu}_x)_{\rm r} = (-2,-4)^T\;,
\\
&\boldsymbol{v}_1^{(\rm r)} = B_{\rm r}^{-1}\cdot (1,0)^T = (-1,-2)^T\;, \; \nu_b =2\;,
\end{split}
     \end{align}
Thus,
\begin{align}
\begin{split}
 &\overline{\CI^{\rm a;{(\rm r)}}_{\rm half}}= \sum_{m_1, m_2 \geq 0} \frac{q^{m_1^2+2m_1 m_2 + 2 m_2^2+ m_1+2m_2}}{(q)_{m_1}(q)_{m_2}} = q^{-17/42} \chi^{\CM(2,7)}_{(1,1)}\;,
 \\
 & \overline{\CI^{\rm b;{(\rm r)}}_{\rm half}}= \sum_{m_1, m_2 \geq 0} \frac{q^{m_1^2+2m_1 m_2 + 2 m_2^2}}{(q)_{m_1}(q)_{m_2}} = q^{1/42} \chi^{\CM(2,7)}_{(1,3)}\;.
 \end{split}
\end{align}
\paragraph{The case
\(\bigl(Q \not\equiv \pm 1 \ (\mathrm{mod}\ P)\bigr)\land
\bigl(P \not\equiv \pm 1 \ (\mathrm{mod}\ Q)\bigr)\).}
In this case, we work in a duality frame (i.e.\ a choice of polarization) in which
the factorization property is manifest.

For the theory \(\mathbb{T}[A,B;\mathbf{Q},\{E_I\}]\), we choose a polarization such that
there exists a charge matrix of the factorized form
\(\mathbf{Q}=\mathbf{Q}^{(1)}\oplus \mathbf{Q}^{(2)}\),
with \(\mathbf{Q}^{(1)}\) and \(\mathbf{Q}^{(2)}\) of sizes
\(r_1\times r_1\) and \(r_2\times r_2\), respectively, and
\(r_1+r_2=r\),
obeying the relation in~\eqref{NQB relation}, for which the resulting data
\((K,\boldsymbol{v}_m,\boldsymbol{v}_1,\boldsymbol{v}_2)\)
exhibit the following factorization property\footnote{One may relax the condition in the second line and consider the most general case,
including the possibility that one of \(r_1\) or \(r_2\) vanishes.
Even in this situation, we expect that the half-indices evaluated at the four twisting points
serve as candidates for r-VOA characters.
However, since we do not yet have a sufficient number of case studies in the case where the IR theory is given by a product of two rank-$0$ SCFTs,
we do not include this possibility among the assumptions of the conjecture
in~\eqref{Main conjecture}.
 Our conjecture is primarily motivated by the case studies presented in this paper
and in~\cite{Gang:2023ggt,Gang:2024loa}, and may therefore require further
refinements or modifications in future work.
 }:
\begin{align}
\begin{split}
&K = K^{(1)} \oplus K^{(2)},\;\;
\boldsymbol{v}_m = \mathbf{0}_{r_1}\oplus \boldsymbol{v}^{(2)}_m
\ \ \text{or}\ \
\boldsymbol{v}^{(1)}_m \oplus \mathbf{0}_{r_2},
\\[2pt]
&\text{and  } \exists \;\alpha, \beta \in \mathbb{Q}\text{ such that }
\bigl(\boldsymbol{v}_1+\alpha\,\boldsymbol{v}_m,\ 
\boldsymbol{v}_2+\beta\,\boldsymbol{v}_m\bigr)
=
\bigl(\boldsymbol{v}^{(1)}\oplus \mathbf{0}_{r_2},\ 
\mathbf{0}_{r_1}\oplus \boldsymbol{v}^{(2)}\bigr),
\\
&\text{with }\boldsymbol{v}^{(1)}\in\mathbb{Z}^{r_1},\;\;
\boldsymbol{v}^{(2)}\in\mathbb{Z}^{r_2}.
\end{split}
\label{factorization polarization}
\end{align}
Here
\(
K:=\mathbf{Q}^T B^{-1} A\,\mathbf{Q}
\),
and the vectors \((\boldsymbol{v}_m,\boldsymbol{v}_1,\boldsymbol{v}_2)\) are defined
as in~\eqref{basis of F[T]} by
\[
\mu_m\boldsymbol{v}_m+\nu_1\boldsymbol{v}_1+\nu_2\boldsymbol{v}_2
:=
\mathbf{Q}^T B^{-1}(\mu_m,\nu_1,\nu_2,\mathbf{0}_{r-3})^T.
\]
Then, in the case
\(\boldsymbol{v}_m=\boldsymbol{v}^{(1)}_m\oplus \mathbf{0}_{r_2}\),
we define
\begin{align}
\begin{split}
&\CI^{a/b,(1)}_{\rm half}[\mu_m]
:=\sum_{\substack{
\mathbf{m}  \in \mathbb{Z}^{r_1} \\
\mathbf{Q}^{(1)}\cdot\mathbf{m} \in (\mathbb{Z}_{\ge 0})^{r_1}
}}
\frac{
q^{\frac12\, \mathbf{m}^T K^{(1)} \mathbf{m}}
\,(-q^{1/2})^{-\mathbf{m}^T \left(\boldsymbol{\mu}_{\rm geo}^{(1)} + \mu_m  \boldsymbol{v}^{(1)}_m+ \nu \boldsymbol{v}^{(1)}\right)}
}{
\prod_{i=1}^{r_1} (q)_{[\mathbf{Q}^{(1)}\cdot\mathbf{m}]_i}
}\bigg{|}_{\nu = \boldsymbol{\nu}_{aa/ba}\cdot (1,0)^T}\;,
\\
&\CI^{a/b,(2)}_{\rm half}
:=\sum_{\substack{
\mathbf{m}  \in \mathbb{Z}^{r_2} \\
\mathbf{Q}^{(1)}\cdot\mathbf{m} \in (\mathbb{Z}_{\ge 0})^{r_2}
}}
\frac{
q^{\frac12\, \mathbf{m}^T K^{(2)} \mathbf{m}}
\,(-q^{1/2})^{-\mathbf{m}^T \left(\boldsymbol{\mu}_{\rm geo}^{(2)} +  \nu \boldsymbol{v}^{(2)}\right)}
}{
\prod_{i=1}^{r_2} (q)_{[\mathbf{Q}^{(2)}\cdot\mathbf{m}]_i}
}\bigg{|}_{\boldsymbol{\nu} = \boldsymbol{\nu}_{aa/ab}\cdot (0,1)^T}\;,
\\
&\overline{\CI^{a/b,(1)}_{\rm half}}[\mu_m]
:=\sum_{\substack{
\mathbf{m}  \in \mathbb{Z}^{r_1} \\
\mathbf{Q}^{(1)}\cdot\mathbf{m} \in (\mathbb{Z}_{\ge 0})^{r_1}
}}
\frac{
q^{\frac12\, \mathbf{m}^T \overline{K^{(1)}} \mathbf{m}}
\,(-q^{1/2})^{-\mathbf{m}^T \left(\overline{\boldsymbol{\mu}}_{\rm geo}^{(1)} - \mu_m  \boldsymbol{v}^{(1)}_m - \nu \boldsymbol{v}^{(1)}\right)}
}{
\prod_{i=1}^{r_1} (q)_{[\mathbf{Q}^{(1)}\cdot\mathbf{m}]_i}
}\bigg{|}_{\nu = \boldsymbol{\nu}_{aa/ba}\cdot (1,0)^T}\;,
\\
&\overline{\CI^{a/b,(2)}_{\rm half}}
:=\sum_{\substack{
\mathbf{m}  \in \mathbb{Z}^{r_2} \\
\mathbf{Q}^{(1)}\cdot\mathbf{m} \in (\mathbb{Z}_{\ge 0})^{r_2}
}}
\frac{
q^{\frac12\, \mathbf{m}^T \overline{K^{(2)}} \mathbf{m}}
\,(-q^{1/2})^{-\mathbf{m}^T \left(\overline{\boldsymbol{\mu}}_{\rm geo}^{(2)} -  \nu \boldsymbol{v}^{(2)}\right)}
}{
\prod_{i=1}^{r_2} (q)_{[\mathbf{Q}^{(2)}\cdot\mathbf{m}]_i}
}\bigg{|}_{\boldsymbol{\nu} = \boldsymbol{\nu}_{aa/ab}\cdot (0,1)^T}\;,
\\
&\textrm{with } \overline{K^{(i)}} := (\mathbf{Q}^{(i)})^T  \mathbf{Q}^{(i)} - K^{(i)}, \quad \overline{\boldsymbol{\mu}}_{\rm geo}^{(i)} := -\mathbf{Q}^{(i)}\cdot (1,\ldots, 1)^T - \boldsymbol{\mu}_{\rm geo}^{(i)}\;.
\label{half-indices at AA/AB/BA/BB-twistings}
\end{split}
\end{align}
Here
\(
\boldsymbol{\mu}_{\rm geo}
:= \mathbf{Q}^T B^{-1} \boldsymbol{\nu}_x
= \boldsymbol{\mu}_{\rm geo}^{(1)} \oplus \boldsymbol{\mu}_{\rm geo}^{(2)}.
\)
Here  \(\boldsymbol{\nu}_{aa}=\mathbf{0}\), while the remaining
\(\boldsymbol{\nu}_*\) are defined as in~\eqref{four special points}.
Similarly, for the case
\(\boldsymbol{v}_m= \mathbf{0}_{r_1} \oplus \boldsymbol{v}^{(2)}_m\),
one can define
\(\CI^{a/b,(1)}_{\rm half}\),
\(\CI^{a/b,(2)}_{\rm half}[\mu_m]\),
\(\overline{\CI^{a/b,(1)}_{\rm half}}\),
and
\(\overline{\CI^{a/b,(2)}_{\rm half}}[\mu_m]\).

For the reduced theory
\(\mathbb{T}[A_{\rm r}, B_{\rm r}; \mathbf{Q}_{\rm r}, \{E_I^{\rm (r)}\}]\),
we choose a polarization such that there exists a factorized charge matrix
\(\mathbf{Q}_{\rm r} = \mathbf{Q}^{(1)}_{\rm r}\oplus \mathbf{Q}^{(2)}_{\rm r}\)
satisfying the relation in~\eqref{NQB relation}, for which the resulting data
\((K_{\rm r},\boldsymbol{v}^{\rm (r)}_1,\boldsymbol{v}^{\rm (r)}_2)\)
obey the following factorization properties ($r-2=r_1+r_2$):
\begin{align}
\begin{split}
&K_{\rm r} = K^{(1)}_{\rm r} \oplus K^{(2)}_{\rm r} \textrm{ and }
\;\;
\\
&(\boldsymbol{v}^{\rm (r)}_1 , \boldsymbol{v}^{\rm (r)}_2)
=
(\boldsymbol{v}^{(1);({\rm r})} \oplus \mathbf{0}_{r_2},
\mathbf{0}_{r_1} \oplus \boldsymbol{v}^{(2);({\rm r})})  \textrm{ with }\boldsymbol{v}^{(1);({\rm r})} \in \mathbb{Z}^{r_1}, \; \boldsymbol{v}^{(2);({\rm r})} \in \mathbb{Z}^{r_2},
\\[2pt]
&\text{where}\;\;
K_{\rm r} := (\mathbf{Q}^T B^{-1} A \mathbf{Q})_{\rm r},
\;\;
\nu_1 \boldsymbol{v}^{\rm (r)}_1+\nu_2\boldsymbol{v}^{\rm (r)}_2
:=
\mathbf{Q}_{\rm r}^T B^{-1}_{\rm r}
(\nu_1,\nu_2,\mathbf{0}_{r-4})^T .
\end{split}
\label{factorization polarization-red}
\end{align}
Then, we define 
\begin{align}
\begin{split}
&\CI^{a/b,(1);({\rm r})}_{\rm half}
:=\sum_{\substack{
\mathbf{m}  \in \mathbb{Z}^{r_1} \\
\mathbf{Q}^{(1)}_{\rm r}\cdot\mathbf{m} \in (\mathbb{Z}_{\ge 0})^{r_1}
}}
\frac{
q^{\frac12\, \mathbf{m}^T K^{(1)}_{\rm r} \mathbf{m}}
\,(-q^{1/2})^{-\mathbf{m}^T \left(\boldsymbol{\mu}_{\rm geo}^{(1),({\rm r})} + \nu \boldsymbol{v}^{(1);({\rm r})}\right)}
}{
\prod_{i=1}^{r_1} (q)_{[\mathbf{Q}^{(1)}\cdot\mathbf{m}]_i}
}\bigg{|}_{\nu = \boldsymbol{\nu}_{aa/ba}\cdot (1,0)^T}\;,
\\
&\CI^{a/b,(2);({\rm r})}_{\rm half}
:=\sum_{\substack{
\mathbf{m}  \in \mathbb{Z}^{r_2} \\
\mathbf{Q}^{(2)}_{\rm r}\cdot\mathbf{m} \in (\mathbb{Z}_{\ge 0})^{r_2}
}}
\frac{
q^{\frac12\, \mathbf{m}^T K^{(1)}_{\rm r} \mathbf{m}}
\,(-q^{1/2})^{-\mathbf{m}^T \left(\boldsymbol{\mu}_{\rm geo}^{(2),({\rm r})} + \nu \boldsymbol{v}^{(2);({\rm r})}\right)}
}{
\prod_{i=1}^{r_1} (q)_{[\mathbf{Q}^{(2)}\cdot\mathbf{m}]_i}
}\bigg{|}_{\nu = \boldsymbol{\nu}_{aa/ab}\cdot (0,1)^T}\;,
\\
&\overline{\CI^{a/b,(1);({\rm r})}_{\rm half}}
:=\sum_{\substack{
\mathbf{m}  \in \mathbb{Z}^{r_1} \\
\mathbf{Q}^{(1)}_{\rm r}\cdot\mathbf{m} \in (\mathbb{Z}_{\ge 0})^{r_1}
}}
\frac{
q^{\frac12\, \mathbf{m}^T \overline{K^{(1)}_{\rm r}} \mathbf{m}}
\,(-q^{1/2})^{-\mathbf{m}^T \left(\overline{\boldsymbol{\mu}}_{\rm geo}^{(1),({\rm r})} - \nu \boldsymbol{v}^{(1);({\rm r})}\right)}
}{
\prod_{i=1}^{r_1} (q)_{[\mathbf{Q}^{(1)}\cdot\mathbf{m}]_i}
}\bigg{|}_{\nu = \boldsymbol{\nu}_{aa/ba}\cdot (1,0)^T}\;,
\\
&\overline{\CI^{a/b,(2);({\rm r})}_{\rm half}}
:=\sum_{\substack{
\mathbf{m}  \in \mathbb{Z}^{r_2} \\
\mathbf{Q}^{(2)}_{\rm r}\cdot\mathbf{m} \in (\mathbb{Z}_{\ge 0})^{r_2}
}}
\frac{
q^{\frac12\, \mathbf{m}^T \overline{K^{(1)}_{\rm r}} \mathbf{m}}
\,(-q^{1/2})^{-\mathbf{m}^T \left(\overline{\boldsymbol{\mu}}_{\rm geo}^{(2),({\rm r})} - \nu \boldsymbol{v}^{(2);({\rm r})}\right)}
}{
\prod_{i=1}^{r_1} (q)_{[\mathbf{Q}^{(2)}\cdot\mathbf{m}]_i}
}\bigg{|}_{\nu = \boldsymbol{\nu}_{aa/ab}\cdot (0,1)^T}\;,
\\
&\textrm{with } \overline{K^{(i)}_{\rm r}} := (\mathbf{Q}^{(i)}_{\rm r})^T  \mathbf{Q}^{(i)}_{\rm r} - K^{(i)}_{\rm r}, \quad \overline{\boldsymbol{\mu}}_{\rm geo}^{(i);({\rm r})} := -\mathbf{Q}^{(i)}\cdot (1,\ldots, 1)^T - \boldsymbol{\mu}_{\rm geo}^{(i);({\rm r})}\;.
\label{half-indices at AA/AB/BA/BB-twistings-red}
\end{split}
\end{align}
We expect the following indices to become characters of rational VOAs up to an overall factor $\pm q^{\mathbb{Q}}$
\begin{align}
\begin{split}
&\CI^{\rm a/b,(i)}_{\rm half}[\mu_m \in 2\mathbb{Z}]  \textrm{ or } \CI^{\rm a/b,(i)}_{\rm half} \textrm{ when }K^{(i)} >0 \;,
\\
&\overline{\CI^{\rm a/b,(i)}_{\rm half}}[\mu_m \in 2\mathbb{Z}]   \textrm{ or } \overline{\CI^{\rm a/b,(i)}_{\rm half}}\textrm{ when }\overline{K^{(i)}} >0 \;,
\\
& \CI^{\rm a/b,(i);({\rm r})}_{\rm half} \textrm{ when }K^{(i)}_{\rm r} >0 \;,
\\
&\overline{\CI^{\rm a/b,(i);({\rm r})}_{\rm half}}\textrm{ when }\overline{K^{(i)}_{\rm r}} >0 \;,\label{half-indices at AA/AB/BA/BB-twistings-2}
\end{split}
\end{align}
\paragraph{Example : $(P,Q)=(5,7)$} Using the \eqref{reduced NZ for (5,7)} and \eqref{reduced K and Q for (5,7)}, we see that
\begin{align}
\begin{split}
&\mathbf{Q}^{(1)}_{\rm r} = (1), \quad  \mathbf{Q}^{(2)}_{\rm r} =  \begin{pmatrix} 1 & 0 \\ 0 & 1 \end{pmatrix}, \quad 
K^{(1)}_{\rm r} = (2), \quad K^{(2)}_{\rm r} = \begin{pmatrix} -3 & -2 \\ -2 & -1 \end{pmatrix},
\\
& \boldsymbol{\mu}^{(1);{\rm (r)}}_{\rm geo} = (-2), \quad  \boldsymbol{\mu}^{(2);{\rm (r)}}_{\rm geo} = (3,1)^T, \quad \boldsymbol{v}^{(1);{\rm (r)}} = (1)\quad   \boldsymbol{v}^{(2);{\rm (r)}} = (-2,-1)^T\;,
\\
&\overline{K^{(1)}_{\rm r}} = (-1), \quad \overline{K^{(2)}_{\rm r}} = \begin{pmatrix} 4 & 2 \\ 2 & 2 \end{pmatrix}, \quad \overline{\boldsymbol{\mu}}^{(1);{\rm (r)}}_{\rm geo} = (1), \quad  \overline{\boldsymbol{\mu}}^{(2);{\rm (r)}}_{\rm geo} = (-4,-2)^T\;.
\end{split}
\end{align}
Note that $K^{(1)}_{\rm r}>0$ and $\overline{K^{(2)}_{\rm r}}>0$ and we have
\begin{align}
\begin{split}
&\CI^{a,(1);(\rm r)}_{\rm half} = \sum_{m\geq 0} \frac{q^{m^2+m}}{(q)_m}=q^{-11/60}  \chi^{\CM(2,5)}_{(1,1)}, \quad \CI^{b,(1);(\rm r)}_{\rm half} = \sum_{m\geq 0} \frac{q^{m^2}}{(q)_m} = q^{1/60} \chi^{\CM(2,5)}_{(1,2)}
\\
&\overline{\CI^{a,(2);(\rm r)}_{\rm half}} = \sum_{m_1, m_2\geq 0} \frac{q^{2m_1^2+2 m_1 m_2 +m_2^2+ 2m_1+m_2}}{(q)_{m_1} (q)_{m_2}} = q^{-17/42} \chi^{\CM(2,7)}_{(1,1)}, 
\\
&\overline{\CI^{b,(2);(\rm r)}_{\rm half}} = \sum_{m_1, m_2\geq 0} \frac{q^{2m_1^2+2 m_1 m_2 +m_2^2}}{(q)_{m_1} (q)_{m_2}} =  q^{1/42} \chi^{\CM(2,7)}_{(1,3)}\;.
\end{split}
\end{align}
\\
\\
Further examples are provided in Appendix~\ref{Appendix Examples}.

%% file: section_Examples.tex
\section{Examples} \label{Appendix Examples}

In this section, we compute the half-indices appearing in
\eqref{half-indices when |P-Q|=1-2},
\eqref{half-indices at A/B-twistings-2}, and
\eqref{half-indices at AA/AB/BA/BB-twistings-2},
and use them to test the conjecture in~\eqref{Main conjecture}.
In the comparison of $q$-series throughout this section, we adopt the following conventions:
\begin{align}
\begin{split}
&``=" \quad \text{equality verified up to } O(q^{100}),\\
&``\simeq" \quad \text{equality up to an overall factor } q^{\mathbb{Q}},
\ \text{verified up to } O(q^{100}).
\end{split}
\end{align}
%

\subsection{The case $(P,Q) = (2,2r-1)$ with $r \ge 2$}

\subsubsection{$(P,Q) = (2,3)$}
The torus knot complement can be ideally triangulated with $r=2$ tetrahedra. Using SnapPy, the gluing data can be obtained by the following input:
\begin{align*}
    \mathtt{Manifold('3\_1').gluing\_equations\_pgl()}\;,
\end{align*}
and the gluing equations are
\begin{align}
    \begin{split}
        & C_1=Z_1+Z_2',\quad C_2=Z_1+2 Z_1'+2 Z_1''+2 Z_2+Z_2'+2 Z_2'',\\
        & M=-Z_1''+Z_2\;.
    \end{split}
\end{align}
In the ideal triangulation, there is an easy internal edge $\{E_I\}=\{C_1\}$.

Among the $3^{2}$ possible polarization choices, there are four choices satisfying
the positivity condition
\[
B^{-1}A > 0 \, , \textrm{ or }(\mathbb{I}-B^{-1}A) > 0 \;,
\]
while all other choices violate this condition.
\\
\paragraph{i)} The first admissible polarization choice is
\begin{align}
    \mathbf{X}=(Z_1'',Z_2'')^T\textrm{ and }\mathbf{P}=(Z_1',Z_2')^T\;.
\end{align}
The NZ matrices in \eqref{NZ matrices} are
\begin{align}
    A=\begin{pmatrix}
        -1 & -1\\
        -1 & 0
    \end{pmatrix},\quad B=\begin{pmatrix}
        0 & -1\\
        -1 & 1
    \end{pmatrix},\quad \boldsymbol{\nu}_x=(1,-1)^T\;.
\end{align}
In the case, $B^{-1} A\in M_2(\BZ)$ and $|\det B|=1$, thus we can choose $\mathbf{Q}=\BI$ and $N=\left|\frac{(\det \mathbf{Q})^2}{\det B}\right|=1$. Therefore,
\begin{align}
    \begin{split}
        & T[S^3\backslash \CK_{(2,3)}]=\BT[A,B;\mathbf{Q}]\\
        & =U(1)_{K-\BI/2}^2\textrm{ coupled to }2\;\F\textrm{s of charge }\mathbf{Q}=\BI\textrm{ with }\CW=\CO_{E_1}\\
        & \textrm{with }K=B^{-1} A=\begin{pmatrix}
            2 & 1\\ 1 & 1
        \end{pmatrix}\textrm{ and }\CO_{E_1}=V_{(1,-1)}\;.
    \end{split}
\end{align}
The affine spaces $\FM[\CT]$ in \eqref{M[T]} and vector space $\FF[\CT]$ in \eqref{F[T]} are
\begin{align}
    \begin{split}
        & \FM[\CT]=\{\boldsymbol{\m}=(\m_1,\m_2)^T: \m_1-\m_2=1\}\;,\\
        & \FF[\CT]=\{\boldsymbol{v}=(v_1,v_2)^T: v_1-v_2=0\}\;.
    \end{split}
\end{align}
As expected in \eqref{mu_geom} and \eqref{basis of F[T]},
\begin{align}
    \begin{split}
        & \boldsymbol{\m}_\mathrm{geo}=B^{-1}\cdot\boldsymbol{\n}_x=(0,-1)^T\in\FM[\CT]\;,\\
        & \FF[\CT]=\mathrm{Span}\{\boldsymbol{v}_m:=B^{-1}\cdot(1,0)^T=(-1,-1)^T\}\;.
    \end{split}
\end{align}
\\
The superconformal index \eqref{SCI for T_{DGG}} for the theory is
\begin{align}
    \begin{split}
        & \CI_\mathrm{sci}^{T[S^3\backslash \CK_{(2,3)}]}[u;\m_m]=\left.\CI_\mathrm{sci}^{T[A,B]}[\boldsymbol{x},\boldsymbol{\m}]\right|_*\textrm{ where}\\
        & \CI_\mathrm{sci}^{T[A,B]}[\boldsymbol{x},\boldsymbol{\m}]=\sum_{\mathbf{m}\in \BZ^2}\paren{\prod_{i=1}^2 \oint\frac{du_i}{2\p i u_i}} u_1^{2 m_1+m_2} u_2^{m_1+m_2}\\
        & \qquad \qquad \qquad \quad \times \prod_{i=1}^2 (x_i (-q^{1/2})^{\m_i})^{m_i} \CI_\D(m_i,u_i)\;,\\
        &\big{|}_* \;:\;(x_1,x_2)\rightarrow \left(\frac{1}{u},\frac{1}{u}\right), \;\;(\mu_1, \mu_2)\rightarrow (-\m_m,-1-\m_m) \;.
    \end{split}
\end{align}
Using the superconformal index, one can explicitly verify that the $U(1)_m$ symmetry decouples in the IR — the dependence on $(u,\m_m)$ identically drops out and the theory flows to a unitary TQFT:
\begin{align}
    \CI_\mathrm{sci}^{T[S^3\backslash \CK_{(2,3)}]}[u;\m_m]=1\;.
\end{align}
\\
The half-index in \eqref{half-indices when |P-Q|=1} is
\begin{align}
    \CI_\mathrm{half}[\mu_m\in 2\mathbb{Z}]=\sum_{\mathbf{m}=(m_1,m_2)^T\in (\BZ_{\ge 0})^2}\frac{q^{\frac{1}{2} \mathbf{m}^T\cdot K \cdot \mathbf{m}}(-q^{1/2})^{m_2+\mu_m (m1+ m2)}}{(q)_{m_1} (q)_{m_2}}=1\;.
\end{align}
\\
\paragraph{ii)} The second admissible polarization choice is
\begin{align}
    \mathbf{X}=(Z_1',Z_2)^T\textrm{ and }\mathbf{P}=(Z_1,Z_2'')^T\;.
\end{align}
The NZ matrices are
\begin{align}
    A=\begin{pmatrix}
        1 & 1\\
        0 & -1
    \end{pmatrix},\quad B=\begin{pmatrix}
        1 & 0\\
        1 & -1
    \end{pmatrix},\quad \boldsymbol{\n}_x=(-1,-1)^T\;.
\end{align}
In the case, we can choose $\mathbf{Q}=\BI$ and $N=1$. $K$, $\boldsymbol{\m}_\mathrm{geo}$ and $\boldsymbol{v}_m$ are
\begin{align}
    K=B^{-1} A=\begin{pmatrix}
        1 & 1\\
        1 & 2
    \end{pmatrix},\quad \boldsymbol{\m}_\mathrm{geo}=(-1,0)^T,\quad \boldsymbol{v}_m=(1,1)^T\;,
\end{align}
so the present choice of polarization gives the same result with the previous case, up to permutation.
\\
\paragraph{iii)} When we choose the following polarizations
\begin{align}
    \mathbf{X}=(Z_1,Z_2'')^T\textrm{ and }\mathbf{P}=(Z_1'',Z_2')^T\;,
\end{align}
the NZ matrices are
\begin{align}
    A=\begin{pmatrix}
        0 & -1\\
        1 & 0
    \end{pmatrix},\quad B=\begin{pmatrix}
        -1 & -1\\
        0 & 1
    \end{pmatrix},\quad \boldsymbol{\n}_x=(1,-2)^T\;.
\end{align}
In the case, $|\det B|=1$, and $B^{-1} A$ is integer valued, but not positive definite. Instead, $\BI-B^{-1} A$ is positive definite. Therefore, we can choose $\mathbf{Q}=\BI$, $N=1$ and 
\begin{equation}
\begin{gathered}
    \overline{K} = \BI-B^{-1} A = \begin{pmatrix} 2 & -1 \\ -1 & 1 \end{pmatrix}\,, \quad 
    \CO_{E_1} = \phi_1 V_{(0,1)}\,, \\
    \overline{\boldsymbol{\m}}_\mathrm{geo}=-(1,1)^T-B^{-1}\cdot\boldsymbol{\n}_x = (-2,1)^T\,, \quad
    \boldsymbol{v}_m=B^{-1}\cdot(1,0)^T = (-1,0)^T\;.
\end{gathered}
\end{equation}
The superconformal index \eqref{SCI for T_{DGG}} for the theory is
\begin{align}
    \begin{split}
        & \CI_\mathrm{sci}^{\overline{T[S^3\backslash\CK_{(2,3)}]}}[u;\m_m]=\left.\CI_\mathrm{sci}^{\overline{T[A,B]}}[\boldsymbol{x},\boldsymbol{\m}]\right|_*\textrm{ where}\\
        & \CI_\mathrm{sci}^{\overline{T[A,B]}}[\boldsymbol{x},\boldsymbol{\m}]=\sum_{\mathbf{m}\in\BZ^2}\paren{\prod_{i=1}^2 \oint\frac{du_i}{2\p i u_i}} u_1^{2 m_1-m_2} u_2^{-m_1+m_2}\\
        & \qquad \qquad \qquad \quad \times \prod_{i=1}^2 (x_i(-q^{1/2})^{\m_i})^{m_i} \CI_\D(m_i,u_i)\;,\\
        & \left.\right|_*:(x_1,x_2)\rightarrow \paren{u,1},\quad (\m_1,\m_2)\rightarrow (-2+\m_m,1)\;.
    \end{split}
\end{align}
The dependence on $(u,\m_m)$ identically drops out, and the theory flows to a unitary TQFT in the IR:
\begin{align}
    \CI_\mathrm{sci}^{\overline{T[S^3\backslash\CK_{(2,3)}]}}[u;\m_m]=1\;.
\end{align}
The half-index in \eqref{half-indices when |P-Q|=1} is
\begin{align}
    \overline{\CI_\mathrm{half}}[\m_m\in 2\mathbb{Z}]=\sum_{\mathbf{m}\in(\BZ_{\geq 0})^2}\frac{q^{\frac{1}{2}\mathbf{m}^T\cdot \overline{K}\cdot\mathbf{m}} (-q^{1/2})^{2 m_1-m_2-\m_m m_1}}{(q)_{m_1} (q)_{m_2}}=0\;.
\end{align}
\\
\paragraph{iv)} When we choose the following polarizations
\begin{align}
    \mathbf{X}=(Z_1',Z_2')^T\textrm{ and }\mathbf{P}=(Z_1,Z_2)^T\;,
\end{align}
the NZ matrices are
\begin{align}
    A=\begin{pmatrix}
        1 & 0\\
        0 & 1
    \end{pmatrix},\quad B=\begin{pmatrix}
        1 & 1\\
        1 & 0
    \end{pmatrix},\quad \boldsymbol{\n}_x=(-1,-2)^T\;.
\end{align}
In the case, we can choose $\mathbf{Q}=\BI$, $N=1$ and $\BI-B^{-1} A$ is positive definite. $\overline{K}$, $\overline{\boldsymbol{\m}}_\mathrm{geo}$ and $\boldsymbol{v}_m$ are
\begin{align}
\begin{split}
    & \overline{K}=\BI-B^{-1} A=\begin{pmatrix}
        1 & -1\\
        -1 & 2
    \end{pmatrix},\quad \overline{\boldsymbol{\m}}_\mathrm{geo}=-(1,1)^T-B^{-1}\cdot\boldsymbol{\n}_x=(1,-2)^T,\\
    & \boldsymbol{v}_m=B^{-1}\cdot (1,0)^T=(0,1)^T\;,
\end{split}
\end{align}
so the present choice of polarization gives the same result with the previous case, up to permutation. We will omit such cases from now on.
\\
\\
\subsubsection{$(P,Q) = (2,5)$} \label{AppEx2,5}
Using SnapPy, the gluing data of the torus knot can be obtained by the following input:
\begin{align*}
    \mathtt{Manifold('5\_1').gluing\_equations\_pgl()}
\end{align*}
The torus knot complement can be ideally triangulated with $r=3$ tetrahedra with following gluing data 
\begin{align}
\begin{split}    &C_1 = Z_1+Z_3, \quad C_2 = 2Z'_1+Z''_1+Z_2+2Z_2''+Z'_3+2 Z_3'',
\\
&C_3 = Z_1+Z_1''+Z_2+2Z_2'+Z_3+Z_3', \quad M = -Z_1''-Z_2'+Z_3''\;.
\end{split}
\label{(2,5) ge}
\end{align}
In the ideal triangulation, there is an easy internal edge $\{E_{I}\}= \{C_1\}$ and we choose $\{H_{I}\}=\{C_2\}$.

Among the $3^3$ possible polarization choices, there are six choices satisfying the positivity condition
\begin{align*}
    B^{-1} A>0,\textrm{ or } (\BI-B^{-1} A)>0\;,
\end{align*}
while all other choices violate this condition. We list some representative cases among them, as well as for the other examples below.
\\
\paragraph{i)} When we choose the following polarizations 
\begin{align}
\mathbf{X} = (Z''_1, Z_2, Z''_3)^T \textrm{ and } \mathbf{P} = (Z'_1, Z_2'', Z'_3)^T\;,
\end{align}
the NZ matrices are
\begin{align}
A= \left(
\begin{array}{ccc}
 -1 & 1 & 1 \\
 1 & 1 & 2 \\
 -1 & 0 & -1 
\end{array}
\right), \quad B = \left(
\begin{array}{ccc}
 0 & 1 & 0 \\
 2 & 2 & 1 \\
-1 & 0 & -1 
\end{array}
\right), \quad {\boldsymbol{\nu}}_{x} = (-1,-2,0)^T\;.
\end{align}
In the case, $B^{-1} A \in M_3 (\mathbb{Z})$ and $|\det B |=1$ and thus we can choose $\mathbf{Q}=\mathbb{I}$. Therefore, 
\begin{align}
\begin{split}
&T[S^3\backslash \mathcal{K}_{(2,5)}] = \mathbb{T}[A, B ;\mathbf{Q}]
\\
&= U(1)^3_{K-\mathbb{I}/2} \textrm{ coupled to } 3\;\Phi \textrm{s of charge }\mathbf{Q} = \mathbb{I} \textrm{ with }\mathcal{W}= \mathcal{O}_{E_1}
\\
&\quad \;\; \textrm{with } K = B^{-1}A = \left(
\begin{array}{ccc}
 2 & -1 & -1 \\
 -1 & 1 & 1 \\
 -1 & 1 & 2 \\
\end{array}
\right) \textrm{ and } \CO_{E_1} =  V_{(1,0,1)}\;.
\end{split}
\end{align}
The affine spaces $\mathfrak{M}[\mathcal{T}]$ in \eqref{M[T]} and vector space in $\mathfrak{F}[\mathcal{T}]$ \eqref{F[T]} are 
\begin{align}
\begin{split}
&\mathfrak{M}[\mathcal{T}]  =\{ \boldsymbol{\mu}= (\mu_1, \mu_2, \mu_3)^T \;:\; \mu_1+\mu_3 =0 \}\;,
\\
&\mathfrak{F}[\mathcal{T}]  =\{ \boldsymbol{v}= (v_1, v_2, v_3)^T \;:\; v_1+v_3 =0 \}\;.
\end{split}
\end{align}
As expected in \eqref{mu_geom} and \eqref{basis of F[T]},
\begin{align}
\begin{split}
&\boldsymbol{\mu}_{\rm geo} = B^{-1}\cdot \boldsymbol{\nu}_x = (0,-1,0)^T \in \mathfrak{M}[\mathcal{T}]\;,
\\
&\mathfrak{F}[\mathcal{T}]  = \textrm{Span} \{   \boldsymbol{v}_m :=B^{-1}\cdot (1,0,0)^T= (-2,1,2)^T, \; \boldsymbol{v}_1 :=B^{-1}\cdot (0,1,0)^T = (1,0,-1)^T\}\;.
\end{split}
\end{align}
\\
The superconformal index \eqref{SCI for T_{DGG}} for the theory is 
\begin{align}
\begin{split}
&\CI^{T[S^3\backslash \CK_{(2,5)}]}_{\rm sci}[u, \eta; \mu_m, \nu] =  \CI^{T[A,B]}_{\rm sci}[\boldsymbol{x},\boldsymbol{\mu}]\big{|}_{ *}\textrm {  where }
\\
&\CI^{T[A,B]}_{\rm sci}[\boldsymbol{x},\boldsymbol{\mu}]= \sum_{\mathbf{m} \in \mathbb{Z}^3} \left( \prod_{i=1}^3\oint \frac{du_i}{2\pi i u_i} \right)   u_1^{2m_1-m_2-m_3} u_2^{-m_1+m_2+m_3}u_3^{-m_1+m_2+2m_3}
\\
& \qquad \qquad \qquad \quad  \times \prod_{i=1}^3 (x_i(-q^{1/2})^{\mu_i})^{m_i}\CI_\Delta (m_i,u_i)\;,
\\
&\big{|}_* \;:\;(x_1,x_2, x_3)\rightarrow \left(\frac{1}{u^2 \eta}, u, \frac{u^2}{\eta}\right), \;\;(\mu_1, \mu_2, \mu_3)\rightarrow (-2 \mu_m-\nu, -1 +\mu_m, 2 \mu_m-\nu) \;.
\end{split}
\end{align}
Using the superconformal index, one can explicitly verify that the $U(1)_m$ symmetry
decouples in the IR — the dependence on $(u,\mu_m)$ identically drops out:
\begin{align} \label{(2,5)index_U1mdecouples}
\CI^{T[S^3\backslash \CK_{(2,5)}]}_{\rm sci}[u,\eta;\mu_m,\nu] =
\CI^{T[S^3\backslash \CK_{(2,5)}]}_{\rm sci}[\eta;\nu]\;.
\end{align}
In $q$-expansion, 
\begin{align} \label{(2,5)index_qexpansion}
\begin{split}
&\CI_{\rm sci}^{T[S^3\backslash \CK_{(2,5)}]} [\eta;\nu] = 1+(\eta-1)q +\left(\eta+\frac{1}\eta -2\right) q^2+ \left(\eta+\frac{1}\eta -2\right) q^3
\\
&+\left(\eta^2+\frac{1}\eta -2\right) q^4+(\eta^2-\eta)q^5+\left(2\eta^2-3\eta- \frac{1}\eta +\frac{1}{\eta^2} +1\right)q^6+\ldots \bigg{|}_{\eta \rightarrow \eta (-q^{1/2})^\nu}\;.
\end{split}
\end{align}
At $\nu_a=0$ and $\nu_b = 2$, as anticipated in \eqref{two special points}, the index becomes
\begin{align} \label{(2,5)indexbecomes1}
\CI_{\rm sci}^{T[S^3\backslash \CK_{(2,5)}]} [\eta=1;\nu=0] = \CI_{\rm sci}^{T[S^3\backslash \CK_{(2,5)}]} [\eta=1;\nu=2]=1\;. 
\end{align}
\\
The next ones to be considered are the twisted partition functions. At the A-twisting point, i.e. $\boldsymbol{\mu} = \boldsymbol{\mu}_{\rm geo} = (0,-1,0)^T$ and $\boldsymbol{\zeta}= \mathbf{0}$, the $\CW_0$ and $\CW_1$ in \eqref{W0 and W1} becomes
\begin{align}
\begin{split}
&\CW^{\rm a}_0 \simeq   Z_1^2-Z_1 Z_2 +\frac{Z_2^2}2-Z_1 Z_3 + Z_2 Z_3 + Z_3^2 - i \pi Z_2 +\sum_{i=1}^3 {\rm Li}_2(e^{-Z_i})\;,
\\
&\CW^{\rm a}_1 \simeq  -\frac{1}2 \left(Z_2 + \sum_{i=1}^3 \log (1-e^{-Z_i}) \right)\;.
\end{split}
\end{align}
Then, $\mathbf{HF}$ at the point is 
\begin{align}
\mathbf{HF}_{\rm a} \simeq  \bigg{\{} \left(\frac{5-\sqrt{5}}2,1\right),\;\left(\frac{5+\sqrt{5}}2,\mathbf{e}(\frac{1}5)\right) \bigg{\}} \;.
\end{align}
At the B-twisting point, i.e. $\boldsymbol{\mu} = \boldsymbol{\mu}_{\rm geo} +2 \boldsymbol{v}_1 = (2,-1,-2)^T$ and $\boldsymbol{\zeta}= \mathbf{0}$, 
\begin{align}
\begin{split}
&\CW^{\rm b}_0 \simeq Z_1^2-Z_1 Z_2 +\frac{Z_2^2}2-Z_1 Z_3 + Z_2 Z_3 + Z_3^2 + i \pi (2Z_1-Z_2-2Z_3) +\sum_{i=1}^3 {\rm Li}_2(e^{-Z_i})\;,
\\
&\CW^{\rm b}_1 \simeq  -\frac{1}2 \left(-2Z_1+Z_2+2 Z_3 + \sum_{i=1}^3 \log (1-e^{-Z_i}) \right)\;.
\end{split}
\end{align}
Then, $\mathbf{HF}$ at the point is 
\begin{align}
\mathbf{HF}_{\rm b} \simeq  \bigg{\{} \left(\frac{5-\sqrt{5}}2,1\right),\;\left(\frac{5+\sqrt{5}}2,\mathbf{e}(-\frac{1}5)\right) \bigg{\}} \;.
\end{align}
It matches the $\mathbf{HF}$ of $\CM(2,5)$.
\\
The half-index in \eqref{half-indices at A/B-twistings} at the A-twisting point is 
\begin{align}
\begin{split}
\CI^{a}_{\rm half} [\m_m\in 2\BZ] &= \sum_{\mathbf{m}=(m_1,m_2,m_3)^T \in (\mathbb{Z}_{\geq 0})^3 } \frac{q^{\frac{1}2 \mathbf{m}^T \cdot K \cdot \mathbf{m} } (-q^{1/2})^{m_2-\m_m (-2 m_1+m_2+2 m_3)}}{(q)_{m_1} (q)_{m_2} (q)_{m_3}} \\
&\simeq 1+q+q^2+q^3+2 q^4+2q^5+3 q^6+3 q^7 + 4 q^8+5 q^9+ \ldots\\
& = q^{1/60} \chi^{\CM(2,5)}_{(1,2)}(q),
\end{split}
\end{align}
The half-index in \eqref{half-indices at A/B-twistings}  at the B-twisting point is 
\begin{align}
\begin{split}
\CI^{b}_{\rm half}[\m_m \in 2 \BZ] &= \sum_{\mathbf{m}=(m_1,m_2,m_3)^T \in (\mathbb{Z}_{\geq 0})^3 } \frac{q^{\frac{1}2 \mathbf{m}^T \cdot K \cdot \mathbf{m} } (-q^{1/2})^{-2m_1+m_2+2m_3-\m_m (-2 m_1+m_2+2 m_3)}}{(q)_{m_1} (q)_{m_2} (q)_{m_3}}
\\
&\simeq 1+q^2+q^3+q^4+q^5+2 q^6+2 q^7+3 q^8+3 q^9+\ldots
\\
& = q^{-11/60} \chi^{\CM(2,5)}_{(1,1)}(q),
\end{split}
\end{align}
%
\\
\paragraph{ii)} When we choose the following polarizations 
\begin{align}
\mathbf{X} = (Z''_1, Z'_2, Z'_3)^T \textrm{ and } \mathbf{P} = (Z'_1, Z_2, Z_3)^T\;,
\end{align}
the NZ matrices are
    \begin{align}
    A= \left(
\begin{array}{ccc}
-1 & -1 & -1 \\
1 & -2 & -1 \\
-1 & 0 & 0 \\
\end{array}
\right), \quad B = \left(
\begin{array}{ccc}
 0 & 0 & -1 \\
 2 & -1 & -2 \\
 -1 & 0 & 1 
\end{array}
\right), \quad \boldsymbol{\nu}_x = (1,2,-1)^T\;.
\end{align}
In the case, $B^{-1} A \in M_3 (\mathbb{Z})$ and $|\det B|=1$ and thus we can choose $\mathbf{Q}=\mathbb{I}$. Therefore, 
\begin{align}
\begin{split}
&T[S^3\backslash \mathcal{K}_{(2,5)}] = \mathbb{T}[A, B ;\mathbf{Q}]
\\
&= U(1)^3_{K-\mathbb{I}/2} \textrm{ coupled to } 3\;\Phi \textrm{s of charge }\mathbf{Q} = \mathbb{I} \textrm{ with }\mathcal{W}= \mathcal{O}_{E_1}
\\
&\quad \;\; \textrm{with } K = B^{-1}A = \left(
\begin{array}{ccc}
 2 & 1 & 1 \\
 1 & 2 & 1 \\
 1 & 1 & 1 \\
\end{array}
\right) \textrm{ and } \mathcal{O}_{E_1} =  V_{(1,0,-1)}\;.
\end{split}
\end{align}
$\FM[\CT]$, $\FF[\CT]$, $\boldsymbol{\mu}_{\rm geo}$, $\boldsymbol{v}_m$, and $\boldsymbol{v}_1$ are
\begin{equation}
\begin{gathered}
    \FM[\CT]  =\{ \boldsymbol{\mu}= (\mu_1, \mu_2, \mu_3)^T :\, \mu_1-\mu_3 = 1 \}\,, \quad
    \FF[\CT]  = \textrm{Span} \{ \boldsymbol{v}_m, \; \boldsymbol{v}_1\}\,, \\
    \boldsymbol{\mu}_{\rm geo} = B^{-1}\cdot \boldsymbol{\nu}_x = (0,0,-1)^T \,, \\
    \boldsymbol{v}_m :=B^{-1}\cdot (1,0,0)^T = (-1,0,-1)^T\,, \quad \boldsymbol{v}_1 :=B^{-1}\cdot (0,1,0)^T = (0,-1,0)^T\,, \\
\end{gathered}
\end{equation}

The superconformal index \eqref{SCI for T_{DGG}} for the theory is 
\begin{align}
\begin{split}
&\CI^{T[S^3\backslash \CK_{(2,5)}]}_{\rm sci}[u, \eta; \mu_m, \nu] =  \CI^{T[A,B]}_{\rm sci}[\boldsymbol{x},\boldsymbol{\mu}]\big{|}_{ *} \,,\,\textrm { where }
\\
&\big{|}_* \;:\;(x_1,x_2, x_3)\rightarrow \left(\frac{1}{u}, \frac{1}{\eta}, \frac{1}{u}\right), \;\;(\mu_1, \mu_2, \mu_3)\rightarrow (-\mu_m,-\nu,-1-\mu_m) \;.
\end{split}
\end{align}
The $q$-expansion and properties of superconformal index \eqref{(2,5)index_U1mdecouples} - \eqref{(2,5)indexbecomes1} are identical to those in {\bf i)}. We have verified that these properties hold for all examples; hence, we will omit the superconformal index analysis in what follows.
\\
At the A/B-twisting point, i.e. $\boldsymbol{\mu} = \boldsymbol{\mu}_{\rm geo} = (0,0,-1)^T$ and $\boldsymbol{\mu} = \boldsymbol{\mu}_{\rm geo} +2 \boldsymbol{v}_1 = (0,-2,-1)^T$, $\mathbf{HF}$ are
\begin{align}
\begin{split}
&\mathbf{HF}_{\rm a} \simeq  \bigg{\{} \left(\frac{5-\sqrt{5}}2,1\right),\;\left(\frac{5+\sqrt{5}}2,\mathbf{e}(\frac{1}5)\right) \bigg{\}} \,, \\
&\mathbf{HF}_{\rm b} \simeq  \bigg{\{} \left(\frac{5-\sqrt{5}}2,1\right),\;\left(\frac{5+\sqrt{5}}2,\mathbf{e}(-\frac{1}5)\right) \bigg{\}} \,.
\end{split}
\end{align}
The $\mathbf{HF}_{\rm b}$ matches the $\mathbf{HF}$ of $\CM(2,5)$.
\\
The half-indices at the A/B-twisting points are 
\begin{align}
\begin{split}
\CI^{a}_{\rm half}[\m_m \in 2\mathbb{Z}] &= 
1+q+q^2+q^3+2 q^4+2q^5+3 q^6+3 q^7 + 4 q^8+5 q^9+\ldots \\
& = q^{1/60} \chi^{\CM(2,5)}_{(1,2)}(q)\,, \\
\CI^{b}_{\rm half}[\m_m \in 2\mathbb{Z}] &= 
1+q^2+q^3+q^4+q^5+2 q^6+2 q^7+3 q^8+3 q^9+ 4 q^{10}+\ldots \\
& = q^{-11/60} \chi^{\CM(2,5)}_{(1,1)}(q)\;.
\end{split}
\end{align}
Note that all BPS partition functions and half-indices match exactly with those of \textbf{i)}. There exist a lot of pairs like \textbf{i)} and \textbf{ii)}, so from now on we will include only one of them in this section.
\\
\paragraph{iii)} When we choose the following polarizations
\begin{align}
    \mathbf{X}=(Z_1,Z_2,Z_3')^T\textrm{ and }\mathbf{P}=(Z_1'',Z_2'',Z_3)^T\;,
\end{align}
the NZ matrices are
\begin{align}
    A=\begin{pmatrix}
        0 & 1 & -1\\
        -2 & 1 & -1\\
        1 & 0 & 0
    \end{pmatrix},\quad B=\begin{pmatrix}
        -1 & 1 & -1\\
        -1 & 2 & -2\\
        0 & 0 & 1
    \end{pmatrix},\quad \boldsymbol{\n}_x=(0,2,-2)^T\;.
\end{align}
In the case, $\BI-B^{-1} A>0$, and we can choose $\mathbf{Q}=\BI$. Then,
\begin{equation}
\begin{gathered}
    \overline{K}=\begin{pmatrix}
        3 & 1 & -1\\
        1 & 1 & 0\\
        -1 & 0 & 1
    \end{pmatrix},\quad \overline{\CO}_{E_1} =  \phi_1 V_{(0,0,1)}\,, \\
    \overline{\boldsymbol{\m}}_\mathrm{geo}=(-3,-1,1)^T, \quad \boldsymbol{v}_m=(2,1,0)^T,\quad \boldsymbol{v}_1=(-1,-1,0)^T\;.
\end{gathered}
\end{equation}
\\
The half-indices at the A/B-twisting points are
\begin{align}
        \overline{\CI_\mathrm{half}^a}[ \m_m \in 2\mathbb{Z}] = \overline{\CI_\mathrm{half}^b}[\m_m \in 2\mathbb{Z}] = 0
\end{align}
\\
\\
\\
By taking the reduction procedure on the gluing equations in \eqref{(2,5) ge}, we have
\begin{align}
    H_1^\mathrm{(r)}:=C_2^\mathrm{(r)}=Z_2+2 Z_2''\;.
    \label{(2,5) reduced gluing}
\end{align}
All 3 choices of polarization satisfy the positivity condition
\begin{align}
    (B^{-1} A)_\mathrm{r}>0\textrm{, or }(\BI-B^{-1} A)_\mathrm{r}>0\;.
\end{align}
\\
\paragraph{reduced-i)}
When we choose the following polarizations
\begin{align}
    X_\mathrm{r}=Z_2'\textrm{ and }P_\mathrm{r}=Z_2\;,
\end{align}
the reduced NZ matrices in \eqref{reduced NZ matrices} are
\begin{align}
    A_\mathrm{r}=(-2),\quad B_\mathrm{r}=(-1),\quad \boldsymbol{\n}_{x;\mathrm{r}}=(0)\;.
\end{align}
In the case, we choose $\mathbf{Q}_\mathrm{r}=\BI$. Then,
\begin{align}
\begin{split}
    K_\mathrm{r}=(2), \quad N_\mathrm{r}=1\,, \quad
    \boldsymbol{\m}_\mathrm{geo}^\mathrm{(r)}=(0),\quad \boldsymbol{v}_1^\mathrm{(r)}=(-1)\,.
\end{split}
\end{align}
\\
The superconformal index for $T^{\rm (r)}[S^3\backslash \CK_{(2,5)}]$ is 
\begin{align}
\begin{split}
    &\CI_{\rm sci}^{T^{\rm (r)}[S^3\backslash \CK_{(2,5)}]} ( \eta ; \nu) = \CI_{\rm sci}^{T[A_{\rm r},B_{\rm r}]}[\boldsymbol{x}; \boldsymbol{\mu}] (q) \bigg{|}_* \,, \textrm{ where } \\
    & \quad\big{|}_* \, : \,
    (\mu_1) \rightarrow \left(-\nu\right), \,\,
    (x_1) \rightarrow \left(\frac{1}{\eta}\right)\,.
\end{split}
\end{align}
Again, the $q$-expansion and properties of superconformal index \eqref{(2,5)index_U1mdecouples} - \eqref{(2,5)indexbecomes1} are identical to those in i). Henceforth, we omit the analysis on superconformal index, including for the reduced examples.
\\
At the A/B-twisting point, i.e. $\boldsymbol{\mu} = \boldsymbol{\mu}_{\rm geo} = (0)$ and $\boldsymbol{\mu} = \boldsymbol{\mu}_{\rm geo} +2 \boldsymbol{v}_1 = (-2)$, $\mathbf{HF}$ are
\begin{align}
\begin{split}
    &\mathbf{HF}_{\rm a} \simeq  \bigg{\{} \left(\frac{5-\sqrt{5}}2,1\right),\,\left(\frac{5+\sqrt{5}}2,\mathbf{e}(\frac{1}{5})\right) \bigg{\}} \,, \\
    &\mathbf{HF}_{\rm b} \simeq  \bigg{\{} \left(\frac{5-\sqrt{5}}2,1\right),\,\left(\frac{5+\sqrt{5}}2,\mathbf{e}(-\frac{1}{5})\right) \bigg{\}} \,.
\end{split}
\end{align}
$\mathbf{HF}_b$ matches the $\mathbf{HF}$ of $\CM(2,5)$.
\\
The half-indices $\CI_\mathrm{half}^{a/b;\mathrm{(r)}}[A,B;\mathbf{Q}]$ in \eqref{half-indices at A/B-twistings} at the A/B-twisting points are
\begin{align}
    \begin{split}
        \CI_\mathrm{half}^{a;\mathrm{(r)}} & =1+q+q^2+q^3+2 q^4+2 q^5+3 q^6+3 q^7+4 q^8+5 q^9+6 q^{10}+\ldots\\
        & =q^{1/60}\c_{(1,2)}^{\CM(2,5)}(q)\;,\\
        \CI_\mathrm{half}^{b;\mathrm{(r)}} & =1+q^2+q^3+q^4+q^5+2 q^6+2 q^7+3 q^8+3 q^9+4 q^{10}+4 q^{11}+\ldots\\
        & =q^{-11/60} \c_{(1,1)}^{\CM(2,5)}(q)\;.
    \end{split}
\end{align}
\\
\paragraph{reduced-ii)} \label{Nahm3,5Ex} When we choose the following polarizations
\begin{align}
    X_\mathrm{r}=Z_2\textrm{ and }P_\mathrm{r}=Z_2''\;,
\end{align}
the reduced NZ matrices are
\begin{align}
    A_\mathrm{r}=(1),\quad B_\mathrm{r}=(2),\quad \boldsymbol{\n}_{x;\mathrm{r}}=(-2)\;.
\end{align}
In the case, we choose $\mathbf{Q}_\mathrm{r}=(2)$. Then,
\begin{equation}
\begin{gathered}
    K_\mathrm{r}=(\mathbf{Q}^T B^{-1} A \mathbf{Q})_\mathrm{r}=(2),
    \quad N_\mathrm{r} = \left|\frac{(\det \mathbf{Q})^2}{\det B}\right|_\mathrm{r}=2\,, \\
    \boldsymbol{\m}_\mathrm{geo}^\mathrm{(r)}=\mathbf{Q}^T B^{-1} \cdot \boldsymbol{\nu}_{x;r} = (-2),\quad
    \boldsymbol{v}_1^\mathrm{(r)}=\mathbf{Q}^T B^{-1} \cdot(1)=(1)\,.
\end{gathered}
\label{K=2 Q=2}
\end{equation}
At the A/B-twisting point, $\mathbf{HF}$ are
\begin{align}
\begin{split}
    &\mathbf{HF}_{\rm a} \simeq  \bigg{\{} \left(\frac{5-\sqrt{5}}2,1\right),\,\left(\frac{5+\sqrt{5}}2,\mathbf{e}(\frac{1}{5})\right) \bigg{\}}
    \times \bigg\{ (2,\mathbf{e}(0)),\, \big(2,\mathbf{e}(\tfrac{1}{4})\big) \bigg\}\,, \\
    &\mathbf{HF}_{\rm b} \simeq  \bigg{\{} \left(\frac{5-\sqrt{5}}2,1\right),\,\left(\frac{5+\sqrt{5}}2,\mathbf{e}(-\frac{1}{5})\right) \bigg{\}}
    \times \bigg\{ (2,\mathbf{e}(0)),\, \big(2,\mathbf{e}(\tfrac{1}{4})\big) \bigg\}\,.
\end{split}
\end{align}
The first factor corresponds to the $\mathbf{HF}$ of $T[S^3\backslash \CK_{(2,5)}]$, while the second factor corresponds to the decoupled Abelian TQFT $\CA^{N=2,p}$.
Therefore, we confirm that
\begin{align}
\begin{split}
\big(\mathbf{HF}_b \text{ of }\, T^{\rm (r)}[S^3\backslash \CK_{(2,5)}]\big) \,\simeq\, \big(\mathbf{HF} \text{ of } \CM(2,5)\big)\,.
\end{split}
\end{align}
\\
The half-indices $\CI_\mathrm{half}^{a/b;\mathrm{(r)}}[A,B;\mathbf{Q}]$ at the A/B-twisting points are
\begin{align}
    \begin{split}
        \CI_\mathrm{half}^{a;\mathrm{(r)}} &= 1+q^2+q^3+2 q^4+2 q^5+4 q^6+4 q^7+6 q^8+7 q^9+10 q^{10}+11 q^{11}+\ldots\\
        & =q^{-1/40} \c_{(1,1)}^{\CM(3,5)}(q)\;,\\
        \CI_\mathrm{half}^{b;\mathrm{(r)}} &= 1+q+q^2+2 q^3+3 q^4+4 q^5+5 q^6+7 q^7+9 q^8+12 q^9+15 q^{10}+\ldots\\
        & =q^{1/40} \c_{(1,2)}^{\CM(3,5)}(q)\;.
    \end{split}
    \label{K=2 Q=2 half index}
\end{align}
It is the Nahm-sum like expression \eqref{Nahmsum3,l} with $l=5$.
\\
\\
\subsubsection{$(P,Q)=(2,7)$} \label{AppEx2,7}
Using the SnapPy, the gluing data of the torus knot can be obtained by the following input:
\begin{align*}
    \mathtt{Manifold('7\_1').gluing\_equations\_pgl()}
\end{align*}
The torus knot complement can be ideally triangulated with $r=4$ tetrahedra with following gluing data
\begin{align}
    \begin{split}
        & C_1=2 Z_1''+Z_2'+Z_3'+Z_4'', \quad C_2=Z_2''+Z_4',\\
        & C_3=2 Z_1+Z_1'+2 Z_2+2 Z_3+2 Z_4,\\
        & C_4=Z_1'+Z_2'+Z_2''+Z_3'+2 Z_3''+Z_4'+Z_4'',\\
        & M=Z_2'+Z_3''-Z_4\;.
    \end{split}
    \label{(2,7) gluing}
\end{align}
In the ideal triangulation, there are two easy internal edges $\{E_I\}=\{C_2,C_1\}$ and we choose $\{H_I\}=\{C_3\}$.

Among the $3^4$ possible polarization choices, there are four choices satisfying the positivity condition.
\\
\paragraph{i)} When we choose the following polarizations
\begin{align}
    \mathbf{X}=(Z_1',Z_2',Z_3'',Z_4'')^T\textrm{ and }\mathbf{P}=(Z_1,Z_2,Z_3',Z_4')^T\;,
\end{align}
the NZ matrices are
\begin{align}
    A=\begin{pmatrix}
        0 & 1 & 1 & 1\\
        1 & 0 & -2 & -2\\
        0 & -1 & 0 & 0\\
        -2 & 1 & 0 & 1
    \end{pmatrix}, \quad B=\begin{pmatrix}
        0 & 0 & 0 & 1\\
        2 & 2 & -2 & -2\\
        0 & -1 & 0 & 1\\
        -2 & 0 & 1 & 0
    \end{pmatrix},\quad {\boldsymbol{\n}}_x=(-1,2,-1,0)^T\;.
\end{align}
In the case, since $B^{-1} A\notin M_4(\BZ)$ , we choose $\mathbf{Q}=\diag(2,1,1,1)$. Then,
\begin{equation}
\begin{gathered}
    K=\begin{pmatrix}
            6 & 0 & 2 & 0\\
            0 & 2 & 1 & 1\\
            2 & 1 & 2 & 1\\
            0 & 1 & 1 & 1  \end{pmatrix}, \quad N=2\,, \quad \CO_{E_1} =  V_{(0,1,0,-1)}, \quad 
            \CO_{E_2} = \phi_2 \phi_4 V_{(1,0,-1,0)}\,, \\
    \boldsymbol{\mu}_{\rm geo}=(0,0,0,-1)^T, \quad \boldsymbol{v}_m=(0,1,0,1)^T, \quad \boldsymbol{v}_1=(-1,0,-1,0)^T .
\end{gathered}
\end{equation}
\\
$\n_a=0$ and $\n_b=2$. The half-indices at the A/B-twisting points are
\begin{align}
\begin{split}
\CI^{a}_{\rm half}[\m_m \in 2\mathbb{Z}] & = 1 + q + q^2 + 2q^3 + 3q^4 + 4q^5 + 6q^6 + 8q^7 + 11q^8 + 15q^9+ 19q^{10}+\ldots \\
& = q^{5/168} \chi^{\CM(3,7)}_{(1,2)}(q)\,, \\
\CI^{b}_{\rm half} [\m_m \in 2\mathbb{Z}] &= 1 + q^2 + q^3 + 2q^4 + 2q^5 + 4q^6 + 4q^7 + 7q^8 + 8q^9+ 12q^{10}+14 q^{11}+\ldots \\
& = q^{-25/168} \chi^{\CM(3,7)}_{(1,1)}(q)\,.
\end{split}
\end{align}
\\
\paragraph{ii)} When we choose the following polarizations
\begin{align}
    \mathbf{X}=(Z_1,Z_2'',Z_3',Z_4'')^T\textrm{ and }\mathbf{P}=(Z_1'',Z_2',Z_3,Z_4')^T\;,
\end{align}
the NZ matrices are
\begin{align}
    A=\begin{pmatrix}
        0 & 0 & -1 & 1 \\
        1 & -2 & 0 & -2 \\
        0 & 1 & 0 & 0 \\
        0 & 0 & 1 & 1 \\
    \end{pmatrix},\quad 
    B=\begin{pmatrix}
        0 & 1 & -1 & 1 \\
        -1 & -2 & 2 & -2 \\
        0 & 0 & 0 & 1 \\
        2 & 1 & 0 & 0 \\
    \end{pmatrix},\quad \boldsymbol{\nu}_x=(0,3,-2,-2)^T\;.
\end{align}
In the case, $\BI-B^{-1} A>0$, and we can choose $\mathbf{Q}=\BI$. Then,
\begin{equation}
\begin{gathered}
    \overline{K}=\begin{pmatrix}
        2 & -2 & -2 & 0 \\
        -2 & 5 & 3 & -1 \\
        -2 & 3 & 3 & 0 \\
        0 & -1 & 0 & 1 \\
    \end{pmatrix},\quad \overline{\CO}_{E_1} =  \phi_2 V_{(0,0,0,1)}\,, \quad \overline{\CO}_{E_2} =  \phi_3 \phi_4 V_{(2,1,0,0)}\,, \\
    \overline{\boldsymbol{\m}}_\mathrm{geo}=(2,-5,-3,1)^T, \quad \boldsymbol{v}_m=(-2,4,3,0)^T, \quad \boldsymbol{v}_1=(-1,2,2,0)^T\,.
\end{gathered}
\end{equation}
\\
$\n_a=0$ and $\n_b=2$. The half-indices at the A/B-twisting points are
\begin{align}
        \overline{\CI_\mathrm{half}^a}[\m_m \in 2\mathbb{Z}] = \overline{\CI_\mathrm{half}^b}[\m_m \in 2\mathbb{Z}] = 0\;.
\end{align}
\\
\\
\\
By taking the reduction procedure on the gluing equations in \eqref{Example : (2,7)}, we have
\begin{align}
H^{(\rm r)}_1 := C^{(\rm r)}_3 = 2Z_1+Z_1'+2Z_3, \quad E^{(\rm r)}_1=C^{(\rm r)}_1= 2Z_1''+Z_3'\;.
\end{align}

Among the $3^2$ possible polarization choices, there are four choices satisfying the positivity condition.
\\
\paragraph{reduced-i)} When we choose the following polarizations
\begin{align}
    \mathbf{X}_{\rm r} = (Z_1',Z_3') \textrm{ and } \mathbf{P}_{\rm r} = (Z_1,Z_3),
\end{align}
the reduced NZ matrices are
\begin{align}
A_{\rm r} = \begin{pmatrix} 1 & 0 \\ -2 & 1 \end{pmatrix}, \quad  B_{\rm r} = \begin{pmatrix} 2 & 2 \\ -2 & 0 \end{pmatrix}, \quad \boldsymbol{\nu}_{x;{\rm r}} = (-2,0)^T\;.
\end{align}
In the case, we choose $\mathbf{Q}_{\rm r} = \textrm{diag} \{1,2 \}$. Then,
\begin{equation}
\begin{gathered}
    K_{\rm r} = \left( \begin{array}{cc} 1 & -1 \\ -1 & 2 \\ \end{array} \right), \quad N_r= 1, \quad \CO_{E_1^{\rm(r)}} = \phi_2 V_{(2,0)}\,, \quad \boldsymbol{\mu}^{\rm (r)}_{\rm geo} =(0,-2)^T, \,\, \boldsymbol{v}^{\rm (r)}_1 =(0,1)^T.
\end{gathered}
\end{equation}
\\
$\n_b=2$, and at the A/B-twisting point, $\mathbf{HF^2}$ are
\begin{align} \label{2,7_HF_fermionic}
\begin{split}
    & \mathbf{HF^2}_{\rm a} \simeq \bigg\{
    \Big(\frac{7}{4\sin^2(\tfrac{2\pi}{7})},\; 1\Big),\;
    \Big(\frac{7}{4\sin^2(\tfrac{4\pi}{7})},\; \mathbf{e}(-\tfrac{3}{7})\Big),\;
    \Big(\frac{7}{4\sin^2(\tfrac{6\pi}{7})},\; \mathbf{e}(-\tfrac{1}{7})\Big)
    \bigg\} \,,\\
    & \mathbf{HF^2}_{\rm b} \simeq \bigg\{
    \Big(\frac{7}{4\sin^2(\tfrac{4\pi}{7})},\; 1\Big),\;
    \Big(\frac{7}{4\sin^2(\tfrac{6\pi}{7})},\; \mathbf{e}(\tfrac{1}{7})\Big),\;
    \Big(\frac{7}{4\sin^2(\tfrac{2\pi}{7})},\; \mathbf{e}(-\tfrac{2}{7})\Big)
    \bigg\} \,.
\end{split}
\end{align}
Note that $\mathbf{HF^2}_{\rm a}$ is equivalent to that of $M(2,7)$ up to complex conjugation.
\\
The half-indices at the A/B-twisting points are
\begin{align}
\begin{split}
\CI^{a;\textrm{(r)}}_{\rm half} &= 
1+q^{1/2}+2q^{3/2}+3q^2+3q^{5/2}+4q^3+6q^{7/2}+10q^4+12q^{9/2}+14q^5+20q^{11/2}\\
&+28q^6+33q^{13/2}+40q^7+54q^{15/2}+70q^{8}+84q^{17/2}+102q^9+130q^{19/2}+163q^{10}+\ldots \\
&= q^{-1/84} \chi_F(q) \, \chi^{SM(3,7)}_{(1,1)}(q)\,, \\
\CI^{b;\textrm{(r)}}_{\rm half} &= 
1+2q^{1/2}+2q+3q^{3/2}+5q^2+8q^{5/2}+10q^3+13q^{7/2}+19q^4+26q^{9/2}+34q^5+42q^{11/2} \\
&+56q^6+74q^{13/2}+92q^7+115q^{15/2}+147q^{8}+186q^{17/2}+228q^9+280q^{19/2}+348q^{10}
+\ldots \\
& =q^{5/84} \chi_F(q) \, \chi^{SM(3,7)}_{(1,3)}(q)\,.
\end{split}
\end{align}
The theory $T^{\rm (r)}[A_r, B_r ; \{E_1^{\rm (r)}\}]$ is $\widetilde{\CT}_{(3,7)}$ in \cite{Baek:2024tuo}, which is the bulk field theory for (supersymmetric minimal model $SM(3,7)$)$\otimes (\textrm{free fermion})$.
\\
\paragraph{reduced-ii)} When we choose the following polarizations
\begin{align}
    \mathbf{X}_{\rm r} = (Z'_1,Z''_3) \textrm{ and } \mathbf{P}_{\rm r} = (Z_1,Z'_3),
\end{align}
the reduced NZ matrices are
\begin{align}
A_{\rm r} = \begin{pmatrix} 1 & -2 \\ -2 & 0 \end{pmatrix}, \quad  B_{\rm r} = \begin{pmatrix} 2 & -2 \\ -2 & 1 \end{pmatrix}, \quad \boldsymbol{\nu}_{x;{\rm r}} = (0,0)^T\;.
\end{align}
In the case, we choose $\mathbf{Q}_{\rm r} = \textrm{diag} \{2,1\}$. Then,
\begin{equation}
\begin{gathered}
    K_{\rm r} = \left( \begin{array}{cc} 6 & 2 \\ 2 & 2 \\ \end{array} \right), \quad N_r = 2, \quad \CO_{E_1^{\rm(r)}} = V_{(1,-1)}\,, \quad \boldsymbol{\mu}^{\rm (r)}_{\rm geo} =(0,0)^T, \quad \boldsymbol{v}^{\rm (r)}_1 =(-1,-1)^T.
\end{gathered}
\end{equation}
\\
$\n_a=0$ and $\n_b=2$. The half-indices at the A/B-twisting points are
\begin{align}
\begin{split}
\CI^{a;\textrm{(r)}}_{\rm half} & = 1 + q + q^2 + 2q^3 + 3q^4 + 4q^5 + 6q^6 + 8q^7 + 11q^8 + 15q^9+ 19q^{10}+\ldots \\
& = q^{5/168} \chi^{\CM(3,7)}_{(1,2)}(q)\,, \\
\CI^{b;\textrm{(r)}}_{\rm half} &= 1 + q^2 + q^3 + 2q^4 + 2q^5 + 4q^6 + 4q^7 + 7q^8 + 8q^9+ 12q^{10}+14 q^{11}+\ldots \\
& = q^{-25/168} \chi^{\CM(3,7)}_{(1,1)}(q)\,.
\end{split}
\end{align}
It is the Nahm-sum like expression \eqref{Nahmsum3,l} with $l=7$.
\\
\paragraph{reduced-iii)} When we choose the following polarizations
\begin{align}
    \mathbf{X}_{\rm r} = (Z_1,Z'_3) \textrm{ and } \mathbf{P}_{\rm r} = (Z''_1,Z_3),
\end{align}
the reduced NZ matrices are
\begin{align}
A_{\rm r} = \begin{pmatrix} 1 & 0 \\ 0 & 1 \end{pmatrix}, \quad  B_{\rm r} = \begin{pmatrix} -1 & 2 \\ 2 & 0 \end{pmatrix}, \quad \boldsymbol{\nu}_{x;{\rm r}} = (-1,-2)^T\;.
\end{align}
In the case, $\BI-B^{-1} A>0$ and we choose $\mathbf{Q}_{\rm r} = \textrm{diag} \{1,2\}$. Then,
\begin{equation}
\begin{gathered}
    \overline{K}_{\rm r} = \left( \begin{array}{cc} 1 & -1 \\ -1 & 3 \\ \end{array} \right), \quad N_r = 2, \quad \overline{\CO}_{E_1^{\rm(r)}} = \phi_2^2 V_{(2,0)}\,, \quad 
    \overline{\boldsymbol{\mu}}^{\rm (r)}_{\rm geo} =(0,0)^T, \,\, \boldsymbol{v}^{\rm (r)}_1 =(0,1)^T.
\end{gathered}
\end{equation}
\\
$\n_a=0$ and $\n_b=2$. The half-indices at the A/B-twisting points are
\begin{align}
\begin{split}
\overline{\CI^{a;\textrm{(r)}}_{\rm half}} &=
1+q^{1/2}+q+3q^{3/2}+3q^2+4q^{5/2}+6q^3+8q^{7/2}+11q^4+14q^{9/2}+18q^5+24q^{11/2} \\
&+31q^6+37q^{13/2}+47q^7+60q^{15/2}+74q^{8}+90q^{17/2}+111q^9+137q^{19/2}+166q^{10}+\ldots \\
&= q^{3/56} \chi_F(q) \left(\chi^{\CM(4,7)}_{(1,2)}(q)+\chi^{\CM(4,7)}_{(1,5)}(q)\right)\,, \\
\overline{\CI^{b;\textrm{(r)}}_{\rm half}} &=
1+q^{1/2}+q^{3/2}+2q^2+3q^{5/2}+3q^3+4q^{7/2}+7q^4+9q^{9/2}+10q^5+12q^{11/2} \\
&+18q^6+23q^{13/2}+26q^7+33q^{15/2}+44q^{8}+53q^{17/2}+62q^9+76q^{19/2}+96q^{10}+\ldots \\
&= q^{-1/56} \chi_F(q) \left(\chi^{\CM(4,7)}_{(1,1)}(q)+\chi^{\CM(4,7)}_{(1,6)}(q)\right)\,.
\end{split}
\end{align}
These are the characters of the RCFT $(\textrm{fermionized } \mathcal{M}(4,7):=  \mathcal{M}(4,7)/\mathbb{Z}_2^{f} )\otimes (\textrm{free fermion})$.
\\
\\
\subsubsection{$(P,Q)=(2,9)$}
There are two ways to obtain the gluing data of the torus knot using the SnapPy. The first way is using the Dowker-Thistlethwaite (DT) code:
\begin{align*}
    \mathtt{Manifold('DT:[(10,12,14,16,18,2,4,6,8)]').gluing\_equations\_pgl()}
\end{align*}
The torus knot complement can be ideally triangulated with $r=5$ tetrahedra with following gluing data
\begin{align}
\begin{split}
& C_1=Z_4'+Z_5'',\quad C_2=Z_1''+2 Z_2'+Z_3,\quad C_3=2 Z_1'+Z_2'',\\
& C_4=2 Z_1+Z_1''+2 Z_2+2 Z_3'+2 Z_4''+2 Z_5',\\
& C_5=Z_2''+Z_3+2 Z_3''+2 Z_4+Z_4'+2 Z_5+Z_5'',\\
& M=2 Z_1+Z_1''+2 Z_2+Z_3'-Z_3''-Z_4'+Z_4''+Z_5'
\end{split}
\label{(2,9) gluing I}
\end{align}
In the ideal triangulation, there are three easy internal edges $\{E_I\}=\{C_1,C_2,C_3\}$, and we choose $\{H_I\}=\{C_4\}$.

Among the $3^5$ possible polarization choices, there are four choices satisfying the positivity condition.
\\
\paragraph{i)} When we choose the following polarizations
\begin{align}
    \mathbf{X}=(Z_1,Z_2,Z_3',Z_4'',Z_5')^T \textrm{ and }\mathbf{P}=(Z_1'',Z_2'',Z_3,Z_4',Z_5)^T\;,
\end{align}
the NZ matrices are
\begin{align}
    A=\begin{pmatrix}
        2 & 2 & 2 & 1 & 1\\
        2 & 2 & 2 & 2 & 2\\
        0 & 0 & 0 & 0 & -1\\
        0 & -2 & 0 & 0 & 0\\
        -2 & 0 & 0 & 0 & 0
    \end{pmatrix}, \quad B=\begin{pmatrix}
        1 & 0 & 1 & -1 & 0\\
        1 & 0 & 0 & 0 & 0\\
        0 & 0 & 0 & 1 & -1\\
        1 & -2 & 1 & 0 & 0\\
        -2 & 1 & 0 & 0 & 0
    \end{pmatrix},\quad {\boldsymbol{\nu}}_x=(-1,-2,-1,0,0)^T\;.
\end{align}
In the case, $B^{-1} A\in M_5(\mathbb{Z})$ and $|\det B|=1$, thus we can choose $\mathbf{Q}=\mathbb{I}$ and $N=1$. Then,
\begin{align}
    \begin{split}
        & K=\begin{pmatrix}
            2 & 2 & 2 & 2 & 2\\
            2 & 4 & 4 & 4 & 4\\
            2 & 4 & 6 & 6 & 6\\
            2 & 4 & 6 & 7 & 7\\
            2 & 4 & 6 & 7 & 8
        \end{pmatrix},\quad \boldsymbol{\m}_\mathrm{geo}=(-2,-4,-6,-7,-6)^T,\\
            & \boldsymbol{v}_m=(0,0,0,-1,-1)^T,\quad \boldsymbol{v}_1=(1,2,3,4,4)^T\;.
        \end{split}
\end{align}
The chiral primary operators corresponding to each easy internal edges are
\begin{align}
    \CO_{E_1}=V_{(0,0,0,-1,1)},\quad \CO_{E_2}=V_{(-1,2,-1,0,0)},\quad \CO_{E_3}=V_{(2,-1,0,0,0)}\;.
\end{align}
\\
$\n_a=0$ and $\n_b=2$. The half-indices at the A/B-twisting points are
\begin{align}
    \begin{split}
        \CI_\mathrm{half}^a[\m_m \in 2\mathbb{Z}] & =1+q^2+q^3+2 q^4+2 q^5+4 q^6+4 q^7+6 q^8+7 q^9+10 q^{10}+12 q^{11}+\ldots \\
        & =q^{-23/36} \c_{(1,1)}^{\CM(2,9)}(q)\;,\\
        \CI_\mathrm{half}^b[\m_m \in 2\mathbb{Z}] & =1+q+2 q^2+3 q^3+4 q^4+5 q^5+8 q^6+10 q^7+14 q^8+18 q^9+24 q^{10}+\ldots\\
        & =q^{1/36} \c_{(1,4)}^{\CM(2,9)}(q)\;.
    \end{split}
\end{align}
\\
\paragraph{ii)} When we choose the following polarizations
\begin{align}
    \mathbf{X}=(Z_1'',Z_2'',Z_3,Z_4'',Z_5'')^T\textrm{ and }\mathbf{P}=(Z_1',Z_2',Z_3'',Z_4',Z_5')^T\;,
\end{align}
the NZ matrices are
\begin{align}
    A=\begin{pmatrix}
        -1 & -2 & -1 & 1 & 0\\
        -1 & -2 & -2 & 2 & 0\\
        0 & 0 & 0 & 0 & 1\\
        1 & 0 & 1 & 0 & 0\\
        0 & 1 & 0 & 0 & 0
    \end{pmatrix},\quad B=\begin{pmatrix}
        -2 & -2 & -2 & -1 & 1\\
        -2 & -2 & -2 & 0 & 2\\
        0 & 0 & 0 & 1 & 0\\
        0 & 2 & 0 & 0 & 0\\
        2 & 0 & 0 & 0 & 0
    \end{pmatrix},\quad \boldsymbol{\n}_x=(5,4,-2,-2,-2)^T\;.
\end{align}
In the case, $B^{-1} A\notin M_5(\BZ)$, $\BI-B^{-1} A>0$ and $|\det B|=8$. If we choose $\mathbf{Q}=\diag (1,2,2,1,1)$, then $N=2$ and
\begin{align}
    \begin{split}
        & \overline{K}=\begin{pmatrix}
            1 & -1 & 0 & 0 & 0\\
            -1 & 4 & -2 & 0 & 0\\
            0 & -2 & 6 & 0 & 2\\
            0 & 0 & 0 & 1 & -1\\
            0 & 0 & 2 & -1 & 2
        \end{pmatrix},\quad \overline{\boldsymbol{\m}}_\mathrm{geo}=(0,0,-4,1,-2)^T,\\
        & \boldsymbol{v}_m=(0,0,-2,0,-1)^T,\quad \boldsymbol{v}_1=(0,0,1,0,1)^T\;.
    \end{split}
\end{align}
\\
$\n_a=0$ and $\n_b=2$. The half-indices at the A/B-twisting points are
\begin{align}
    \overline{\CI_\mathrm{half}^a}[\m_m \in 2\mathbb{Z}] = \overline{\CI_\mathrm{half}^b}[\m_m \in 2\mathbb{Z}] =0\,.
\end{align}
\\
\\
By taking the reduction procedure on the gluing equations in \eqref{(2,9) gluing I}, we have
\begin{align}
\begin{split}
    & H_1^{\mathrm{(r)}}:=C_4^{\mathrm{(r)}}=2 Z_1+Z_1''+2 Z_2+2 Z_3',\\
    & E_1^{\mathrm{(r)}}=C_1^{\mathrm{(r)}} = Z_1''+2 Z_2'+Z_3, \quad E_2^{\mathrm{(r)}}=C_2^{\mathrm{(r)}} = 2 Z_1'+Z_2''
    \end{split}
\end{align}

Among the $3^3$ possible polarization choices, there are four choices satisfying the positivity condition.
\\
\paragraph{reduced-i)} When we choose the following polarizations
\begin{align}
    \mathbf{X}_\mathrm{r}=(Z_1,Z_2,Z_3')\textrm{ and }\mathbf{P}_\mathrm{r}=(Z_1'',Z_2'',Z_3),
\end{align}
the reduced NZ matrices are
\begin{align}
    A_\mathrm{r}=\begin{pmatrix}
        2 & 2 & 2\\
        0 & -2 & 0\\
        -2 & 0 & 0
    \end{pmatrix},\quad B_\mathrm{r}=\begin{pmatrix}
        1 & 0 & 0\\
        1 & -2 & 1\\
        -2 & 1 & 0
    \end{pmatrix},\quad \boldsymbol{\n}_{x;\mathrm{r}}=(-2,0,0)^T\;.
\end{align}
In the case, we choose $\mathbf{Q}_\mathrm{r}=\BI$. Then, $N_\mathrm{r}=1$ and 
\begin{align}
    K_\mathrm{r}=(B^{-1} A)_\mathrm{r}=\begin{pmatrix}
        2 & 2 & 2\\
        2 & 4 & 4\\
        2 & 4 & 6
    \end{pmatrix},\quad \boldsymbol{\m}_\mathrm{geo}^\mathrm{(r)}=(-2,-4,-6)^T\,, \quad \boldsymbol{v}_1^\mathrm{(r)}=(1,2,3)^T\,.
\end{align}
The chiral primary operators corresponding to each easy internal edges are
\begin{align}
    \CO_{E_1^\mathrm{(r)}}=V_{(-1,2,-1)},\quad \CO_{E_2^\mathrm{(r)}}=V_{(2,-1,0)}\;.
\end{align}
In the present choice of polarization, the Nahm-sum like expression \eqref{Nahmsum2,k} with $k=3$ is reproduced.
\\
\\
\\
The second way to obtain the gluing data of the torus knot is using the following input:
\begin{align*}
    \mathtt{Manifold('9\_1').gluing\_equations\_pgl()}
\end{align*}
The torus knot complement can be ideally triangulated with $r=5$ tetrahedra with following gluing data
\begin{align}
    \begin{split}
        &C_1=Z_1'+Z_2'', \quad C_2=Z_1+Z_2+Z_3''+2 Z_4+Z_5, \quad C_3=Z_4''+2 Z_5', \\
        &C_4=2 Z_1''+2 Z_2'+2 Z_3'+2 Z_4'+Z_5+2 Z_5'',\\ &C_5=Z_1+Z_1'+Z_2+Z_2''+2 Z_3+Z_3''+Z_4'',\quad M=Z_1-Z_2'+Z_3\;.
    \end{split}
    \label{(2,9) gluing II}
\end{align}
In the ideal triangulation, there are three easy internal edges $\{E_I\}=\{C_1,C_2,C_3\}$, and we choose $\{H_I\}=\{C_4\}$.

Among the $3^5$ possible polarization choices, there are four choices satisfying the positivity condition.
\\
\paragraph{iii)} When we choose the quad structure as follows
\begin{align}
    \mathbf{X}=(Z_1'',Z_2',Z_3,Z_4'',Z_5)^T \textrm{ and } \mathbf{P}=(Z_1',Z_2,Z_3'',Z_4',Z_5'')\;,
\end{align}
the NZ matrices are
\begin{align}
    A=\begin{pmatrix}
        -1 & -1 & 1 & 0 & 0\\
        2 & 2 & -2 & 0 & 1\\
        0 & -1 & 0 & 0 & 0\\
        -1 & 0 & 0 & -2 & 1\\
        0 & 0 & 0 & 1 & -2
    \end{pmatrix}, \quad B=\begin{pmatrix}
        -1 & 0 & 0 & 0 & 0\\
        0 & 0 & -2 & 2 & 2\\
        1 & -1 & 0 & 0 & 0\\
        -1 & 1 & 1 & -2 & 0\\
        0 & 0 & 0 & 0 & -2
    \end{pmatrix},\quad {\boldsymbol{\n}}_x=(1,0,-1,1,0)^T\;.
\end{align}
In the case, $B^{-1} A\notin M_5(\BZ)$ and $|\det B|=4$. If we choose $\mathbf{Q}=\diag (1,1,1,2,1)$, $N=1$ and $\mathbf{Q}^T B^{-1} A \mathbf{Q}\in M_5(\BZ)$. Then,
\begin{align}
    \begin{split}
       &  K=\begin{pmatrix}
            1 & 1 & -1 & 0 & 0\\
            1 & 2 & -1 & 0 & 0\\
            -1 & -1 & 2 & 2 & 0\\
            0 & 0 & 2 & 6 & -1\\
            0 & 0 & 0 & -1 & 1
        \end{pmatrix},\quad {\boldsymbol{\m}}_{\rm geo}=(-1,0,0,0,0)^T,\\
        & \boldsymbol{v}_m=(-1,-1,0,0,0)^T,\quad \boldsymbol{v}_1=(0,0,-1,-1,0)^T\;.
    \end{split}
\end{align}
The chiral primary operators corresponding to each easy internal edges are
\begin{align}
    \CO_{E_1}=V_{(-1,1,0,0,0)}, \quad \CO_{E_2}=\phi_5 V_{(1,-1,-1,1,0)},\quad \CO_{E_3}=\phi_4 V_{(0,0,0,0,2)}
\end{align}
\\
$\n_a=0$ and $\n_b=2$. The half-indices at the A/B-twisting points are
\begin{align}
    \begin{split}
        \CI_\mathrm{half}^a[\m_m \in 2\mathbb{Z}] & =1+q^{1/2}+q+2 q^{3/2}+2 q^2+4 q^{5/2}+5 q^3+6 q^{7/2}+9 q^4+\ldots\\
        & =q^\frac{1}{18} \c_F(q) \paren{\c_{(1,2)}^{\CM(4,9)}(q)+\c_{(3,2)}^{\CM(4,9)}(q)}\;,\\
        \CI_\mathrm{half}^b[\m_m \in 2\mathbb{Z}] & =1+q^{1/2}+q^{3/2}+2 q^2+2 q^{5/2}+2 q^3+4 q^{7/2}+6 q^4+7 q^{9/2}+\ldots\\
        & =q^{-\frac{1}{9}}\c_F(q)\paren{\c_{(1,1)}^{\CM(4,9)}(q)+\c_{(3,1)}^{\CM(4,9)}(q)}\;.
    \end{split}
\end{align}
These are the characters of the RCFT
\[
\bigl(\text{fermionized } \mathcal{M}(4,9):=\mathcal{M}(4,9)/\mathbb{Z}_2^{f}\bigr)
\otimes (\text{free fermion}) \, .
\]
The fermionic $\mathbb{Z}_2$ symmetry $\mathbb{Z}_2^{f}$ is generated by the
Verlinde loop operator associated with the primary $(3,1)$, whose conformal
dimension is $h_{3,1}=\tfrac{7}{2}$.
\\
\paragraph{iv)} When we choose the quad structure as follows
\begin{align}
    \mathbf{X}=(Z_1'',Z_2'',Z_3'',Z_4',Z_5'')^T\textrm{ and }\mathbf{P}=(Z_1',Z_2',Z_3',Z_4,Z_5')^T\;,
\end{align}
the NZ matrices are
\begin{align}
    A=\begin{pmatrix}
        -1 & 0 & -1 & 0 & 0\\
        2 & 0 & 0 & 2 & 1\\
        0 & 1 & 0 & 0 & 0\\
        -1 & -1 & 1 & 0 & -1\\
        0 & 0 & 0 & -1 & 0
    \end{pmatrix},\quad B=\begin{pmatrix}
        -1 & -1 & -1 & 0 & 0\\
        0 & 2 & 2 & 0 & -1\\
        1 & 0 & 0 & 0 & 0\\
        -1 & -1 & 0 & 2 & -1\\
        0 & 0 & 0 & -1 & 2
    \end{pmatrix},\quad \boldsymbol{\n}_x=(2,-1,-2,1,-1)^T\;.
\end{align}
In the case, $B^{-1} A\in M_5(\BZ)$, $\BI-B^{-1} A>0$, and $|\det B|=1$. If we choose $\mathbf{Q}=\BI$, then $N=1$ and
\begin{align}
    \begin{split}
        & \overline{K}=\begin{pmatrix}
            1 & -1 & 0 & 0 & 0\\
            -1 & 7 & -5 & 4 & 2\\
            0 & -5 & 5 & -4 & -2\\
            0 & 4 & -4 & 4 & 2\\
            0 & 2 & -2 & 2 & 2
        \end{pmatrix},\quad \overline{\boldsymbol{\m}}_\mathrm{geo}=(1,-7,5,-4,-2)^T,\\
        & \boldsymbol{v}_m=(0,-6,5,-4,-2)^T,\quad \boldsymbol{v}_1=(0,-3,3,-2,-1)^T\;.
    \end{split}
\end{align}
\\
$\n_a=0$ and $\n_b=2$. The half-indices at the A/B-twisting points are
\begin{align}
    \overline{\CI_\mathrm{half}^a}[\m_m \in 2\mathbb{Z}] = \overline{\CI_\mathrm{half}^b}[\m_m \in 2\mathbb{Z}] = 0 \,.
\end{align}
\\
\\
\\
By taking the reduction procedure on the gluing equations in \eqref{(2,9) gluing II}, we have
\begin{align}
    \begin{split}
        & H_1^\mathrm{(r)}:=C_4^\mathrm{(r)}=2 Z_3'+2 Z_4'+Z_5+2 Z_5'',\\
        & E_1^\mathrm{(r)}=C_2^\mathrm{(r)}=Z_3''+2 Z_4+Z_5,\quad E_2^\mathrm{(r)}=C_3^\mathrm{(r)}=Z_4''+2 Z_5'\;.
    \end{split}
\end{align}
\\
Among the $3^3$ possible polarization choices, there are four choices satisfying the positivity condition.
\\
\paragraph{reduced-ii)} When we choose the following polarizations
\begin{align}
    \mathbf{X}_\mathrm{r}=(Z_3'',Z_4'',Z_5)\textrm{ and }\mathbf{P}_\mathrm{r}=(Z_3',Z_4',Z_5'')\;,
\end{align}
the reduced NZ matrices are
\begin{align}
    A_\mathrm{r}=\begin{pmatrix}
        0 & 0 & 1\\
        1 & -2 & 1\\
        0 & 1 & -2
    \end{pmatrix},\quad B_\mathrm{r}=\begin{pmatrix}
        2 & 2 & 2\\
        0 & -2 & 0\\
        0 & 0 & -2
    \end{pmatrix},\quad \boldsymbol{\n}_{x;\mathrm{r}}=(-2,0,0)^T\;.
\end{align}
In the case, we choose $\mathbf{Q}_\mathrm{r}=\diag (2,2,1)$. Then, $N_\mathrm{r}=2$ and
\begin{align}
    \begin{split}
        & K_\mathrm{r}=\begin{pmatrix}
            2 & -2 & 0\\
            -2 & 4 & -1\\
            0 & -1 & 1
        \end{pmatrix},\quad \boldsymbol{\m}_\mathrm{geo}^\mathrm{(r)}=(-2,0,0)^T,\\
        & \boldsymbol{v}_1^\mathrm{(r)}=(1,0,0)^T\;.
    \end{split}
\end{align}
The chiral primary operators corresponding to each easy internal edges are
\begin{align}
    \CO_{E_1^{\mathrm{(r)}}}=\f_1 \f_3 V_{(0,1,0)},\quad \CO_{E_2^{\mathrm{(r)}}}=\f_2 V_{(0,0,2)}\;.
\end{align}
$\n_a=0$ and $\n_b=2$. The half-indices at the A/B-twisting points are
\begin{align}
    \begin{split}
        \CI_\mathrm{half}^{a;\mathrm{(r)}} & =1+q^{1/2}+3 q^{3/2}+6 q^2+7 q^{5/2}+10 q^3+18 q^{7/2}+30 q^4+\ldots\;,\\
        \CI_\mathrm{half}^{b;\mathrm{(r)}} & =1+2 q^{1/2}+3 q+7 q^{3/2}+12 q^2+19 q^{5/2}+31 q^3+48 q^{7/2}+...\;.
    \end{split}
\end{align}
We could not identify these half-indices with previously known RCFT characters.
Instead, we find that they can be completed into a full set of eight characters
\(\{\chi_\alpha(q)\}_{\alpha=0}^{7}\) given by 
\begin{align}
    \begin{split}
        \c_0(q)& =q^{-1/72} \sum_{\mathbf{m}\in (\BZ_{\geq 0})^3} \frac{q^{\frac{1}{2}\mathbf{m}^T\cdot K_\mathrm{r}\cdot \mathbf{m}} (-q^{1/2})^{2 m_1}}{(q)_{2 m_1} (q)_{2 m_2} (q)_{m_3}}=q^{-1/72} \CI_\mathrm{half}^{a;\mathrm{(r)}}\,,\\
        \c_1(q) &=q^{41/72}\sum_{\mathbf{m}\in (\BZ_{\geq 0})^3} \frac{q^{\frac{1}{2}\mathbf{m}^T\cdot K_\mathrm{r}\cdot \mathbf{m}} (-q^{1/2})^{2 m_1}}{(q)_{2 m_1} (q)_{2 m_2} (q)_{m_3}}\paren{q^{m_1-2 m_2}-q^{m_1}},\\
        \c_2(q)& =q^{1/24}\sum_{\mathbf{m}\in (\BZ_{\geq 0})^3} \frac{q^{\frac{1}{2}\mathbf{m}^T\cdot K_\mathrm{r}\cdot \mathbf{m}} (-q^{1/2})^{2 m_1}}{(q)_{2 m_1} (q)_{2 m_2} (q)_{m_3}} q^{-m_2},\\
        \c_3(q)& =q^{-7/72}\sum_{\mathbf{m}\in (\BZ_{\geq 0})^3} \frac{q^{\frac{1}{2}\mathbf{m}^T\cdot K_\mathrm{r}\cdot \mathbf{m}} (-q^{1/2})^{2 m_1}}{(q)_{2 m_1} (q)_{2 m_2} (q)_{m_3}} q^{-m_1}=q^{-7/72} \CI_\mathrm{half}^{b;\mathrm{(r)}}\,,\\
        \c_4(q) &=q^{11/72}\sum_{\mathbf{m}\in (\BZ_{\geq 0})^3} \frac{q^{\frac{1}{2}\mathbf{m}^T\cdot K_\mathrm{r}\cdot \mathbf{m}} (-q^{1/2})^{2 m_1}}{(q)_{2 m_1} (q)_{2 m_2} (q)_{m_3}}\paren{q^{-2 m_1+m_2}-q^{m_2}},\\
        \c_5(q) &=q^{-5/24}\sum_{\mathbf{m}\in (\BZ_{\geq 0})^3} \frac{q^{\frac{1}{2}\mathbf{m}^T\cdot K_\mathrm{r}\cdot \mathbf{m}} (-q^{1/2})^{2 m_1}}{(q)_{2 m_1} (q)_{2 m_2} (q)_{m_3}}\paren{q^{-m_1}-q^{m_1}},\\
        \c_6(q) &=q^{-13/72}\sum_{\mathbf{m}\in (\BZ_{\geq 0})^3} \frac{q^{\frac{1}{2}\mathbf{m}^T\cdot K_\mathrm{r}\cdot \mathbf{m}} (-q^{1/2})^{2 m_1}}{(q)_{2 m_1} (q)_{2 m_2} (q)_{m_3}}\paren{q^{m_2}-q^{2 m_1-m_2}-q^{m_2}+q^{2 m_1+m_2}},\\
        \c_7(q)& =q^{-19/72}\sum_{\mathbf{m}\in (\BZ_{\geq 0})^3} \frac{q^{\frac{1}{2}\mathbf{m}^T\cdot K_\mathrm{r}\cdot \mathbf{m}} (-q^{1/2})^{2 m_1}}{(q)_{2 m_1} (q)_{2 m_2} (q)_{m_3}} \paren{q^{-m_1+m_2}-q^{m_1+m_2}}\;.
    \end{split}
\end{align}
These $q$-series satisfy the modular transformation property
\begin{align}
\chi_{\alpha}\!\left(q=e^{-2\pi i/\tau}\right)
= S_{\alpha\beta}\,
\chi_{\beta}\!\left(q=e^{2\pi i\tau}\right),
\end{align}
with
\begin{align}
S_{\alpha\beta}
= (-1)^{\alpha\beta}\,\frac{\sqrt{2}}{3}\,
\sin\!\left(\frac{2(\alpha+1)(\beta+1)\pi}{9}\right).
\end{align}
We have numerically checked this modular relation for various values of \(\tau\)
satisfying \(|e^{2\pi i\tau}|<1\) and \(|e^{2\pi i(-1/\tau)}|<1\), and we expect that
\(\{\chi_{\alpha}(q)\}\) arise as characters of a rational VOA.
\\
\paragraph{reduced-iii)} When we choose the following polarizations
\begin{align}
    \mathbf{X}_\mathrm{r}=(Z_3,Z_4'',Z_5)\textrm{ and }\mathbf{P}_\mathrm{r}=(Z_3'',Z_4',Z_5'')\;,
\end{align}
the reduced NZ matrices are
\begin{align}
    A_\mathrm{r}=\begin{pmatrix}
        -2 & 0 & 1\\
        0 & -2 & 1\\
        0 & 1 & -2
    \end{pmatrix},\quad B_\mathrm{r}=\begin{pmatrix}
        -2 & 2 & 2\\
        1 & -2 & 0\\
        0 & 0 & -2
    \end{pmatrix},\quad \boldsymbol{\n}_{x;\mathrm{r}}=(0,0,0)^T\;.
\end{align}
In the case, we choose $\mathbf{Q}_\mathrm{r}=\diag (1,2,1)$. Then, $N_\mathrm{r}=1$ and
\begin{align}
    \begin{split}
        & K_\mathrm{r}=(\mathbf{Q}^T B^{-1} A \mathbf{Q})_\mathrm{r}=\begin{pmatrix}
            2 & 2 & 0\\
            2 & 6 & -1\\
            0 & -1 & 1
        \end{pmatrix},\quad \boldsymbol{\m}_\mathrm{geo}^\mathrm{(r)}=(0,0,0)^T, \quad \boldsymbol{v}_1^\mathrm{(r)}=(-1,-1,0)^T\;.
    \end{split}
\end{align}
The chiral primary operators corresponding to each easy internal edges are
\begin{align}
    \CO_{E_1^\mathrm{(r)}}=\f_3 V_{(-1,1,0)},\quad \CO_{E_2^\mathrm{(r)}}=\f_2 V_{(0,0,2)}\;.
\end{align}
\\
$\n_a=0$ and $\n_b=2$. The half-indices at the A/B-twisting points are
\begin{align}
    \begin{split}
        & \CI_\mathrm{half}^{a;{\rm (r)}} = q^\frac{1}{18} \c_F(q) \paren{\c_{(1,2)}^{\CM(4,9)}(q)+\c_{(3,2)}^{\CM(4,9)}(q)}\;,\\
        & \CI_\mathrm{half}^{b;{\rm (r)}} = q^{-\frac{1}{9}}\c_F(q)\paren{\c_{(1,1)}^{\CM(4,9)}(q)+\c_{(3,1)}^{\CM(4,9)}(q)}\;.
    \end{split}
\end{align}
These are the characters of the RCFT $(\textrm{fermionized } \mathcal{M}(4,9):=  \mathcal{M}(4,9)/\mathbb{Z}_2^{f} )\otimes (\textrm{free fermion})$.
\\
\paragraph{reduced-iv)} When we choose the following polarizations
\begin{align}
    \mathbf{X}_\mathrm{r}=(Z_3',Z_4',Z_5'')\textrm{ and }\mathbf{P}_\mathrm{r}=(Z_3,Z_4,Z_5'),
\end{align}
the reduced NZ matrices are
\begin{align}
    A_\mathrm{r}=\begin{pmatrix}
        2 & 2 & 1\\
        -1 & 0 & -1\\
        0 & -1 & 0
    \end{pmatrix},\quad B_\mathrm{r}=\begin{pmatrix}
        0 & 0 & -1\\
        -1 & 2 & -1\\
        0 & -1 & 2
    \end{pmatrix},\quad \boldsymbol{\n}_{x;\mathrm{r}}=(-1,0,-1)^T\;.
\end{align}
In the case, $B^{-1} A\in M_3(\BZ)$, $\BI-B^{-1} A>0$ and $|\det B|=1$. Then, as we choose $\mathbf{Q}_\mathrm{r}=\BI$, and
\begin{align}
    \begin{split}
        & \overline{K}_\mathrm{r}=\begin{pmatrix}
            6 & 4 & 2\\
            4 & 4 & 2\\
            2 & 2 & 2
        \end{pmatrix},\quad N_\mathrm{r}=1, \quad \overline{\boldsymbol{\m}}_\mathrm{geo}^\mathrm{(r)}=(-6,-4,-2)^T, \quad \boldsymbol{v}_1^\mathrm{(r)}=(-3,-2,-1)^T\;.
    \end{split}
\end{align}
Compared with {\bf reduced-i)}, although two cases come from the different ideal triangulations, one can obtain the same field theory description.
\\
\\
\subsubsection{$(P,Q)=\left.(2,2r-1)\right|_{r\geq 6}$} From the results of $(P,Q)=\left.(2,2r-1)\right|_{6\leq r\leq 9}$, we guess the general forms of gluing data for all $r\geq 6$. We guess that the torus knot complement can be ideally triangulated with $r$ tetrahedra with the following gluing data
\begin{align}
    \begin{split}
        & C_1=2 Z_1'+Z_2'', \quad \left.C_I\right|_{2\leq I \leq r-3}=Z_{I-1}''+2 Z_I'+Z_{I+1}'', \quad C_{r-2}=Z_{r-1}'+Z_r'', \\
        & C_{r-1}=2 Z_1+Z_1''+\sum_{i=2}^{r-2} 2 Z_i+2 Z_{r-1}''+2 Z_r',\\
        & C_r=Z_{r-3}''+2 Z_{r-2}'+Z_{r-2}''+2 Z_{r-1}+Z_{r-1}'+2 Z_r+Z_r'', \\
        & M=-Z_{r-3}''-2 Z_{r-2}'+Z_{r-1}''-Z_r\;.
    \end{split}
    \label{(2,2r-1) gluing equations}
\end{align}
In the ideal triangulation, there are $r-2$ easy internal edges $\{E_I\}=\{C_{r-2}, C_1,\cdots,C_{r-3}\}$, and we choose $\{H_I\}=\{C_{r-1}\}$.
\\
\paragraph{i)} When we choose the quad structure as follows
\begin{align}
    \mathbf{X}=(Z_1,\cdots,Z_{r-2},Z_{r-1}'',Z_r')^T, \quad \mathbf{P}=(Z_1'',\cdots,Z_{r-2}'',Z_{r-1}',Z_r)\;,
\end{align}
the NZ matrices are
\begin{align}
    \begin{split}
        & A=\begin{pmatrix}
        0 & 0 & \cdots & 0 & 2 & 1 & 0\\
        2 & 2 & \cdots & 2 & 2 & 2 & 2\\
        0 & 0 & \cdots & 0 & 0 & 0 & -1\\
        -2 & 0 & \cdots & 0 & 0 & 0 & 0\\
        0 & -2 & \cdots & 0 & 0 & 0 & 0\\
        \vdots & \vdots & \ddots & \vdots & \vdots & \vdots & \vdots\\
        0 & 0 & \cdots & -2 & 0 & 0 & 0
    \end{pmatrix}, \quad B=\begin{pmatrix}
        0 & 0 & 0 & 0 & \cdots & 0 & -1 & 2 & 0 & -1\\
        1 & 0 & 0 & 0 & \cdots & 0& 0 & 0 & 0 & 0\\
        0 & 0 & 0 & 0 & \cdots & 0 & 0 & 0 & 1 & -1\\
        -2 & 1 & 0 & 0 & \cdots & 0 & 0 & 0 & 0 & 0\\
        1 & -2 & 1 & 0 & \cdots & 0 & 0 & 0 & 0 & 0\\
        \vdots & \vdots & \vdots & \vdots & \ddots & \vdots & \vdots & \vdots & \vdots & \vdots \\
        0 & 0 & 0 & 0 & \cdots & 1 & -2 & 1 & 0 & 0
    \end{pmatrix},\\
    & {\boldsymbol{\nu}}_x=(-2,-2,-1,0,\cdots,0)^T\;.
    \end{split}
\end{align}
In the cases, $B^{-1} A\in M_{r}(\BZ)$ and $|\det B|=1$, thus we can choose $\mathbf{Q}=\BI$ and $N=1$. Then,
\begin{align}
    \begin{split}
        & K=\begin{pmatrix}
            2 & 2 & 2 & \cdots & 2 & 2 & 2\\
            2 & 4 & 4 & \cdots & 4 & 4 & 4\\
            2 & 4 & 6 & \cdots & 6 & 6 & 6\\
            \vdots & \vdots & \vdots & \ddots & \vdots & \vdots & \vdots\\
            2 & 4 & 6 & \cdots & 2r-4 & 2r-4 & 2r-4\\
            2 & 4 & 6 & \cdots & 2r-4 & 2r-3 & 2r-3\\
            2 & 4 & 6 & \cdots & 2r-4 & 2r-3 & 2r-2
        \end{pmatrix},\\
        & {\boldsymbol{\m}}_{\rm geo}=(-2,-4,\cdots,-2r+4,-2r+3,-2r+4)^T,\\
        & \boldsymbol{v}_m=(0,\cdots,0,-1,-1)^T,\\
        & \boldsymbol{v}_1=(1,2,\cdots,r-2,r-1,r-1)^T\;.
    \end{split}
\end{align}
The chiral primary operators corresponding to each easy internal edges are
\begin{align}
    \begin{split}
        & \CO_{E_1}=V_{(0,\cdots,0,-1,1)},\quad \CO_{E_2}=V_{(2,-1,0,\cdots,0)},\quad \CO_{E_3}=V_{(-1,2,-1,0,\cdots,0)},\\
        & \cdots,\quad \CO_{E_{r-2}}=V_{(0,\cdots,0,-1,2,-1,0,0)}\;.
    \end{split}
\end{align}
\\
\\
\\
By taking the reduction procedure on the gluing equations in \eqref{(2,2r-1) gluing equations}, we have
\begin{align}
    \begin{split}
        & H_1^\mathrm{(r)}:=C_{r-1}^\mathrm{(r)}=2 Z_1+Z_1''+\sum_{i=2}^{r-2} 2 Z_i,\\
        & E_1^\mathrm{(r)}=C_1^{(r)}=2 Z_1'+Z_2'',\\
        & \left.E_I^\mathrm{(r)}\right|_{2\leq I\leq r-3}=\left.C_I^\mathrm{(r)}\right|_{2\leq I\leq r-3}=Z_{I-1}''+2 Z_I'+Z_{I+1}''\;.
    \end{split}
\end{align}
\\
\paragraph{reduced-i)}
When we choose the quad structure as follows
\begin{align}
    \mathbf{X}_\mathrm{r}=(Z_1,Z_2,\cdots,Z_{r-2})^T\textrm{ and }\mathbf{P}_\mathrm{r}=(Z_1'',Z_2'',\cdots,Z_{r-2}'')^T\;,
\end{align}
the reduced NZ matrices are
\begin{align}
    \begin{split}
        & A_\mathrm{r}=\begin{pmatrix}
            2 & 2 & \cdots & 2 & 2 & 2\\
            -2 & 0 & \cdots & 0 & 0 & 0\\
            0 & -2 & \cdots & 0 & 0 & 0\\
            \vdots & \vdots & \ddots & \vdots & \vdots & \vdots\\
            0 & 0 & \cdots & -2 & 0 & 0\\
            0 & 0 & \cdots & 0 & -2 & 0
        \end{pmatrix},\quad B_\mathrm{r}=\begin{pmatrix}
            1 & 0 & 0 & 0 & \cdots & 0 & 0 & 0 & 0\\
            -2 & 1 & 0 & 0 & \cdots & 0 & 0 &  0 & 0\\
            1 & -2 & 1 & 0 & \cdots & 0 & 0 & 0 & 0\\
            0 & 1 & -2 & 1 & \cdots & 0 & 0 & 0 & 0\\
            \vdots & \vdots & \vdots & \vdots & \ddots & \vdots & \vdots & \vdots & \vdots\\
            0 & 0 & 0 & 0 & \cdots & 0 & 1 & -2 & 1
        \end{pmatrix},\\
        & \boldsymbol{\n}_{x;\mathrm{r}}=(-2,0,\cdots,0)^T\;.
    \end{split}
\end{align}
In the cases, we choose $\mathbf{Q}_\mathrm{r}=\BI$. Then, $N_\mathrm{r}=1$ and
\begin{equation}
\begin{gathered}
    K_\mathrm{r}=\begin{pmatrix}
            2 & 2 & 2 & \cdots & 2 & 2\\
            2 & 4 & 4 & \cdots & 4 & 4\\
            2 & 4 & 6 & \cdots & 6 & 6\\
            \vdots & \vdots & \vdots & \ddots & \vdots & \vdots\\
            2 & 4 & 6 & \cdots & 2 (r-3) & 2 (r-3)\\
            2 & 4 & 6 & \cdots & 2 (r-3) & 2 (r-2)
        \end{pmatrix}, \\
    \boldsymbol{\m}_\mathrm{geo}^\mathrm{(r)}=(-2,-4,\cdots,-2 (r-2))^T, \quad
    \boldsymbol{v}_1^\mathrm{(r)}=(1,2,\cdots,r-2)^T\;.
\end{gathered}
\end{equation}
The chiral primary operators corresponding to each easy internal edges are
\begin{align}
    \CO_{E_1^\mathrm{(r)}}=V_{(2,-1,0,\cdots,0)},\quad \CO_{E_2^\mathrm{(r)}}=V_{(-1,2,-1,0,\cdots,0)},\quad \cdots,\quad \CO_{E_{r-3}^\mathrm{(r)}}=V_{(0,\cdots,0,-1,2,-1)}\;.
\end{align}
The half-indices at the A- and B-twisting points, obtained using
$K_{\rm r}$, $\boldsymbol{\mu}_{\mathrm{geo}}^{(\mathrm r)}$, and
$\boldsymbol{v}^{(\mathrm r)}_{1}$,
reproduce the Nahm-sum  expressions for the characters of
$\mathcal{M}(2,2r-1)$ given in \eqref{Nahmsum2,k}.
\\
\\
\subsection{The case $(P,Q) = (3,*)$}

\subsubsection{$(P,Q)=(3,4)$}
Using SnapPy, the gluing data of the torus knot can be obtained by the following inputs:
\begin{align*}
    \mathtt{Manifold('8\_19').gluing\_equations\_pgl()}
\end{align*}
or
\begin{align*}
    \mathtt{Manifold('DT:[(6,-8,10,-12,14,-16,2,-4)]').gluing\_equations\_pgl()}
\end{align*}
Both inputs give the same gluing data, up to cyclic choice of $(Z_i,Z_i',Z_i'')$ and labelling of the internal edges. The torus knot complement can be ideally triangulated with $r=3$ tetrahedra with following gluing data
\begin{align} \label{(3,4) gluing equations}
\begin{split}
    & C_1=Z_1+Z_2+2 Z_3,\quad C_2=Z_1'+Z_2'',\\
    & C_3=Z_1+Z_1'+2 Z_1''+Z_2+2 Z_2'+Z_2''+2 Z_3'+2 Z_3'',\\
    & M=Z_1'-Z_1''-Z_2'-Z_2''-2 Z_3'-Z_3''\;.
\end{split}
\end{align}
In the ideal triangulation, there are two easy internal edges $\{E_I\}=\{C_2,C_1\}$.
\\
\\
Among the $3^3$ possible polarization choices, there are two choices satisfying the positivity condition.
\\
\paragraph{i)} When we choose the quad structure as follows
\begin{align}
    \mathbf{X}=(Z_1,Z_2',Z_3'')^T \textrm{ and }\mathbf{P}=(Z_1'',Z_2,Z_3')^T\;,
\end{align}
the NZ matrices are
\begin{align}
    A=\begin{pmatrix}
        -1 & 0 & -1\\
        -1 & -1 & 0\\
        1 & 0 & -2
    \end{pmatrix},\quad B=\begin{pmatrix}
        -2 & 1 & -2\\
        -1 & -1 & 0\\
        0 & 1 & -2
    \end{pmatrix},\quad {\boldsymbol{\n}}_x=(0,0,0)^T\;.
\end{align}
In the case, $B^{-1} A\notin  M_3(\BZ)$ and $|\det B|=4$. If we choose $\mathbf{Q}=\diag (1,1,2)$, 
\begin{align}
    \begin{split}
        & K=\begin{pmatrix}
            1 & 0 & -1\\
            0 & 1 & 1\\
            -1 & 1 & 5
        \end{pmatrix}, \quad N=1, \quad {\boldsymbol{\m}}_{\rm geo}=(0,0,0)^T, \quad \boldsymbol{v}_m=\frac{1}{2}\paren{-1,1,1}^T\;.
    \end{split}
\end{align}
The chiral primary operators corresponding to each easy internal edges are
\begin{align}
    \CO_{E_1}=V_{(1,1,0)}, \quad \CO_{E_2}=\f_1 V_{(0,-1,1)}\;.
\end{align}
The half-index in \eqref{half-indices when |P-Q|=1} is
\begin{align}
    \begin{split}
        \CI_\mathrm{half}[\mu_m \in 4\BZ] &\simeq 1+2q^{1/2}+q+2q^{3/2}+5q^2+6q^{5/2}+7q^3+10q^{7/2}+16q^4+22q^{9/2} \\
        & +26q^5+34q^{11/2}+48q^6+62q^{13/2}+75q^7+96q^{15/2}+128q^8+\ldots\\
        & =q^{1/16} \paren{\c_F(q)}^2 \c_{(1,1)}^{\CM(3,4)}\,, \\
        \CI_\mathrm{half}[\mu_m \in 4\BZ+2] &= 0\,.
    \end{split}
\end{align}
\\
\paragraph{ii)} When we choose the quad structure as follows
\begin{align}
    \mathbf{X}=(Z_1'',Z_2,Z_3')^T \textrm{ and }\mathbf{P}=(Z_1',Z_2'',Z_3)^T\;,
\end{align}
the NZ matrices are
\begin{align}
    A=\begin{pmatrix}
        -1 & 1 & -1\\
        0 & 0 & 0\\
        -1 & 1 & 0
    \end{pmatrix},\quad B=\begin{pmatrix}
        1 & 0 & 1\\
        1 & 1 & 0\\
        -1 & 0 & 2 \end{pmatrix}, \quad {\boldsymbol{\n}}_x=(-2,-2,-1)^T\;.
\end{align}
In the case, $B^{-1} A\notin  M_3(\BZ)$ and $|\det B|=3$. If we choose
\begin{align}
    \mathbf{Q}=\begin{pmatrix}
            1 & 0 & 0 \\
            1 & 1 & -1 \\
            0 & 2 & 1\end{pmatrix}\,,
\end{align}
$N=3$ and $\mathbf{Q}^T B^{-1} A \mathbf{Q}\in M_3(\BZ)$. Then,
\begin{align}
    \begin{split}
        & \overline{K}=\begin{pmatrix}
            2 & 1 & -1 \\
            1 & 4 & 2 \\
            -1 & 2 & 4
        \end{pmatrix}, \quad N=3, \quad \overline{{\boldsymbol{\m}}}_{\rm geo}=(0,0,0)^T, \quad \boldsymbol{v}_m=\paren{0,0,1}^T\;.
    \end{split}
\end{align}
The chiral primary operators corresponding to each easy internal edges are
\begin{align}
    \CO_{E_1} = V_{(1,0,0)},\quad \CO_{E_2} = \phi_2 V_{(-1,1,0)}\;.
\end{align}
\\
The half-index is
\begin{align}
\begin{split}
    \CI_\mathrm{half}[\mu_m] & =q^{1/16-\mu_m^2/24}\paren{\chi^{\CM(3,4)}_{(1,1)}(q)\,\chi^{U(1)_{12}}_{\mu_m}(q)
    +(-1)^{\mu_m} \chi^{\CM(3,4)}_{(1,3)}(q)\,\chi^{U(1)_{12}}_{6+\mu_m}(q)} \,.
\end{split}
\end{align}
When $\mu_m \in 2\mathbb{Z}$, the half-indices coincide with the characters of the RCFT
\[
\frac{\CM(3,4)\otimes U(1)_{12}}{\mathbb{Z}_2^{\rm diag}} \, .
\]
Both $\CM(3,4)$ and the $U(1)_{12}$ WZW model possess a fermionic $\mathbb{Z}_2$
symmetry, while their diagonal $\mathbb{Z}_2^{\rm diag}$ is bosonic.
\\
\\
\\
By taking the reduction procedure on the gluing equations in \eqref{(3,4) gluing equations}, we have
\begin{align}
    E_1^\mathrm{(r)}=C_1^\mathrm{(r)}=2 Z_3\;.
\end{align}

Among the 3 possible polarization choices, there are two choices satisfying the positivity condition.
\\
\paragraph{reduced-i)}
When we choose the quad structure as follows
\begin{align}
    \mathbf{X}_\mathrm{r}=(Z_3'') \textrm{ and }\mathbf{P}_\mathrm{r}=(Z_3')\;,
\end{align}
the reduced NZ matrices are
\begin{align}
    A_\mathrm{r}=(-2),\quad B_\mathrm{r}=(-2),\quad \boldsymbol{\n}_{x;\mathrm{r}}=(0)\;.
\end{align}
In the case, $(B^{-1}A)_\mathrm{r}\in \BZ$, and $|\det B|=2$. To make the $N_\mathrm{r}$ be a square-free integer, we should choose $\mathbf{Q}_\mathrm{r}=(2)$. Then, $N_\mathrm{r}=2$ and
\begin{align}
    K_\mathrm{r}=(4),\quad \boldsymbol{\m}_\mathrm{geo}^\mathrm{(r)}=(0).
\end{align}
The half-index is
\begin{align}
    \begin{split}
        \CI_\mathrm{half}^\mathrm{(r)} &= 1+q^2+q^3+2 q^4+2 q^5+3 q^6+3 q^7+5 q^8+5 q^9+7 q^{10}+\ldots\\
        & =q^{1/48} \c_{(1,1)}^{\CM(3,4)}(q)\;.
    \end{split}
\end{align}
It is the Nahm-sum like expression \eqref{Nahmsum3,l} with $l=4$.
\\
\\
\subsubsection{$(P,Q)=(3,5)$} \label{AppEx3,5}
There are two ways to obtain the gluing data of the torus knot using the SnapPy. The first way is using the DT code:
\begin{align*}
    \mathtt{Manifold('DT:[(14,-16,18,-20,2,-4,6,-8,10,-12)]').gluing\_equations\_pgl()}
\end{align*}
The torus knot complement can be ideally triangulated with $r=3$ tetrahedra with following gluing data
\begin{align}
    \begin{split}
        & C_1=Z_2+Z_3'',\quad C_2=Z_1+2 Z_1''+Z_2'+Z_3',\\
        & C_3=Z_1+2 Z_1'+Z_2+Z_2'+2 Z_2''+2 Z_3+Z_3'+Z_3'',\\
        & M=-Z_1-Z_2+Z_3''\;.
    \end{split}
    \label{(3,5) gluing I}
\end{align}
In the ideal triangulation, there is an easy internal edge $\{E_I\}=\{C_1\}$ and we choose $\{H_I\}=\{C_2\}$.

Among the $3^3$ possible polarization choices, there are four choices satisfying the positivity condition.
\\
\paragraph{i)} When we choose the quad structure as follows
\begin{align}
    \mathbf{X}=(Z_1,Z_2'',Z_3')^T\textrm{ and }\mathbf{P}=(Z_1'',Z_2',Z_3)^T\;,
\end{align}
the NZ matrices are
\begin{align}
    A=\begin{pmatrix}
        -1 & 1 & -1\\
    1 & 0 & 1\\
    0 & -1 & -1
    \end{pmatrix}, \quad B=\begin{pmatrix}
        0 & 1 & -1\\
        2 & 1 & 0\\
        0 & -1 & -1
    \end{pmatrix},\quad {\boldsymbol{\n}}_x=(0,-2,0)^T\;.
\end{align}
In the case, $B^{-1} A\notin  M_3(\BZ)$ and $|\det B|=4$. If we choose $\mathbf{Q}=\diag (2,1,1)$, $N=1$ and $\mathbf{Q}^T B^{-1} A \mathbf{Q}\in M_3(\BZ)$. Then,
\begin{align}
    \begin{split}
        & K=\begin{pmatrix}
            3 & -1 & 1\\
            -1 & 1 & 0\\
            1 & 0 & 1
        \end{pmatrix},\quad {\boldsymbol{\m}}_{\rm geo}=(-2,0,0)^T,\\
        & \boldsymbol{v}_m=\paren{-\frac{1}{2},\frac{1}{2},-\frac{1}{2}}^T,\quad \boldsymbol{v}_1=(1,0,0)^T\;.
    \end{split}
\end{align}
The chiral primary operator corresponding to the easy internal edge is
\begin{align}
    \CO_{E_1}=V_{(0,1,1)}\;.
\end{align}
\\
$\n_a=0$ and $\n_b=2$. The half-indices at the A/B-twisting points are
\begin{align}
    \begin{split}
        \CI_\mathrm{half}^a[\mu_m \in 4\mathbb{Z}]& \simeq 1+2 q^{1/2}+q+2 q^{3/2}+5 q^2+6 q^{5/2}+7 q^3+10 q^{7/2}+16 q^4+\ldots\\
        & =q^{1/60} \paren{\c_F(q)}^2 \c_{(1,1)}^{\CM(3,5)}(q),\\
        \CI_\mathrm{half}^b[\mu_m \in 4\mathbb{Z}]&\simeq 1+2 q^{1/2}+2 q+4 q^{3/2}+6 q^2+8 q^{5/2}+12 q^3+16 q^{7/2}+23 q^4+\ldots\\
        & =q^{1/15} \paren{\c_F(q)}^2 \c_{(1,2)}^{\CM(3,5)}(q)\;,
        \\
         \CI_\mathrm{half}^a[\mu_m \in 4\mathbb{Z}+2] &= \CI_\mathrm{half}^b[\mu_m \in 4\mathbb{Z}+2]=0\;.
    \end{split}
\end{align}
\\
\paragraph{ii)} When we choose the quad structure as follows
\begin{align}
    \mathbf{X}=(Z_1',Z_2'',Z_3')^T\textrm{ and }\mathbf{P}=(Z_1,Z_2',Z_3)^T\;,
\end{align}
the NZ matrices are
\begin{align}
    A=\begin{pmatrix}
        0 & 1 & -1\\
        -2 & 0 & 1\\
        0 & -1 & -1
    \end{pmatrix}, \quad B=\begin{pmatrix}
        -1 & 1 & -1\\
        -1 & 1 & 0\\
        0 & -1 & -1
    \end{pmatrix},\quad {\boldsymbol{\n}}_x=(0,0,0)^T
\end{align}
In the case, $B^{-1} A\in M_3(\BZ)$ and $|\det B|=1$, thus we can choose $\mathbf{Q}=\BI$. Then, $N=1$ and
\begin{align}
\begin{split}
    & K=\left(
        \begin{array}{ccc}
         4 & 2 & -2 \\
         2 & 2 & -1 \\
         -2 & -1 & 2 \\
        \end{array}
\right),\quad \boldsymbol{\m}_\mathrm{geo}=(0,0,0)^T,\\
& \boldsymbol{v}_m=(1,1,-1)^T,\quad \boldsymbol{v}_1=(-2,-1,1)^T\;.
\end{split}
\end{align}
The chiral primary operator corresponding to the easy internal edge is
\begin{align}
    \CO_{E_1}=V_{(0,1,1)}\;.
\end{align}
\\
$\n_a=0$ and $\n_b=2$. The half-indices at the A/B-twisting points are
\begin{align}
    \begin{split}
        \CI_\mathrm{half}^a[\mu_m\in 4\mathbb{Z}] &\simeq 1+4 q+8 q^2+15 q^3+29 q^4+51 q^5+85 q^6+138 q^7+\ldots\\
        & =q^{7/120} \c_{(1,2)}^{\CM(2,5)}(q) \, \c_0^{U(1)_2} (q),\\
        \CI_\mathrm{half}^b[\mu_m \in 4\mathbb{Z}] &\simeq 2+2 q+8 q^2+12 q^3+24 q^4+38 q^5+68 q^6+102 q^7+\ldots\\
        & =q^{-47/120} \c_{(1,1)}^{\CM(2,5)}(q) \, \c_1^{U(1)_2}(q)\;,\\
        \CI_\mathrm{half}^a[\mu_m\in 4\mathbb{Z}+2] &\simeq 2+4 q+10 q^2+20 q^3+38 q^4+68 q^5+120 q^6+200 q^7+\ldots\\
        & =q^{-23/120} \c_{(1,2)}^{\CM(2,5)}(q) \, \c_1^{U(1)_2} (q),\\
        \CI_\mathrm{half}^b[\mu_m\in 4\mathbb{Z}+2] &\simeq 1+3 q+5 q^2+11 q^3+22 q^4+38 q^5+67 q^6+113 q^7+\ldots\\
        & =q^{-17/120} \c_{(1,1)}^{\CM(2,5)}(q) \, \c_0^{U(1)_2} (q)\;.
    \end{split}
\end{align}
These are the characters of $\CM(2,5)\otimes U(1)_2$.
\\
\\
\\
By taking the reduction procedure on the gluing equations in \eqref{(3,5) gluing I}, we have
\begin{align}
    H_1^\mathrm{(r)}:=C_2^\mathrm{(r)}=Z_1+2 Z_1''\;.
\end{align}
The reduced gluing equation is the same as \eqref{(2,5) reduced gluing}. 
\\
\\
\\
The second way to obtain the gluing data of the torus knot is using the following input:
\begin{align*}
    \mathtt{Manifold('10\_124').gluing\_equations\_pgl()}
\end{align*}
The torus knot complement can be ideally triangulated with $r=3$ tetrahedra with following gluing data
\begin{align}
    \begin{split}
        & C_1=Z_1+Z_2,\quad C_2=Z_1'+Z_1''+Z_2'+Z_2''+2 Z_3'+Z_3'',\\
        & C_3=Z_1+Z_1'+Z_1''+Z_2+Z_2'+Z_2''+2 Z_3+Z_3'',\\
        & M=Z_1-Z_1''+Z_2''+Z_3-Z_3'\;.
    \end{split}
\end{align}
In the ideal triangulation, there is an easy internal edge $\{E_I\}=\{C_1\}$ and we choose $\{H_I\}=\{C_2\}$.

Among the $3^3$ possible polarization choices, there are 12 choices satisfying the positivity condition.
\\
\paragraph{iii)} When we choose the quad structure as follows
\begin{align}
    \mathbf{X}=(Z_1'',Z_2'',Z_3'')^T\textrm{ and }\mathbf{P}=(Z_1',Z_2',Z_3')^T\;,
\end{align}
the NZ matrices are
\begin{align}
    A=\begin{pmatrix}
        -2 & 1 & -1\\
        1 & 1 & 1\\
        -1  & -1 & 0
    \end{pmatrix},\quad B=\begin{pmatrix}
        -1 & 0 & -2\\
        1 & 1 & 2\\
        -1 & -1 & 0
    \end{pmatrix},\quad \boldsymbol{\n}_x=(2,-2,0)^T\;.
\end{align}
In the case, $B^{-1} A\notin M_3(\BZ)$ and $|\det B|=2$. If we choose $\mathbf{Q}=\diag (1,1,2)$, $N=2$ and
\begin{align}
    \begin{split}
        & K=\begin{pmatrix}
            2 & -1 & 0\\
            -1 & 2 & 0\\
            0 & 0 & 2
        \end{pmatrix},\quad \boldsymbol{\m}_\mathrm{geo}=(0,0,-2)^T,\\
        & \boldsymbol{v}_m=(-1,1,0)^T,\quad \boldsymbol{v}_1=(-1,1,1)^T\;.
    \end{split}
\end{align}
The chiral primary operator corresponding to the easy internal edge is
\begin{align}
    \CO_{E_1}=V_{(1,1,0)}\;.
\end{align}
$\n_a=0$ and $\n_b=2$. The half-indices at the A/B-twisting points are
\begin{align}
    \begin{split}
        \CI_\mathrm{half}^a[\mu_m \in 4\BZ] &\simeq q^{1/60} \c_{(1,1)}^{\CM(3,5)}(q) \c_0^{U(1)_2}(q)\,,\\
        \CI_\mathrm{half}^b[\mu_m \in 4\BZ] &\simeq q^{-11/60} \c_{(1,2)}^{\CM(3,5)}(q) \c_1^{U(1)_2}(q)\,, \\
        \CI_\mathrm{half}^a[\mu_m \in 4\BZ+2] &\simeq q^{-7/30} \c_{(1,1)}^{\CM(3,5)}(q) \c_1^{U(1)_2}(q)\,,\\
        \CI_\mathrm{half}^b[\mu_m \in 4\BZ+2] &\simeq q^{1/15} \c_{(1,2)}^{\CM(3,5)}(q) \c_0^{U(1)_2}(q)\,.
    \end{split}
\end{align}
The result is consistent with the combined results of \textbf{reduced-ii)} in \ref{AppEx2,5} and \eqref{U(1) Nahm sum}.
\\
\paragraph{iv)} When we choose the following polarizations
\begin{align}
    \mathbf{X}=(Z_1'',Z_2',Z_3'')^T\textrm{ and }\mathbf{P}=(Z_1',Z_2,Z_3')^T,
\end{align}
the NZ matrices are
\begin{align}
    A=\begin{pmatrix}
        -2 & -1 & -1\\
        1 & 0 & 1\\
        -1 & 0 & 0
    \end{pmatrix},\quad B=\begin{pmatrix}
        -1 & -1 & -2\\
        1 & -1 & 2\\
        -1 & 1 & 0
    \end{pmatrix},\quad \boldsymbol{\n}_x=(3,-1,-1)^T\;.
\end{align}
In the case, we choose
\begin{equation}
    \mathbf{Q}=\begin{pmatrix}
        1 & 1 & 1\\
        1 & 0 & 0\\
        0 & 1 & -1
    \end{pmatrix}\;.
\end{equation}
Then, $N=1$ and
\begin{align}
    \begin{split}
        & K=\begin{pmatrix}
            3 & 2 & 2\\
            2 & 2 & 1\\
            2 & 1 & 2
        \end{pmatrix},\quad \boldsymbol{\m}_\mathrm{geo}=(-1,-1,1)^T,\\
        & \boldsymbol{v}_m=\paren{-1,-\frac{1}{2},-\frac{1}{2}}^T,\quad \boldsymbol{v}_1=(-1,0,-1)^T\;.
    \end{split}
\end{align}
The chiral primary operator corresponding to the easy internal edge is
\begin{align}
    \CO_{E_1}=V_{(-1,1,1)}\;.
\end{align}
\\
$\n_a=0$ and $\n_b=2$. The half-indices at the A/B-twisting points are
\begin{align}
    \begin{split}
        & \CI_\mathrm{half}^a[\mu_m \in 2\mathbb{Z}]=q^{-1/40} \c_{(1,1)}^{\CM(3,5)}(q),\\
        & \CI_\mathrm{half}^b[\mu_m \in 2\mathbb{Z}]=q^{1/40} \c_{(1,2)}^{\CM(3,5)}(q)\,.
    \end{split}
\end{align}
\paragraph{v)} When we choose the following polarizations
\begin{align}
    \mathbf{X}=(Z_1,Z_2',Z_3'')^T\textrm{ and }\mathbf{P}=(Z_1'',Z_2,Z_3')^T\;,
\end{align}
the NZ matrices are
\begin{align}
    A=\begin{pmatrix}
        1 & -1 & -1\\
        -1 & 0 & 1\\
        1 & 0 & 0
    \end{pmatrix},\quad B=\begin{pmatrix}
        -1 & -1 & -2\\
        0 & -1 & 2\\
        0 & 1 & 0
    \end{pmatrix},\quad \boldsymbol{\n}_x=(2,0,-2)^T\;.
\end{align}
In the case, $\BI-B^{-1} A>0$. If we choose $\mathbf{Q}=\diag (1,1,2)$, $N=2$ and
\begin{align}
    \begin{split}
        & \overline{K}=\begin{pmatrix}
            3 & -1 & 0\\
            -1 & 1 & 0\\
            0 & 0 & 2
        \end{pmatrix},\quad \overline{\boldsymbol{\m}}_\mathrm{geo}=(-3,1,0)^T,\\
        & \boldsymbol{v}_m=(-1,0,0)^T,\quad \boldsymbol{v}_1=(-1,0,1)^T\;.
    \end{split}
\end{align}
\\
$\n_a=0$ and $\n_b=2$. The half-indices are
\begin{align}
\overline{\CI_\mathrm{half}^a}[\m_m \in 2\mathbb{Z}]=\overline{\CI_\mathrm{half}^b}[\m_m \in 2\mathbb{Z}]=0\;.
\end{align}
\\
\\
\subsubsection{$(P,Q)=(3,7)$} \label{AppEx3,7}
Using SnapPy, the gluing data of the torus knot can be obtained by the following input:
\begin{align}
\begin{split}
    & \mathtt{Manifold('DT:[(10,-12,14,-16,18,-20,22,-24,}\\
    & \mathtt{26,-28,2,-4,6,-8)]').gluing\_equations\_pgl()}
\end{split}
\end{align}
The torus knot complement can be ideally triangulated with $r=4$ tetrahedra with following gluing data
\begin{align}
    \begin{split}
        & C_1=2 Z_1'+Z_2+Z_3'+Z_4',\quad C_2=Z_3+Z_4'',\quad C_3=Z_1+Z_2+2 Z_2',\\
        & C_4=Z_1+2 Z_1''+2 Z_2''+Z_3+Z_3'+2 Z_3''+2 Z_4+Z_4'+Z_4'',\\
        & M=-Z_1+2 Z_1''-Z_2'+3 Z_2''+Z_3-Z_3'+Z_4\;.
    \end{split}
\end{align}
In the ideal triangulation, there are two easy internal edges $\{E_I\}=\{C_2,C_1\}$, and we choose $\{H_I\}=\{C_3\}$.
\\
\\
Among the $3^4$ possible polarization choices, there are four choices satisfying the positivity condition.
\\
\paragraph{i)} When we choose the following polarizations
\begin{align}
    \mathbf{X}=(Z_1,Z_2',Z_3'',Z_4')^T\textrm{ and }\mathbf{P}=(Z_1'',Z_2,Z_3',Z_4)^T\;,
\end{align}
the NZ matrices are
\begin{align}
    A=\begin{pmatrix}
        -1 & -4 & -1 & 0 \\
        1 & 2 & 0 & 0 \\
        0 & 0 & -1 & -1 \\
        -2 & 0 & 0 & 1 \\
    \end{pmatrix}, \quad 
    B=\begin{pmatrix}
        2 & -3 & -2 & 1 \\
        0 & 1 & 0 & 0 \\
        0 & 0 & -1 & -1 \\
        -2 & 1 & 1 & 0 \\
    \end{pmatrix},\quad {\boldsymbol{\n}}_x=(4,-2,0,0)^T\;.
\end{align}
In the case, since $B^{-1} A\notin M_4(\BZ)$ , we choose $\mathbf{Q}=\diag(2,1,1,1)$. Then,
\begin{equation}
\begin{gathered}
    K=\begin{pmatrix}
        7 & 2 & 1 & -1 \\
        2 & 2 & 0 & 0 \\
        1 & 0 & 1 & 0 \\
        -1 & 0 & 0 & 1 \\  \end{pmatrix}, \quad N=1, \quad \CO_{E_1} =  V_{(0,0,1,1)}\,, \quad \CO_{E_2} = \phi_4 V_{(1,-1,-1,0)}\,, \\
    \boldsymbol{\mu}_{\rm geo}=(-2,-2,0,0)^T, \quad \boldsymbol{v}_m = \frac12(-1,0,-1,1)^T, \quad \boldsymbol{v}_1 = (0,1,-1,1)^T .
\end{gathered}
\end{equation}
\\
$\n_a=0$ and $\n_b=2$. The half-indices at the A/B-twisting points are
\begin{align}
\begin{split}
\CI^{a}_{\rm half} [\mu_m \in 4\mathbb{Z}] &\simeq  q^{-3/28} \paren{\chi_F(q)}^2\,\chi^{\CM(3,7)}_{(1,1)}(q)\,, \\
\CI^{b}_{\rm half} [\mu_m \in 4\mathbb{Z}] &\simeq q^{-3/7} \paren{\chi_F(q)}^2\,\chi^{\CM(3,7)}_{(1,2)}(q)\,, \\
\CI^{a}_{\rm half} [\mu_m \in 4\mathbb{Z}+2] &= \CI^{b}_{\rm half} [\mu_m \in 4\mathbb{Z}+2]=0\;.
\end{split}
\end{align}
\\
\paragraph{ii)} When we choose the following polarizations
\begin{align}
    \mathbf{X}=(Z_1,Z_2,Z_3'',Z_4')^T\textrm{ and }\mathbf{P}=(Z_1'',Z_2'',Z_3',Z_4)^T\;,
\end{align}
the NZ matrices are
\begin{align}
    A=\begin{pmatrix}
        -1 & 1 & -1 & 0\\
        1 & -1 & 0 & 0\\
        0 & 0 & -1 & -1 \\
        -2 & 1 & 0 & 1
    \end{pmatrix},\quad B=\begin{pmatrix}
        2 & 4 & -2 & 1\\
        0 & -2 & 0 & 0\\
        0 & 0 & -1 & -1\\
        -2 & 0 & 1 & 0
    \end{pmatrix},\quad \boldsymbol{\n}_x=(0,0,0,0)^T\;.
\end{align}
In the case, since $B^{-1} A\notin M_4(\BZ)$, we choose $\mathbf{Q}=\diag (2,2,1,1)$. Then,
\begin{align}
    \begin{split}
        & K=\begin{pmatrix}
            5 & -2 & 1 & -1\\
            -2 & 2 & 0 & 0\\
            1 & 0 & 1 & 0\\
            -1 & 0 & 0 & 1
        \end{pmatrix},\quad N=2, \quad \CO_{E_1}=V_{(0,0,1,1)},\quad \CO_{E_2}=\f_2 \f_4 V_{(1,0,-1,0)},\\
        & \boldsymbol{\m}_\mathrm{geo}=\frac{1}{2}(-1,0,-1,1)^T, \quad \boldsymbol{v}_m=\frac{1}{2} (-1,0,-1,1)^T,\quad \boldsymbol{v}_1=(-1,-1,-1,1)^T\;.
    \end{split}
\end{align}
$\n_a=0$ and $\n_b=2$. The half-indices at the A/B-twisting points are
\begin{align}
\begin{split}
    \CI^{a}_{\rm half} [\mu_m \in 4\mathbb{Z}] \simeq  q^{17/168} \paren{\chi_F(q)}^2 \, 
    & \bigg(\chi^{\CM(3,4)}_{(1,1)}(q)\paren{\chi^{\CM(7,12)}_{(3,5)}(q)+\chi^{\CM(7,12)}_{(3,11)}(q)} \\
    & +\chi^{\CM(3,4)}_{(1,3)}(q)\paren{\chi^{\CM(7,12)}_{(3,1)}(q)+\chi^{\CM(7,12)}_{(3,7)}(q)} \bigg)\,, \\
    \CI^{b}_{\rm half} [\mu_m \in 4\mathbb{Z}] \simeq  q^{-79/168} \paren{\chi_F(q)}^2 \, 
    & \bigg(\chi^{\CM(3,4)}_{(1,1)}(q)\paren{\chi^{\CM(7,12)}_{(1,1)}(q)+\chi^{\CM(7,12)}_{(1,7)}(q)} \\
    & +\chi^{\CM(3,4)}_{(1,3)}(q)\paren{\chi^{\CM(7,12)}_{(1,5)}(q)+\chi^{\CM(7,12)}_{(1,11)}(q)} \bigg)\,, \\
    \CI^{a}_{\rm half} [\mu_m \in 4\mathbb{Z}+2] &= \CI^{b}_{\rm half} [\mu_m \in 4\mathbb{Z}+2]=0\;.
\end{split}
\end{align}
The half-indices at $\mu_m \in 4\mathbb{Z}$ coincide with the characters of the RCFT
\begin{align}
(\textrm{two free fermions}) \otimes \left(\left(\frac{\CM(7,12)}{\mathbb{Z}_2}\right)\otimes \CM(3,4)\right)\Big/\mathbb{Z}_2^{\rm diag}\,.
\end{align}
Here $\CM(7,12)$ admits a \emph{bosonic} $\mathbb{Z}_2$ symmetry generated by the Verlinde loop associated with the primary $(1,7)$; this is the $\mathbb{Z}_2$ by which we quotient in $\CM(7,12)/\mathbb{Z}_2$.
In addition, $\CM(7,12)$ also has a \emph{fermionic} $\mathbb{Z}_2^{f}$ generated by the primary $(1,5)$.
Likewise, $\CM(3,4)$ has a fermionic $\mathbb{Z}_2^{f}$ generated by $(1,3)$.
The final quotient $\mathbb{Z}_2^{\rm diag}$ is the diagonal subgroup of these two fermionic symmetries, i.e.
\begin{align}
\mathbb{Z}_2^{\rm diag}\;\subset\; \mathbb{Z}_2^{f}(\CM(7,12))\times \mathbb{Z}_2^{f}(\CM(3,4))\,,
\end{align}
generated by the simultaneous action of $(1,5)$ in $\CM(7,12)$ and $(1,3)$ in $\CM(3,4)$.
\\
\paragraph{iii)} When we choose the following polarizations
\begin{align}
    \mathbf{X}=(Z_1'',Z_2'',Z_3',Z_4)^T\textrm{ and }\mathbf{P}=(Z_1',Z_2',Z_3,Z_4'')^T\;,
\end{align}
the NZ matrices are
\begin{align}
    A=\begin{pmatrix}
        3 & 3 & -1 & 1 \\
        -1 & -1 & 0 & 0 \\
        0 & 0 & 0 & 0 \\
        0 & -1 & 1 & -1 \\
    \end{pmatrix},\quad B=\begin{pmatrix}
        1 & -1 & 1 & 0 \\
        -1 & 1 & 0 & 0 \\
        0 & 0 & 1 & 1 \\
        2 & -1 & 0 & -1 \\
    \end{pmatrix},\quad \boldsymbol{\n}_x=(-1,0,-2,0)^T\;.
\end{align}
In the case, $B^{-1}A \in M_4(\BZ)$ and $\BI-B^{-1} A>0$. Thus we can choose $\mathbf{Q}=\BI$, then
\begin{equation}
\begin{gathered}
    \overline{K}=\begin{pmatrix}
        4 & 4 & -2 & 2 \\
        4 & 6 & -2 & 2 \\
        -2 & -2 & 2 & -1 \\
        2 & 2 & -1 & 2 \\
    \end{pmatrix}, \quad \overline{\CO}_{E_1} = V_{(0,0,1,1)}, \quad \overline{\CO}_{E_2} = \phi_3 V_{(2,-1,0,-1)}, \\
    \overline{\boldsymbol{\m}}_\mathrm{geo}=(0,0,0,0)^T, \quad
    \boldsymbol{v}_m=(-1,-1,1,-1)^T,\quad \boldsymbol{v}_1=(0,1,1,-1)^T\;.
\end{gathered}
\end{equation}
\\
$\n_a=0$ and $\n_b=2$. The half-indices are
\begin{align}
\begin{split}
    \overline{\CI_\mathrm{half}^a}[\mu_m \in 4\BZ] &\simeq q^{11/168}\chi^{U(1)_2}_0(q) \, \chi^{\CM(2,7)}_{(1,3)}(q)\,, \\
    \overline{\CI_\mathrm{half}^b}[\mu_m \in 4\BZ] &\simeq q^{-103/168}\chi^{U(1)_2}_1(q) \, \chi^{\CM(2,7)}_{(1,1)}(q)\,, \\
    \overline{\CI_\mathrm{half}^a}[\mu_m \in 4\BZ+2] &\simeq q^{-31/168}\chi^{U(1)_2}_1(q) \, \chi^{\CM(2,7)}_{(1,3)}(q)\,, \\
    \overline{\CI_\mathrm{half}^b}[\mu_m \in 4\BZ+2] &\simeq q^{-61/168}\chi^{U(1)_2}_0(q) \, \chi^{\CM(2,7)}_{(1,1)}(q)\,.
\end{split}
\end{align}
\\
\\
\subsubsection{$(P,Q)=(3,8)$}
Using SnapPy, the gluing data of the torus knot can be obtained by the following input:
\begin{align}
    \begin{split}
        & \mathtt{Manifold('DT:[(22,-24,26,-28,30,-32,2,-4,6,-8,}\\
        & \mathtt{10,-12,14,-16,18,-20)]').gluing\_equations\_pgl()}
    \end{split}
\end{align}
The torus knot complement can be ideally triangulated with $r=4$ tetrahedra with following gluing data
\begin{align}
    \begin{split}
        & C_1=Z_1+Z_2''+2 Z_3+2 Z_4'',\quad C_2=Z_1''+Z_2,\quad C_3=Z_3'+2 Z_3''+Z_4,\\
        & C_4=Z_1+2 Z_1'+Z_1''+Z_2+2 Z_2'+Z_2''+Z_3'+Z_4+2 Z_4',\\
        & M=-Z_1''+Z_2-Z_4\;.
    \end{split}
    \label{(3,8) gluing equations}
\end{align}
In the ideal triangulation, there are two easy internal edges $\{C_2,C_1\}$, and we choose $\{H_I\}=\{C_3\}$.

Among the $3^4$ possible polarization choices, there are four choices satisfying the positivity condition.
\\
\paragraph{i)} When we choose the following polarizations
\begin{align}
    \mathbf{X}=(Z_1',Z_2'',Z_3'',Z_4)^T\textrm{ and }\mathbf{P}=(Z_1,Z_2',Z_3',Z_4'')^T\;,
\end{align}
the NZ matrices are
\begin{align}
    A=\begin{pmatrix}
        1 & -1 & 0 & -1 \\
        0 & 0 & 2 & 1 \\
        -1 & -1 & 0 & 0 \\
        0 & 1 & -2 & 0 \\
    \end{pmatrix}, \quad 
    B=\begin{pmatrix}
        1 & -1 & 0 & 0 \\
        0 & 0 & 1 & 0 \\
        -1 & -1 & 0 & 0 \\
        1 & 0 & -2 & 2 \\
    \end{pmatrix},\quad {\boldsymbol{\n}}_x=(0,-2,0,0)^T\;.
\end{align}
In the case, since $B^{-1} A\notin M_4(\BZ)$ , we choose $\mathbf{Q}=\diag(1,1,1,2)$. Then,
\begin{equation}
\begin{gathered}
    K=\begin{pmatrix}
        1 & 0 & 0 & -1 \\
        0 & 1 & 0 & 1 \\
        0 & 0 & 2 & 2 \\
        -1 & 1 & 2 & 5 \\  \end{pmatrix}, \quad N=1, \quad \CO_{E_1} =  V_{(1,1,0,0)}\,, \quad \CO_{E_2} = \phi_2 V_{(-1,0,2,-1)}\,, \\
    \boldsymbol{\mu}_{\rm geo}=(0,0,-2,-4)^T, \quad \boldsymbol{v}_m = \frac12(1,-1,0,-1)^T, \quad \boldsymbol{v}_1 = (0,0,1,2)^T .
\end{gathered}
\end{equation}
\\
$\n_a=0$ and $\n_b=2$. The half-indices at the A/B-twisting points are
\begin{align}
\begin{split}
\CI^{a}_{\rm half}[\m_m\in 4\BZ] & \simeq q^{-17/96} \paren{\chi_F(q)}^2\,\chi^{\CM(3,8)}_{(1,1)}(q)\,, \\
\CI^{b}_{\rm half}[\m_m\in 4\BZ] & \simeq q^{7/96} \paren{\chi_F(q)}^2\,\chi^{\CM(3,8)}_{(1,3)}(q)\,,\\
\CI^{a}_{\rm half}[\m_m\in 4\BZ+2] &= \CI^{b}_{\rm half}[\m_m\in 4\BZ+2] = 0\,.
\end{split}
\end{align}
\\
\paragraph{ii)} When we choose the following polarizations
\begin{align}
    \mathbf{X}=(Z_1',Z_2'',Z_3',Z_4')^T\textrm{ and }\mathbf{P}=(Z_1,Z_2',Z_3,Z_4)^T\;,
\end{align}
the NZ matrices are
\begin{align}
    A=\begin{pmatrix}
        1 & -1 & 0 & 0\\
        0 & 0 & -1 & 0\\
        -1 & -1 & 0 & 0\\
        0 & 1 & 0 & -2
    \end{pmatrix},\quad B=\begin{pmatrix}
        1 & -1 & 0 & -1\\
        0 & 0 & -2 & 1\\
        -1 & -1 & 0 & 0\\
        1 & 0 & 2 & -2
    \end{pmatrix},\quad \boldsymbol{\n}_x=(0,0,0,0)^T\;.
\end{align}
In the case, since $B^{-1}A\notin M_4(\BZ)$, we choose $\mathbf{Q}=\diag (1,1,2,1)$. Then,
\begin{align}
    \begin{split}
        & K=\begin{pmatrix}
            2 & -1 & 2 & 2\\
            -1 & 2 & -2 & -2\\
            2 & -2 & 6 & 4\\
            2 & -2 & 4 & 4
        \end{pmatrix},\quad N=2,\quad \CO_{E_1}=V_{(1,1,0,0)},\quad \CO_{E_2}=\f_2 V_{(-1,0,-1,2)},\\
        & \boldsymbol{\mu}_\mathrm{geo}=(0,0,0,0)^T,\quad \boldsymbol{v}_m=(1,-1,1,1)^T,\quad \boldsymbol{v}_1=(-1,1,-3,-2)^T\;.
    \end{split}
\end{align}
$\n_a=0$ and $\n_b=2$. The half-indices at the A/B-twisting points are
\begin{align}
    \begin{split}
        \CI_\mathrm{half}^a[\m_m\in 4\BZ] &\simeq \c_{(1,3)}^{\CM(3,8)}(q) \c_0^{U(1)_2}(q)\,,\\
        \CI_\mathrm{half}^b[\m_m\in 4\BZ] &\simeq \c_{(1,1)}^{\CM(3,8)}(q) \c_1^{U(1)_2}(q)\,,\\
        \CI_\mathrm{half}^a[\m_m\in 4\BZ+2] &\simeq \c_{(1,3)}^{\CM(3,8)}(q) \c_1^{U(1)_2}(q)\,,\\
        \CI_\mathrm{half}^b[\m_m\in 4\BZ+2] &\simeq \c_{(1,1)}^{\CM(3,8)}(q) \c_0^{U(1)_2}(q)\,.
    \end{split}
\end{align}
\\
\\
By taking the reduction procedure on the gluing equations in \eqref{(3,8) gluing equations}, we have
\begin{align}
    H_1^\mathrm{(r)}:=C_3^\mathrm{(r)}=Z_3'+2 Z_3''+Z_4, \quad E_1^\mathrm{(r)}=C_1^\mathrm{(r)}=2 Z_3+2 Z_4''
\end{align}
Among the $3^2$ possible polarization choices, there are four choices satisfying the positivity condition.
\\
\paragraph{reduced-i)} When we choose the following polarizations
\begin{align}
    \mathbf{X}_{\rm r} = (Z_3'',Z_4)^T \textrm{ and } \mathbf{P}_{\rm r} = (Z_3',Z_4'')^T,
\end{align}
the reduced NZ matrices are
\begin{align}
A_{\rm r} = \begin{pmatrix} 2 & 1 \\ -2 & 0 \end{pmatrix}, \quad  
B_{\rm r} = \begin{pmatrix} 1 & 0 \\ -2 & 2 \end{pmatrix}, \quad \boldsymbol{\nu}_{x;{\rm r}} = (-2,0)^T\;.
\end{align}
In the case, we choose $\mathbf{Q}_{\rm r} = \textrm{diag} \{1,2\}$. Then, $N_r = 2$ and
\begin{equation}
\begin{gathered}
    K_{\rm r} = \left( \begin{array}{cc} 2 & 2 \\ 2 & 4 \\ \end{array} \right), \quad \CO_{E_1^{\rm(r)}} = V_{(2,-1)}\,, \quad \boldsymbol{\mu}^{\rm (r)}_{\rm geo} =(-2,-4)^T, \quad \boldsymbol{v}^{\rm (r)}_1 =(1,2)^T.
\end{gathered}
\end{equation}
\\
$\n_a=0$ and $\n_b=2$. The half-indices at the A/B-twisting points are
\begin{align}
\begin{split}
\CI^{a;\textrm{(r)}}_{\rm half} & = q^{-7/32} \chi^{\CM(3,8)}_{(1,1)}(q)\,, \\
\CI^{b;\textrm{(r)}}_{\rm half} & = q^{1/32} \chi^{\CM(3,8)}_{(1,2)}(q)\,.
\end{split}
\end{align}
It is the Nahm-sum like expression \eqref{Nahmsum3,l} with $l=8$.
\\
\\
\subsection{The case $(P,Q) = (4,5)$}   

Using SnapPy, the gluing data of the torus knot can be obtained by the following input:
\begin{align*}
\begin{split}
    & \mathtt{Manifold('DT:[(16,-26,-12,22,-2,-18,28,-8,-24,}\\
    & \mathtt{4,-14,-30,10,-20,-6)]').gluing\_equations\_pgl()}
\end{split}
\end{align*}
The torus knot complement can be ideally triangulated with $r=5$ tetrahedra with following gluing data
\begin{align}
    \begin{split}
        & C_1=Z_1+Z_5'',\quad C_2=Z_1''+Z_2'+2 Z_3'+Z_5,\\
        & C_3=Z_1''+2 Z_4'+Z_5,\quad C_4=Z_1'+2 Z_2''+Z_3+Z_5',\\
        & C_5=Z_1+Z_1'+2 Z_2+Z_2'+Z_3+2 Z_3''+2 Z_4+2 Z_4''+Z_5'+Z_5'',\\
        & M=Z_1-Z_2'+Z_3-Z_4''-Z_5''\;.
    \end{split}
    \label{(4,5) gluing}
\end{align}
In the ideal triangulation, there are four easy internal edges $\{E_I\}=\{C_1,C_2,C_3,C_4\}$.

Among the $3^5$ possible polarization choices, there are two choices satisfying the positivity condition.
\\
\paragraph{i)} When we choose the quad structure as follows
\begin{align}
    \mathbf{X}=(Z_1'',Z_2',Z_3,Z_4,Z_5')^T\textrm{ and }\mathbf{P}=(Z_1',Z_2,Z_3'',Z_4'',Z_5)^T\;,
\end{align}
the NZ matrices are
\begin{align}
    A = \begin{pmatrix}
        -1 & -1 & 1 & 0 & 1 \\
        -1 & 0 & 0 & 0 & -1 \\
        1 & 1 & -2 & 0 & 0 \\
        1 & 0 & 0 & -2 & 0 \\
        0 & -2 & 1 & 0 & 1 \\ \end{pmatrix}, \quad 
    B = \begin{pmatrix}
        -1 & 0 & 0 & -1 & 1 \\
        -1 & 0 & 0 & 0 & -1 \\
        0 & 0 & -2 & 0 & 1 \\
        0 & 0 & 0 & -2 & 1 \\
        1 & -2 & 0 & 0 & 0 \\
    \end{pmatrix},\quad {\boldsymbol{\nu}}_x=(0,0,0,0,0)^T\;.
\end{align}
In the case, $B^{-1} A\notin  M_5(\BZ)$ and $|\det B|=12$. If we choose
\begin{align}
        \mathbf{Q} = \begin{pmatrix}
        1 & 0 & 0 & 0 & 0 \\
        -2 & 2 & 0 & 0 & 0 \\
        0 & 2 & 1 & 0 & 0 \\
        0 & 0 & 2 & 1 & 1 \\
        0 & 0 & 0 & -2 & 1 \\ \end{pmatrix}\;,
\end{align}
then, $N=3$ and $\mathbf{Q}^T B^{-1} A \mathbf{Q}\in M_5(\BZ)$. 
Then,
\begin{align}
\begin{split}
    & K=\begin{pmatrix}
        4 & -2 & 1 & -2 & 1 \\
        -2 & 4 & 1 & 0 & 0 \\
        1 & 1 & 8 & -1 & 5 \\
        -2 & 0 & -1 & 4 & -2 \\
        1 & 0 & 5 & -2 & 4 \\ \end{pmatrix},\quad \boldsymbol{\m}_\mathrm{geo}=(0,0,0,0,0)^T, \quad \boldsymbol{v}_m=(0,0,1,-1,1)^T\;.
\end{split}
\end{align}
The chiral primary operators corresponding to each easy internal edges are
\begin{align}
\begin{split}
    \CO_{E_1} &= V_{(1,1,-2,1,3)},\quad \CO_{E_2} =\f_1\f_2 V_{(0,0,2,-1,-3)}, \\
    \CO_{E_3} &=\f_1 V_{(0,0,0,1,1)},\quad \CO_{E_4} =\f_3\f_5 V_{(-1,0,0,0,0)}\;.
\end{split}
\end{align}
\\
The half-index in \eqref{half-indices when |P-Q|=1} is
\begin{align}
\begin{split}
    \CI_\mathrm{half}[\m_m\in 12\BZ]& \simeq q^{9/80}\paren{\c_{(1,1)}^{\CM(4,5)}(q) \c_{(0,0)}^{U(1)^2_J}(q)+\c_{(3,1)}^{\CM(4,5)}(q) \c_{(1,2)}^{U(1)^2_J}(q)},\\
    \CI_\mathrm{half}[\m_m\in 12\BZ\pm 2]& \simeq q^{-13/240}\paren{\c_{(1,1)}^{\CM(4,5)}(q) \c_{(1,0)}^{U(1)^2_J}(q)+\c_{(3,1)}^{\CM(4,5)}(q) \c_{(2,2)}^{U(1)^2_J}(q)},\\
    \CI_\mathrm{half}[\m_m\in 12\BZ\pm 4]& \simeq q^{-133/240}\paren{\c_{(1,1)}^{\CM(4,5)}(q) \c_{(2,0)}^{U(1)^2_J}(q)+\c_{(3,1)}^{\CM(4,5)}(q) \c_{(3,2)}^{U(1)^2_J}(q)},\\
    \CI_\mathrm{half}[\m_m\in 12\BZ+6]& \simeq q^{-111/80}\paren{\c_{(1,1)}^{\CM(4,5)}(q) \c_{(3,0)}^{U(1)^2_J}(q)+\c_{(3,1)}^{\CM(4,5)}(q) \c_{(4,2)}^{U(1)^2_J}(q)}\;,\\
    \textrm{with } J &= \begin{pmatrix}
    4 & 2\\
    2 & 4
\end{pmatrix}\;.
\end{split}
\end{align}
These are   characters of the RCFT $\frac{\CM(4,5)\otimes U(1)^2_J}{\mathbb{Z}_2^{\rm diag}}$
\\
\\
\\
By taking the reduction procedure on the gluing equations in \eqref{(4,5) gluing}, we have
\begin{align}
\begin{split}
    & E_1^\mathrm{(r)}=C_2^\mathrm{(r)}=Z_2'+2 Z_3',\quad E_2^\mathrm{(r)}=C_3^\mathrm{(r)}=2 Z_4',\\
    & E_3^\mathrm{(r)}=C_4^\mathrm{(r)}=2 Z_2''+Z_3\;.
\end{split}
\end{align}

Among the $3^3$ possible polarization choices, there are two choices satisfying the positivity condition.
\\
\paragraph{reduced-i)} When we choose the following polarizations
\begin{align}
    \mathbf{X}_\mathrm{r}=(Z_2',Z_3,Z_4)^T\textrm{ and }\mathbf{P}_\mathrm{r}=(Z_2,Z_3'',Z_4'')^T\;,
\end{align}
the reduced NZ matrices are
\begin{align}
    A_\mathrm{r}=\begin{pmatrix}
        1 & -2 & 0\\
        0 & 0 & -2\\
        -2 & 1 & 0
    \end{pmatrix},\quad B_\mathrm{r}=\begin{pmatrix}
        0 & -2 & 0 \\
        0 & 0 & -2\\
        -2 & 0 & 0
    \end{pmatrix},\quad \boldsymbol{\n}_{x;\mathrm{r}}=(0,0,0)^T\;.
\end{align}
In the case, we choose $\mathbf{Q}_\mathrm{r}=\diag (1,2,2)$. Then, $N_\mathrm{r}=2$ and
\begin{align}
    K_\mathrm{r}=\begin{pmatrix}
        1 & -1 & 0\\
        -1 & 4 & 0\\
        0 & 0 & 4
    \end{pmatrix},\quad \boldsymbol{\m}_\mathrm{geo}^\mathrm{(r)}=(0,0,0)^T\;.
\end{align}
The chiral primary operators corresponding to each easy internal edges are
\begin{align}
    \CO_{E_1^\mathrm{(r)}}=\f_1 V_{(0,1,0)},\quad \CO_{E_2^\mathrm{(r)}}=V_{(0,0,1)},\quad \CO_{E_3^\mathrm{(r)}}=\f_2 V_{(2,0,0)}\;.
\end{align}
The half-index in \eqref{half-indices when |P-Q|=1} is
\begin{align}
    \begin{split}
        \CI_\mathrm{half}^\mathrm{(r)}& =1+q^{1/2}+2 q^{3/2}+4 q^2+4 q^{5/2}+5 q^3+9 q^{7/2}+14 q^4+17 q^{9/2}+\ldots\\
        & =q^{17/240} \c_F(q) \c_{(1,1)}^{\CM(3,4)}(q) \paren{\c_{(1,1)}^{\CM(4,5)}(q)+\c_{(3,1)}^{\CM(4,5)}(q)}\\
        & =q^{17/240} \c_F(q) \c_{(1,1)}^{\CM(3,4)}(q) \c_{(1,1)}^{SM(3,5)}(q)\;.
    \end{split}
\end{align}
\\
\paragraph{reduced-ii)} In the same choice of polarization with \textbf{reduced-i)}, there is another possible choice of the charge matrix: $\mathbf{Q}_\mathrm{r}=\diag (2,2,1)$. Then, $N_\mathrm{r}=2$ and
\begin{align}
    K_\mathrm{r}=\begin{pmatrix}
        4 & -2 & 0\\
        -2 & 4 & 0\\
        0 & 0 & 1
    \end{pmatrix},\quad \boldsymbol{\m}_\mathrm{geo}^\mathrm{(r)}=(0,0,0)^T\;.
\end{align}
The chiral primary operators corresponding to each easy internal edges are
\begin{align}
    \CO_{E_1^\mathrm{(r)}}=\phi_1^2 V_{(0,1,0)},\quad \CO_{E_2^\mathrm{(r)}}=V_{(0,0,2)},\quad \CO_{E_3^\mathrm{(r)}} = \phi_2^2 V_{(1,0,0)}\;.
\end{align}
The half-index in \eqref{half-indices when |P-Q|=1} is
\begin{align}
        \CI_\mathrm{half}^\mathrm{(r)}& = q^{17/240} \c_F(q) \paren{\c_{(1,1)}^{\CM(3,4)}(q)\c_{(1,1)}^{\CM(4,5)}(q) + \c_{(1,3)}^{\CM(3,4)}(q)\c_{(1,4)}^{\CM(4,5)}(q)}\,.
\end{align}
This is the vacuum character of the RCFT $\frac{\CM(3,4)\otimes \CM(4,5)}{\mathbb{Z}_2^{\rm diag}} \otimes (\textrm{free fermion})$.
\\
\paragraph{reduced-iii)} When we choose the following polarizations
\begin{align}
    \mathbf{X}_{\rm r} = (Z_2,Z_3'',Z_4'')^T \textrm{ and } \mathbf{P}_{\rm r} = (Z_2'',Z_3',Z_4')^T,
\end{align}
the reduced NZ matrices are
\begin{align}
    A_{\rm r} = \begin{pmatrix} 
        -1 & 0 & 0 \\
        0 & 0 & 0 \\
        0 & -1 & 0 \\ \end{pmatrix}, \quad  
    B_{\rm r} = \begin{pmatrix} 
        -1 & 2 & 0 \\
        0 & 0 & 2 \\
        2 & -1 & 0 \\ \end{pmatrix}, \quad 
    \boldsymbol{\nu}_{x;{\rm r}} = (-1,-2,-1)^T\;.
\end{align}
In the case, $\BI-B^{-1} A>0$ and if we choose
\begin{align}
    \mathbf{Q}_{\rm r} = \begin{pmatrix}
        1 & -1 & 0 \\
        1 & 2 & 0 \\
        0 & 0 & 2 \\ \end{pmatrix}\,,
\end{align}
then
\begin{equation}
\begin{gathered}
    \overline{K}_{\rm r} = \begin{pmatrix}
        4 & 2 & 0 \\
        2 & 4 & 0 \\
        0 & 0 & 4 \\
    \end{pmatrix}, \quad N_r = 6, \quad \overline{\boldsymbol{\mu}}^{\rm (r)}_{\rm geo} =(0,0,0)^T\,.
\end{gathered}
\end{equation}
The chiral primary operators corresponding to each easy internal edges are
\begin{align}
    \overline{\CO}_{E_1^{\rm(r)}} = V_{(0,1,0)}\,, \quad
    \overline{\CO}_{E_2^{\rm(r)}} = V_{(0,0,1)}\,, \quad
    \overline{\CO}_{E_3^{\rm(r)}} = V_{(1,-1,0)}\,.
\end{align}
The half-index in \eqref{half-indices when |P-Q|=1} is
\begin{align}
\overline{\CI^{\textrm{(r)}}_{\rm half}} = q^{13/240} \chi^{\CM(3,4)}_{(1,1)} \left(\chi^{\CM(5,6)}_{(1,1)}+\chi^{\CM(5,6)}_{(1,5)}\right)\,.
\end{align}
This is the vacuum character of the RCFT $\CM(3,4) \otimes \frac{ \CM(5,6)}{\mathbb{Z}_2}$.
\\
\\
\subsection{The case $(P,Q)=(5,7)$}
Using the SnapPy, the gluing data of the torus knot can be obtained by the following input:
\begin{align*}
    \begin{split}
        & \mathtt{Manifold('DT:[(34,-48,-38,52,42,-56,-46,4,50,-8,-54,12,2,-16,-6,20,10,}\\
        & \mathtt{-24,-14,28,18,-32,-22,36,26,-40,-30,44)]').gluing\_equations\_pgl()}
    \end{split}
\end{align*}
The knot complement can be ideally triangulated using 5 tetrahedra and gluing equations are
\begin{align}
\begin{split}
&C_1 = 2Z_1+Z_1''+Z_3+Z_3'+Z_4'+Z_4'', \quad C_2 = Z_3''+Z_4, \quad C_3= 2Z_2+Z_3'+Z_4'+Z_5+2Z_5''
\\
&C_4 = Z_2'+Z_3+Z_4''+2Z_5', \quad C_5 = 2Z_1'+Z_1''+Z_2'+2Z_2''+Z_3''+Z_4+Z_5,
\\
& M= -Z_1+Z_1'-Z_2+Z_2''+2Z_3''\;. \label{(5,7) ge}
\end{split}
\end{align}
In the ideal triangulation, there are two easy internal edges $\{E_I\}=\{C_2,C_4\}$, and we choose $\{H_I\}=\{C_1,C_3\}$.
\\
\\
Among the $3^5$ possible polarization choices, there are eight choices satisfying the positivity condition.
\\
\paragraph{i)} When we choose the following polarizations
\begin{align}
    \mathbf{X}=(Z_1'',Z_2'',Z_3',Z_4'',Z_5)^T \textrm{ and }\mathbf{P}=(Z_1',Z_2',Z_3,Z_4',Z_5'')^T\;,
\end{align}
the NZ matrices are
\begin{align}
    A=\begin{pmatrix}
        1 & 2 & -2 & 0 & 0 \\
        -1 & 0 & 1 & 1 & 0 \\
        0 & -2 & 1 & 0 & 1 \\
        0 & 0 & -1 & -1 & 0 \\
        0 & 0 & 0 & 1 & -2 \\ \end{pmatrix}, \quad 
        B=\begin{pmatrix}
        2 & 1 & -2 & 0 & 0 \\
        -2 & 0 & 1 & 1 & 0 \\
        0 & -2 & 0 & 1 & 2 \\
        0 & 0 & -1 & -1 & 0 \\
        0 & 1 & 1 & 0 & -2 \\ \end{pmatrix},\quad {\boldsymbol{\nu}}_x=(0,0,0,0,0)^T\;.
\end{align}
In the case, $B^{-1} A\notin M_5(\mathbb{Z})$ and $|\det B|=8$. If we choose $\mathbf{Q}=\mathrm{diag}(2,1,1,1,2)$, $N=2$ and
\begin{equation}
\begin{gathered}
    K = \begin{pmatrix}
        2 & 0 & 0 & 0 & 0 \\
        0 & 2 & 0 & 0 & 2 \\
        0 & 0 & 1 & 0 & 1 \\
        0 & 0 & 0 & 1 & -1 \\
        0 & 2 & 1 & -1 & 7 \\
        \end{pmatrix}, \quad \boldsymbol{\m}_\mathrm{geo}=(0,0,0,0,0)^T,\\
    \boldsymbol{v}_m = \frac12(0,0,-1,1,-1)^T, \quad \boldsymbol{v}_1 = \frac12(-2,0,-1,1,-1)^T\,, \quad
    \boldsymbol{v}_2 = \frac12(0,-2,-1,1,-3)^T\,.
\end{gathered}
\end{equation}
The chiral primary operators corresponding to each easy internal edges are
\begin{align}
    \CO_{E_1} = V_{(0,0,1,1,0)}\,,\quad \CO_{E_2} = \phi_4 V_{(0,-1,-1,0,1)}\,.
\end{align}
The half-indices in \eqref{half-indices at AA/AB/BA/BB-twistings} are
\begin{align}
    \begin{split}
        \CI_\mathrm{half}^{aa}[\mu_m \in 4\mathbb{Z}] & \simeq q^{27/280} \paren{\chi_F(q)}^2 \, \chi^{M(3,5)}_{(1,2)} \chi^{M(3,7)}_{(1,2)} \,, \\
        \CI_\mathrm{half}^{ba}[\mu_m \in 4\mathbb{Z}+2] &\simeq   q^{13/280} \paren{\chi_F(q)}^2 \, \chi^{M(3,5)}_{(1,1)} \chi^{M(3,7)}_{(1,2)} \,, \\
        \CI_\mathrm{half}^{ab}[\mu_m \in 4 \mathbb{Z}+2] &\simeq   q^{-23/280} \paren{\chi_F(q)}^2 \, \chi^{M(3,5)}_{(1,2)} \chi^{M(3,7)}_{(1,1)} \,, \\
        \CI_\mathrm{half}^{bb}[\mu_m \in 4\mathbb{Z}] & \simeq  q^{-37/280} \paren{\chi_F(q)}^2 \, \chi^{M(3,5)}_{(1,1)} \chi^{M(3,7)}_{(1,1)} \,, \\
        \CI_\mathrm{half}^{aa}[\mu_m \in 4\mathbb{Z}+2] = \CI_\mathrm{half}^{ab}[\mu_m \in 4\mathbb{Z}] &= \CI_\mathrm{half}^{ba}[\mu_m \in 4\mathbb{Z}] = \CI_\mathrm{half}^{bb}[\mu_m \in 4\mathbb{Z}+2]=0\;.
    \end{split}
\end{align}
Note that $K$ is block-diagonal; the result is consistent with the combined results of {\bf reduced-ii)} in \ref{AppEx2,5} and {\bf i)} in \ref{AppEx3,7}.
\\
\\
\\
Note that the $C_2$ is the special easy internal edge. After taking the reduction procedure, we have
\begin{align}
\begin{split}
&H_1^{(r)} = C_1^{(r)} = 2Z_1+Z_1'', \quad H_2^{(r)} = C_3^{(r)} = 2Z_2+Z_5+2Z_5'',
\quad E_1^{(r)} = C_4^{(r)} = Z_2'+2Z_5'\;.
\end{split}
\end{align}

Among the $3^3$ possible polarization choices, there are eight choices satisfying the positivity condition.
\\
\paragraph{reduced-i)} When we choose the following polarizations
\begin{align}
    \mathbf{X}_\mathrm{r}=(Z_1'',Z_2',Z_5)^T\textrm{ and }\mathbf{P}_\mathrm{r}=(Z_1',Z_2,Z_5'')^T\;,
\end{align}
the reduced NZ matrices are
\begin{align}
    A_\mathrm{r}=\begin{pmatrix}
        -1 & 0 & 0\\
        0 & 0 & 1\\
        0 & 1 & -2
    \end{pmatrix},\quad B_\mathrm{r}=\begin{pmatrix}
        -2 & 0 & 0\\
        0 & 2 & 2\\
        0 & 0 & -2
    \end{pmatrix},\quad \boldsymbol{\n}_{x;\mathrm{r}}=(0,-2,0)^T\;.
\end{align}
In the case, we choose $\mathbf{Q}_\mathrm{r}=\diag(2,2,1)$. Then, $N_\mathrm{r}=2$ and
\begin{align}
    \begin{split}
        & K_\mathrm{r}=\begin{pmatrix}
            2 & 0 & 0\\
            0 & 2 & -1\\
            0 & -1 & 1
        \end{pmatrix},\quad \boldsymbol{\m}_\mathrm{geo}^\mathrm{(r)}=(0,-2,0)^T,\\
        & \CO_{E_1^{\mathrm{(r)}}}=\f_2 V_{(0,0,2)},\quad \boldsymbol{v}_1^\mathrm{(r)}=(-1,0,0)^T,\quad \boldsymbol{v}_2^\mathrm{(r)}=(0,1,0)^T\;.
    \end{split}
\end{align}
$\boldsymbol{\n}_{aa}=(0,0)^T$, $\boldsymbol{\n}_{ab}=(0,2)^T$, $\boldsymbol{\n}_{ba}=(2,0)^T$, and $\boldsymbol{\n}_{bb}=(2,2)^T$. The half-indices in \eqref{half-indices at AA/AB/BA/BB-twistings} are
\begin{align}
    \begin{split}
        \CI_\mathrm{half}^{aa,\mathrm{(r)}}& =q^{11/840} \c_F(q) \c_{(1,2)}^{\CM(3,5)}(q) \c_{(1,1)}^{SM(3,7)}(q),\\
        \CI_\mathrm{half}^{ab,\mathrm{(r)}}& =q^{71/840} \c_F(q) \c_{(1,2)}^{\CM(3,5)}(q) \c_{(1,3)}^{SM(3,7)}(q),\\
        \CI_\mathrm{half}^{ba,\mathrm{(r)}}& =q^{-31/840} \c_F(q) \c_{(1,1)}^{\CM(3,5)}(q) \c_{(1,1)}^{SM(3,7)}(q),\\
        \CI_\mathrm{half}^{bb,\mathrm{(r)}}& =q^{29/840} \c_F(q) \c_{(1,1)}^{\CM(3,5)}(q) \c_{(1,3)}^{SM(3,7)}(q)\;.
    \end{split}
\end{align}
The result is consistent with the combined results of \textbf{reduced-ii)} in \ref{AppEx2,5} and \textbf{reduced-i)} in \ref{AppEx2,7}.

For the other polarization choices that can be manifestly factorized as \eqref{factorization polarization} or \eqref{factorization polarization-red}, one can check that the factors $(K^{(i)},\,Q^{(i)})$ are that of \ref{AppEx2,5} or \ref{AppEx3,5}, and \ref{AppEx2,7} or \ref{AppEx3,7}.

%% file: draft.bbl
\providecommand{\href}[2]{#2}\begingroup\raggedright\begin{thebibliography}{10}

\bibitem{Hikami:2003tb}
K.~Hikami and A.~N. Kirillov, ``{Torus knot and minimal model},''
  \href{http://dx.doi.org/10.1016/j.physletb.2003.09.007}{{\em Phys. Lett. B}
  {\bfseries 575} (2003) 343--348},
  \href{http://arxiv.org/abs/hep-th/0308152}{{\ttfamily arXiv:hep-th/0308152}}.

\bibitem{zagier1999modular}
D.~Zagier and R.~Lawrence, ``Modular forms and quantum invariants of
  3-manifolds,''.

\bibitem{Terashima:2011qi}
Y.~Terashima and M.~Yamazaki, ``{SL(2,R) Chern-Simons, Liouville, and Gauge
  Theory on Duality Walls},''
  \href{http://dx.doi.org/10.1007/JHEP08(2011)135}{{\em JHEP} {\bfseries 08}
  (2011) 135}, \href{http://arxiv.org/abs/1103.5748}{{\ttfamily arXiv:1103.5748
  [hep-th]}}.

\bibitem{Dimofte:2011ju}
T.~Dimofte, D.~Gaiotto, and S.~Gukov, ``{Gauge Theories Labelled by
  Three-Manifolds},'' \href{http://dx.doi.org/10.1007/s00220-013-1863-2}{{\em
  Commun. Math. Phys.} {\bfseries 325} (2014) 367--419},
  \href{http://arxiv.org/abs/1108.4389}{{\ttfamily arXiv:1108.4389 [hep-th]}}.

\bibitem{Kim:2009wb}
S.~Kim, ``{The Complete superconformal index for N=6 Chern-Simons theory},''
  \href{http://dx.doi.org/10.1016/j.nuclphysb.2009.06.025}{{\em Nucl. Phys. B}
  {\bfseries 821} (2009) 241--284},
  \href{http://arxiv.org/abs/0903.4172}{{\ttfamily arXiv:0903.4172 [hep-th]}}.
  [Erratum: Nucl.Phys.B 864, 884 (2012)].

\bibitem{Kapustin:2009kz}
A.~Kapustin, B.~Willett, and I.~Yaakov, ``{Exact Results for Wilson Loops in
  Superconformal Chern-Simons Theories with Matter},''
  \href{http://dx.doi.org/10.1007/JHEP03(2010)089}{{\em JHEP} {\bfseries 03}
  (2010) 089}, \href{http://arxiv.org/abs/0909.4559}{{\ttfamily arXiv:0909.4559
  [hep-th]}}.

\bibitem{Imamura:2011su}
Y.~Imamura and S.~Yokoyama, ``{Index for three dimensional superconformal field
  theories with general R-charge assignments},''
  \href{http://dx.doi.org/10.1007/JHEP04(2011)007}{{\em JHEP} {\bfseries 04}
  (2011) 007}, \href{http://arxiv.org/abs/1101.0557}{{\ttfamily arXiv:1101.0557
  [hep-th]}}.

\bibitem{Jafferis:2010un}
D.~L. Jafferis, ``{The Exact Superconformal R-Symmetry Extremizes Z},''
  \href{http://dx.doi.org/10.1007/JHEP05(2012)159}{{\em JHEP} {\bfseries 05}
  (2012) 159}, \href{http://arxiv.org/abs/1012.3210}{{\ttfamily arXiv:1012.3210
  [hep-th]}}.

\bibitem{Hama:2010av}
N.~Hama, K.~Hosomichi, and S.~Lee, ``{Notes on SUSY Gauge Theories on
  Three-Sphere},'' \href{http://dx.doi.org/10.1007/JHEP03(2011)127}{{\em JHEP}
  {\bfseries 03} (2011) 127}, \href{http://arxiv.org/abs/1012.3512}{{\ttfamily
  arXiv:1012.3512 [hep-th]}}.

\bibitem{Hama:2011ea}
N.~Hama, K.~Hosomichi, and S.~Lee, ``{SUSY Gauge Theories on Squashed
  Three-Spheres},'' \href{http://dx.doi.org/10.1007/JHEP05(2011)014}{{\em JHEP}
  {\bfseries 05} (2011) 014}, \href{http://arxiv.org/abs/1102.4716}{{\ttfamily
  arXiv:1102.4716 [hep-th]}}.

\bibitem{HIKAMI_2001}
K.~HIKAMI, ``{HYPERBOLIC} {STRUCTURE} {ARISING} {FROM} a {KNOT} {INVARIANT},''
  \href{https://doi.org/10.1142%2Fs0217751x0100444x}{{\em International Journal
  of Modern Physics A} {\bfseries 16} no.~19, (Jul, 2001) 3309--3333}.

\bibitem{Dimofte:2009yn}
T.~Dimofte, S.~Gukov, J.~Lenells, and D.~Zagier, ``{Exact Results for
  Perturbative Chern-Simons Theory with Complex Gauge Group},''
  \href{http://dx.doi.org/10.4310/CNTP.2009.v3.n2.a4}{{\em Commun. Num. Theor.
  Phys.} {\bfseries 3} (2009) 363--443},
  \href{http://arxiv.org/abs/0903.2472}{{\ttfamily arXiv:0903.2472 [hep-th]}}.

\bibitem{Dimofte:2011gm}
T.~Dimofte, ``{Quantum Riemann Surfaces in Chern-Simons Theory},''
  \href{http://dx.doi.org/10.4310/ATMP.2013.v17.n3.a1}{{\em Adv. Theor. Math.
  Phys.} {\bfseries 17} no.~3, (2013) 479--599},
  \href{http://arxiv.org/abs/1102.4847}{{\ttfamily arXiv:1102.4847 [hep-th]}}.

\bibitem{EllegaardAndersen:2011vps}
J.~Ellegaard~Andersen and R.~Kashaev, ``{A TQFT from Quantum Teichm\"uller
  Theory},'' \href{http://dx.doi.org/10.1007/s00220-014-2073-2}{{\em Commun.
  Math. Phys.} {\bfseries 330} (2014) 887--934},
  \href{http://arxiv.org/abs/1109.6295}{{\ttfamily arXiv:1109.6295 [math.QA]}}.

\bibitem{Dimofte:2012qj}
T.~D. Dimofte and S.~Garoufalidis, ``{The Quantum content of the gluing
  equations},'' {\em Geom. Topol.} {\bfseries 17} (2013) 1253--1316,
  \href{http://arxiv.org/abs/1202.6268}{{\ttfamily arXiv:1202.6268 [math.GT]}}.

\bibitem{Dimofte:2014zga}
T.~Dimofte, ``{Complex Chern\textendash{}Simons Theory at Level k via the
  3d\textendash{}3d Correspondence},''
  \href{http://dx.doi.org/10.1007/s00220-015-2401-1}{{\em Commun. Math. Phys.}
  {\bfseries 339} no.~2, (2015) 619--662},
  \href{http://arxiv.org/abs/1409.0857}{{\ttfamily arXiv:1409.0857 [hep-th]}}.

\bibitem{zbMATH04092352}
E.~Witten, ``Quantum field theory and the {Jones} polynomial,''
  \href{http://dx.doi.org/10.1007/BF01217730}{{\em Commun. Math. Phys.}
  {\bfseries 121} no.~3, (1989) 351--399}.

\bibitem{Gang:2018huc}
D.~Gang and M.~Yamazaki, ``{Three-dimensional gauge theories with supersymmetry
  enhancement},'' \href{http://dx.doi.org/10.1103/PhysRevD.98.121701}{{\em
  Phys. Rev. D} {\bfseries 98} no.~12, (2018) 121701},
  \href{http://arxiv.org/abs/1806.07714}{{\ttfamily arXiv:1806.07714
  [hep-th]}}.

\bibitem{Gang:2021hrd}
D.~Gang, S.~Kim, K.~Lee, M.~Shim, and M.~Yamazaki, ``{Non-unitary TQFTs from 3D
  $ \mathcal{N} $ = 4 rank 0 SCFTs},''
  \href{http://dx.doi.org/10.1007/JHEP08(2021)158}{{\em JHEP} {\bfseries 08}
  (2021) 158}, \href{http://arxiv.org/abs/2103.09283}{{\ttfamily
  arXiv:2103.09283 [hep-th]}}.

\bibitem{Gang:2023rei}
D.~Gang, H.~Kim, and S.~Stubbs, ``{Three-Dimensional Topological Field Theories
  and Nonunitary Minimal Models},''
  \href{http://dx.doi.org/10.1103/PhysRevLett.132.131601}{{\em Phys. Rev.
  Lett.} {\bfseries 132} no.~13, (2024) 131601},
  \href{http://arxiv.org/abs/2310.09080}{{\ttfamily arXiv:2310.09080
  [hep-th]}}.

\bibitem{Gang:2022kpe}
D.~Gang and D.~Kim, ``{Generalized non-unitary Haagerup-Izumi modular data from
  3D S-fold SCFTs},'' \href{http://dx.doi.org/10.1007/JHEP03(2023)185}{{\em
  JHEP} {\bfseries 03} (2023) 185},
  \href{http://arxiv.org/abs/2211.13561}{{\ttfamily arXiv:2211.13561
  [hep-th]}}.

\bibitem{Ferrari:2023fez}
A.~E.~V. Ferrari, N.~Garner, and H.~Kim, ``{Boundary vertex algebras for 3d
  $\mathcal{N}=4$ rank-0 SCFTs},''
  \href{http://dx.doi.org/10.21468/SciPostPhys.17.2.057}{{\em SciPost Phys.}
  {\bfseries 17} no.~2, (2024) 057},
  \href{http://arxiv.org/abs/2311.05087}{{\ttfamily arXiv:2311.05087
  [hep-th]}}.

\bibitem{Costello:2018fnz}
K.~Costello and D.~Gaiotto, ``{Vertex Operator Algebras and 3d $ \mathcal{N} $
  = 4 gauge theories},'' \href{http://dx.doi.org/10.1007/JHEP05(2019)018}{{\em
  JHEP} {\bfseries 05} (2019) 018},
  \href{http://arxiv.org/abs/1804.06460}{{\ttfamily arXiv:1804.06460
  [hep-th]}}.

\bibitem{Costello:2020ndc}
K.~Costello, T.~Dimofte, and D.~Gaiotto, ``{Boundary Chiral Algebras and
  Holomorphic Twists},''
  \href{http://dx.doi.org/10.1007/s00220-022-04599-0}{{\em Commun. Math. Phys.}
  {\bfseries 399} no.~2, (2023) 1203--1290},
  \href{http://arxiv.org/abs/2005.00083}{{\ttfamily arXiv:2005.00083
  [hep-th]}}.

\bibitem{dedushenko:2018bpp}
M.~Dedushenko, S.~Gukov, H.~Nakajima, D.~Pei, and K.~Ye, ``{3d TQFTs from
  Argyres{\textendash}Douglas theories},''
  \href{http://dx.doi.org/10.1088/1751-8121/abb481}{{\em J. Phys. A} {\bfseries
  53} no.~43, (2020) 43LT01}, \href{http://arxiv.org/abs/1809.04638}{{\ttfamily
  arXiv:1809.04638 [hep-th]}}.

\bibitem{Gang:2023ggt}
D.~Gang, D.~Kim, and S.~Lee, ``{A non-unitary bulk-boundary correspondence:
  Non-unitary Haagerup RCFTs from S-fold SCFTs},''
  \href{http://dx.doi.org/10.21468/SciPostPhys.17.2.064}{{\em SciPost Phys.}
  {\bfseries 17} no.~2, (2024) 064},
  \href{http://arxiv.org/abs/2310.14877}{{\ttfamily arXiv:2310.14877
  [hep-th]}}.

\bibitem{Dedushenko:2023cvd}
M.~Dedushenko, ``{On the 4d/3d/2d view of the SCFT/VOA correspondence},''
  \href{http://arxiv.org/abs/2312.17747}{{\ttfamily arXiv:2312.17747
  [hep-th]}}.

\bibitem{Creutzig:2024ljv}
T.~Creutzig, N.~Garner, and H.~Kim, ``{Mirror symmetry and level-rank duality
  for 3d $\mathcal {N} = 4$ rank 0 SCFTs},''
  \href{http://dx.doi.org/10.1007/s11005-025-02015-x}{{\em Lett. Math. Phys.}
  {\bfseries 115} no.~6, (2025) 123},
  \href{http://arxiv.org/abs/2406.00138}{{\ttfamily arXiv:2406.00138
  [hep-th]}}.

\bibitem{ArabiArdehali:2024ysy}
A.~Arabi~Ardehali, M.~Dedushenko, D.~Gang, and M.~Litvinov, ``{Bridging 4D QFTs
  and 2D VOAs via 3D high-temperature EFTs},''
  \href{http://arxiv.org/abs/2409.18130}{{\ttfamily arXiv:2409.18130
  [hep-th]}}.

\bibitem{Gaiotto:2024ioj}
D.~Gaiotto and H.~Kim, ``{3D TFTs from 4d $ \mathcal{N} $ = 2 BPS particles},''
  \href{http://dx.doi.org/10.1007/JHEP03(2025)173}{{\em JHEP} {\bfseries 03}
  (2025) 173}, \href{http://arxiv.org/abs/2409.20393}{{\ttfamily
  arXiv:2409.20393 [hep-th]}}.

\bibitem{ArabiArdehali:2024vli}
A.~Arabi~Ardehali, D.~Gang, N.~J. Rajappa, and M.~Sacchi, ``{3d SUSY
  enhancement and non-semisimple TQFTs from four dimensions},''
  \href{http://dx.doi.org/10.1007/JHEP09(2025)179}{{\em JHEP} {\bfseries 09}
  (2025) 179}, \href{http://arxiv.org/abs/2411.00766}{{\ttfamily
  arXiv:2411.00766 [hep-th]}}.

\bibitem{Kim:2024dxu}
H.~Kim and J.~Song, ``{A Family of Vertex Algebras from Argyres-Douglas
  Theory},'' \href{http://dx.doi.org/10.21468/SciPostPhys.19.6.144}{{\em
  SciPost Phys.} {\bfseries 19} (2025) 144},
  \href{http://arxiv.org/abs/2412.20015}{{\ttfamily arXiv:2412.20015
  [hep-th]}}.

\bibitem{Gang:2024loa}
D.~Gang, H.~Kim, B.~Park, and S.~Stubbs, ``{Three Dimensional Topological Field
  Theories and Nahm Sum Formulas},''
  \href{http://dx.doi.org/10.21468/SciPostPhys.19.5.128}{{\em SciPost Phys.}
  {\bfseries 19} (2025) 128}, \href{http://arxiv.org/abs/2411.06081}{{\ttfamily
  arXiv:2411.06081 [hep-th]}}.

\bibitem{Go:2025ixu}
B.~Go, Q.~Jia, H.~Kim, and S.~Kim, ``{From BPS spectra of Argyres-Douglas
  theories to families of 3d TFTs},''
  \href{http://dx.doi.org/10.1007/JHEP08(2025)012}{{\em JHEP} {\bfseries 08}
  (2025) 012}, \href{http://arxiv.org/abs/2502.15133}{{\ttfamily
  arXiv:2502.15133 [hep-th]}}.

\bibitem{Jeong:2025xid}
K.~Jeong and S.~Lee, ``{QFT Realization of Non-Unitary
  $\mathfrak{sl}(2,\mathbb{C})$ WRT Invariants and Their Galois
  Conjugations},'' \href{http://arxiv.org/abs/2511.16380}{{\ttfamily
  arXiv:2511.16380 [hep-th]}}.

\bibitem{Kim:2025klh}
H.~Kim, H.~Kim, and J.~Song, ``{Macdonald index from 3d TQFT},''
  \href{http://arxiv.org/abs/2511.11186}{{\ttfamily arXiv:2511.11186
  [hep-th]}}.

\bibitem{Kim:2025rog}
M.~Kim and S.~Kim, ``{3D TFTs and boundary VOAs from BPS spectra of $(G,G')$
  Argyres-Douglas theories},''
  \href{http://arxiv.org/abs/2511.23194}{{\ttfamily arXiv:2511.23194
  [hep-th]}}.

\bibitem{Nishinaka:2025ytu}
T.~Nishinaka and Y.~Yoshida, ``{3d Chern--Simons matter theories from
  generalized Argyres--Douglas theories},''
  \href{http://arxiv.org/abs/2512.15201}{{\ttfamily arXiv:2512.15201
  [hep-th]}}.

\bibitem{Gadde:2013wq}
A.~Gadde, S.~Gukov, and P.~Putrov, ``{Walls, Lines, and Spectral Dualities in
  3d Gauge Theories},'' \href{http://dx.doi.org/10.1007/JHEP05(2014)047}{{\em
  JHEP} {\bfseries 05} (2014) 047},
  \href{http://arxiv.org/abs/1302.0015}{{\ttfamily arXiv:1302.0015 [hep-th]}}.

\bibitem{Yoshida:2014ssa}
Y.~Yoshida and K.~Sugiyama, ``{Localization of three-dimensional
  $\mathcal{N}=2$ supersymmetric theories on $S^1 \times D^2$},''
  \href{http://dx.doi.org/10.1093/ptep/ptaa136}{{\em PTEP} {\bfseries 2020}
  no.~11, (2020) 113B02}, \href{http://arxiv.org/abs/1409.6713}{{\ttfamily
  arXiv:1409.6713 [hep-th]}}.

\bibitem{Dimofte:2017tpi}
T.~Dimofte, D.~Gaiotto, and N.~M. Paquette, ``{Dual boundary conditions in 3d
  SCFT\textquoteright{}s},''
  \href{http://dx.doi.org/10.1007/JHEP05(2018)060}{{\em JHEP} {\bfseries 05}
  (2018) 060}, \href{http://arxiv.org/abs/1712.07654}{{\ttfamily
  arXiv:1712.07654 [hep-th]}}.

\bibitem{Witten:2003ya}
E.~Witten, ``{SL(2,Z) action on three-dimensional conformal field theories with
  Abelian symmetry},'' in {\em {From Fields to Strings: Circumnavigating
  Theoretical Physics: A Conference in Tribute to Ian Kogan}}, pp.~1173--1200.
\newblock 7, 2003.
\newblock \href{http://arxiv.org/abs/hep-th/0307041}{{\ttfamily
  arXiv:hep-th/0307041}}.

\bibitem{Aharony:1997bx}
O.~Aharony, A.~Hanany, K.~A. Intriligator, N.~Seiberg, and M.~J. Strassler,
  ``{Aspects of N=2 supersymmetric gauge theories in three-dimensions},''
  \href{http://dx.doi.org/10.1016/S0550-3213(97)00323-4}{{\em Nucl. Phys. B}
  {\bfseries 499} (1997) 67--99},
  \href{http://arxiv.org/abs/hep-th/9703110}{{\ttfamily arXiv:hep-th/9703110}}.

\bibitem{Cordova:2013cea}
C.~Cordova and D.~L. Jafferis, ``{Complex Chern-Simons from M5-branes on the
  Squashed Three-Sphere},''
  \href{http://dx.doi.org/10.1007/JHEP11(2017)119}{{\em JHEP} {\bfseries 11}
  (2017) 119}, \href{http://arxiv.org/abs/1305.2891}{{\ttfamily arXiv:1305.2891
  [hep-th]}}.

\bibitem{Gang:2018wek}
D.~Gang and K.~Yonekura, ``{Symmetry enhancement and closing of knots in 3d/3d
  correspondence},'' \href{http://dx.doi.org/10.1007/JHEP07(2018)145}{{\em
  JHEP} {\bfseries 07} (2018) 145},
  \href{http://arxiv.org/abs/1803.04009}{{\ttfamily arXiv:1803.04009
  [hep-th]}}.

\bibitem{thurston2022geometry}
W.~P. Thurston, {\em The geometry and topology of three-manifolds: With a
  preface by Steven P. Kerckhoff}, vol.~27.
\newblock American Mathematical Society, 2022.

\bibitem{neumann1985volumes}
W.~D. Neumann and D.~Zagier, ``Volumes of hyperbolic three-manifolds,'' {\em
  Topology} {\bfseries 24} no.~3, (1985) 307--332.

\bibitem{Hsin:2018vcg}
P.-S. Hsin, H.~T. Lam, and N.~Seiberg, ``{Comments on One-Form Global
  Symmetries and Their Gauging in 3d and 4d},''
  \href{http://dx.doi.org/10.21468/SciPostPhys.6.3.039}{{\em SciPost Phys.}
  {\bfseries 6} no.~3, (2019) 039},
  \href{http://arxiv.org/abs/1812.04716}{{\ttfamily arXiv:1812.04716
  [hep-th]}}.

\bibitem{SnapPy}
M.~Culler, N.~M. Dunfield, M.~Goerner, and J.~R. Weeks, ``Snap{P}y, a computer
  program for studying the geometry and topology of $3$-manifolds.'' Available
  at \url{http://snappy.computop.org} (30/03/2025).

\bibitem{Baek:2024tuo}
S.~Baek and D.~Gang, ``{3D bulk field theories for 2D non-unitary $ \mathcal{N}
  $ = 1 supersymmetric minimal models},''
  \href{http://dx.doi.org/10.1007/JHEP01(2025)027}{{\em JHEP} {\bfseries 01}
  (2025) 027}, \href{http://arxiv.org/abs/2405.05746}{{\ttfamily
  arXiv:2405.05746 [hep-th]}}.

\bibitem{Gang:2024tlp}
D.~Gang, H.~Kang, and S.~Kim, ``{Non-hyperbolic 3-manifolds and 3D field
  theories for 2D Virasoro minimal models},''
  \href{http://arxiv.org/abs/2405.16377}{{\ttfamily arXiv:2405.16377
  [hep-th]}}.

\bibitem{Dimofte:2011py}
T.~Dimofte, D.~Gaiotto, and S.~Gukov, ``{3-Manifolds and 3d Indices},''
  \href{http://dx.doi.org/10.4310/ATMP.2013.v17.n5.a3}{{\em Adv. Theor. Math.
  Phys.} {\bfseries 17} no.~5, (2013) 975--1076},
  \href{http://arxiv.org/abs/1112.5179}{{\ttfamily arXiv:1112.5179 [hep-th]}}.

\bibitem{Lee:2013ida}
S.~Lee and M.~Yamazaki, ``{3d Chern-Simons Theory from M5-branes},''
  \href{http://dx.doi.org/10.1007/JHEP12(2013)035}{{\em JHEP} {\bfseries 12}
  (2013) 035}, \href{http://arxiv.org/abs/1305.2429}{{\ttfamily arXiv:1305.2429
  [hep-th]}}.

\bibitem{Yagi:2013fda}
J.~Yagi, ``{3d TQFT from 6d SCFT},''
  \href{http://dx.doi.org/10.1007/JHEP08(2013)017}{{\em JHEP} {\bfseries 08}
  (2013) 017}, \href{http://arxiv.org/abs/1305.0291}{{\ttfamily arXiv:1305.0291
  [hep-th]}}.

\bibitem{Cordova:2016emh}
C.~Cordova, T.~T. Dumitrescu, and K.~Intriligator, ``{Multiplets of
  Superconformal Symmetry in Diverse Dimensions},''
  \href{http://dx.doi.org/10.1007/JHEP03(2019)163}{{\em JHEP} {\bfseries 03}
  (2019) 163}, \href{http://arxiv.org/abs/1612.00809}{{\ttfamily
  arXiv:1612.00809 [hep-th]}}.

\bibitem{Evtikhiev:2017heo}
M.~Evtikhiev, ``{Studying superconformal symmetry enhancement through
  indices},'' \href{http://dx.doi.org/10.1007/JHEP04(2018)120}{{\em JHEP}
  {\bfseries 04} (2018) 120}, \href{http://arxiv.org/abs/1708.08307}{{\ttfamily
  arXiv:1708.08307 [hep-th]}}.

\bibitem{Gang:2019jut}
D.~Gang and M.~Yamazaki, ``{Expanding 3d $ \mathcal{N} $ = 2 theories around
  the round sphere},'' \href{http://dx.doi.org/10.1007/JHEP02(2020)102}{{\em
  JHEP} {\bfseries 02} (2020) 102},
  \href{http://arxiv.org/abs/1912.09617}{{\ttfamily arXiv:1912.09617
  [hep-th]}}.

\bibitem{Benini:2015noa}
F.~Benini and A.~Zaffaroni, ``{A topologically twisted index for
  three-dimensional supersymmetric theories},''
  \href{http://dx.doi.org/10.1007/JHEP07(2015)127}{{\em JHEP} {\bfseries 07}
  (2015) 127}, \href{http://arxiv.org/abs/1504.03698}{{\ttfamily
  arXiv:1504.03698 [hep-th]}}.

\bibitem{Benini:2016hjo}
F.~Benini and A.~Zaffaroni, ``{Supersymmetric partition functions on Riemann
  surfaces},'' {\em Proc. Symp. Pure Math.} {\bfseries 96} (2017) 13--46,
  \href{http://arxiv.org/abs/1605.06120}{{\ttfamily arXiv:1605.06120
  [hep-th]}}.

\bibitem{Closset:2016arn}
C.~Closset and H.~Kim, ``{Comments on twisted indices in 3d supersymmetric
  gauge theories},'' \href{http://dx.doi.org/10.1007/JHEP08(2016)059}{{\em
  JHEP} {\bfseries 08} (2016) 059},
  \href{http://arxiv.org/abs/1605.06531}{{\ttfamily arXiv:1605.06531
  [hep-th]}}.

\bibitem{Closset:2017zgf}
C.~Closset, H.~Kim, and B.~Willett, ``{Supersymmetric partition functions and
  the three-dimensional A-twist},''
  \href{http://dx.doi.org/10.1007/JHEP03(2017)074}{{\em JHEP} {\bfseries 03}
  (2017) 074}, \href{http://arxiv.org/abs/1701.03171}{{\ttfamily
  arXiv:1701.03171 [hep-th]}}.

\bibitem{Closset:2018ghr}
C.~Closset, H.~Kim, and B.~Willett, ``{Seifert fibering operators in 3d
  $\mathcal{N}=2$ theories},''
  \href{http://dx.doi.org/10.1007/JHEP11(2018)004}{{\em JHEP} {\bfseries 11}
  (2018) 004}, \href{http://arxiv.org/abs/1807.02328}{{\ttfamily
  arXiv:1807.02328 [hep-th]}}.

\bibitem{Jockers:2021omw}
H.~Jockers, P.~Mayr, U.~Ninad, and A.~Tabler, ``{BPS indices, modularity and
  perturbations in quantum K-theory},''
  \href{http://dx.doi.org/10.1007/JHEP02(2022)044}{{\em JHEP} {\bfseries 02}
  (2022) 044}, \href{http://arxiv.org/abs/2106.07670}{{\ttfamily
  arXiv:2106.07670 [hep-th]}}.

\bibitem{Choi:2022dju}
S.~Choi, D.~Gang, and H.-C. Kim, ``{Infrared phases of 3D class R theories},''
  \href{http://dx.doi.org/10.1007/JHEP11(2022)151}{{\em JHEP} {\bfseries 11}
  (2022) 151}, \href{http://arxiv.org/abs/2206.11982}{{\ttfamily
  arXiv:2206.11982 [hep-th]}}.

\bibitem{Cho:2020ljj}
G.~Y. Cho, D.~Gang, and H.-C. Kim, ``{M-theoretic Genesis of Topological
  Phases},'' \href{http://dx.doi.org/10.1007/JHEP11(2020)115}{{\em JHEP}
  {\bfseries 11} (2020) 115}, \href{http://arxiv.org/abs/2007.01532}{{\ttfamily
  arXiv:2007.01532 [hep-th]}}.

\bibitem{Dimofte:2013iv}
T.~Dimofte, M.~Gabella, and A.~B. Goncharov, ``{K-Decompositions and 3d Gauge
  Theories},'' \href{http://dx.doi.org/10.1007/JHEP11(2016)151}{{\em JHEP}
  {\bfseries 11} (2016) 151}, \href{http://arxiv.org/abs/1301.0192}{{\ttfamily
  arXiv:1301.0192 [hep-th]}}.

\bibitem{Kanade:2022wtm}
S.~Kanade, ``{Coloured $\mathfrak {sl}_r$ invariants of torus knots and
  characters of ${\mathcal {W}}_r$ algebras},''
  \href{http://dx.doi.org/10.1007/s11005-022-01628-w}{{\em Lett. Math. Phys.}
  {\bfseries 113} no.~1, (2023) 5},
  \href{http://arxiv.org/abs/2207.03685}{{\ttfamily arXiv:2207.03685
  [math.QA]}}.

\bibitem{Baek:2025uev}
S.~Baek and H.~Kang, ``{Non-hyperbolic 3-manifolds and bulk field theories for
  supersymmetric/$W_N$ minimal models},''
  \href{http://arxiv.org/abs/2511.04524}{{\ttfamily arXiv:2511.04524
  [hep-th]}}.

\bibitem{Garoufalidis:2015}
S.~Garoufalidis, M.~Goerner, and C.~Zickert, ``Gluing equations for
  {PGL}$(n,\mathbb{C})$ representations of $3$-manifolds,''
  \href{http://dx.doi.org/10.2140/agt.2015.15.565}{{\em Algebraic \& Geometric
  Topology} {\bfseries 15} no.~1, (March, 2015) 565--622}.

\bibitem{Gang:2013sqa}
D.~Gang, E.~Koh, S.~Lee, and J.~Park, ``{Superconformal Index and 3d-3d
  Correspondence for Mapping Cylinder/Torus},''
  \href{http://dx.doi.org/10.1007/JHEP01(2014)063}{{\em JHEP} {\bfseries 01}
  (2014) 063}, \href{http://arxiv.org/abs/1305.0937}{{\ttfamily arXiv:1305.0937
  [hep-th]}}.

\bibitem{Evans:2010yr}
D.~E. Evans and T.~Gannon, ``{The exoticness and realisability of twisted
  Haagerup-Izumi modular data},''
  \href{http://dx.doi.org/10.1007/s00220-011-1329-3}{{\em Commun. Math. Phys.}
  {\bfseries 307} (2011) 463--512},
  \href{http://arxiv.org/abs/1006.1326}{{\ttfamily arXiv:1006.1326 [math.QA]}}.

\bibitem{feller2015optimal}
P.~Feller, ``Optimal cobordisms between torus knots,'' {\em arXiv preprint
  arXiv:1501.00483} (2015) .

\bibitem{feller2021genus}
P.~Feller and J.~Park, ``Genus one cobordisms between torus knots,'' {\em
  International Mathematics Research Notices} {\bfseries 2021} no.~1, (2021)
  521--548.

\bibitem{baader2024minimal}
S.~Baader, L.~Lewark, F.~Misev, and P.~Tru{\"o}l, ``Minimal cobordisms between
  thin and thick torus knots,'' {\em arXiv preprint arXiv:2405.13719} (2024) .

\bibitem{Garoufalidis:2016ckn}
S.~Garoufalidis, C.~Hodgson, N.~Hoffman, and H.~Rubinstein, ``{The 3D-index and
  normal surfaces},'' \href{http://arxiv.org/abs/1604.02688}{{\ttfamily
  arXiv:1604.02688 [math.GT]}}.

\bibitem{Gang:2018gyt}
D.~Gang, ``{Quantum Approach to Dehn Surgery Problem},''
  \href{http://arxiv.org/abs/1803.11143}{{\ttfamily arXiv:1803.11143
  [math.GT]}}.

\bibitem{Garoufalidis:2020xec}
S.~Garoufalidis, J.~Gu, and M.~Marino, ``{Peacock patterns and resurgence in
  complex Chern-Simons theory.},''
  \href{http://dx.doi.org/10.1007/s40687-023-00391-1}{{\em Res. Math. Sci.}
  {\bfseries 10} no.~3, (2023) 29},
  \href{http://arxiv.org/abs/2012.00062}{{\ttfamily arXiv:2012.00062
  [math.GT]}}.

\bibitem{Celoria:2025lqm}
D.~Celoria, C.~D. Hodgson, and J.~H. Rubinstein, ``{The 3D index and Dehn
  filling},'' \href{http://arxiv.org/abs/2509.09886}{{\ttfamily
  arXiv:2509.09886 [math.GT]}}.

\bibitem{Agarwal:2022cdm}
P.~Agarwal, D.~Gang, S.~Lee, and M.~Romo, ``{Quantum trace map for 3-manifolds
  and a 'length conjecture'},''
  \href{http://arxiv.org/abs/2203.15985}{{\ttfamily arXiv:2203.15985
  [hep-th]}}.

\bibitem{Duan:2022ijp}
Z.~Duan, S.~Garoufalidis, and J.~Gu, ``{The descendants of the 3d-index},''
  \href{http://arxiv.org/abs/2301.00098}{{\ttfamily arXiv:2301.00098
  [math.GT]}}.

\bibitem{Garoufalidis:2024hoe}
S.~Garoufalidis and T.~Yu, ``{The 3d-index of the 3d-skein module via the
  quantum trace map},'' \href{http://arxiv.org/abs/2406.04918}{{\ttfamily
  arXiv:2406.04918 [math.GT]}}.

\bibitem{Refined3DIndex:ToAppear}
D.~Gang, K.~Jeong, J.~Jung, T.~Kim, and S.~Lee, ``Refined 3d index,'' 2025.
\newblock to appear.

\bibitem{Cheng:2018vpl}
M.~C.~N. Cheng, S.~Chun, F.~Ferrari, S.~Gukov, and S.~M. Harrison, ``{3d
  Modularity},'' \href{http://dx.doi.org/10.1007/JHEP10(2019)010}{{\em JHEP}
  {\bfseries 10} (2019) 010}, \href{http://arxiv.org/abs/1809.10148}{{\ttfamily
  arXiv:1809.10148 [hep-th]}}.

\bibitem{Cheng:2024vou}
M.~C.~N. Cheng, I.~Coman, P.~Kucharski, D.~Passaro, and G.~Sgroi, ``{3d
  Modularity Revisited},'' \href{http://arxiv.org/abs/2403.14920}{{\ttfamily
  arXiv:2403.14920 [hep-th]}}.

\bibitem{Gukov:2016gkn}
S.~Gukov, P.~Putrov, and C.~Vafa, ``{Fivebranes and 3-manifold homology},''
  \href{http://dx.doi.org/10.1007/JHEP07(2017)071}{{\em JHEP} {\bfseries 07}
  (2017) 071}, \href{http://arxiv.org/abs/1602.05302}{{\ttfamily
  arXiv:1602.05302 [hep-th]}}.

\bibitem{Gukov:2017kmk}
S.~Gukov, D.~Pei, P.~Putrov, and C.~Vafa, ``{BPS spectra and 3-manifold
  invariants},'' \href{http://dx.doi.org/10.1142/S0218216520400039}{{\em J.
  Knot Theor. Ramifications} {\bfseries 29} no.~02, (2020) 2040003},
  \href{http://arxiv.org/abs/1701.06567}{{\ttfamily arXiv:1701.06567
  [hep-th]}}.

\bibitem{Gukov:2019mnk}
S.~Gukov and C.~Manolescu, ``{A two-variable series for knot complements},''
  \href{http://dx.doi.org/10.4171/qt/145}{{\em Quantum Topol.} {\bfseries 12}
  no.~1, (2021) 1--109}, \href{http://arxiv.org/abs/1904.06057}{{\ttfamily
  arXiv:1904.06057 [math.GT]}}.

\bibitem{Faddeev:1993rs}
L.~D. Faddeev and R.~M. Kashaev, ``{Quantum Dilogarithm},''
  \href{http://dx.doi.org/10.1142/S0217732394000447}{{\em Mod. Phys. Lett. A}
  {\bfseries 9} (1994) 427--434},
  \href{http://arxiv.org/abs/hep-th/9310070}{{\ttfamily arXiv:hep-th/9310070}}.

\bibitem{Nahm:2004ch}
W.~Nahm, \href{http://dx.doi.org/10.1007/978-3-540-30308-4_2}{``{Conformal
  field theory and torsion elements of the Bloch group},''} in {\em {Les
  Houches School of Physics: Frontiers in Number Theory, Physics and
  Geometry}}, pp.~67--132.
\newblock 2007.
\newblock \href{http://arxiv.org/abs/hep-th/0404120}{{\ttfamily
  arXiv:hep-th/0404120}}.

\bibitem{Zagier:2007knq}
D.~Zagier, \href{http://dx.doi.org/10.1007/978-3-540-30308-4_1}{``{The
  Dilogarithm Function},''} in {\em {Les Houches School of Physics: Frontiers
  in Number Theory, Physics and Geometry}}, pp.~3--65.
\newblock 2007.

\bibitem{Berkovich_1996}
A.~Berkovich and B.~M. McCoy, ``Continued fractions and fermionic
  representations for characters of m(p,{p'}) minimal models,''
  \href{http://dx.doi.org/10.1007/BF00400138}{{\em Letters in Mathematical
  Physics} {\bfseries 37} no.~1, (May, 1996) 49–66}.

\bibitem{melzer1994susyanalogGAid}
E.~Melzer, ``Supersymmetric analogs of the gordon-andrews identities, and
  related tba systems,'' 1994.
\newblock \url{https://arxiv.org/abs/hep-th/9412154}.

\bibitem{Bytsko_1999}
A.~G. Bytsko, ``Fermionic representations for characters of calm(3,t),
  calm(4,5), calm(5,6) and calm(6,7) minimal models and related dilogarithm and
  rogers-ramanujan-type identities,''
  \href{http://dx.doi.org/10.1088/0305-4470/32/46/305}{{\em Journal of Physics
  A: Mathematical and General} {\bfseries 32} no.~46, (Nov., 1999)
  8045–8058}.

\bibitem{Kedem:1993ze}
R.~Kedem, T.~R. Klassen, B.~M. McCoy, and E.~Melzer, ``{Fermionic sum
  representations for conformal field theory characters},''
  \href{http://dx.doi.org/10.1016/0370-2693(93)90194-M}{{\em Phys. Lett. B}
  {\bfseries 307} (1993) 68--76},
  \href{http://arxiv.org/abs/hep-th/9301046}{{\ttfamily arXiv:hep-th/9301046}}.

\bibitem{Berkovich:1994es}
A.~Berkovich and B.~M. McCoy, ``{Continued fractions and Fermionic
  representations for characters of M(p,p-prime) minimal models},''
  \href{http://dx.doi.org/10.1007/BF00400138}{{\em Lett. Math. Phys.}
  {\bfseries 37} (1996) 49--66},
  \href{http://arxiv.org/abs/hep-th/9412030}{{\ttfamily arXiv:hep-th/9412030}}.

\bibitem{welsh2005fermionic}
T.~A. Welsh, {\em Fermionic expressions for minimal model Virasoro characters}.
\newblock American Mathematical Soc., 2005.

\bibitem{Delmastro:2019vnj}
D.~Delmastro and J.~Gomis, ``{Symmetries of Abelian Chern-Simons Theories and
  Arithmetic},'' \href{http://dx.doi.org/10.1007/JHEP03(2021)006}{{\em JHEP}
  {\bfseries 03} (2021) 006}, \href{http://arxiv.org/abs/1904.12884}{{\ttfamily
  arXiv:1904.12884 [hep-th]}}.

\bibitem{DiFrancesco:1997nk}
P.~Di~Francesco, P.~Mathieu, and D.~Senechal,
  \href{http://dx.doi.org/10.1007/978-1-4612-2256-9}{{\em {Conformal Field
  Theory}}}.
\newblock Graduate Texts in Contemporary Physics. Springer-Verlag, New York,
  1997.

\end{thebibliography}\endgroup
